\documentclass[aps,prd,showpacs,letterpaper,reprint,superscriptaddress,notitlepage,nofootinbi,floatfix,nofootinbib,longbibliography]{revtex4-1}

\usepackage[caption=false]{subfig}
\usepackage{graphicx}
\usepackage{bm}
\usepackage{amssymb}
\usepackage{amsmath}
\usepackage{amsfonts}
\usepackage[utf8]{inputenc}
\usepackage{booktabs}
\usepackage{gensymb}
\usepackage[colorlinks=true,allcolors=blue]{hyperref}
\usepackage{microtype}
\usepackage{multirow, makecell}
\usepackage[capitalise]{cleveref}
\usepackage{longtable}
\usepackage{placeins}

\usepackage[utf8]{inputenc}

\usepackage{listings}
\usepackage{siunitx}
\usepackage{xspace}
\usepackage{comment}
\usepackage{floatrow}
\usepackage{afterpage}

\usepackage{ragged2e}
\usepackage{stackengine}
\usepackage{needspace}

\usepackage[T5,T1]{fontenc}

\DeclareSIUnit \s {\second}
\DeclareSIUnit \ns {\nano\second}
\DeclareSIUnit \mus {\micro\second}
\DeclareSIUnit \ms {\milli\second}
\DeclareSIUnit \MB {\mega\byte}
\DeclareSIUnit \GB {\giga\byte}
\DeclareSIUnit \TB {\tera\byte}
\DeclareSIUnit \PB {\peta\byte}
\DeclareSIUnit \Mbps {\mega\bit/\s}
\DeclareSIUnit \Gbps {\giga\bit/\s}
\DeclareSIUnit \Tbps {\tera\bit/\s}
\DeclareSIUnit \Pbps {\peta\bit/\s}
\DeclareSIUnit \kton {\kilo\tonne}
\DeclareSIUnit \kt {\kilo\tonne}
\DeclareSIUnit \Mt {\mega\tonne}
\DeclareSIUnit \eV {\electronvolt}
\DeclareSIUnit \keV {\kilo\electronvolt}
\DeclareSIUnit \MeV {\mega\electronvolt}
\DeclareSIUnit \GeV {\giga\electronvolt}
\DeclareSIUnit \PeV {\peta\electronvolt}
\DeclareSIUnit \EeV {\exa\electronvolt}
\DeclareSIUnit \m {\meter}
\DeclareSIUnit \cm {\centi\meter}
\DeclareSIUnit \in {\inchcommand}
\DeclareSIUnit \km {\kilo\meter}
\DeclareSIUnit \kV {\kilo\volt}
\DeclareSIUnit \kW {\kilo\watt}
\DeclareSIUnit \MW {\mega\watt}
\DeclareSIUnit \MHz {\mega\hertz}
\DeclareSIUnit \mrad {\milli\radian}
\DeclareSIUnit \year {years}
\DeclareSIUnit \POT {POT}
\DeclareSIUnit \sig {$\sigma$}
\DeclareSIUnit\parsec{pc}
\DeclareSIUnit\lightyear{ly}
\DeclareSIUnit\foot{ft}
\DeclareSIUnit\ft{ft}
\DeclareSIUnit \ppb{ppb}
\DeclareSIUnit \ppt{ppt}
\DeclareSIUnit \samples{S}
\DeclareSIUnit \pe{PE}
\DeclareSIUnit \sr{\steradian}

\newcommand\SigmaOne{\SI{68.3}\percent}
\newcommand\SigmaTwo{\SI{95.4}\percent}

\newcommand\EnergyCut{\SI{60}\TeV\xspace}

\newcommand\Livetime{\SI{7.5}\year\xspace}

\newcommand\astronorm{\Phi_\texttt{astro}}
\newcommand\astrodeltagamma{\gamma_\texttt{astro}}
\newcommand\convnorm{\Phi_\texttt{conv}}
\newcommand\promptnorm{\Phi_\texttt{prompt}}
\newcommand\pik{R_{K/\pi}}
\newcommand\atmonunubar{{2\nu/\left(\nu+\bar{\nu}\right)}_\texttt{atmo}}
\newcommand\crdeltagamma{\Delta\gamma_\texttt{CR}}
\newcommand\muonnorm{\Phi_\mu}
\newcommand\domeff{\epsilon_\texttt{DOM}}
\newcommand\holeice{\epsilon_\texttt{head-on}}
\newcommand\anisotropy{a_\texttt{s}}

\usepackage{xparse}

\ExplSyntaxOn
\DeclareExpandableDocumentCommand{\eval}{m}{\fp_eval:n {#1}}
\ExplSyntaxOff

\newcommand\CutoffBFCutoff{\SI{5.0}\PeV}
\newcommand\CutoffWilksLowerCutoff{\SI{1.2}\PeV}
\newcommand\CutoffWilksLowerCutoffNinety{\SI{670}\TeV}
\newcommand\CutoffPValue{0.71}

\newcommand\BayesFactorStrongCutoff{\SI{370}\TeV}

\newcommand\BayesFactorEqualCutoff{\SI{1.6}\PeV}

\newcommand\NoAstroPromptNorm{21.56} 
\newcommand\ICFiftyNinePromptUpperLimit{3.80} 
\newcommand\SPLFreqWilksNinetyUpperLimit{9.82}

\newcommand\BayesFactorStrongPromptNorm{13.29}

\newcommand\SPLFreqBFCRDeltaGamma{-0.053}
\newcommand\SPLFreqWilksUpperCRDeltaGamma{-0.005}
\newcommand\SPLFreqWilksLowerCRDeltaGamma{-0.184}

\newcommand\SPLFreqBFAtmoNuRatio{1.002}
\newcommand\SPLFreqWilksUpperAtmoNuRatio{1.102}
\newcommand\SPLFreqWilksLowerAtmoNuRatio{0.902}

\newcommand\SPLFreqBFAniScale{1.00}
\newcommand\SPLFreqWilksUpperAniScale{1.20}
\newcommand\SPLFreqWilksLowerAniScale{0.80}

\newcommand\SPLFreqBFIndex{2.87}
\newcommand\SPLFreqWilksUpperIndex{3.08}
\newcommand\SPLFreqWilksLowerIndex{2.68}
\newcommand\SPLFreqWilksUpperIndexDelta{0.20}
\newcommand\SPLFreqWilksLowerIndexDelta{0.19}
\newcommand\SPLFreqWilksIndexSummary{2.87^{+0.20}_{-0.19}}
\newcommand\SPLFreqBFNorm{6.37}
\newcommand\SPLFreqWilksUpperNorm{7.83}
\newcommand\SPLFreqWilksLowerNorm{4.75}

\newcommand\SPLFreqBFConvNorm{1.01}
\newcommand\SPLFreqWilksUpperConvNorm{1.35}
\newcommand\SPLFreqWilksLowerConvNorm{0.67}

\newcommand\SPLFreqBFDOMEff{0.952}
\newcommand\SPLFreqWilksUpperDOMEff{1.045}
\newcommand\SPLFreqWilksLowerDOMEff{0.886}

\newcommand\SPLFreqBFHoleIce{-0.06}
\newcommand\SPLFreqWilksUpperHoleIce{0.45}
\newcommand\SPLFreqWilksLowerHoleIce{-0.54}

\newcommand\SPLFreqBFMuonNorm{1.19}
\newcommand\SPLFreqWilksUpperMuonNorm{1.64}
\newcommand\SPLFreqWilksLowerMuonNorm{0.75}

\newcommand\SPLFreqBFKPi{1.000}
\newcommand\SPLFreqWilksUpperKPi{1.100}
\newcommand\SPLFreqWilksLowerKPi{0.901}

\newcommand\SPLFreqBFPromptNorm{0.00}
\newcommand\SPLFreqWilksUpperPromptNorm{5.34}
\newcommand\SPLFreqWilksLowerPromptNorm{0.00}

\newcommand\SPLBayesMAPCRDeltaGamma{-0.036}
\newcommand\SPLBayesHPDUpperCRDeltaGamma{0.010}
\newcommand\SPLBayesHPDLowerCRDeltaGamma{-0.088}

\newcommand\SPLBayesMAPAtmoNuRatio{0.986}
\newcommand\SPLBayesHPDUpperAtmoNuRatio{1.100}
\newcommand\SPLBayesHPDLowerAtmoNuRatio{0.901}

\newcommand\SPLBayesMAPAniScale{1.01}
\newcommand\SPLBayesHPDUpperAniScale{1.20}
\newcommand\SPLBayesHPDLowerAniScale{0.80}

\newcommand\SPLBayesMAPIndex{2.89}
\newcommand\SPLBayesHPDUpperIndex{3.12}
\newcommand\SPLBayesHPDLowerIndex{2.69}

\newcommand\SPLBayesIndexSummary{2.89^{+0.23}_{-0.20}}\newcommand\SPLBayesMAPNorm{5.68}
\newcommand\SPLBayesHPDUpperNorm{7.24}
\newcommand\SPLBayesHPDLowerNorm{4.13}

\newcommand\SPLBayesMAPConvNorm{0.93}
\newcommand\SPLBayesHPDUpperConvNorm{1.29}
\newcommand\SPLBayesHPDLowerConvNorm{0.61}

\newcommand\SPLBayesMAPDOMEff{0.935}
\newcommand\SPLBayesHPDUpperDOMEff{1.002}
\newcommand\SPLBayesHPDLowerDOMEff{0.848}

\newcommand\SPLBayesMAPHoleIce{-0.07}
\newcommand\SPLBayesHPDUpperHoleIce{0.39}
\newcommand\SPLBayesHPDLowerHoleIce{-0.63}

\newcommand\SPLBayesMAPMuonNorm{1.20}
\newcommand\SPLBayesHPDUpperMuonNorm{1.61}
\newcommand\SPLBayesHPDLowerMuonNorm{0.73}

\newcommand\SPLBayesMAPKPi{0.993}
\newcommand\SPLBayesHPDUpperKPi{1.095}
\newcommand\SPLBayesHPDLowerKPi{0.894}

\newcommand\SPLBayesMAPPromptNorm{0.54}
\newcommand\SPLBayesHPDUpperPromptNorm{6.15}
\newcommand\SPLBayesHPDLowerPromptNorm{0.00}

\newcommand\SPLBayesPromptNormSummary{0.54^{+5.62}_{-0.54}}

\newcommand\Stecker{Stecker~\cite{Stecker:2013fxa}}
\newcommand\SteckerBayes{7.26\times10^{-13}}
\newcommand\SteckerSPLBayes{8.27\times10^{-11}}
\newcommand\SteckerSPLPhiLower{2.95}
\newcommand\SteckerSPLPhiMode{4.24}
\newcommand\SteckerSPLPhiUpper{5.83}
\newcommand\SteckerSPLPhiLowerDelta{\eval{\SteckerSPLPhiMode - \SteckerSPLPhiLower}}
\newcommand\SteckerSPLPhiUpperDelta{\eval{\SteckerSPLPhiUpper - \SteckerSPLPhiMode}}
\newcommand\SteckerSPLPhiSummary{\SteckerSPLPhiMode^{+\SteckerSPLPhiUpperDelta}_{-\SteckerSPLPhiLowerDelta}}
\newcommand\SteckerSPLGammaLower{3.45}
\newcommand\SteckerSPLGammaMode{3.9}
\newcommand\SteckerSPLGammaUpper{4.48}
\newcommand\SteckerSPLGammaLowerDelta{\eval{\SteckerSPLGammaMode - \SteckerSPLGammaLower}}
\newcommand\SteckerSPLGammaUpperDelta{\eval{\SteckerSPLGammaUpper - \SteckerSPLGammaMode}}
\newcommand\SteckerSPLGammaSummary{\SteckerSPLGammaMode^{+\SteckerSPLGammaUpperDelta}_{-\SteckerSPLGammaLowerDelta}}
\newcommand\SteckerTableSummary{AGN core & \Stecker & $\SteckerBayes$ & $\SteckerSPLBayes$ & $\SteckerSPLGammaSummary$ & $\SteckerSPLPhiSummary$}
\newcommand\Fang{Fang et al.~\cite{Fang:2017zjf}}
\newcommand\FangBayes{0.379}
\newcommand\FangSPLBayes{0.153}
\newcommand\FangSPLPhiLower{0.94}
\newcommand\FangSPLPhiMode{2.1}
\newcommand\FangSPLPhiUpper{3.6}
\newcommand\FangSPLPhiLowerDelta{\eval{\FangSPLPhiMode - \FangSPLPhiLower}}
\newcommand\FangSPLPhiUpperDelta{\eval{\FangSPLPhiUpper - \FangSPLPhiMode}}
\newcommand\FangSPLPhiSummary{\FangSPLPhiMode^{+\FangSPLPhiUpperDelta}_{-\FangSPLPhiLowerDelta}}
\newcommand\FangSPLGammaLower{3.38}
\newcommand\FangSPLGammaMode{4.07}
\newcommand\FangSPLGammaUpper{4.71}
\newcommand\FangSPLGammaLowerDelta{\eval{\FangSPLGammaMode - \FangSPLGammaLower}}
\newcommand\FangSPLGammaUpperDelta{\eval{\FangSPLGammaUpper - \FangSPLGammaMode}}
\newcommand\FangSPLGammaSummary{\FangSPLGammaMode^{+\FangSPLGammaUpperDelta}_{-\FangSPLGammaLowerDelta}}
\newcommand\FangTableSummary{AGN & \Fang & $\FangBayes$ & $\FangSPLBayes$ & $\FangSPLGammaSummary$ & $\FangSPLPhiSummary$}
\newcommand\KimuraBOne{Kimura et al. (B1)~\cite{Kimura:2014jba}}
\newcommand\KimuraBOneBayes{7.12\times10^{-6}}
\newcommand\KimuraBOneSPLBayes{5.20\times10^{-7}}
\newcommand\KimuraBOneSPLPhiLower{0.0}
\newcommand\KimuraBOneSPLPhiMode{1.03}
\newcommand\KimuraBOneSPLPhiUpper{1.92}
\newcommand\KimuraBOneSPLPhiLowerDelta{\eval{\KimuraBOneSPLPhiMode - \KimuraBOneSPLPhiLower}}
\newcommand\KimuraBOneSPLPhiUpperDelta{\eval{\KimuraBOneSPLPhiUpper - \KimuraBOneSPLPhiMode}}
\newcommand\KimuraBOneSPLPhiSummary{\KimuraBOneSPLPhiMode^{+\KimuraBOneSPLPhiUpperDelta}_{-\KimuraBOneSPLPhiLowerDelta}}
\newcommand\KimuraBOneSPLGammaLower{3.88}
\newcommand\KimuraBOneSPLGammaMode{4.74}
\newcommand\KimuraBOneSPLGammaUpper{5.0}
\newcommand\KimuraBOneSPLGammaLowerDelta{\eval{\KimuraBOneSPLGammaMode - \KimuraBOneSPLGammaLower}}
\newcommand\KimuraBOneSPLGammaUpperDelta{\eval{\KimuraBOneSPLGammaUpper - \KimuraBOneSPLGammaMode}}
\newcommand\KimuraBOneSPLGammaSummary{\KimuraBOneSPLGammaMode^{+\KimuraBOneSPLGammaUpperDelta}_{-\KimuraBOneSPLGammaLowerDelta}}
\newcommand\KimuraBOneTableSummary{LLAGN & \KimuraBOne & $\KimuraBOneBayes$ & $\KimuraBOneSPLBayes$ & $\KimuraBOneSPLGammaSummary$ & $\KimuraBOneSPLPhiSummary$}
\newcommand\KimuraBFour{Kimura et al. (B4)~\cite{Kimura:2014jba}}
\newcommand\KimuraBFourBayes{3.74\times10^{-4}}
\newcommand\KimuraBFourSPLBayes{0.219}
\newcommand\KimuraBFourSPLPhiLower{0.62}
\newcommand\KimuraBFourSPLPhiMode{1.57}
\newcommand\KimuraBFourSPLPhiUpper{2.55}
\newcommand\KimuraBFourSPLPhiLowerDelta{\eval{\KimuraBFourSPLPhiMode - \KimuraBFourSPLPhiLower}}
\newcommand\KimuraBFourSPLPhiUpperDelta{\eval{\KimuraBFourSPLPhiUpper - \KimuraBFourSPLPhiMode}}
\newcommand\KimuraBFourSPLPhiSummary{\KimuraBFourSPLPhiMode^{+\KimuraBFourSPLPhiUpperDelta}_{-\KimuraBFourSPLPhiLowerDelta}}
\newcommand\KimuraBFourSPLGammaLower{2.16}
\newcommand\KimuraBFourSPLGammaMode{2.43}
\newcommand\KimuraBFourSPLGammaUpper{2.73}
\newcommand\KimuraBFourSPLGammaLowerDelta{\eval{\KimuraBFourSPLGammaMode - \KimuraBFourSPLGammaLower}}
\newcommand\KimuraBFourSPLGammaUpperDelta{\eval{\KimuraBFourSPLGammaUpper - \KimuraBFourSPLGammaMode}}
\newcommand\KimuraBFourSPLGammaSummary{\KimuraBFourSPLGammaMode^{+\KimuraBFourSPLGammaUpperDelta}_{-\KimuraBFourSPLGammaLowerDelta}}
\newcommand\KimuraBFourTableSummary{LLAGN & \KimuraBFour & $\KimuraBFourBayes$ & $\KimuraBFourSPLBayes$ & $\KimuraBFourSPLGammaSummary$ & $\KimuraBFourSPLPhiSummary$}
\newcommand\KimuraTwoComp{Kimura et al. (two component)~\cite{Kimura:2014jba}}
\newcommand\KimuraTwoCompBayes{1.87\times10^{-4}}
\newcommand\KimuraTwoCompSPLBayes{3.63\times10^{-6}}
\newcommand\KimuraTwoCompSPLPhiLower{0.0}
\newcommand\KimuraTwoCompSPLPhiMode{0.0}
\newcommand\KimuraTwoCompSPLPhiUpper{0.62}
\newcommand\KimuraTwoCompSPLPhiLowerDelta{\eval{\KimuraTwoCompSPLPhiMode - \KimuraTwoCompSPLPhiLower}}
\newcommand\KimuraTwoCompSPLPhiUpperDelta{\eval{\KimuraTwoCompSPLPhiUpper - \KimuraTwoCompSPLPhiMode}}
\newcommand\KimuraTwoCompSPLPhiSummary{\KimuraTwoCompSPLPhiMode^{+\KimuraTwoCompSPLPhiUpperDelta}_{-\KimuraTwoCompSPLPhiLowerDelta}}
\newcommand\KimuraTwoCompSPLGammaLower{3.39}
\newcommand\KimuraTwoCompSPLGammaMode{4.25}
\newcommand\KimuraTwoCompSPLGammaUpper{5.0}
\newcommand\KimuraTwoCompSPLGammaLowerDelta{\eval{\KimuraTwoCompSPLGammaMode - \KimuraTwoCompSPLGammaLower}}
\newcommand\KimuraTwoCompSPLGammaUpperDelta{\eval{\KimuraTwoCompSPLGammaUpper - \KimuraTwoCompSPLGammaMode}}
\newcommand\KimuraTwoCompSPLGammaSummary{\KimuraTwoCompSPLGammaMode^{+\KimuraTwoCompSPLGammaUpperDelta}_{-\KimuraTwoCompSPLGammaLowerDelta}}
\newcommand\KimuraTwoCompTableSummary{LLAGN & \KimuraTwoComp & $\KimuraTwoCompBayes$ & $\KimuraTwoCompSPLBayes$ & $\KimuraTwoCompSPLGammaSummary$ & $\KimuraTwoCompSPLPhiSummary$}
\newcommand\MariaBLLacs{Padovani et al.~\cite{Padovani:2015mba}}
\newcommand\MariaBLLacsBayes{1.08\times10^{-10}}
\newcommand\MariaBLLacsSPLBayes{1.75\times10^{-7}}
\newcommand\MariaBLLacsSPLPhiLower{3.48}
\newcommand\MariaBLLacsSPLPhiMode{4.92}
\newcommand\MariaBLLacsSPLPhiUpper{6.56}
\newcommand\MariaBLLacsSPLPhiLowerDelta{\eval{\MariaBLLacsSPLPhiMode - \MariaBLLacsSPLPhiLower}}
\newcommand\MariaBLLacsSPLPhiUpperDelta{\eval{\MariaBLLacsSPLPhiUpper - \MariaBLLacsSPLPhiMode}}
\newcommand\MariaBLLacsSPLPhiSummary{\MariaBLLacsSPLPhiMode^{+\MariaBLLacsSPLPhiUpperDelta}_{-\MariaBLLacsSPLPhiLowerDelta}}
\newcommand\MariaBLLacsSPLGammaLower{3.26}
\newcommand\MariaBLLacsSPLGammaMode{3.66}
\newcommand\MariaBLLacsSPLGammaUpper{4.17}
\newcommand\MariaBLLacsSPLGammaLowerDelta{\eval{\MariaBLLacsSPLGammaMode - \MariaBLLacsSPLGammaLower}}
\newcommand\MariaBLLacsSPLGammaUpperDelta{\eval{\MariaBLLacsSPLGammaUpper - \MariaBLLacsSPLGammaMode}}
\newcommand\MariaBLLacsSPLGammaSummary{\MariaBLLacsSPLGammaMode^{+\MariaBLLacsSPLGammaUpperDelta}_{-\MariaBLLacsSPLGammaLowerDelta}}
\newcommand\MariaBLLacsTableSummary{BLLac & \MariaBLLacs & $\MariaBLLacsBayes$ & $\MariaBLLacsSPLBayes$ & $\MariaBLLacsSPLGammaSummary$ & $\MariaBLLacsSPLPhiSummary$}
\newcommand\MurasechockedJets{Senno et al.~\cite{Senno:2015tsn}}
\newcommand\MurasechockedJetsBayes{0.353}
\newcommand\MurasechockedJetsSPLBayes{2.31}
\newcommand\MurasechockedJetsSPLPhiLower{1.97}
\newcommand\MurasechockedJetsSPLPhiMode{3.16}
\newcommand\MurasechockedJetsSPLPhiUpper{4.97}
\newcommand\MurasechockedJetsSPLPhiLowerDelta{\eval{\MurasechockedJetsSPLPhiMode - \MurasechockedJetsSPLPhiLower}}
\newcommand\MurasechockedJetsSPLPhiUpperDelta{\eval{\MurasechockedJetsSPLPhiUpper - \MurasechockedJetsSPLPhiMode}}
\newcommand\MurasechockedJetsSPLPhiSummary{\MurasechockedJetsSPLPhiMode^{+\MurasechockedJetsSPLPhiUpperDelta}_{-\MurasechockedJetsSPLPhiLowerDelta}}
\newcommand\MurasechockedJetsSPLGammaLower{3.06}
\newcommand\MurasechockedJetsSPLGammaMode{3.7}
\newcommand\MurasechockedJetsSPLGammaUpper{4.24}
\newcommand\MurasechockedJetsSPLGammaLowerDelta{\eval{\MurasechockedJetsSPLGammaMode - \MurasechockedJetsSPLGammaLower}}
\newcommand\MurasechockedJetsSPLGammaUpperDelta{\eval{\MurasechockedJetsSPLGammaUpper - \MurasechockedJetsSPLGammaMode}}
\newcommand\MurasechockedJetsSPLGammaSummary{\MurasechockedJetsSPLGammaMode^{+\MurasechockedJetsSPLGammaUpperDelta}_{-\MurasechockedJetsSPLGammaLowerDelta}}
\newcommand\MurasechockedJetsTableSummary{GRB choked jet & \MurasechockedJets & $\MurasechockedJetsBayes$ & $\MurasechockedJetsSPLBayes$ & $\MurasechockedJetsSPLGammaSummary$ & $\MurasechockedJetsSPLPhiSummary$}
\newcommand\SBGminBmodel{Bartos et al.~\cite{Bartos:2015xpa}}
\newcommand\SBGminBmodelBayes{1.16\times10^{-14}}
\newcommand\SBGminBmodelSPLBayes{1.62\times10^{-16}}
\newcommand\SBGminBmodelSPLPhiLower{0.0}
\newcommand\SBGminBmodelSPLPhiMode{0.0}
\newcommand\SBGminBmodelSPLPhiUpper{0.51}
\newcommand\SBGminBmodelSPLPhiLowerDelta{\eval{\SBGminBmodelSPLPhiMode - \SBGminBmodelSPLPhiLower}}
\newcommand\SBGminBmodelSPLPhiUpperDelta{\eval{\SBGminBmodelSPLPhiUpper - \SBGminBmodelSPLPhiMode}}
\newcommand\SBGminBmodelSPLPhiSummary{\SBGminBmodelSPLPhiMode^{+\SBGminBmodelSPLPhiUpperDelta}_{-\SBGminBmodelSPLPhiLowerDelta}}
\newcommand\SBGminBmodelSPLGammaLower{3.38}
\newcommand\SBGminBmodelSPLGammaMode{4.43}
\newcommand\SBGminBmodelSPLGammaUpper{4.99}
\newcommand\SBGminBmodelSPLGammaLowerDelta{\eval{\SBGminBmodelSPLGammaMode - \SBGminBmodelSPLGammaLower}}
\newcommand\SBGminBmodelSPLGammaUpperDelta{\eval{\SBGminBmodelSPLGammaUpper - \SBGminBmodelSPLGammaMode}}
\newcommand\SBGminBmodelSPLGammaSummary{\SBGminBmodelSPLGammaMode^{+\SBGminBmodelSPLGammaUpperDelta}_{-\SBGminBmodelSPLGammaLowerDelta}}
\newcommand\SBGminBmodelTableSummary{SBG & \SBGminBmodel & $\SBGminBmodelBayes$ & $\SBGminBmodelSPLBayes$ & $\SBGminBmodelSPLGammaSummary$ & $\SBGminBmodelSPLPhiSummary$}
\newcommand\TavecchilowPower{Tavecchio et al.~\cite{Tavecchio:2014eia}}
\newcommand\TavecchilowPowerBayes{0.104}
\newcommand\TavecchilowPowerSPLBayes{0.703}
\newcommand\TavecchilowPowerSPLPhiLower{2.16}
\newcommand\TavecchilowPowerSPLPhiMode{3.63}
\newcommand\TavecchilowPowerSPLPhiUpper{5.04}
\newcommand\TavecchilowPowerSPLPhiLowerDelta{\eval{\TavecchilowPowerSPLPhiMode - \TavecchilowPowerSPLPhiLower}}
\newcommand\TavecchilowPowerSPLPhiUpperDelta{\eval{\TavecchilowPowerSPLPhiUpper - \TavecchilowPowerSPLPhiMode}}
\newcommand\TavecchilowPowerSPLPhiSummary{\TavecchilowPowerSPLPhiMode^{+\TavecchilowPowerSPLPhiUpperDelta}_{-\TavecchilowPowerSPLPhiLowerDelta}}
\newcommand\TavecchilowPowerSPLGammaLower{3.36}
\newcommand\TavecchilowPowerSPLGammaMode{3.79}
\newcommand\TavecchilowPowerSPLGammaUpper{4.53}
\newcommand\TavecchilowPowerSPLGammaLowerDelta{\eval{\TavecchilowPowerSPLGammaMode - \TavecchilowPowerSPLGammaLower}}
\newcommand\TavecchilowPowerSPLGammaUpperDelta{\eval{\TavecchilowPowerSPLGammaUpper - \TavecchilowPowerSPLGammaMode}}
\newcommand\TavecchilowPowerSPLGammaSummary{\TavecchilowPowerSPLGammaMode^{+\TavecchilowPowerSPLGammaUpperDelta}_{-\TavecchilowPowerSPLGammaLowerDelta}}
\newcommand\TavecchilowPowerTableSummary{LLBLLac & \TavecchilowPower & $\TavecchilowPowerBayes$ & $\TavecchilowPowerSPLBayes$ & $\TavecchilowPowerSPLGammaSummary$ & $\TavecchilowPowerSPLPhiSummary$}
\newcommand\TDEWinterBiehl{Biehl et al.~\cite{Biehl:2017hnb}}
\newcommand\TDEWinterBiehlBayes{1.31\times10^{-6}}
\newcommand\TDEWinterBiehlSPLBayes{0.145}
\newcommand\TDEWinterBiehlSPLPhiLower{3.97}
\newcommand\TDEWinterBiehlSPLPhiMode{5.36}
\newcommand\TDEWinterBiehlSPLPhiUpper{7.08}
\newcommand\TDEWinterBiehlSPLPhiLowerDelta{\eval{\TDEWinterBiehlSPLPhiMode - \TDEWinterBiehlSPLPhiLower}}
\newcommand\TDEWinterBiehlSPLPhiUpperDelta{\eval{\TDEWinterBiehlSPLPhiUpper - \TDEWinterBiehlSPLPhiMode}}
\newcommand\TDEWinterBiehlSPLPhiSummary{\TDEWinterBiehlSPLPhiMode^{+\TDEWinterBiehlSPLPhiUpperDelta}_{-\TDEWinterBiehlSPLPhiLowerDelta}}
\newcommand\TDEWinterBiehlSPLGammaLower{2.97}
\newcommand\TDEWinterBiehlSPLGammaMode{3.32}
\newcommand\TDEWinterBiehlSPLGammaUpper{3.75}
\newcommand\TDEWinterBiehlSPLGammaLowerDelta{\eval{\TDEWinterBiehlSPLGammaMode - \TDEWinterBiehlSPLGammaLower}}
\newcommand\TDEWinterBiehlSPLGammaUpperDelta{\eval{\TDEWinterBiehlSPLGammaUpper - \TDEWinterBiehlSPLGammaMode}}
\newcommand\TDEWinterBiehlSPLGammaSummary{\TDEWinterBiehlSPLGammaMode^{+\TDEWinterBiehlSPLGammaUpperDelta}_{-\TDEWinterBiehlSPLGammaLowerDelta}}
\newcommand\TDEWinterBiehlTableSummary{GRB & \TDEWinterBiehl & $\TDEWinterBiehlBayes$ & $\TDEWinterBiehlSPLBayes$ & $\TDEWinterBiehlSPLGammaSummary$ & $\TDEWinterBiehlSPLPhiSummary$}

\numberwithin{figure}{section}
\numberwithin{table}{section}
\numberwithin{equation}{section}

\lstdefinestyle{mystyle}{basicstyle=\ttfamily,
    breakatwhitespace=false,
    breaklines=true,
    captionpos=b,
    keepspaces=true,
    showspaces=false,
    showstringspaces=false,
    showtabs=false,
    tabsize=4
}
\AtBeginDocument{\heavyrulewidth=.08em
\lightrulewidth=.05em
\cmidrulewidth=.03em
\belowrulesep=.65ex
\belowbottomsep=0pt
\aboverulesep=.4ex
\abovetopsep=0pt
\cmidrulesep=\doublerulesep{}
\cmidrulekern=.5em
\defaultaddspace=.5em
}

\newcommand{\CORSIKA}{\texttt{CORSIKA}}
\newcommand{\MUONGUN}{\texttt{MUONGUN}}
\newcommand{\PHOTOSPLINE}{\texttt{PHOTOSPLINE}}
\newcommand{\like}{\mathcal{L}}
\newcommand{\likeSAY}{\mathcal{L}_{\rm Eff}}
\newcommand{\pdfSxi}{S(\vec{x}_i, \vec{\psi}_s)}
\newcommand{\TS}{\mathrm{TS}}
\newcommand{\Fermi}{\textit{Fermi}}
\newcommand{\nuveto}{{\LARGE $\nu$}\texttt{eto}}

\DeclareMathOperator*{\argmax}{arg\,max}

\newcommand{\reffig}[1]{Fig.~\ref{#1}}

\newcommand{\refsec}[1]{Section~\ref{#1}}

\newcommand{\refappsec}[1]{Appendix Section~\ref{#1}}
\newcommand{\refapp}[1]{Appendix~\ref{#1}}
\newcommand{\reftab}[1]{Table~\ref{#1}}

\allowdisplaybreaks

\begin{document}

\collaboration{IceCube Collaboration}
\noaffiliation{}

\title{The IceCube high-energy starting event sample: Description and flux characterization with \Livetime{} of data}

\begin{abstract}
    The IceCube Neutrino Observatory has established the existence of a high-energy all-sky neutrino flux of astrophysical origin.
    This discovery was made using events interacting within a fiducial region of the detector surrounded by an active veto and with reconstructed energy above \EnergyCut, commonly known as the high-energy starting event sample, or HESE\@.
    We revisit the analysis of the HESE sample with an additional $\SI{4.5}\year$ of data, newer glacial ice models, and improved systematics treatment.
    This paper describes the sample in detail, reports on the latest astrophysical neutrino flux measurements, and presents a source search for astrophysical neutrinos.
    We give the compatibility of these observations with specific isotropic flux models proposed in the literature as well as generic power-law-like scenarios.
    Assuming $\nu_e:\nu_\mu:\nu_\tau=1:1:1$, and an equal flux of neutrinos and antineutrinos, we find that the astrophysical neutrino spectrum is compatible with an unbroken power law, with a preferred spectral index of  $\SPLFreqBFIndex^{+\SPLFreqWilksUpperIndexDelta}_{-\SPLFreqWilksLowerIndexDelta}$ for the $\SigmaOne$ confidence interval.
\end{abstract}

\affiliation{III. Physikalisches Institut, RWTH Aachen University, D-52056 Aachen, Germany}
\affiliation{Department of Physics, University of Adelaide, Adelaide, 5005, Australia}
\affiliation{Dept. of Physics and Astronomy, University of Alaska Anchorage, 3211 Providence Dr., Anchorage, AK 99508, USA}
\affiliation{Dept. of Physics, University of Texas at Arlington, 502 Yates St., Science Hall Rm 108, Box 19059, Arlington, TX 76019, USA}
\affiliation{CTSPS, Clark-Atlanta University, Atlanta, GA 30314, USA}
\affiliation{School of Physics and Center for Relativistic Astrophysics, Georgia Institute of Technology, Atlanta, GA 30332, USA}
\affiliation{Dept. of Physics, Southern University, Baton Rouge, LA 70813, USA}
\affiliation{Dept. of Physics, University of California, Berkeley, CA 94720, USA}
\affiliation{Lawrence Berkeley National Laboratory, Berkeley, CA 94720, USA}
\affiliation{Institut f{\"u}r Physik, Humboldt-Universit{\"a}t zu Berlin, D-12489 Berlin, Germany}
\affiliation{Fakult{\"a}t f{\"u}r Physik {\&} Astronomie, Ruhr-Universit{\"a}t Bochum, D-44780 Bochum, Germany}
\affiliation{Universit{\'e} Libre de Bruxelles, Science Faculty CP230, B-1050 Brussels, Belgium}
\affiliation{Vrije Universiteit Brussel (VUB), Dienst ELEM, B-1050 Brussels, Belgium}
\affiliation{Department of Physics and Laboratory for Particle Physics and Cosmology, Harvard University, Cambridge, MA 02138, USA}
\affiliation{Dept. of Physics, Massachusetts Institute of Technology, Cambridge, MA 02139, USA}
\affiliation{Dept. of Physics and Institute for Global Prominent Research, Chiba University, Chiba 263-8522, Japan}
\affiliation{Department of Physics, Loyola University Chicago, Chicago, IL 60660, USA}
\affiliation{Dept. of Physics and Astronomy, University of Canterbury, Private Bag 4800, Christchurch, New Zealand}
\affiliation{Dept. of Physics, University of Maryland, College Park, MD 20742, USA}
\affiliation{Dept. of Astronomy, Ohio State University, Columbus, OH 43210, USA}
\affiliation{Dept. of Physics and Center for Cosmology and Astro-Particle Physics, Ohio State University, Columbus, OH 43210, USA}
\affiliation{Niels Bohr Institute, University of Copenhagen, DK-2100 Copenhagen, Denmark}
\affiliation{Dept. of Physics, TU Dortmund University, D-44221 Dortmund, Germany}
\affiliation{Dept. of Physics and Astronomy, Michigan State University, East Lansing, MI 48824, USA}
\affiliation{Dept. of Physics, University of Alberta, Edmonton, Alberta, Canada T6G 2E1}
\affiliation{Erlangen Centre for Astroparticle Physics, Friedrich-Alexander-Universit{\"a}t Erlangen-N{\"u}rnberg, D-91058 Erlangen, Germany}
\affiliation{Physik-department, Technische Universit{\"a}t M{\"u}nchen, D-85748 Garching, Germany}
\affiliation{D{\'e}partement de physique nucl{\'e}aire et corpusculaire, Universit{\'e} de Gen{\`e}ve, CH-1211 Gen{\`e}ve, Switzerland}
\affiliation{Dept. of Physics and Astronomy, University of Gent, B-9000 Gent, Belgium}
\affiliation{Dept. of Physics and Astronomy, University of California, Irvine, CA 92697, USA}
\affiliation{Karlsruhe Institute of Technology, Institute for Astroparticle Physics, D-76021 Karlsruhe, Germany }
\affiliation{Dept. of Physics and Astronomy, University of Kansas, Lawrence, KS 66045, USA}
\affiliation{SNOLAB, 1039 Regional Road 24, Creighton Mine 9, Lively, ON, Canada P3Y 1N2}

\affiliation{King’s College London, London WC2R 2LS, UK}
\affiliation{School of Physics and Astronomy, Queen Mary University of London, London E1 4NS, UK}

\affiliation{Department of Physics and Astronomy, UCLA, Los Angeles, CA 90095, USA}
\affiliation{Department of Physics, Mercer University, Macon, GA 31207-0001, USA}
\affiliation{Dept. of Astronomy, University of Wisconsin{\textendash}Madison, Madison, WI 53706, USA}
\affiliation{Dept. of Physics and Wisconsin IceCube Particle Astrophysics Center, University of Wisconsin{\textendash}Madison, Madison, WI 53706, USA}
\affiliation{Institute of Physics, University of Mainz, Staudinger Weg 7, D-55099 Mainz, Germany}
\affiliation{Department of Physics, Marquette University, Milwaukee, WI, 53201, USA}
\affiliation{Institut f{\"u}r Kernphysik, Westf{\"a}lische Wilhelms-Universit{\"a}t M{\"u}nster, D-48149 M{\"u}nster, Germany}
\affiliation{Bartol Research Institute and Dept. of Physics and Astronomy, University of Delaware, Newark, DE 19716, USA}
\affiliation{Dept. of Physics, Yale University, New Haven, CT 06520, USA}
\affiliation{Dept. of Physics, University of Oxford, Parks Road, Oxford OX1 3PU, UK}
\affiliation{Dept. of Physics, Drexel University, 3141 Chestnut Street, Philadelphia, PA 19104, USA}
\affiliation{Physics Department, South Dakota School of Mines and Technology, Rapid City, SD 57701, USA}
\affiliation{Dept. of Physics, University of Wisconsin, River Falls, WI 54022, USA}
\affiliation{Dept. of Physics and Astronomy, University of Rochester, Rochester, NY 14627, USA}
\affiliation{Oskar Klein Centre and Dept. of Physics, Stockholm University, SE-10691 Stockholm, Sweden}
\affiliation{Dept. of Physics and Astronomy, Stony Brook University, Stony Brook, NY 11794-3800, USA}
\affiliation{Dept. of Physics, Sungkyunkwan University, Suwon 16419, Korea}
\affiliation{Institute of Basic Science, Sungkyunkwan University, Suwon 16419, Korea}
\affiliation{Dept. of Physics and Astronomy, University of Alabama, Tuscaloosa, AL 35487, USA}
\affiliation{Dept. of Astronomy and Astrophysics, Pennsylvania State University, University Park, PA 16802, USA}
\affiliation{Dept. of Physics, Pennsylvania State University, University Park, PA 16802, USA}
\affiliation{Dept. of Physics and Astronomy, Uppsala University, Box 516, S-75120 Uppsala, Sweden}
\affiliation{Dept. of Physics, University of Wuppertal, D-42119 Wuppertal, Germany}
\affiliation{DESY, D-15738 Zeuthen, Germany}

\author{R. Abbasi}
\affiliation{Department of Physics, Loyola University Chicago, Chicago, IL 60660, USA}
\author{M. Ackermann}
\affiliation{DESY, D-15738 Zeuthen, Germany}
\author{J. Adams}
\affiliation{Dept. of Physics and Astronomy, University of Canterbury, Private Bag 4800, Christchurch, New Zealand}
\author{J. A. Aguilar}
\affiliation{Universit{\'e} Libre de Bruxelles, Science Faculty CP230, B-1050 Brussels, Belgium}
\author{M. Ahlers}
\affiliation{Niels Bohr Institute, University of Copenhagen, DK-2100 Copenhagen, Denmark}
\author{M. Ahrens}
\affiliation{Oskar Klein Centre and Dept. of Physics, Stockholm University, SE-10691 Stockholm, Sweden}
\author{C. Alispach}
\affiliation{D{\'e}partement de physique nucl{\'e}aire et corpusculaire, Universit{\'e} de Gen{\`e}ve, CH-1211 Gen{\`e}ve, Switzerland}
\author{A. A. Alves Jr.}
\affiliation{Karlsruhe Institute of Technology, Institute for Astroparticle Physics, D-76021 Karlsruhe, Germany }
\author{N. M. Amin}
\affiliation{Bartol Research Institute and Dept. of Physics and Astronomy, University of Delaware, Newark, DE 19716, USA}
\author{K. Andeen}
\affiliation{Department of Physics, Marquette University, Milwaukee, WI, 53201, USA}
% ...

\maketitle
\tableofcontents
\section{Introduction\label{sec:introduction}}
During the last few decades, observations of low-energy extraterrestrial neutrinos have pushed forward our understanding of astrophysical environments and elucidated properties of neutrinos.
This is because neutrinos only interact via the weak force, allowing them to escape dense astrophysical environments where they are produced and travel long distances unperturbed to us.
These properties have enabled measurements of $\si\MeV$ neutrinos from the Sun, the closest detected extraterrestrial neutrino source, improving our understanding of the Sun's inner workings~\cite{Raffelt:1999tx,Bahcall:2004pz}, and have been pivotal in resolving the neutrino flavor-changing puzzle~\cite{McDonald:2016ixn}.
Similarly, the observation of neutrinos from supernova 1987A~\cite{Hirata:1987hu,Bratton:1988ww}, approximately 168,000 light-years away in the Large Magellanic Cloud, has provided invaluable information for supernova physics~\cite{Bethe:1990mw}, characterization of neutrino properties~\cite{Arnett:1987iz,Goldman:1987fg,Manohar:1987ec,Lattimer:1988mf}, and fundamental physics~\cite{Raffelt:1987yt,Turner:1987by}.

\begin{figure}
    \centering
    \includegraphics[width=\linewidth]{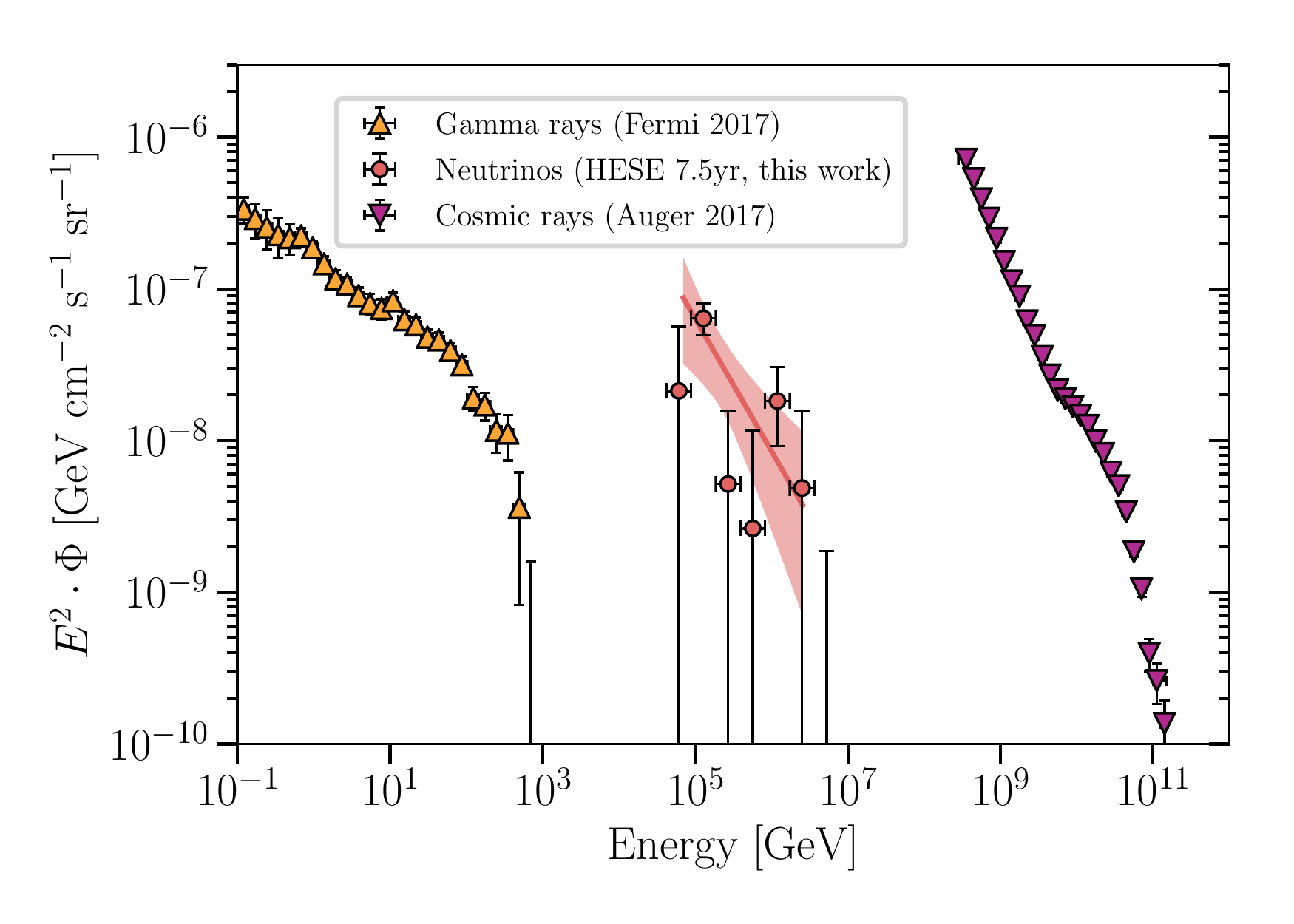}
    \caption{\textbf{\textit{High-energy fluxes of gamma rays, neutrinos, and cosmic rays.}}
    The segmented power-law neutrino flux, described in \refsec{sec:unfolding}, obtained in the analysis described in this paper, is shown with red circles.
    The single power-law assumption, described in \refsec{sec:spl}, is shown with the light red region.
    The high-energy gamma-ray measurements by Fermi~\cite{Ackermann:2014usa} are shown in orange, while the extremely-high-energy cosmic-ray measurements by the Pierre Auger Observatory~\cite{Fenu:2017hlc} are shown as purple data points.
    The comparable energy content of these three fluxes is of particular interest in the investigation of cosmic-ray origin.}\label{fig:all_cosmic}
\end{figure}

Enabled by these unique neutrino properties, the study of high-energy extraterrestrial neutrinos is revealing many new opportunities for discovery.
Before reaching Earth, these neutrinos have likely travelled distances that far exceed the those traversed by neutrinos observed in 1987.
They are expected to be produced in high-energy hadronic processes in our Universe either directly from decaying hadrons or from decaying charged leptons produced in the hadronic interactions~\cite{Gaisser:1994yf}.
Regions of charged-particle acceleration are prime candidates for high-energy neutrino sources.
The observation of $\si\EeV$ cosmic rays indicates that objects of large size or high magnetic field strength are accelerating charged particles to high energies, narrowing the search for neutrino sources to a subclass of objects~\cite{gaisser1990cosmic,LetessierSelvon:2011dy}.
The diffuse cosmic ray, gamma ray, and neutrino fluxes show similar energy content despite their disparate energy regimes, as recent data demonstrates (\reffig{fig:all_cosmic}).
Despite this information and a wealth of cosmic-ray observations, the sources of ultra-high-energy cosmic rays are an unresolved mystery~\cite{Kotera:2011cp}.
Thus, much like solar neutrinos, which can escape their birthplace, high energy astrophysical neutrinos are an indispensable probe for cosmic-ray sources, providing insight into the long-standing problem of the origin of cosmic-rays, as they can escape dense environments and reach us unperturbed.
By studying their flux and energy spectrum, constraints can be placed on the acceleration environments that produce these neutrinos.

High-energy astrophysical neutrinos are also powerful probes of new physics~\cite{Ahlers:2018mkf}.
This is in large part because neutrinos are charged under flavor~\cite{Gaisser:1994yf,Ackermann:2019cxh,Arguelles:2019rbn}, unlike other cosmic messengers.
New nontrivial flavor interactions can arise from a breaking of space-time symmetries~\cite{Diaz:2013wia,Arguelles:2015dca}, secret neutrino interactions with the cosmic-neutrino background~\cite{Davis:2015rza,Cherry:2016jol,Arguelles:2017atb,Kelly:2018tyg}, flavored dark-matter neutrino interactions~\cite{Capozzi:2018bps,Farzan:2018pnk,Choi:2019zxy}, or other nonstandard interactions~\cite{Rasmussen:2017ert}.
Beyond flavor, the very long distances traversed by high-energy astrophysical neutrinos can be used for accurate time-of-flight~\cite{Ellis:2018ogq} and neutrino-flux spectral distortion~\cite{Liao:2017yuy} measurements.
High-energy astrophysical neutrinos can probe very heavy decaying and annihilating dark matter, whose other Standard Model products will not reach Earth~\cite{Aartsen:2018mxl}.
Finally, these neutrinos can also probe the high-energy neutrino-nucleon cross section~\cite{Aartsen:2017kpd,Bustamante:2017xuy,Xu:2018zyc,Bertone:2018dse,Anchordoqui:2019ufu,yiqianxuthesis}.
Such a measurement is of interest due to the possibility of observing gluon screening~\cite{Henley:2005ms}, which could reduce the cross section at the highest energies~\cite{Block:2014kza,Goncalves:2015fua,Arguelles:2015wba}, or of uncovering new physics phenomena, e.g., low-scale quantum gravity~\cite{AlvarezMuniz:2001mk}, leptoquarks~\cite{Anchordoqui:2006wc,Dey:2015eaa,Dutta:2015dka,Dey:2016sht,Dey:2017ede,Chauhan:2017ndd,Becirevic:2018uab}, sphalerons~\cite{Cohen:1993nk,Ellis:2016dgb}, and micro black hole production~\cite{AlvarezMuniz:2002ga,Mack:2019bps}; see~\cite{Klein:2019nbu} for a recent review.

The IceCube Neutrino Observatory has firmly established the existence of high-energy astrophysical neutrinos.
Northern sky measurements of through-going muon tracks~\cite{Aartsen:2015rwa,Aartsen:2016xlq}, all-sky measurements using events with interaction vertices contained in the detector fiducial volume~\cite{Aartsen:2013jdh,Aartsen:2014gkd,Aartsen:2015zva,Aartsen:2017mau} such as high-energy starting events (HESE), and additional studies extending to lower energies with contained cascades~\cite{Aartsen:2014muf,Aartsen:2020aqd} have all contributed to the characterization of the astrophysical neutrino flux.
Archival and real-time directional searches have found an excess with respect to background from a starburst galaxy~\cite{Aartsen:2019fau} and evidence of neutrino emission associated with a blazar~\cite{IceCube:2018dnn,IceCube:2018cha}.
However, the energy spectrum, directional distribution, and composition of this neutrino flux are still too poorly constrained to differentiate between many astrophysical scenarios.
This work focuses on measuring the astrophysical neutrino spectrum using events with their interaction vertex contained inside a fiducial volume; see~\cite{schneider2020precision} for additional details.
The astrophysical flux measurement assumes that the flux is isotropic and equal in composition between all neutrino species, whose end result is shown in \reffig{fig:all_cosmic}.
We also present a directional search for neutrino sources in \refapp{sec:sources}.
Other work with this sample includes the measurement of the neutrino flavor composition~\cite{HESETAU}, the search for additional neutrino interactions~\cite{Farrag:2019jty,Katori:2019xpc} and dark matter in the galactic core~\cite{Arguelles:2019boy}, and the measurement of the neutrino cross section~\cite{HESEXS}.

This paper is organized as follows.
In the first sections,~\ref{sec:selection},~\ref{sec:reconstruction},~\ref{sec:backgrounds}, and~\ref{sec:uncertainties} the detector is described, the event selection is defined, and relevant backgrounds, systematics, and statistical methodology are discussed.
In \refsec{sec:diffuse}, the results of this work concerning the isotropic astrophysical flux are presented.
Each of the results subsections begins with a brief summary in italics, followed by detailed discussions.
Finally, \refsec{sec:conclusion} summarizes the main conclusions of this work.

\section{Detector and event selection\label{sec:selection}}
IceCube is a gigaton-scale Cherenkov detector embedded in the Antarctic ice~\cite{Aartsen:2013rt} at the geographical South Pole~\cite{Aartsen:2016nxy}.
The detector consists of photomultiplier tubes (PMTs) and digitization electronics contained within glass pressure housings.
Each of these units is referred to as a digital optical module, or ``DOM''~\cite{Abbasi:2008aa}.
DOMs are mounted onto 86 vertical cables forming ``strings''.
Of these strings, 78 are arranged in a hexagonal grid with $\sim\SI{125}\meter$ spacing, where DOMs are spaced $\sim\SI{17}\meter$ apart vertically.
The remaining eight strings comprise the DeepCore sub-array, which has DOMs arranged with smaller vertical and horizontal spacing within a hexagonal cell of the main array and has PMTs with higher quantum efficiency~\cite{Collaboration:2011ym}.
The DOMs record discretized charge and timing information if a PMT readout voltage corresponding to at least $0.25$ photo-electrons ($\si\pe$) is observed.
The discretized information is referred to as ``hits,'' where each hit has a recorded charge and time.

IceCube detects neutrinos by observing the Cherenkov light emitted by relativistic charged particles that are produced by neutrino interactions in the ice or bedrock below the detector.
Neutrino neutral-current (NC) interactions initiate a hadronic shower that appears as a ``cascade''-like morphology in the detector and produces an outgoing neutrino that is not observable.
Here, ``cascade''-like refers to a highly localized energy deposition and roughly spherical light emission.
Charged-current (CC) interactions produce a hadronic shower at the site of the neutrino interaction and an outgoing charged lepton.
When the CC interaction is triggered by an electron neutrino ($\nu_e$), an electron ($e$) is produced, and its subsequent interaction starts an electromagnetic shower.
This type of event is observable as a cascade-like morphology, and is indistinguishable from a NC interaction.
In the case the incident particle is a muon neutrino ($\nu_\mu$), a muon ($\mu$) is produced in the interaction, which will generally traverse several kilometers and exit the kilometer-scale detection volume while depositing energy stochastically~\cite{PhysRevD.98.030001}.
IceCube observes this muon as a ``track''-like morphology, where ``track''-like refers roughly to a long and narrow trail of energy depositions and light emission pointed in the same direction.
Finally, tau neutrino ($\nu_\tau$) charged-current interactions produce a hadronic shower and tau ($\tau$) which has a mean decay length of $\sim \SI{50}\m/\si\PeV$.
When the initial interaction and subsequent decay of the tau can be distinguished from each other such events are classified as ``double cascades''~\cite{Learned:1994wg,Cowen:2007ny}.
The morphologies induced by a tau neutrino interaction are often trackless.
About $\sim \SI{17}\percent$~\cite{PDG2016} of the time the tau decays to $\nu_\tau \mu \bar\nu_\mu$ producing an observable track.
By distinguishing between cascades, tracks, and double cascades, IceCube is sensitive to the neutrino flux's flavor composition.
Astrophysical neutrinos are expected to arrive in roughly equal amounts of the three neutrino flavors~\cite{Farzan:2002ct,Palladino:2015vna,Arguelles:2015dca,Bustamante:2015waa,Ahlers:2018yom,Arguelles:2019tum}, be distributed isotropically across the sky~\cite{Gaisser:1994yf}, and dominate the observed neutrino flux above $\sim\SI{100}\TeV$~\cite{Aartsen:2016xlq}.
At energies above $\sim\SI{100}\TeV$, IceCube has obtained results consistent with expectations~\cite{Aartsen:2015ivb,Aartsen:2015knd,Palladino:2015zua,Mena:2014sja,Palomares-Ruiz:2015mka}.

\begin{figure}
    \centering
    \includegraphics[width=\textwidth,height=\textheight,keepaspectratio]{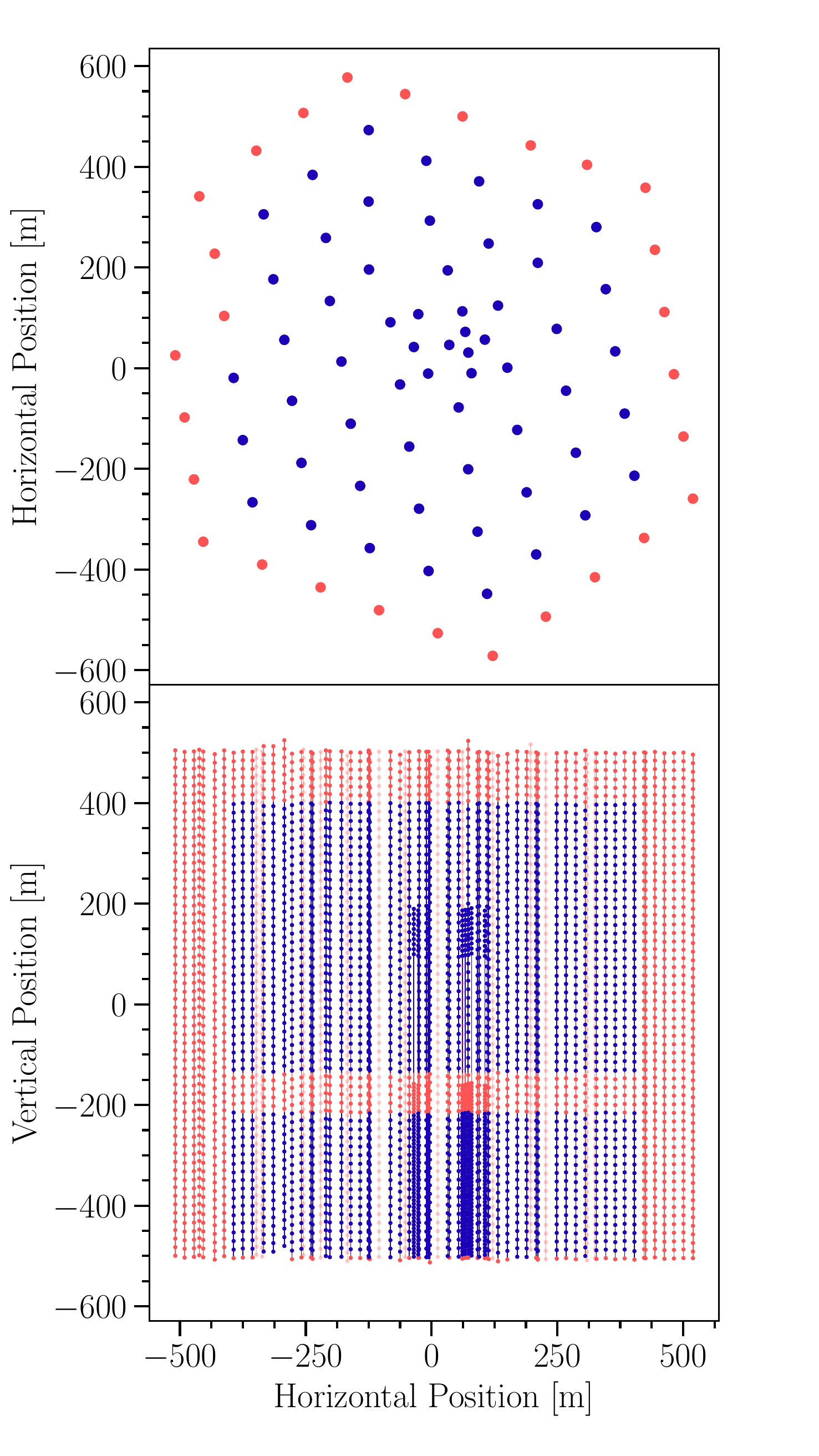}
    \caption{\textbf{\textit{HESE veto.}}
    Diagram of the IceCube detector indicating veto DOMs.
    Top panel: an overhead view of the IceCube detector.
    Positions of strings with only veto DOMs are shown in red, while those with at least one non-veto DOM are shown in blue.
    Bottom panel: a side view of the IceCube detector strings and DOMs.
    Veto DOMs are indicated with red circles and non-veto DOMs with blue circles.
    Strings in front of or behind the region without veto DOMs are semi-transparent.}\label{fig:veto}
\end{figure}

\begin{figure}
    \centering
    \includegraphics[width=\linewidth]{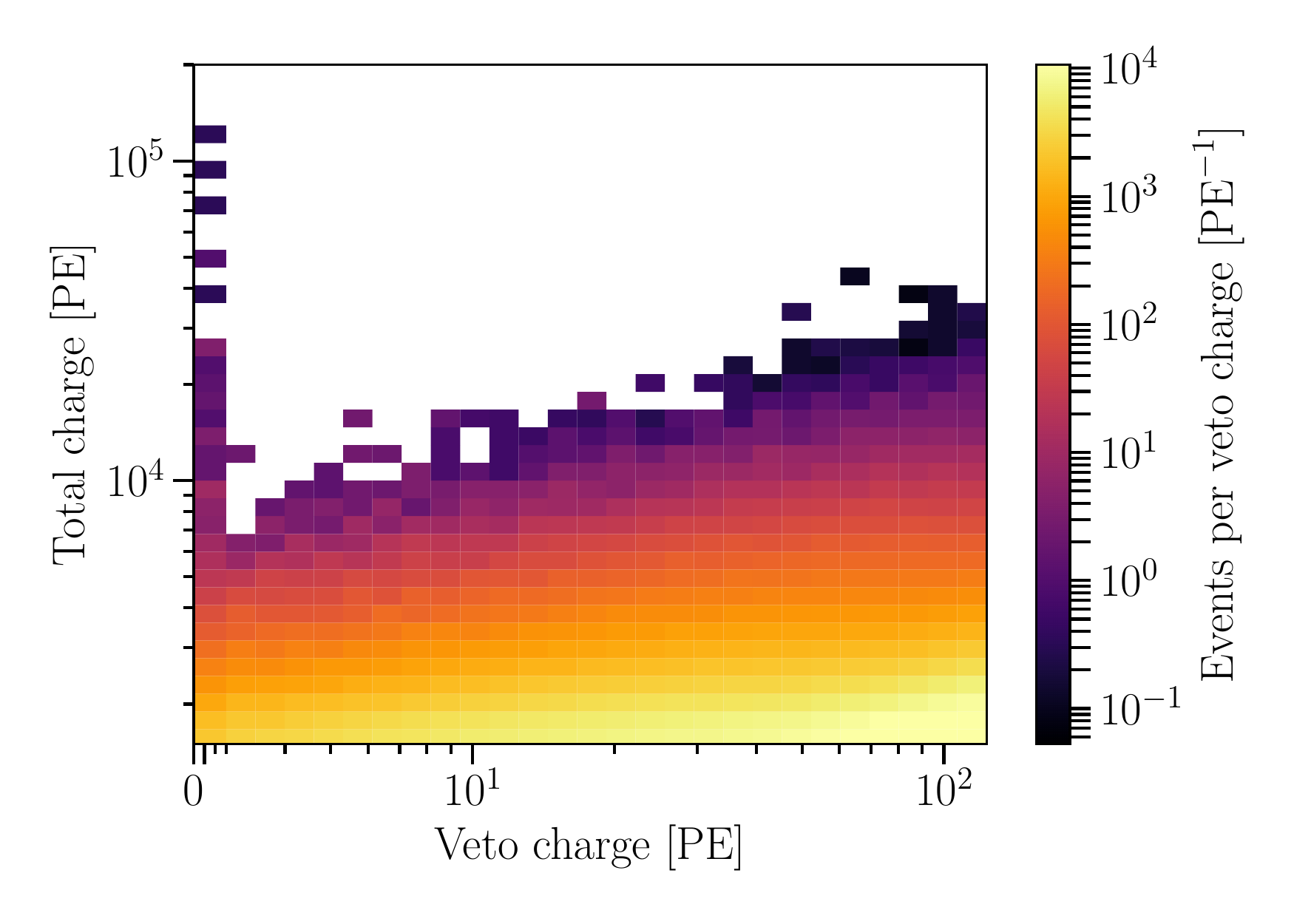}
    \caption{\textbf{\textit{Veto charge vs. total charge.}} The color scale shows the event density with respect to the veto charge, prior to the veto cuts and charge cuts, where darker color represents higher event density and lighter represents lower event density.
    The vertical axis is the total charge, in photo-electrons, deposited in the detector, while the horizontal axis is the veto charge as defined in \refsec{sec:selection}.
    The horizontal axis is plotted in linear scale from $0-\SI{3}\pe$ and in log scale from $\SI{3}\pe$ upwards.
    The high-charge population of low-veto-charge events (between 0 and $\SI{3}\pe$) can be clearly seen against the background of higher-veto-charge events.}\label{fig:charge_veto}
\end{figure}

The high-energy starting-event (HESE) sample aims to isolate astrophysical neutrinos by reducing the background of not only atmospheric muons but also atmospheric neutrinos.
In order to do so, the outer parts of the detector are used as a veto layer, aiming to select only events with a contained interaction vertex, here referred to as ``starting events''.
The fiducial volume excludes approximately the topmost $\SI{90}\meter$ of the detector, $\SI{90}\meter$ from the outer layer of DOMs on the detector sides, the bottommost $\SI{10}\meter$ of DOMs, and a $\SI{60}\meter$ thick horizontal layer directly below the region of ice with the largest dust concentration.
The veto then consists of the DOMs excluded from the fiducial volume.
\reffig{fig:veto} shows a schematic of the veto.

We first define the approximate trigger time ($t_0$) and vertex position ($\vec{x}_0$).
The event trigger time is defined as when the integrated charge deposition in the detector reaches $\SI{250}\pe$, excluding the DOMs in DeepCore, and considering only charges with hard-local-coincidence (HLC) triggers~\cite{Achterberg:2006md,Abbasi:2008aa,Aartsen:2016nxy}.
On each DOM, at least one ``hit'' is recorded if the module detects a voltage corresponding to at least $\SI{0.25}\pe$.
HLC triggered hits are DOM hits that are in coincidence ($\pm\SI{1}\mus$) with another hit on any of the two to four nearest or next-nearest neighbor DOMs on the same string.
Choosing only HLC triggered hits reduces the noise, and excluding DeepCore makes the detector response more uniform.
To further limit the contribution of random noise to the threshold, only a three-microsecond time window is considered when computing the event start time.
This time window encompasses the time a high-energy muon takes to traverse the detector.
The event's interaction vertex is approximated by the charge-averaged position of the event's first $\SI{250}\pe$ from HLC hits.
To reduce the background of muons entering from outside the detector, we select only events with three $\si\pe{}$s or less from veto hits and with veto hits on fewer than three DOMs.
A veto hit with time $t_1$ and position $\vec{x}_1$ is defined as an HLC triggered hit that meets the conditions summarized in \reftab{tbl:veto_hits}.
As the first condition, the hit must be on a DOM within the veto region.
Hits are required to arrive before $t_0+\SI{50}\ns$ to select only hits from light originating outside the fiducial volume; this cut is sufficient as the minimum distance from the fiducial volume to a veto DOM is $\SI{17}\meter$.
The time of veto hits is required to be within $\SI{3}\mus$ of the trigger time to reduce the contribution of noise.
Finally, to select only hits that may be related to the event vertex, veto hits are required to be on DOMs within $\SI{3}\mus \cdot c$ ($\sim\SI{899}\meter$) of the event vertex position.
The $\SI{3}\pe$ veto cut removes atmospheric events that are likely to deposit charge in the veto region but is not efficient at removing lower energy atmospheric events.
The distribution of data events with respect to veto charge and total charge is shown in \reffig{fig:charge_veto}.
Atmospheric muon events comprise the bulk of the distribution above a veto charge of $\SI{3}\pe$, whereas, below this threshold, the distribution is dominated by a population of neutrinos at a higher total charge.
In the Northern sky, both atmospheric and astrophysical neutrinos can contribute significantly to this low-veto-charge population.
However, in the Southern sky, atmospheric neutrinos are accompanied by muons from the same air-shower, whereas astrophysical neutrinos are not.
Thus, in the Southern sky, we expect astrophysical neutrinos to be the dominant component in this population.

The veto allows astrophysical neutrino events in the Southern sky to be separated from the vast majority of atmospheric muons and from a proportion of atmospheric neutrinos in a manner that provides some advantages with respect to non-veto methods~\cite{Aartsen:2020aqd,Stettner:2019tok}.
The physical separation of the veto and non-veto regions allows us to separately characterize the veto's response to incoming atmospheric muons independently of the non-veto portions of the detector.
This characterization is a key ingredient to calculations of the event selection response to atmospheric neutrino events with accompanying muons, which subverts the need for expensive air-shower simulations.
Improved background rejection can also be achieved through other methods~\cite{Aartsen:2020aqd}, but without the same physical separation.
The clear separation of the detector into fiducial volume and veto region allows for a specific type of background estimation.
A data-driven estimate of the atmospheric muon background is obtained using the outermost layer to identify muons, and performing the event selection in a reduced volume.
This method is described in more detail in~\refsec{sec:backgrounds}.
The event selection effective area is also approximately isotropic for astrophysical neutrinos (without accounting for absorption in the Earth), but the background has a highly zenith-dependent acceptance.

To minimize the number of atmospheric events in the selection, we select only events with at least $\SI{6000}\pe$ deposited in the detector.
This cut was determined using a testing sample, equivalent to $\SI{10}\percent$ of two years of detector operation, by requiring that no identified muons pass the charge cut.
This charge cut keeps only events guaranteed to be high energy, removing events with downward fluctuations in muon energy losses or overall light yield.
A charge cut is preferred over an energy cut, as it is more closely related to the observed event light yield and thus is a more robust estimator of expected veto charge.
For the analyses in~\refsec{sec:diffuse}, a reconstructed deposited energy cut is placed at $\SI{60}\TeV$ to reduce muon contamination further and limit the impact of normalization uncertainties and unknown shape uncertainties for this background component.
As shown in \refsec{sec:backgrounds}, the muon component does not significantly contribute to the sample above $\SI{60}\TeV$.

\begin{table}[t!]
    \centering
    \begin{tabular}{l}
        \toprule
        Veto-hit conditions \\ [0.5ex]
        \midrule
        1) Hit on DOM within veto region \\
        2) $t_1 \leq t_0+\SI{50}\ns$  \\
        3) $t_1 \geq t_0-\SI{3}\mus$  \\
        4) $\left|\vec{x}_0 - \vec{x}_1\right| \leq \SI{3}\mus \cdot c$ \\
        \bottomrule
    \end{tabular}
    \caption{\textbf{\textit{Summary of the veto-hit definition.}} This table contains the criteria a hit must satisfy to be considered a veto hit, where $t_0$ is the approximate trigger time and $\vec{x}_0$ is the approximate vertex position.
    If the veto hits constitute more than three photo-electrons or are distributed over more than two DOMs, the event is rejected from the sample.}\label{tbl:veto_hits}
\end{table}

\reftab{tbl:hese_cuts} summarizes the event selection criteria; these are limited to cuts on the total charge, veto charge, and veto hit multiplicity.
The combination of these cuts and the definition of the approximate event interaction vertex fully specifies the event selection.
In this work, the selection is applied to approximately seven and a half years of data, corresponding to a detector livetime of approximately 2635 days once offline periods are accounted for.
\reftab{tbl:observed_events} summarizes the number of observed events that pass these criteria from the chosen data taking period.
A total of 102 events were observed in this time.
Of these events, 60 have deposited energies above $\SI{60}\TeV$: 41 cascades, 17 tracks, and 2 double cascades.
Even though the event selection has not changed with respect to previously reported results~\cite{Aartsen:2013jdh,Aartsen:2014gkd,Aartsen:2015zva,Aartsen:2017mau} the event properties and the selected events themselves have changed due to a re-calibration of the single photo-electron (SPE) charge distributions of each digitizer in the detector.
This re-calibration is described in more detail in \refsec{sec:charge_calibration}.
The net effect of this re-calibration is a $\sim\SI{4}\percent$ decrease in average charge, resulting in some events dropping below the total charge cut, which was not changed correspondingly.
Seven events were removed because their total charge is now less than $\SI{6000}\pe$, with event numbers: 5, 6, 42, 53, 63, 69, and 73.
Of these, three -- one cascade and two tracks -- have deposited energy above $\SI{60}\TeV$ where the astrophysical component measurements are performed.
An additional track event, event 61, was also removed as it now fails the veto criterion: the time to accumulate $\SI{250}\pe$ increased, changing the vertex position, and allowing more hits to meet the veto hit criteria.
Finally, a ninth event, event 62, is not included in the sample due to a loss of low-level data required for the re-calibration.

Of the two double cascades above $\SI{60}\TeV$, one has a high probability of originating from a $\nu_\tau$ interaction~\cite{HESETAU}, while for the other event, the $\nu_\tau$ origin is simply favored with respect to a $\nu_e$ or $\nu_\mu$ origin.
Below $\SI{60}\TeV$ the flavor content of the three morphologies shifts; in particular, the identification of tau neutrinos is no longer robust.
In this lower energy region, we classify $41$ cascades, $10$ tracks, and $2$ double cascades.
Unlike the events examined in~\cite{HESETAU}, the two double cascade events below $\SI{60}\TeV$ are in a region of reconstructed parameter space with higher contamination from other neutrino flavors and larger background uncertainties.
These events are likely neutrinos as indicated by their up-going direction, and have a similar likelihood of astrophysical or atmospheric origin as their flavor is unknown and energy is lower.
Although higher muon contamination is expected below $\SI{60}\TeV$, such events appear in a limited range of reconstructed directions.

\begin{table}[t!]
    \centering
    \begin{tabular}{l r}
        \toprule
        Parameter & Value \\ [0.5ex]
        \midrule
        Event start time charge threshold & $\SI{250}\pe$  \\
        Maximum veto charge & $\SI{3.0}\pe$  \\
        Maximum DOMs with veto hits & $2$ \\
        Minimum total charge & $\SI{6000}\pe$  \\
        Trigger time window & $\SI{3}\mus$  \\
        \bottomrule
    \end{tabular}
    \caption{\textbf{\textit{Summary of the HESE cuts and definitions.}} Table contains the cuts and thresholds of the HESE sample.
    PE stands for photo-electron.}\label{tbl:hese_cuts}
\end{table}

\begin{table}[h!]
	\centering
    \begin{tabular}{l c c c}
        \toprule
        Category & $E < \SI{60}\TeV$ & $E > \SI{60}\TeV$ & Total \\
        \midrule
        Total Events & 42 & 60 & 102 \\
        \midrule
        Up	    & 19 & 21 & 40 \\
        Down	& 23 & 39 & 62 \\
        \midrule
        Cascade	& 30 & 41 & 71 \\
        Track	& 10 & 17 & 27 \\
        Double Cascade & 2 & 2 & 4 \\
        \bottomrule
    \end{tabular}
    \caption{\textbf{\textit{Observed events by category.}}
    The left-most column indicates the event category, which may correspond to a particular choice of morphology or direction.
    The right-most column shows the total number of data events observed in a given category.
    Intermediate columns split events into those with less than or greater than $\SI{60}\TeV$ reconstructed deposited energy.
    }
    \label{tbl:observed_events}
\end{table}

\section{Reconstruction and simulation\label{sec:reconstruction}}
Reconstruction of the neutrino events involves determining the interaction vertex, the incident direction, and the energy depositions -- positions and magnitudes -- in the detector.
The interaction vertex determined as discussed in \refsec{sec:selection} is used only in the event selection, while the more sophisticated reconstructions described in this section introduce the interaction vertex as a free parameter. 
We separately consider hypotheses formed according to the three morphologies: track, cascade, and double cascade.
For each of these hypotheses, we determine the expected light arrival time distribution on a selected subset of DOMs and maximize the likelihood of these light distributions given the data with respect to the direction, vertex, and energy depositions, as described in sections 6 and 8.2 of~\cite{Aartsen:2013vja}.

To incorporate information about neutrino flavor, we assign each event a reconstructed morphology according to the classification algorithm described in~\cite{Usner2018Search,HESETAU}. 
This method should produce a majority sample of tau neutrinos in the double cascade category above $\SI{60}\TeV$ and discriminates well between muon neutrino CC and other events.
However, there remain contributions from all flavors in all three morphological categories, which are outlined in \reftab{tbl:misid}.
Despite the non-negligible rate of misidentification, the asymmetry in the contributions can be used to constrain the flavor composition of the neutrino events.
A more detailed analysis of the astrophysical neutrino flavor content is presented in~\cite{HESETAU}.

\begin{table}[h!]
	\centering
    \begin{tabular}{l c c c}
        \toprule
        Morphology & Cascade & Track & Double Cascade \\ 
        \midrule
        Total & $\SI{72.7}\percent$ & $\SI{23.4}\percent$ & $\SI{3.9}\percent$ \\
        \bottomrule
        \multicolumn{4}{c}{}\\
        \toprule
        Morphology & Cascade & Track & Double Cascade \\ 
        \midrule
        $\nu_e$ & $\SI{56.7}\percent$ & $\SI{9.8}\percent$ & $\SI{21.1}\percent$ \\ 
        $\nu_\mu$ & $\SI{15.7}\percent$ & $\SI{72.8}\percent$ & $\SI{14.2}\percent$ \\ 
        $\nu_\tau$ & $\SI{27.6}\percent$ & $\SI{10.5}\percent$ & $\SI{64.7}\percent$ \\ 
        \midrule
        $\nu_e$ CC & $\SI{51.9}\percent$ & $\SI{8.8}\percent$ & $\SI{18.2}\percent$ \\ 
        $\nu_\mu$ CC & $\SI{8.7}\percent$ & $\SI{71.6}\percent$ & $\SI{10.9}\percent$ \\ 
        $\nu_\tau$ CC & $\SI{23.6}\percent$ & $\SI{9.8}\percent$ & $\SI{62.9}\percent$ \\ 
        \midrule
        $\nu$ CC & $\SI{84.3}\percent$ & $\SI{90.2}\percent$ & $\SI{92.0}\percent$ \\ 
        $\nu$ NC & $\SI{14.8}\percent$ & $\SI{2.6}\percent$ & $\SI{6.9}\percent$ \\ 
        $\nu$ GR & $\SI{0.9}\percent$ & $\SI{0.3}\percent$ & $\SI{1.2}\percent$ \\
        \midrule
        $\mu$ & $\SI{0.0}\percent$ & $\SI{6.9}\percent$ & $\SI{0.0}\percent$ \\
        \bottomrule
    \end{tabular}
    \caption{\textbf{\textit{Expected events by category for best-fit parameters above $\SI{60}\TeV$.}} Each column specifies the morphology of reconstructed events.
    Each row specifies a particle type, interaction type, or combination thereof.
    The top table provides the percentage of events expected in each morphology with respect to the total number of events.
    The bottom table provides the percentage of events in each category for a particular morphology, where percentages are computed with respect to the total number of expected events of the specified morphology.
    Here, CC stands for deep inelastic charged-current scattering, NC for its neutral-current counterpart, and GR for Glashow resonance.
    The percentages have been rounded to one decimal point.}
    \label{tbl:misid}
\end{table}

\begin{figure}
    \centering
    \includegraphics[width=\linewidth]{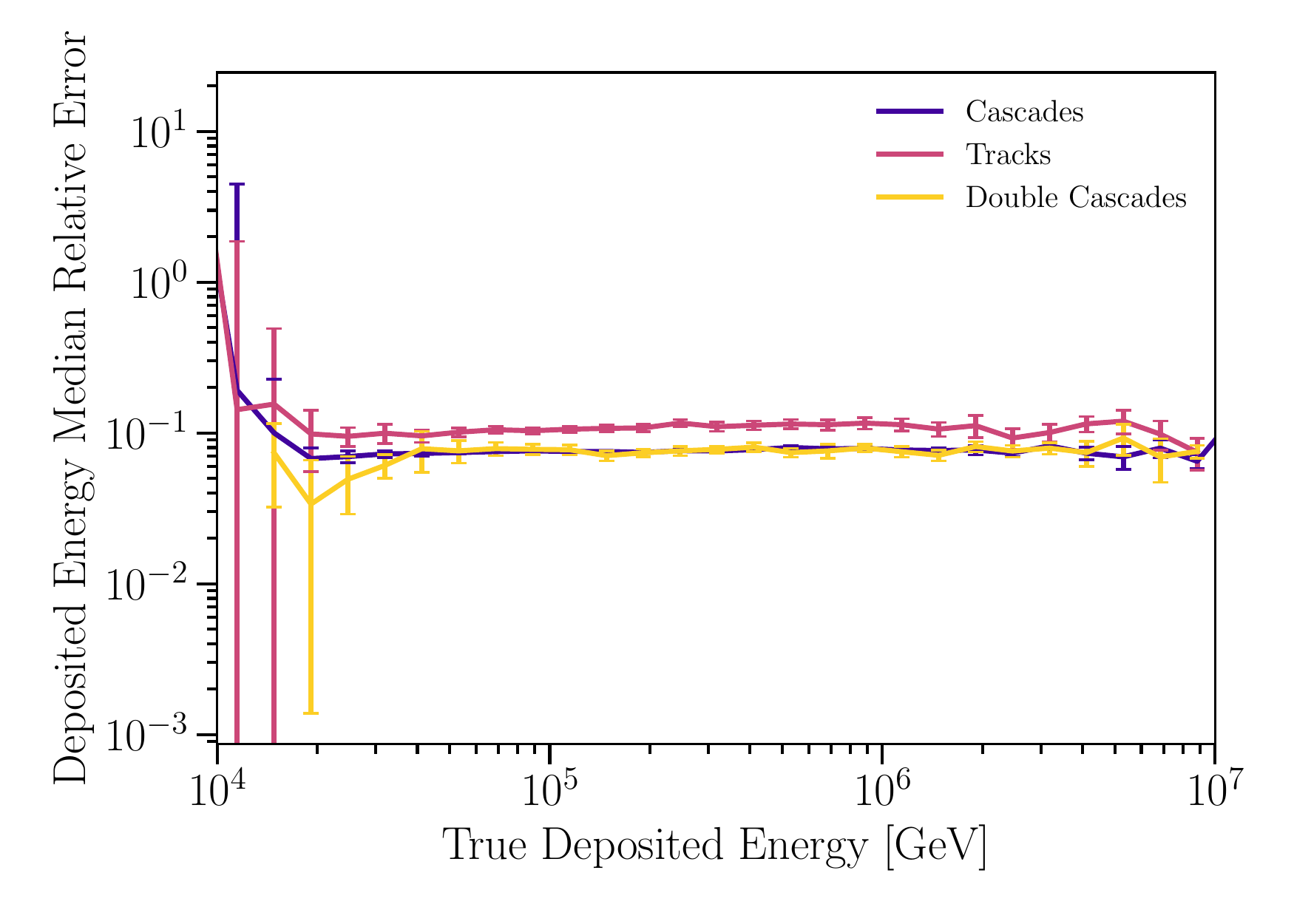}
    \caption{\textbf{\textit{Deposited energy resolution.}} Each line shows the median energy resolution plotted as a function of the true deposited energy in the detector for a reconstructed morphology.
    The error bars indicate the statistical uncertainty in the calculation of the median.
    At these energies the uncertainty of the cascade deposited energy is affected by the spatial extension of the showers not modelled by the reconstruction, whereas the uncertainty in track deposited energy is dominated by the stochasticity of the losses.
    }\label{fig:energy_resolution}
\end{figure}

\begin{figure*}
    \centering
    \subfloat{\includegraphics[width=0.5\linewidth]{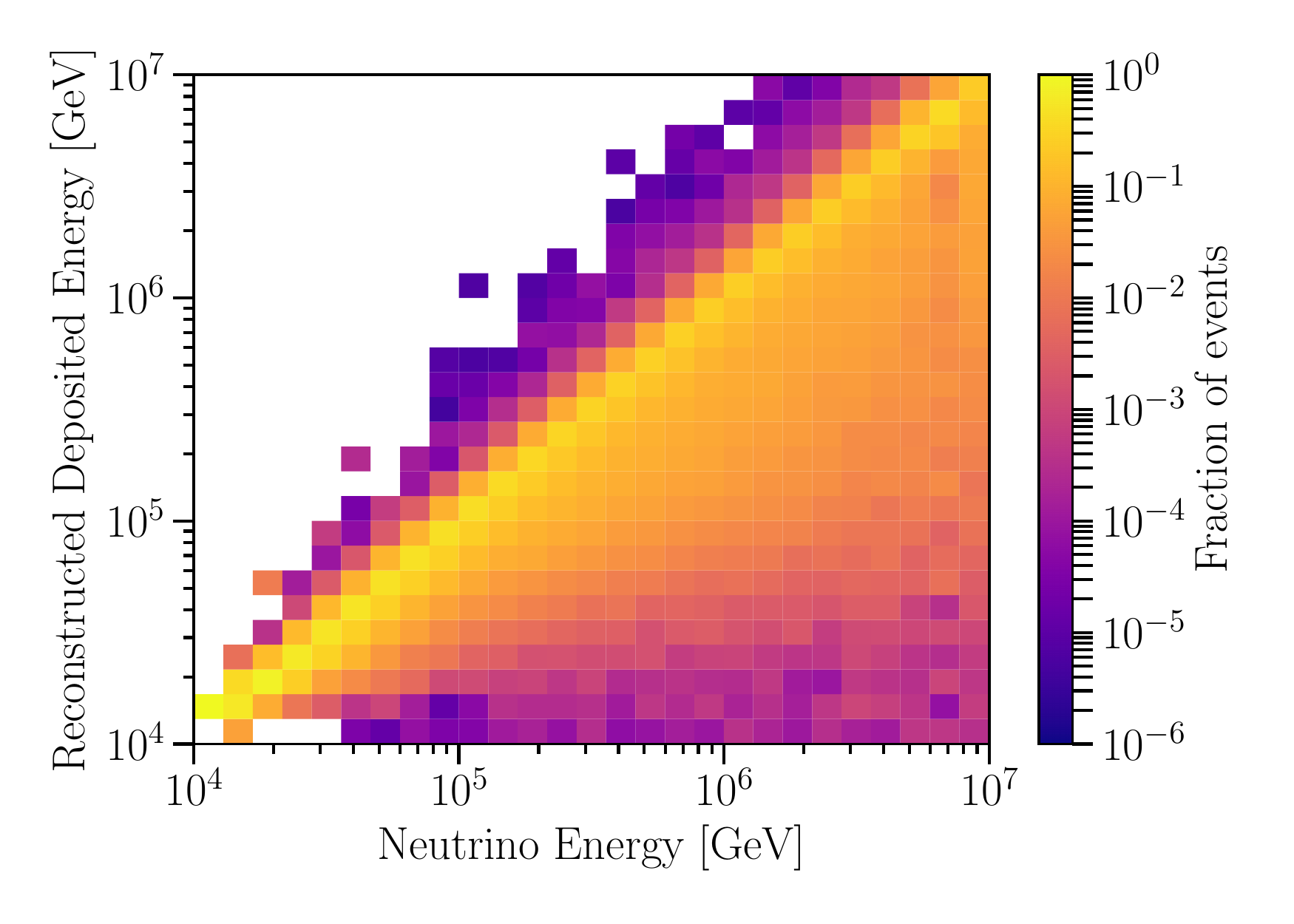}}
    \subfloat{\includegraphics[width=0.5\linewidth]{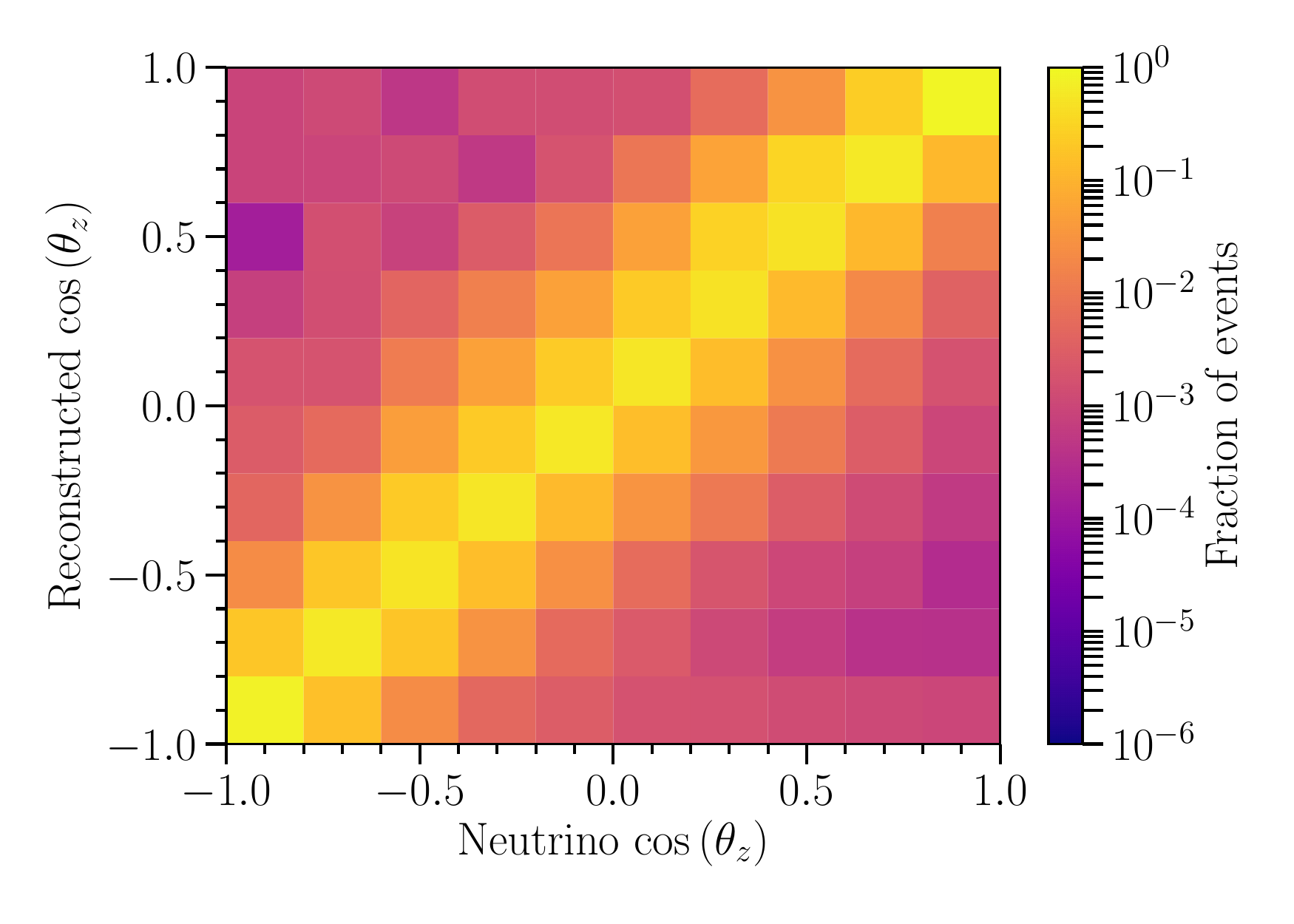}}
    \caption{\textbf{\textit{Distribution of expected reconstruction quantities as a function of true parameters.}} Transfer matrices, evaluated using simulation weighted to the single power-law best-fit parameters, are shown for all morphologies combined.
    The probability of a reconstructed deposited energy for a given neutrino energy (left) and the probability of a reconstructed cosine of the zenith angle for a particular cosine of the neutrino zenith angle (right) are shown.
    The matrices are column normalized.
    The asymmetry of the energy transfer matrix (left) is due to energy conservation, preventing large over fluctuations in reconstructed energy, and the wide range of visible energies possible for NC events which can lose large fractions of energy to the outgoing neutrino.}\label{fig:transfer_matrices}
\end{figure*}

\begin{figure}
    \centering
    \includegraphics[width=\linewidth]{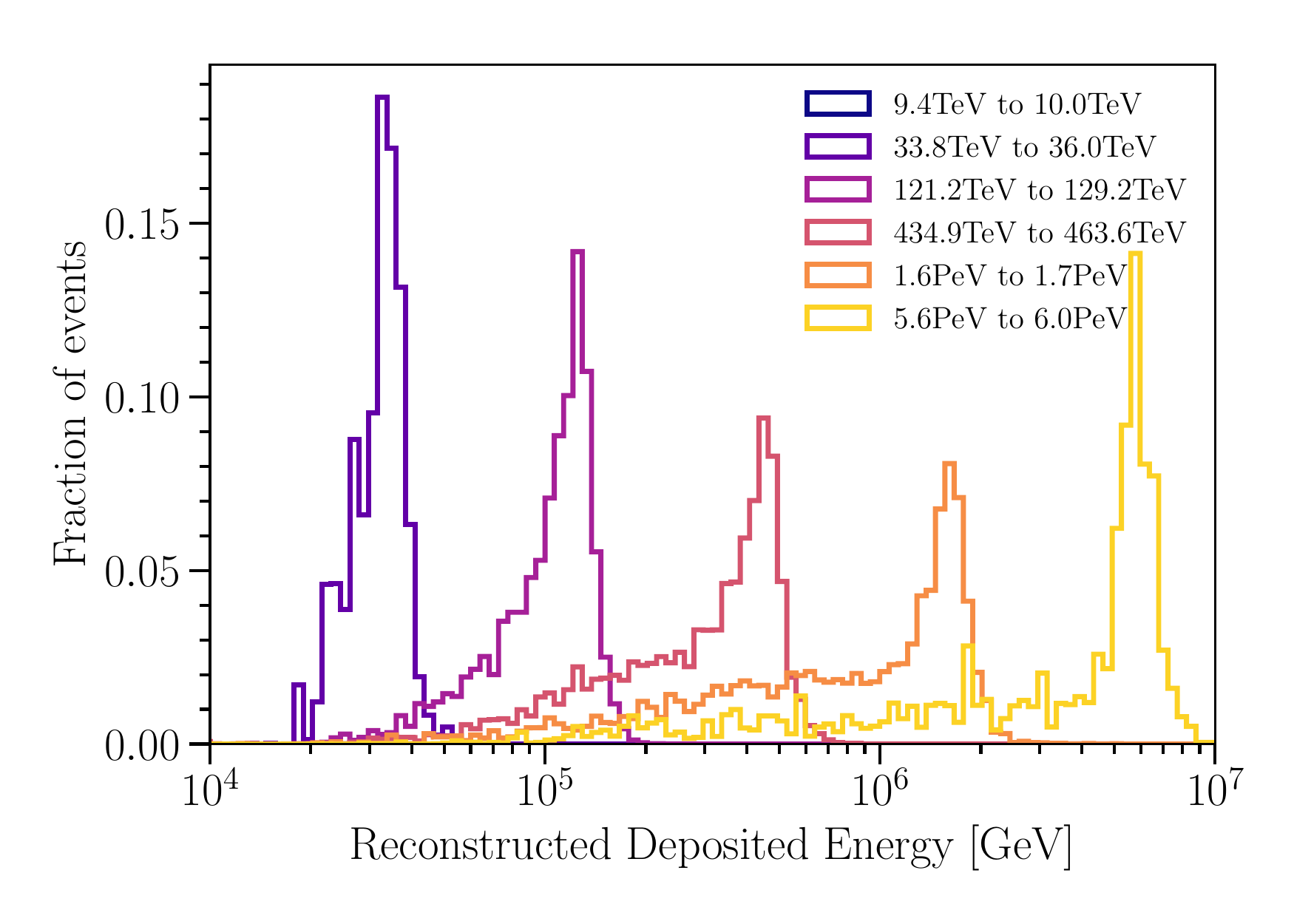}
    \caption{\textbf{\textit{Deposited energy probability distributions.}} Probability distribution of reconstructed deposited energies for slices of true neutrino energy for all morphologies weighted to the best-fit parameters.
    The lowest energy bin in the legend does not contribute to the sample.}\label{fig:energy_spread}
\end{figure}

The magnitude of energy depositions can be reconstructed to $\sim\SI{10}\percent$ accuracy if they are contained within the detector~\cite{Aartsen:2013vja} and barring systematic uncertainties of the ice.
In this sample, the median deposited energy resolution is $\sim\SI{7.9}\percent$, $\sim\SI{11}\percent$, and $\sim\SI{7.8}\percent$ for cascades, tracks, and double cascades, respectively.
\reffig{fig:energy_resolution} shows the median resolution of the reconstructed electromagnetic-equivalent deposited energy as a function of the simulated true electromagnetic-equivalent deposited energy within the detector for the three reconstructed morphologies.
Some reconstruction uncertainty stems from hadronic cascades, which have more variability in Cherenkov light yield than their electromagnetic counterpart~\cite{Aartsen:2013vja}.
The deposited energy is correlated with the neutrino energy, and so can be used to constrain the neutrino energy spectrum.
However, for some interactions, the outgoing neutrino can take away a large fraction of the initial energy, reducing the deposited energy.
The left panel of \reffig{fig:transfer_matrices} shows this behavior.
The deposited energy is peaked close to the neutrino energy, and the reconstructed distributions have long tails.
To visualize this effect more clearly, \reffig{fig:energy_spread} shows the distribution of reconstructed deposited energy for slices in true neutrino energy, where the selection truncates the tail of the distribution for lower neutrino energies.
If we use the deposited energy as a proxy for the neutrino energy, then we obtain a median neutrino energy resolution of $\sim\SI{11}\percent$, $\sim\SI{30}\percent$, and $\sim\SI{18}\percent$ for reconstructed cascades, tracks, and double cascades respectively.
We can compare these to the resolutions of the deposited energy to see the impact of other effects.
Cascades have additional uncertainty that stems from the neutrino interactions' kinematics and from differences in light yield for electromagnetic and hadronic showers as IceCube is unable to differentiate between these types of showers.
Tracks also suffer from the same kinematics issues but lack complete information in $\nu_\mu$ CC events where the resulting muon exits the detector.
Finally, double cascades have more uncertainty than cascades because of the additional degrees of freedom in the reconstruction hypothesis associated with the production and decay of a tau.

The angular reconstruction is more straightforward by comparison, as the average angle between the primary neutrino and secondary particles of the interaction is smaller than $\SI{0.25}\degree$ above $\SI{10}\TeV$.
This separation is negligible compared to reconstruction uncertainties and is even smaller at the energy scale we are concerned with.
Cascades, tracks, and, double cascades in this sample have a median zenith resolution of $\sim\SI{6.3}\degree$, $\sim\SI{1.5}\degree$, and $\sim\SI{5.0}\degree$ respectively.
The analysis of the astrophysical flux in \refsec{sec:diffuse} does not use the azimuthal directional information.
Azimuthal resolution of the different event categories is worse than the zenith resolution by $0.3-\SI{0.6}\degree$ because the inter-DOM spacing is smaller along the vertical axis than the horizontal.
The track angular resolution in this sample is worse than the resolution in $\nu_\mu$ dominated samples~\cite{Stettner:2019tok} due to the non-negligible contamination from $\nu_e$ and $\nu_\tau$ given in \reftab{tbl:misid}, the shorter length of tracks that start within the detector, and the presence of an initial hadronic cascade for $\nu_\mu$ CC events.
The angular resolution of cascades is limited by the large separation between the DOMs, the limited number of unscattered photons that are detected, and our modelling of photon propagation in ice ~\cite{Aartsen:2013rt,Aartsen:2016nxy}.
Tracks and double cascades have better angular resolution than cascades because of their longer path length.
The right panel of \reffig{fig:transfer_matrices} summarizes the angular resolution, showing the distribution of reconstructed zenith angles as a function of the true neutrino zenith angle.
The large smearing in this matrix arises from the cascades that dominate the data sample.

\begin{figure}
    \centering
    \includegraphics[width=\linewidth]{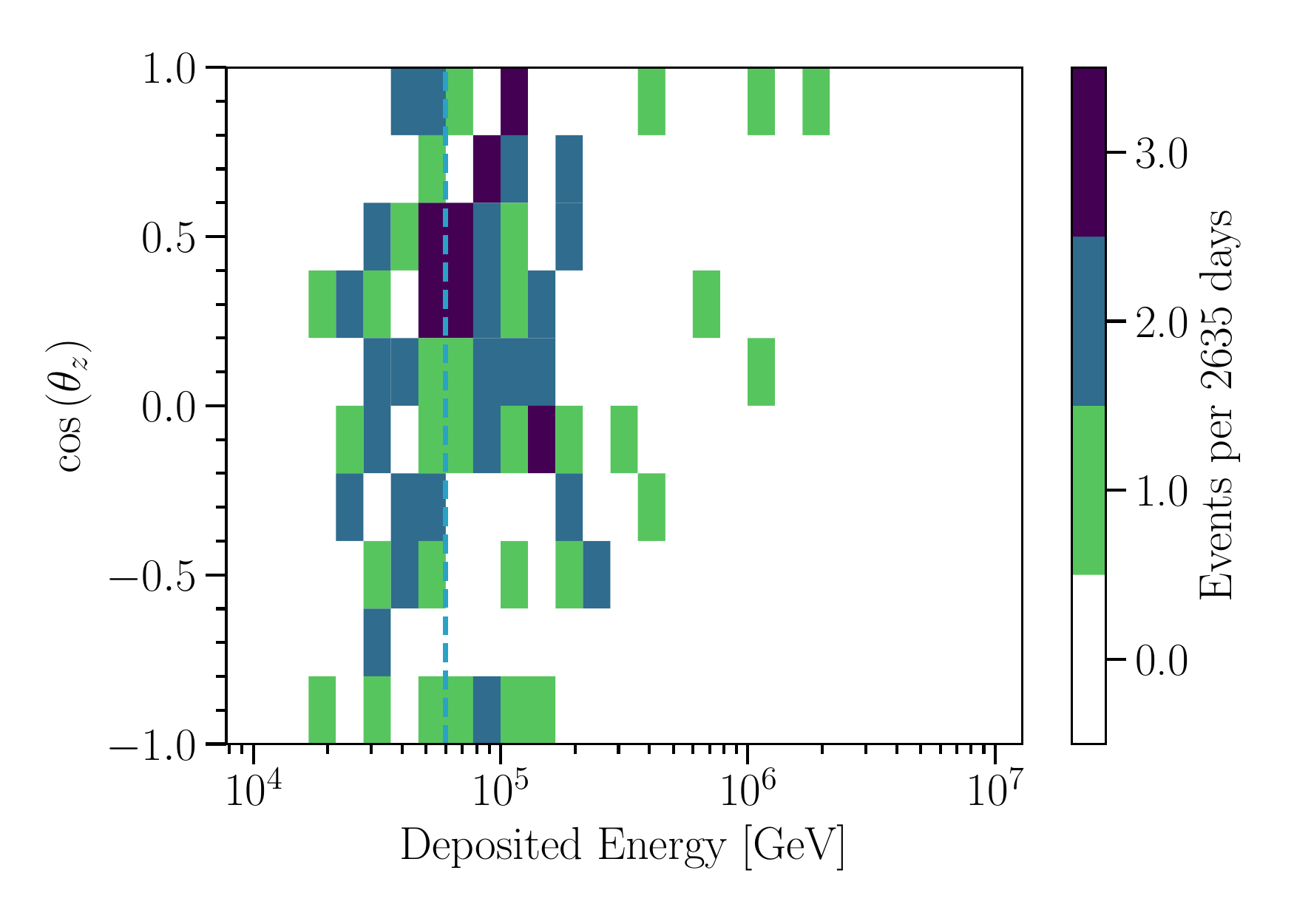}
    \caption{\textbf{\textit{HESE events observed in $\SI{7.5}\year$.}} Histogram of the observed events as a function of their inferred deposited energy and cosine of the reconstructed zenith angle.
    The dashed line indicates the low-energy threshold of $\SI{60}\TeV$.}\label{fig:hese_events}
\end{figure}

The reconstruction method used in this analysis has been changed compared to previous iterations~\cite{Aartsen:2013jdh,Aartsen:2014gkd,Aartsen:2015zva,Aartsen:2017mau} to enable better treatment of reconstruction uncertainties, which improves the accuracy of the analysis.
Previously, progressively narrower brute-force scans of the neutrino direction were used for data.
We now use a minimizer to determine the best-fit neutrino direction of data events, with a significant computational speed improvement.
Additionally, the morphology determination is now performed algorithmically, whereas previous analyses performed morphology identification by hand.
Although these changes may worsen the reconstruction's accuracy for individual events, they also enable us to run the reconstruction and classification on simulation events.
By using the same algorithmic procedure for simulated events, we now account for reconstruction and classification uncertainties on an event-by-event basis, as opposed to using average uncertainties.
Finally, the third morphological category (double cascades) was added to the previous two category classification scheme, which adds additional flavor information to the fit.

The distribution of events in the sample after reconstruction is shown in \reffig{fig:hese_events}.
The deficit of up-going high-energy events is due to the absorption of neutrinos in the Earth.
For atmospheric neutrinos, a similar deficit is expected for down-going high-energy events due to rejection by the veto.
However, such a deficit is not present for data in the down-going region.
These properties of the data are further investigated in later sections.

\section{Determination of atmospheric neutrino and muon backgrounds\label{sec:backgrounds}}
The backgrounds in measuring the astrophysical neutrino flux are atmospheric neutrinos and muons.
Atmospheric neutrinos are predominantly produced by the decay of pions and kaons, which we shall call the ``conventional'' component.
Above $\SI{1}\TeV$, the conventional neutrino spectrum is softer than the incident cosmic-ray spectrum by one unit in the spectral index due to the interactions of these mesons in the atmosphere.
This flux is also peaked at the horizon, $\cos\theta_z=0$, because of a larger path length through the atmosphere~\cite{Gaisser:2002jj,Barr:2004br,Honda:2006qj,Petrova:2012qf}.
A sub-leading -- yet unobserved -- contribution due to charmed hadron decays is expected to be important above $\sim\SI{100}\TeV$~\cite{Bhattacharya:2015jpa}.
Since the charmed hadrons decay promptly and do not interact in the atmosphere at the energies relevant for this analysis, we call this the ``prompt'' component.
Thus, at these energies, the prompt component has a spectral index close to the incident cosmic-ray spectrum and is constant with respect to the cosine of the zenith angle.

The angular and energy distribution of the initial atmospheric neutrino flux is modified after propagation through the Earth since it is not transparent to neutrinos at these energies.
This effect is accounted for with a dedicated Monte Carlo, similar to the one described in~\cite{Gazizov:2004va}.
The simulation uses the isoscalar neutrino cross sections given in~\cite{CooperSarkar:2011pa} for the neutrino-nucleon interactions and the Earth density model described in~\cite{Dziewonski:1981xy}.
Neutrino-electron scattering can be safely neglected except for resonant $W$-boson production~\cite{Glashow:1960zz}, which is included.
Uncertainties on the Earth opacity~\cite{Gandhi:1995tf,CooperSarkar:2011pa,Vincent:2017svp} and the neutrino cross section~\cite{Vincent:2017svp,Garcia:2020jwr} are ignored as they should be sub-leading in this energy range.
In order to account for uncertainties in the cosmic-ray flux~\cite{Dembinski:2017zsh} and hadronic interactions~\cite{Fedynitch:2012fs} the atmospheric neutrino flux is parameterized as
\begin{equation}
\begin{split}
    \phi_\nu^\texttt{atm} =& \promptnorm \bigg(\phi^\pi_\nu + R_{K/\pi} \phi^K_\nu\bigg) {\bigg(\frac{E_\nu}{E_0^c} \bigg)}^{-\Delta \gamma_{CR}} \\ &+ \convnorm \phi^p_\nu {\bigg(\frac{E_\nu}{E_0^p} \bigg)}^{-\Delta \gamma_{CR}},
\end{split}
\label{eq:atm_flux_equation}
\end{equation}
\noindent
where $\phi^\pi_\nu$, $\phi^K_\nu$, and $\phi^p_\nu$ are the conventional pion, kaon, and prompt atmospheric neutrino fluxes at a neutrino energy $E_\nu$ respectively as given in the Honda {\it{}et al.} and BERSS flux calculations~\cite{Honda:2006qj,Bhattacharya:2015jpa}
\footnote{The baseline conventional component here uses the parameterization of the Honda {\it{}et al.} 2006 flux~\cite{Honda:2006qj} given in~\cite{Montaruli:2011as}, which at the highest energies uses the analytic parameterization of the neutrino flux in~\cite{Gaisser:2002jj}.
This does not account for the contribution of $K_s$~\cite{Gaisser:2014pda}, which is $\sim \SI{10}\percent$ at $\SI{100}\TeV$ and well-within our uncertainties.}.
The parameters $\convnorm$ and $\promptnorm$ are normalizations for the conventional and prompt neutrino fluxes, respectively.
The value of $\pik$ modifies the relative kaon and pion contributions, where $\pik=1$ corresponds to the baseline contributions.
The $\crdeltagamma$ parameter allows for the hardening or softening of the atmospheric neutrino components to account for uncertainties in the cosmic-ray flux slope.
Although not shown in \eqref{eq:atm_flux_equation}, the relative contribution of atmospheric neutrinos and antineutrinos is also allowed to vary.
An additional parameter, $\atmonunubar$, is introduced where $\atmonunubar=1$ corresponds to the baseline contributions, $\atmonunubar=0$ is zero neutrinos, and $\atmonunubar=2$ is zero antineutrinos.
These parameters are incorporated as analysis nuisance parameters with priors as summarized in \reftab{tbl:priors}.
The roles these nuisance parameters play in the simulation weighting are presented in \refappsec{sec:likelihood}.
Priors are selected either to be Gaussian or uniform distributions if otherwise unspecified.
This analysis refrains from using prior information from other IceCube neutrino studies in order to provide independent results.
The width of the Gaussian prior for the conventional flux normalization is motivated by studies of the total uncertainty due to cosmic-ray and high-energy hadronic processes~\cite{Fedynitch:2012fs}.
A width of $0.05$ is chosen for the cosmic-ray slope parameter prior in order to accommodate values measured at intermediate~\cite{Karelin:2011zz} and high~\cite{Bartoli:2015fhw,Yoon:2017qjx,Alfaro:2017cwx} energies.
Uncertainties in the correction to the ratio of atmospheric neutrinos-to-anti-neutrinos ($\atmonunubar$) and the correction to the relative kaon and pion yields in air showers ($\pik$) were estimated by comparing the expectation of different atmospheric neutrino calculations and picking a width that encompasses their predictions~\cite{CollinFluxes,Jones:2015bya}.
The prior on the atmospheric muon rate is chosen to be Gaussian with a $\SI{50}\percent$ standard deviation; this encompasses the statistical uncertainty of the muon background measurement.
Uncertainties in the detector efficiency parameters have been found by studying dedicated calibration source data and low-energy muons~\cite{Aartsen:2016nxy}.
These systematic parameters are described in more detail in \refsec{sec:detector_systematics}.
Finally, the parameters $E_0^c=\SI{2020}\GeV$ and $E_0^p=\SI{7887}\GeV$ are pivot energies for the conventional and prompt components where the differential flux is fixed with respect to $\crdeltagamma$.

\begin{table*}[thb]
    \centering
    \begin{tabular}{l rrr}
        \toprule
        Parameter & Prior (constraint) & Range & Description \\
        \midrule
        \multicolumn{1}{l }{\textbf{Astrophysical neutrino flux:}} & & & \\
        $\astronorm$ & - & $[0,\infty)$ & Normalization scale\\
        $\astrodeltagamma$ & - &  $(-\infty,\infty)$ & Spectral index\\
        &&\\
        \midrule
        \multicolumn{1}{l }{\textbf{Atmospheric neutrino flux:}} & & &\\
        $\convnorm$ & $1.0\pm0.4$ & $[0, \infty)$ & Conventional normalization scale\\
        $\promptnorm$ & - & $[0, \infty)$ & Prompt normalization scale\\
        $\pik$ & $1.0\pm0.1$ & $[0, \infty)$ & Kaon-Pion ratio correction\\
        $\atmonunubar$ & $1.0\pm0.1$ & $[0,2]$ & Neutrino-anti-neutrino ratio correction\\
        &&\\
        \midrule
        \multicolumn{1}{l }{\textbf{Cosmic-ray flux:}} & & &\\
        $\crdeltagamma$ & $0.0\pm 0.05$ & $(-\infty,\infty)$ & Cosmic-ray spectral index modification\\
        $\muonnorm$ & $1.0\pm 0.5$ & $[0,\infty)$ & Muon normalization scale\\
        &&\\
        \midrule
        \multicolumn{1}{l }{\textbf{Detector:}} & & &\\
        $\domeff$ & $0.99 \pm 0.1$ & $[0.80, 1.25]$ & Absolute energy scale\\
        $\holeice$ & $0.0 \pm 0.5$ & $[-3.82, 2.18]$ & DOM angular response\\
        $\anisotropy$ & $1.0 \pm 0.2$ & $[0.0, 2.0]$ & Ice anisotropy scale\\
        \bottomrule
    \end{tabular}
    \caption{\textbf{\textit{Analysis model parameters for the single power-law astrophysical model.}} Prior probabilities and constraints for analysis parameters used in Bayesian and frequentist analyses respectively are shown above.
    These priors and constraints on the parameters are either uniform or Gaussian.
    The mean and standard deviation are given for Gaussian priors and constraints, while uniform priors and constraint-free parameters are denoted with a -.
    Bounds are given for all parameters.}\label{tbl:priors}
\end{table*}

\begin{figure}
    \centering
    \includegraphics[width=\linewidth]{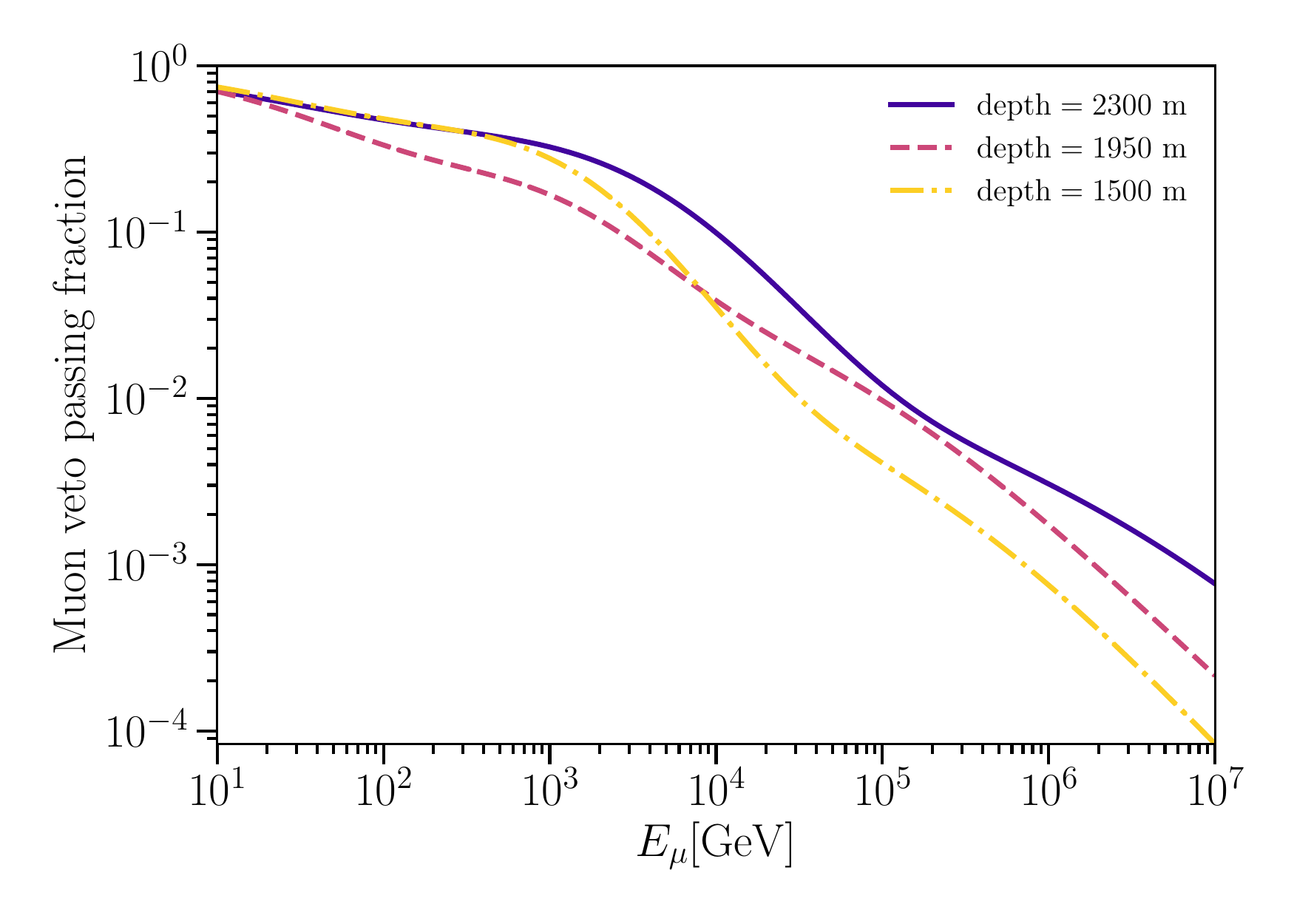}
    \caption{\textbf{\textit{Muon veto passing fraction.}} Each line shows the fraction of muons of given energy at the detector edge, $E_\mu$, that pass without triggering the veto when entering the detector at a particular depth.
    Three depths are shown: 1500, 1950, and 2300 meters from the surface, with lines of darkening color as the depth increases.
    The veto efficiency increases with the muon energy.
    Differences at various depths are due to the changing ice properties and varying acceptance as a function of depth due to the veto region's asymmetric structure.
    At the top of the detector, the veto region is larger, so it is more efficient at rejecting muons.
    Additionally, the horizontal veto layer just above the dust region provides more rejection power for muons intersecting with it.
    Finally, the expected angular distribution of incident muons is dominated by vertical events which are more easily rejected by the upper veto region.
    Above $\sim\SI{10}\TeV$, limited simulation samples are available to assess the response, and above $\sim\SI{100}\TeV$ the function is entirely extrapolated.
    At all depths the sum of a sigmoid function and a Gaussian distribution is fit to the results of muon simulation.}\label{fig:P_light}
\end{figure}

As noted in~\cite{Schonert:2008is}, muons produced in the same air-shower may trigger the detector veto in coincidence with the neutrino interaction.
To account for this, when weighting the neutrino-only simulation, each atmospheric neutrino flux component, $i$, is multiplied by the veto passing fraction, $\mathcal{P}^{i,\alpha}_{passing}$, which depends on the neutrino flavor $\alpha$.
The passing fraction depends on the neutrino energy, the cosine of the zenith angle, and the incident depth in the detector.
In previous analyses, the passing fractions were calculated using an extension of the method described in~\cite{Schonert:2008is} and bounded at $\SI{10}\percent$; details of the method are provided in~\cite{Aartsen:2013jdh}.
Cosmic-ray simulations remain a computationally prohibitive way of accounting for the effects of accompanying muons, so we still rely on calculations of the average passing rate. 
In this analysis, a new calculation given in~\cite{Arguelles:2018awr} is used that allows for cosmic-ray and hadronic models to be changed easily.
More importantly, for this analysis, any parameterization of the detector veto response to muons can be used in the calculation instead of just an energy threshold.
This capability allows us to model the detector response to atmospheric neutrinos more accurately.
Figure~\ref{fig:P_light} shows the probability that a muon will pass the veto as a function of the true muon incident energy for different detector depths.

\begin{figure*}
    \centering
    \subfloat{\includegraphics[width=0.3\linewidth]{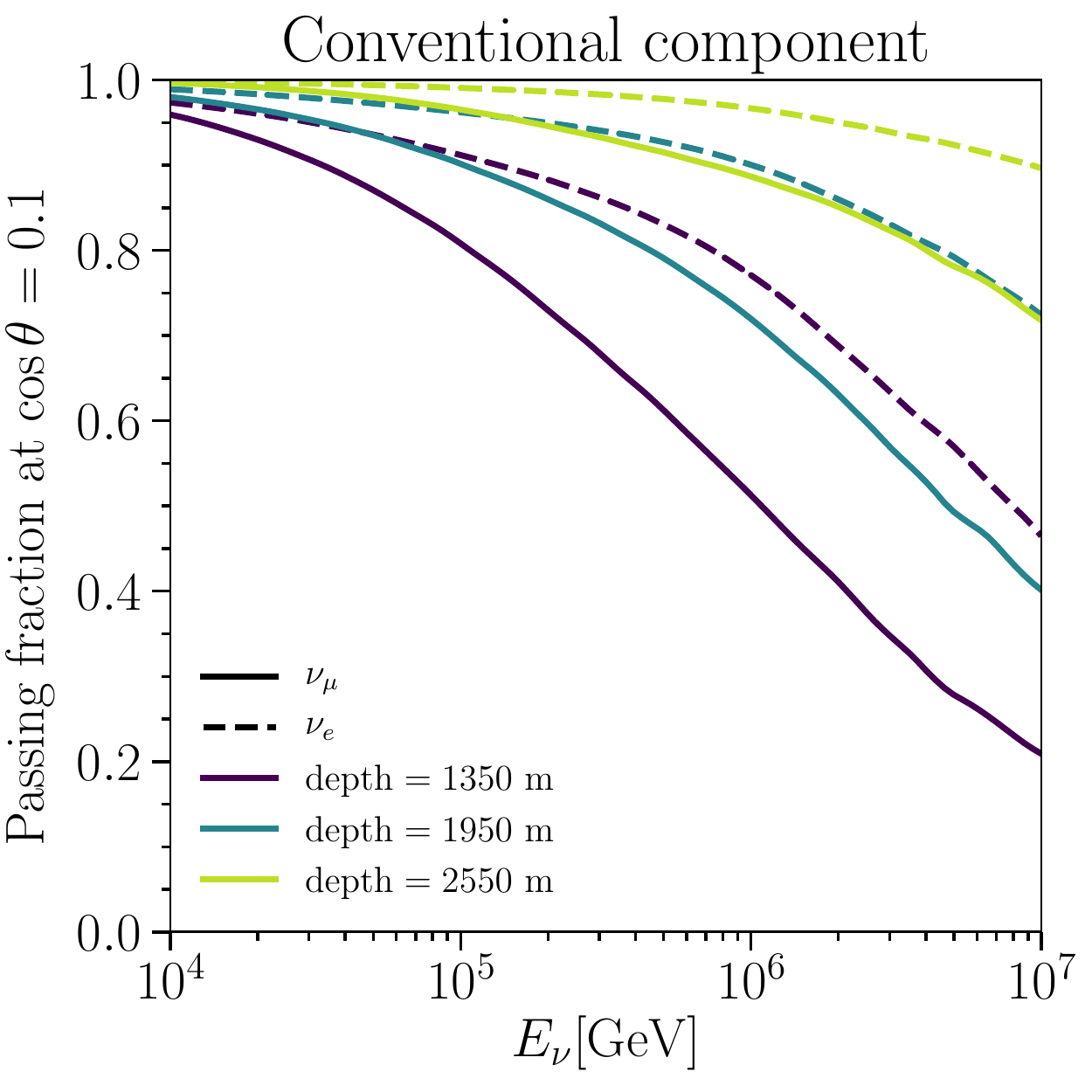}}
    \subfloat{\includegraphics[width=0.3\linewidth]{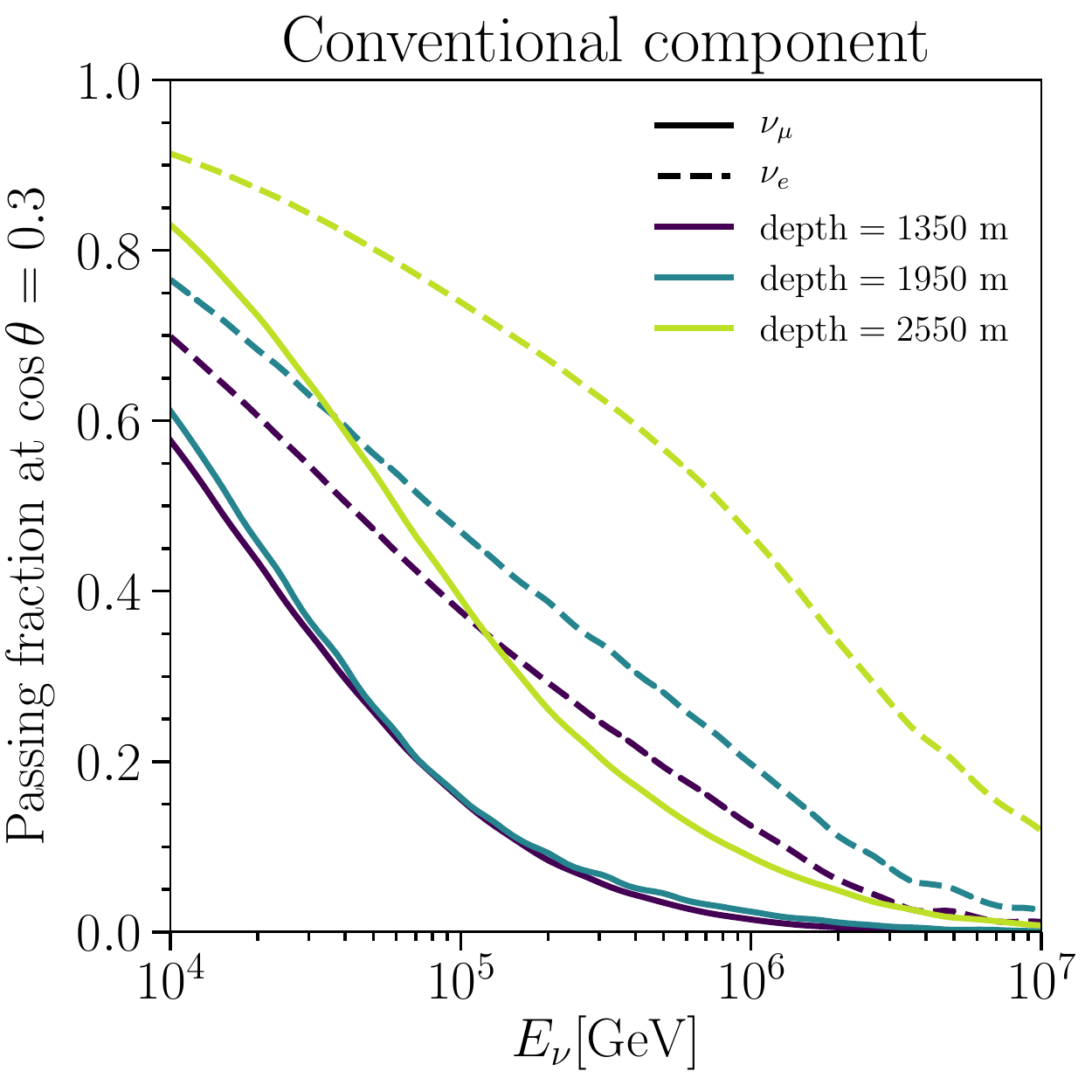}}
    \subfloat{\includegraphics[width=0.3\linewidth]{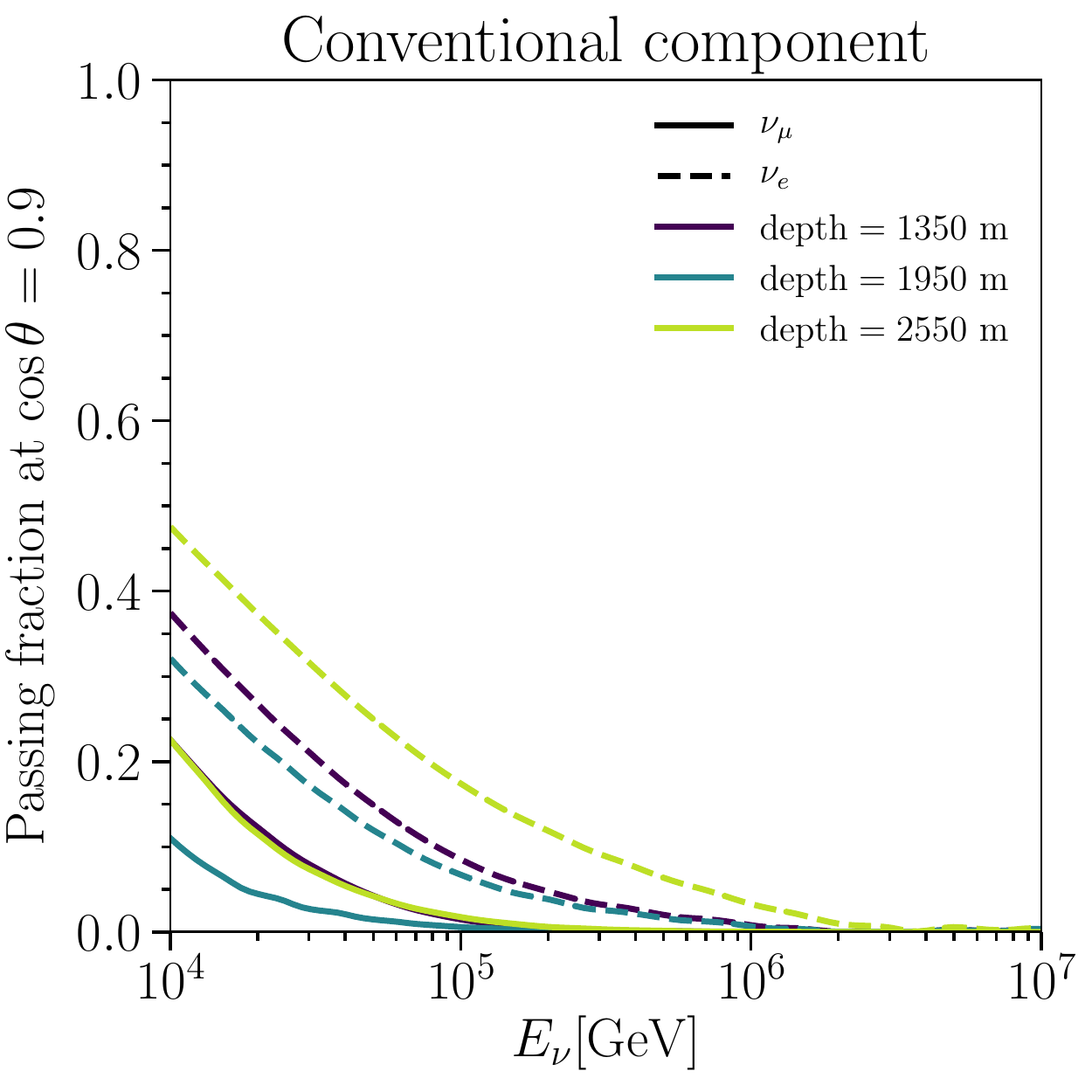}} \\
    \subfloat{\includegraphics[width=0.3\linewidth]{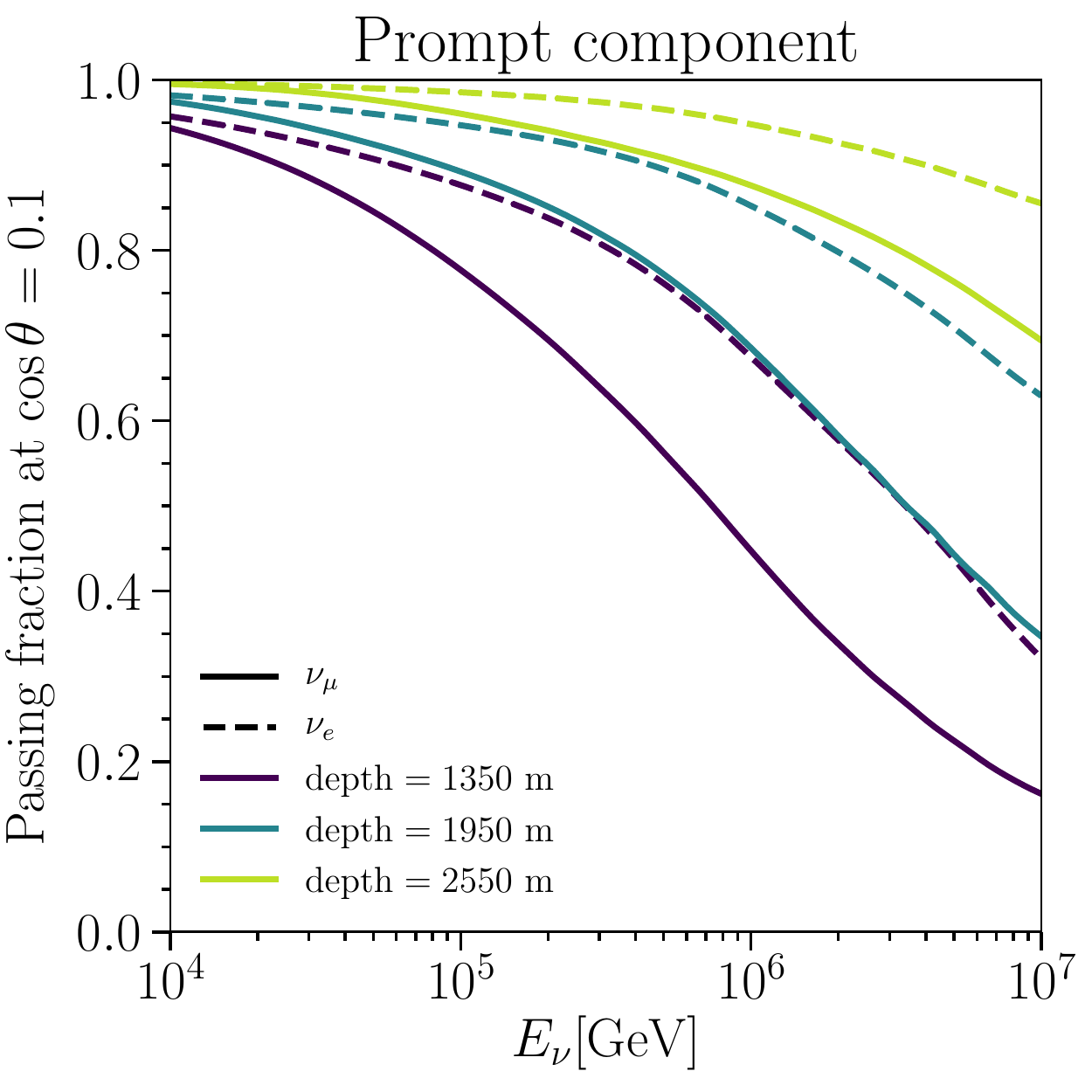}}
    \subfloat{\includegraphics[width=0.3\linewidth]{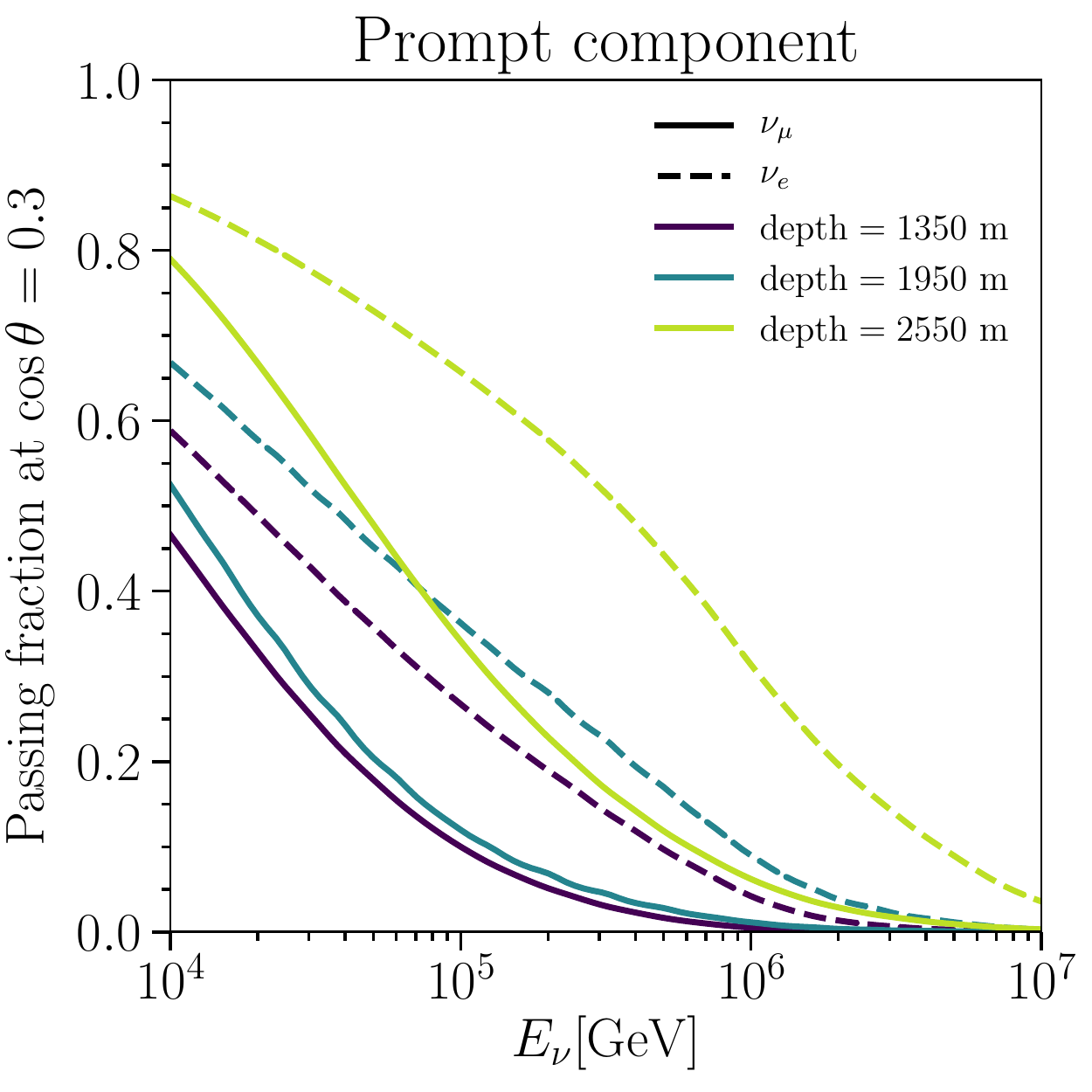}}
    \subfloat{\includegraphics[width=0.3\linewidth]{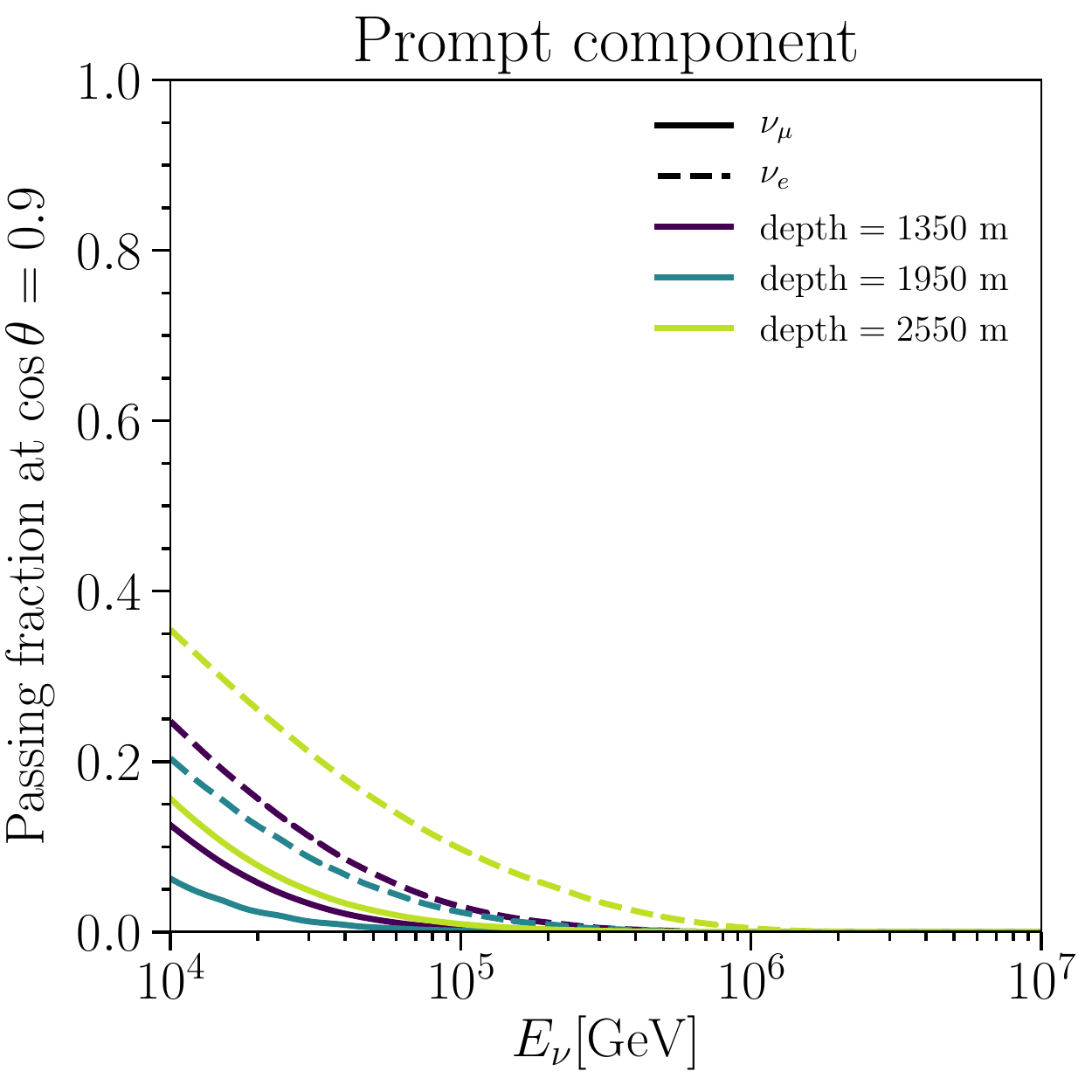}}
    \caption{\textbf{\textit{Conventional and prompt atmospheric component passing fraction.}}
    The top row of plots shows the atmospheric neutrino passing fraction as a function of the neutrino energy for a flux of neutrinos originating from pions and kaons, assuming the Hillas-Gaisser H3a~\cite{Gaisser:2013bla,Gaisser:2011cc,Hillas:2006ms} cosmic-ray model and SIBYLL 2.3c~\cite{Riehn:2017mfm} hadronic interaction model.
    While the bottom row of plots shows the atmospheric neutrino passing fraction for a flux of neutrinos originating from charmed hadrons under the same assumptions.
    Solid lines correspond to muon neutrinos and dashed lines to electron neutrinos.
    The different colors, from darkest to lightest, are for three different detector depths: 1350, 1950, and 2550 meters below the surface.
    The left, center, and right panel correspond to cosine of the zenith angles 0.1, 0.3, and 0.9 respectively (or zenith angles of $\SI{84.3}\degree$, $\SI{72.5}\degree$, and $\SI{25.8}\degree$).}\label{fig:passingfraction}
\end{figure*}

\begin{figure}
    \centering
    \includegraphics[width=\linewidth]{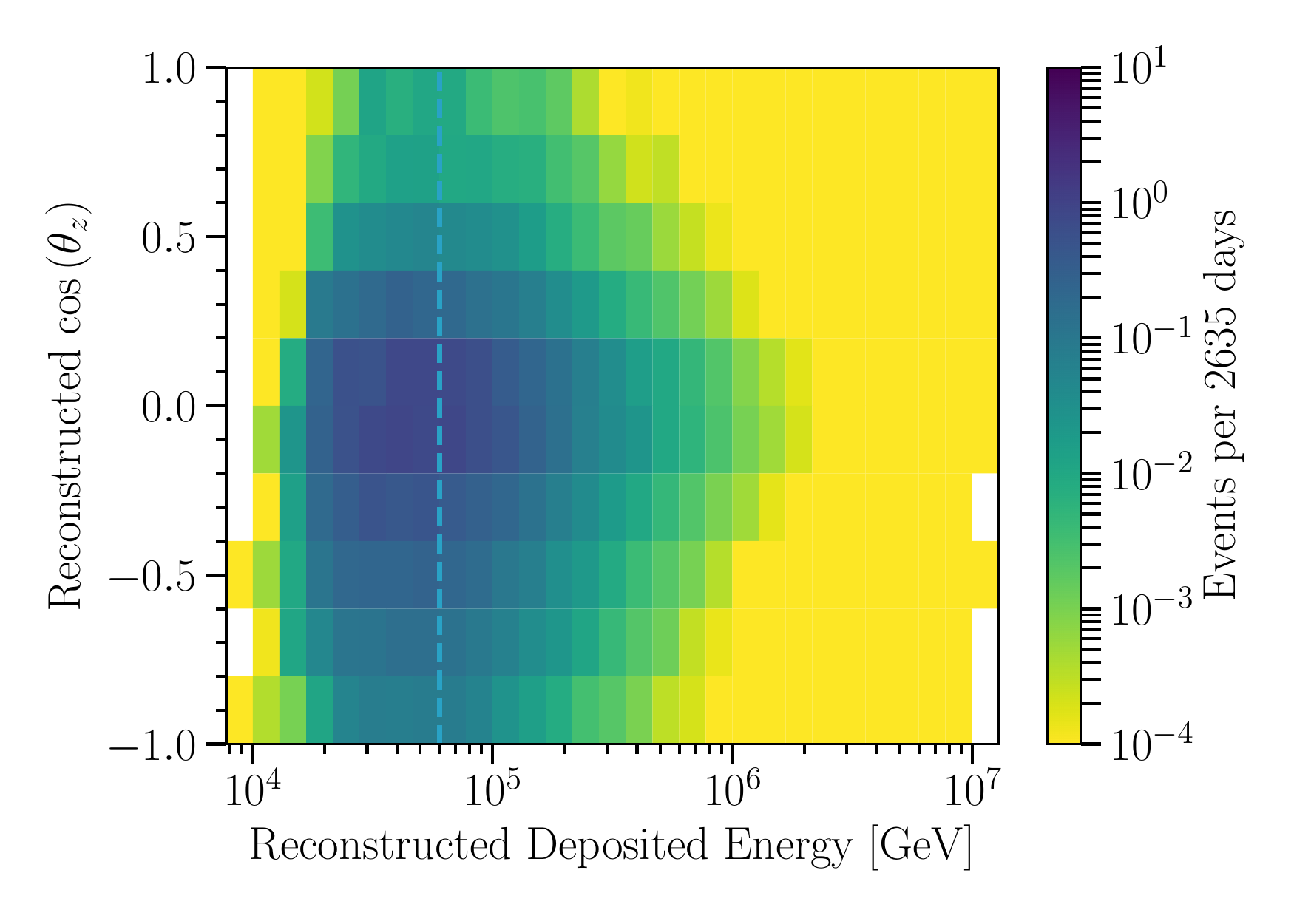}
    \caption{\textbf{\textit{Expected distribution of atmospheric neutrinos produced by pions and kaons in the sample.}} Distribution of neutrinos that pass the veto as a function of the deposited energy and the cosine of the zenith angle assuming nominal values for the nuisance parameters.
    The dashed line at $\SI{60}\TeV$ marks the low energy cut of the analysis.
    Suppression in the down-going region is due to the veto.
    Suppression in the up-going region is due to absorption of neutrinos in the Earth.}\label{fig:conventional_distribution}
\end{figure}

\begin{figure}
    \centering
    \includegraphics[width=\linewidth]{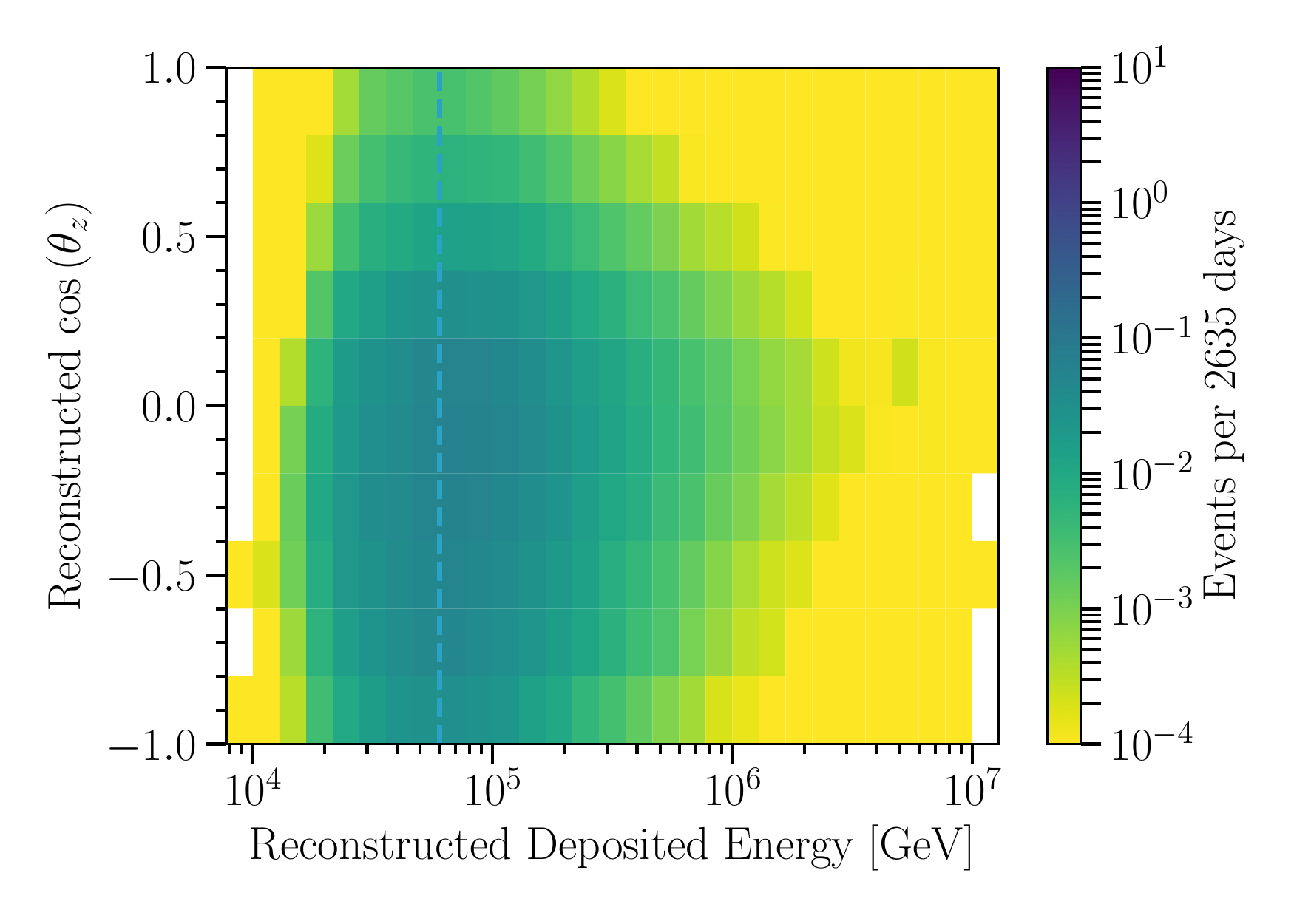}
    \caption{\textbf{\textit{Expected distribution of atmospheric neutrinos produced by charmed hadrons in the sample.}} Displays the same information as~\reffig{fig:conventional_distribution} but for the BERSS flux calculation for neutrinos from charmed hadrons~\cite{Bhattacharya:2015jpa}.
    }\label{fig:prompt_distribution}
\end{figure}

\begin{figure}
    \centering
    \includegraphics[width=\linewidth]{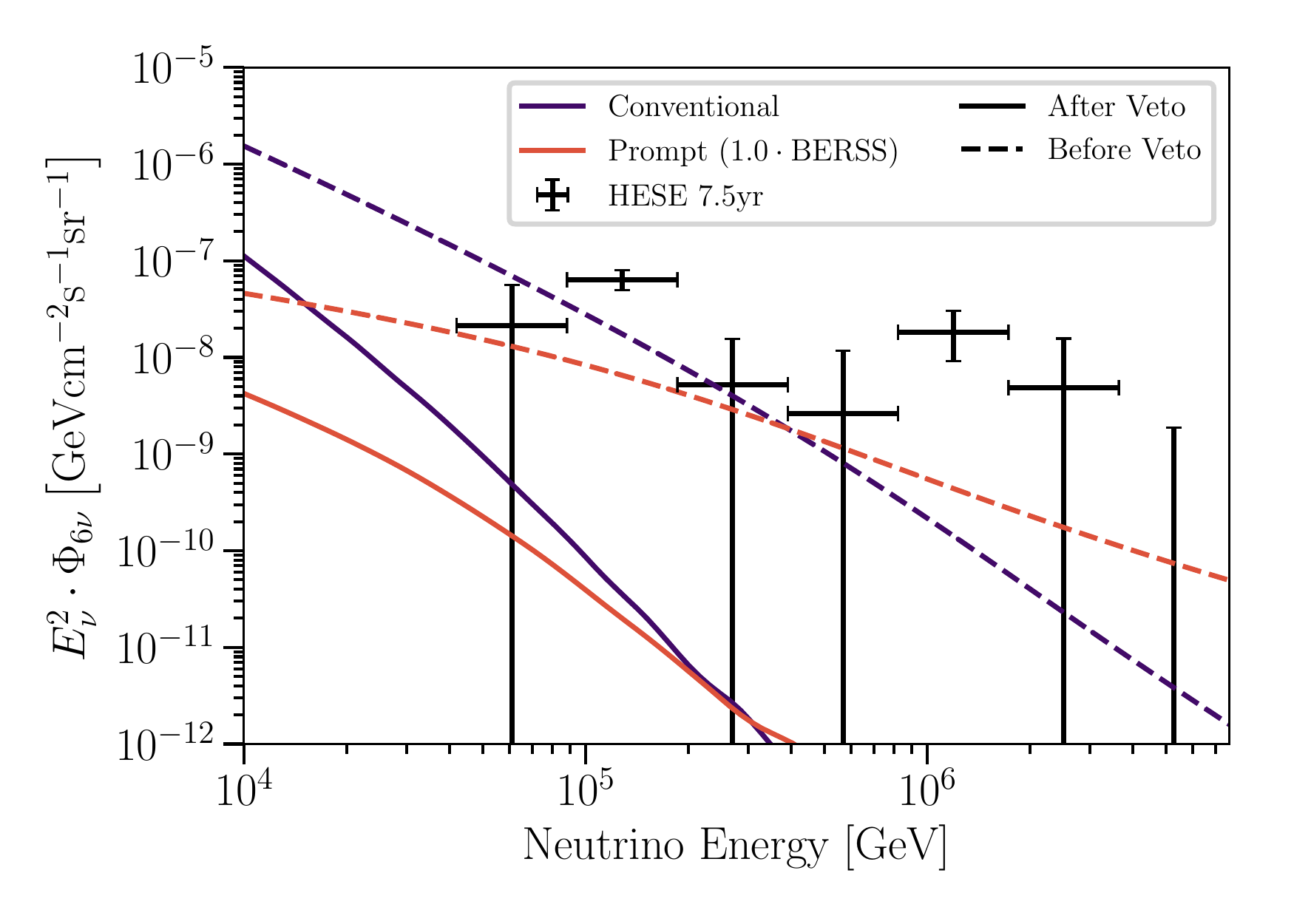}
    \caption{\textbf{\textit{All-sky average astrophysical neutrino flux and atmospheric neutrino fluxes in the vertical direction compared at $\cos\theta_z=1$ before and after the veto.}}
    The atmospheric neutrino fluxes considered in this analysis are shown as dashed lines.
    The solid lines show the product of the atmospheric flux with the passing fraction averaged over depth at a zenith angle of $\SI{0}\degree$.
    The frequentist segmented power-law fit of the astrophysical flux assumes isotropy, as described in \refsec{sec:unfolding} is shown in black.
    This comparison demonstrates the effect of the veto in the down-going region, where it is strongest.
    The atmospheric flux suppression becomes weaker towards the horizon and is not present in the up-going region.
    The dashed lines labeled ``before-veto'' are equivalent to the up-going atmospheric fluxes, with or without the veto, neglecting Earth absorption effects.}
    \label{fig:neutrino_spectrum}
\end{figure}

We calculate the atmospheric neutrino passing fractions for each component and flavor, using the muon passing fractions in \reffig{fig:P_light} as input and the \nuveto{} code provided in~\cite{Arguelles:2018awr}.
This calculation is performed assuming the Hillas-Gaisser H3a~\cite{Gaisser:2013bla,Gaisser:2011cc,Hillas:2006ms} model for the incident cosmic-ray spectra and SIBYLL2.3c~\cite{Riehn:2017mfm} for the hadronic interactions in the air shower.
Using passing fractions derived from alternative cosmic-ray and hadronic interaction models has sub-leading effects in determining the astrophysical flux~\cite{Arguelles:2018awr}.
These effects were studied by repeating the analysis for different passing fractions that arise from a given combination of cosmic-ray spectrum and hadronic model for various spectra and models available in the literature.
The inclusion of these effects, in addition to other discrete ice choices, mentioned later in \refsec{sec:detector_systematics}, increases the reported uncertainty of the astrophysical parameters by at most $\SI{20}\percent$ with respect to errors computed without these effects.
For this reason, these effects are not included in the analysis or reflected in the reported errors of any model parameters.
Figure~\ref{fig:passingfraction} shows the passing fractions for the conventional and prompt neutrino components.
The left, center, and right panels correspond to $\cos\theta_z$ values of 0.1, 0.3, and 0.9, respectively; the solid lines correspond to muon neutrinos and the dashed lines to electron neutrinos.
The passing fractions become smaller as one approaches vertical directions, as seen in the progression from left to right.
Vertical muons have the highest probability of reaching the detector because the overburden they pass through is the smallest.
Though not shown in this figure, the conventional passing fractions differ from neutrinos to anti-neutrinos~\cite{Arguelles:2018awr}; the appropriate passing fractions are used in this analysis.
Figures~\ref{fig:conventional_distribution} and~\ref{fig:prompt_distribution} show the distributions of conventional and prompt neutrinos, respectively, after this correction is applied.
This reduction in atmospheric background accounts for much of the sensitivity of this analysis to the astrophysical neutrino flux, as the observed down-going atmospheric fluxes in IceCube would otherwise be comparable in magnitude and remain similar in their angular distribution.
This is best seen when comparing the atmospheric fluxes before and after the veto to the measured astrophysical flux.
Figure~\ref{fig:neutrino_spectrum} shows the veto suppression effect for straight down-going atmospheric neutrinos.

\begin{figure}
    \centering
    \includegraphics[width=\linewidth]{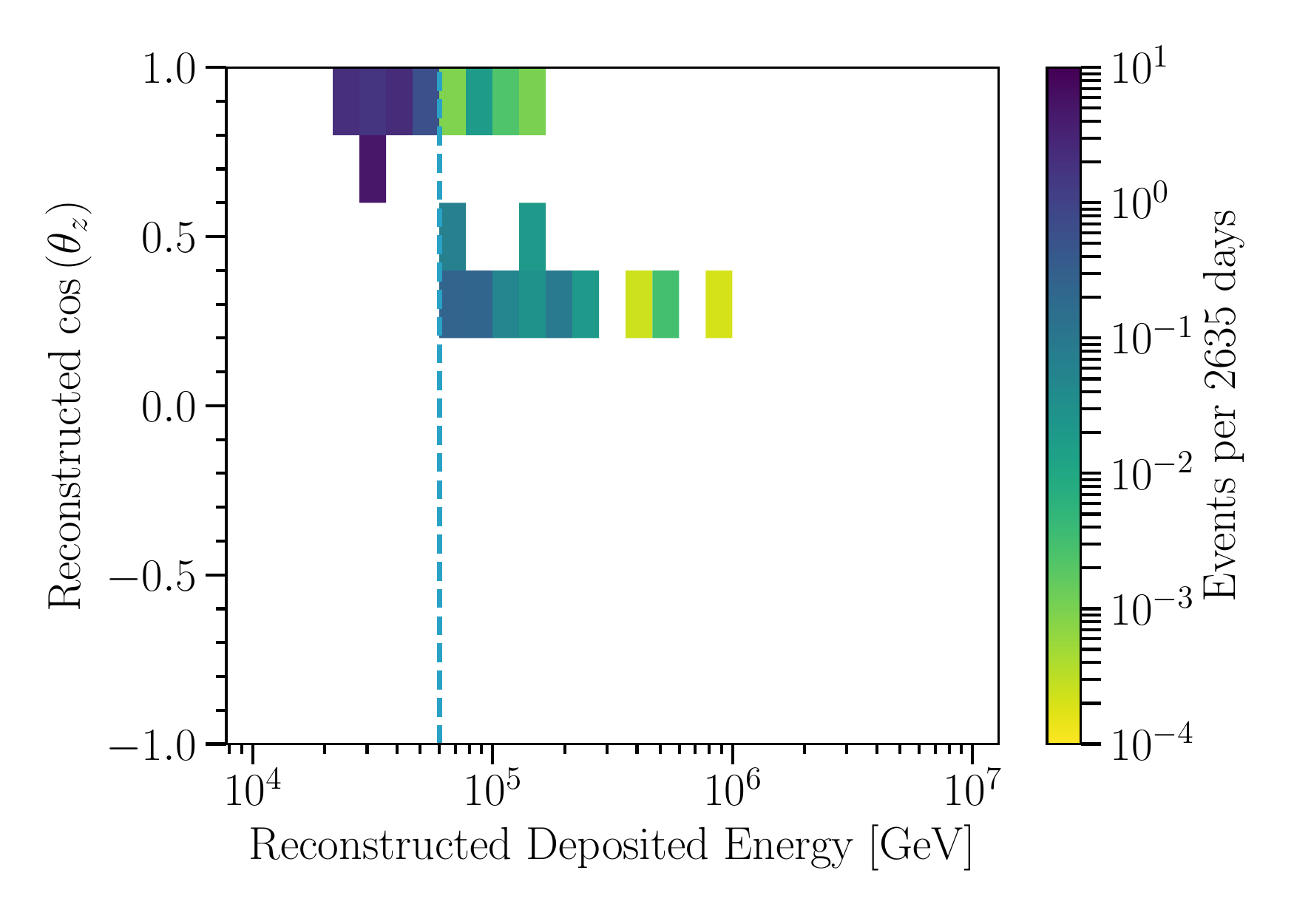}
    \caption{\textbf{\textit{Expected distribution of atmospheric muons in the sample.}} Distribution of muons that pass the veto as calculated with \MUONGUN~as a function of the deposited energy and the cosine of the zenith angle.
    The normalization is set to match the data driven sub-detector study.
    The dashed line at $\SI{60}\TeV$ marks the low energy cut of the analysis.}\label{fig:muons}
\end{figure}

Finally, there is also the possibility of single muons that trigger the event selection without a neutrino interaction in the detector and still pass the veto.
The shape of the atmospheric muon and neutrino fluxes are closely related to each other, and bounded by the cosmic-ray flux so that they must be steeply falling.
The energy losses of muons in the atmosphere and ice further soften the muon spectrum from that of cosmic rays.
Although there is uncertainty in the shape of the muon spectrum, the yield of muons from cosmic-ray air showers has more significant modeling uncertainties that stem from uncertainties in the hadronic interaction cross sections~\cite{Pierog:2017nes} and the cosmic-ray composition~\cite{Bluemer:2009zf}.
As we lack the capability to parameterize both the uncertainty in shape and normalization from first principles, we turn to data-driven techniques to constrain the size of this background.
Unfortunately, the available data-driven techniques do not provide us with enough events to determine the muon background's shape.
For this reason, we take a pragmatic approach to treat the muon component.
A dedicated muon simulation, called \MUONGUN~\cite{vanSanten:2014kqa}, provides a reasonable estimate for the shape of a steeply falling muon spectrum but neglects shape uncertainties.
The input spectrum of atmospheric muons is modeled by a parameterization of muons from air showers simulated with the \CORSIKA~\cite{Heck:1998vt} package assuming the Hillas-Gaisser H4a~\cite{Gaisser:2013bla} cosmic-ray flux model and SIBYLL 2.1~\cite{Ahn:2009wx} hadronic model.
The normalization is then constrained using a procedure that tags background muons in data.
A second veto layer inside the original outer veto layer is introduced to construct the data based prior.
Events that trigger the outer veto layer, but do not trigger this second inner veto layer, are tagged as muons that pass the inner veto.
The muon normalization from simulation is re-scaled from $N_\MUONGUN$ to $2.1\cdot N^\mu_\texttt{tagged}$ to match the number of tagged muons while accounting for the relative size of the fiducial volumes.
Thus, the baseline expected muon flux is given by
\begin{equation}
\begin{split}
        \frac{d^3\Phi}{d E_\mu d \theta_{z,\mu} d D_\mu} ={}& \frac{d^3\Phi_\texttt{GaisserH4a}}{d E_\mu d \theta_{z,\mu} d D_\mu}(E_\mu, \theta_{z,\mu},D_\mu)\\* & \cdot \frac{2.1 \cdot N^\mu_\texttt{tagged}}{N_\MUONGUN},
\end{split}
\label{eq:muon_scaling}
\end{equation}
where $\Phi_\texttt{GaisserH4a}$ is the aforementioned parameterization; and $E_\mu$, $\theta_{z,\mu}$, and $D_\mu$ are the muon energy, zenith, and depth at injection, respectively.
Table~\ref{tbl:tag_muons} lists the number of tagged muons observed per year; in total, 17 muons were observed.
The expected distribution of passing atmospheric muon events is shown in \reffig{fig:muons} as a function of the deposited energy and reconstructed cosine of the zenith angle.

\begin{table}
    \centering
    \begin{tabular}{l r}
        \toprule
        Season & $N^\mu_{tagged}$ \\
        \midrule
        2010 & 2 \\
        2011 & 1 \\
        2012 & 1 \\
        2013 & 1 \\
        2014 & 2 \\
        2015 & 6 \\
        2016 & 2 \\
        2017 & 2 \\
        \midrule
        Total & 17 \\
        \bottomrule
    \end{tabular}
    \caption{\textbf{\textit{Number of tagged muons per season.}}
    The table above shows the number of tagged muons used to construct the muon normalization prior.
    The first season, 2010, used a partial IceCube configuration with 79 strings, while the rest of the seasons took data with the full configuration of 86 strings.
    The larger number of tagged muons in the 2015 season is consistent with a statistical fluctuation.
    The last season, 2017, represents only a partial year of data taking in this paper as the 2017 data processing was not yet completed at the time of this analysis.}\label{tbl:tag_muons}
\end{table}

\section{Systematic uncertainties and statistical treatment\label{sec:uncertainties}}
\subsection{Detector systematic uncertainties\label{sec:detector_systematics}}
The primary detector systematic uncertainties can be organized as arising from either incomplete knowledge of the ice properties or detector response.
The ice properties can, in turn, be separated into global ice effects -- such as anisotropy, scattering length, and absorption of photons in the bulk ice -- and local ice properties, {\it{}i.e.}\ effects of the re-frozen ice surrounding the DOMs previously melted during deployment~\cite{Karle:1994eua}.
Additional air bubbles introduced in the drilling process and concentrated in the center of the hole during re-freezing increase the scattering of light, particularly in the vertical direction~\cite{Aartsen:2016nxy}.

Uncertainties in the optical module light acceptance and local ice effects are modeled with three parameters: DOM efficiency ($\domeff$), head-on efficiency ($\holeice$), and lateral efficiency ($\epsilon_\texttt{lateral}$).
The first parameter is an overall change in the efficiency of all the DOMs in the detector, with respect to the individual baseline of each DOM.
The latter two parameters are part of a parameterization of the efficiency's angular dependence, which depends most strongly on local effects~\cite{Aartsen:2016nxy, Aartsen:2014yll, Aartsen:2017nmd}.
The $\holeice$ parameter modifies the photon efficiency in the vertical direction, while the $\epsilon_\texttt{lateral}$ parameter modifies the lateral direction efficiency.
Of these three parameters, only $\domeff$ and $\holeice$ have a significant effect on the observable distributions in this analysis, and so $\epsilon_\texttt{lateral}$ is fixed to a nominal value obtained from calibration data in the simulation used for this analysis.
Dedicated simulations are run for different values of relative $\domeff$ to incorporate uncertainties that stem from these parameters into the analysis.
From these simulations interpolating b-splines are constructed with \PHOTOSPLINE~\cite{Whitehorn:2013nh,photospline} to describe the ratio between expected event distributions.
It is apparent from these observable distributions that $\domeff$ primarily changes the overall normalization of the event rates.
This systematic correction is applied multiplicatively to the expectation of the sample.
The result of applying this correction is shown in Appendix \reffig{fig:domeff}.

A similar procedure is performed to include the effect of changing the head-on efficiency.
Again, a dedicated simulation is run for several values of $\holeice$ to compute the systematic correction.
This correction is also applied multiplicatively to the expectation resulting in the distributions in Appendix \reffig{fig:holeice}, where this parameter is varied within one standard deviation.
The $\holeice$ parameter primarily modifies the relative rate of observed up-going and down-going events.

The global properties of the ice are taken into account in different ways for different effects.
The scattering and absorption of photons in the ice is azimuthally anisotropic because of the ice flow~\cite{Aartsen:2013rt}.
This azimuthal anisotropy is modelled as the effect is described in~\cite{Williams:2014era}\footnote{Newer modeling of the anisotropy through the birefringent properties of ice~\cite{Chirkin:2019vyq,Chirkin:2013lpu} will supersede this in the future and are expected to produce some changes to the reconstructed event directions.}.
For the double cascade morphology, changes to the scattering and absorption lengths can alter the event's apparent length.
Therefore, this effect can bias the double cascade length reconstruction if the orientation of the anisotropy axis and the strength of the anisotropy are not well modelled.
Calibration measurements well constrain the anisotropy axis, however, the strength of this effect is more uncertain.
Uncertainties of this effect are incorporated by parameterizing the length reconstruction bias with an analytic function; see~\cite{HESETAU} for details.
The effect of this bias on the distribution of event observables is then parameterized with splines in the same way as the previously mentioned ice and detector systematics.
The result of applying this change in anisotropy to this sample is shown in Appendix \reffig{fig:anisotropy}.
Other observables used in this analysis are not strongly affected by this systematic uncertainty, so the effect on the energy and zenith observables is neglected.

The bulk ice scattering and absorption uncertainties are sub-leading, and their impact is evaluated by repeating the analysis with three different ice variants.
The three ice variants used are a $\SI{10}\percent$ increase in overall light scattering, a $\SI{10}\percent$ increase in overall light absorption, and a simultaneous $\SI{7.5}\percent$ reduction of both absorption and scattering.
We found that the inclusion of the effects of bulk ice scattering, bulk ice absorption, and the discrete atmospheric flux choices previously mentioned in \refsec{sec:backgrounds} increases the reported uncertainty of the astrophysical parameters by at most $\SI{20}\percent$ with respect to errors computed without these effects.
For this reason, these effects are not included in the analysis and are not reflected in the reported errors of any model parameters.

\subsection{Statistical treatment\label{sec:statistics}}
The model parameters described in this section fall into two categories: parameters of interest ($\vec\theta$) and nuisance parameters ($\vec\eta$).
The former depend on the analysis, and the latter include parameters that modify the systematic effects discussed in \refsec{sec:detector_systematics}, as well as physics parameters not being examined.
The physics model parameters of interest often refer solely to the astrophysical model parameters and are discussed in greater detail in \refsec{sec:diffuse}.
In the case of a single power-law astrophysical flux hypothesis, these are the astrophysical neutrino flux normalization ($\Phi_{\rm astro}$) and the spectral index of the power-law flux ($\gamma_{\rm astro}$).
Different parameters of interest are given for other generic astrophysical models in \refsec{sec:generic_models} and source-specific models in \refsec{sec:specific_models}.
In the case of searches for new physics, more terms are incorporated into the model parameters; {\it{}e.g.} for the dark matter decay search: the dark matter mass and its lifetime.
In all cases, the systematic treatment of the relevant uncertainties described in \refsec{sec:detector_systematics} applies.

Results are presented using both frequentist and Bayesian statistical methodologies in this work and associated analyses~\cite{HESEFLV,HESEDM, HESEXS}.
These two methodologies provide distinct information~\cite{PhysRevD.98.030001}.
In the frequentist framework, we report the parameters that most likely explain the data.
We also report intervals in parameter space constructed such that they contain the true value of the parameter some fraction of the time for repeated experiments.
These constructions free us from dependence on priors, but do not make probabilistic statements about the model parameters; see~\cite{stuart1963advanced,James:2000et,Biller:2014eya,Cousins:2018tiz,Barlow:2019svl} for further discussion on confidence intervals.
On the other hand, the Bayesian framework makes statements about the model by invoking Bayes' theorem, at the cost of dependence on prior choice.
In this framework, we report the most probable model parameters given the observed data and the preferred parameter space regions.
Frequentist and Bayesian methods can both be applied and provide complementary information about the model and the data~\cite{Heinrich:2007zza,Cousins:1994yw}.

Here both of these approaches make use of the likelihood function, which reflects the plausibility of model parameters given observed data and is defined as $\like(\vec\theta, \vec\eta|\texttt{data}) = p(\texttt{data}|\vec\theta, \vec\eta)$, where $p(\texttt{data}|\vec\theta, \vec\eta)$ is the probability of the data given the model parameters.
External knowledge of the model parameters is also included with the term $\Pi(\vec\theta, \vec\eta)$, which is the constraint (prior) on the parameters in the frequentist (Bayesian) interpretation.
Details of $\Pi(\vec\theta,\vec\eta)$ are given in \reftab{tbl:priors}.

Frequentist results are presented with the best-fit parameters and their errors using the profile likelihood technique.
Dropping the explicit notational dependence on data, the profile likelihood function is defined as
\begin{equation}
    \tilde{\like}^\texttt{profile}(\vec\theta) = \max_{\vec\eta} \like(\vec\theta,\vec\eta)\cdot\Pi(\vec\theta, \vec\eta),
    \label{eq:likelihood_freq}
\end{equation}
where the negative log of the function is minimized in place of maximizing the function.
This minimization is performed over continuous nuisance parameters using the L-BFGS-B algorithm~\cite{doi:10.1137/0916069}.
In the frequentist statistical treatment, the constraint $\Pi(\vec\theta, \vec\eta)$ is the likelihood of the model parameters given external data.
For some parameters, no external data is available, and so the constraint $\Pi(\vec\theta, \vec\eta)$ is constant with respect to those parameters.
Maximizing $\tilde{\like}^\texttt{profile}$ over all parameters defines the best-fit point $\hat{\vec\theta}$.
Frequentist results are then presented assuming Wilks' theorem~\cite{wilks1938}, with the appropriate degrees of freedom; see~\cite{Algeri:2019arh} for a recent summary of the conditions under which this theorem holds.
Although analyses presented with the asymptotic approximation described in~\cite{wilks1938} violate some conditions of Wilks' theorem, the introduction of many nuisance parameters helps to alleviate differences between the real and approximated test-statistic distributions as demonstrated in~\cite{Algeri:2019arh}.
As a result, this asymptotic approximation is sufficient for presentation here.
The required model parameter test-statistic ($\TS$) is defined as
\begin{equation}
    \TS(\vec\theta) = -2\log{\left(\frac{\tilde{\like}^\texttt{profile}(\vec\theta)}{\tilde{\like}^\texttt{profile}(\hat{\vec\theta})}\right)}.
\end{equation}

The product of $\like(\vec\theta, \vec\eta)$ and $\Pi(\vec\theta, \vec\eta)$ can be normalized to form a probability distribution of the model parameters known as the posterior distribution.
The posterior distribution encodes model parameter information after being updated by the observed data~\cite{RevModPhys.83.943}, and is used to present many of the Bayesian results of this work.
This probabilistic interpretation allows one to determine the regions of parameter space with the largest probability of containing the parameter~\cite{laplace1820theorie}.
Integrating the posterior over the nuisance parameters, we obtain the marginal posterior
\begin{equation}
    \mathcal{P}(\vec\theta) = \frac{\int d\vec\eta~\like(\vec\theta,\vec\eta) \Pi(\vec\theta,\vec\eta)}{\int d\vec\theta d\vec\eta~\like(\vec\theta,\vec\eta) \Pi(\vec\theta,\vec\eta)}.
    \label{eq:likelihood_bayes}
\end{equation}
Practically, this is achieved using a Markov Chain Monte Carlo (MCMC) based on the \texttt{emcee} package~\cite{ForemanMackey:2012ig} to sample the posterior distribution, and examining the distribution of samples in $\vec\theta$~\cite{RevModPhys.83.943}.

In the case that the model parameter posterior is confined to a compact region, we report the highest-posterior-density (HPD) credible region of that parameter and its maximum {\it{}a posteriori} (MAP) estimation.
However, there are some cases where credible regions are ill-defined, or a natural choice of prior is not immediately apparent.
In these scenarios we report our results using the Bayes factor as a function of the model parameter~\cite{Trotta:2017wnx,Gariazzo:2019xhx}, {\it i.e.}\ the ratio of the evidence between the alternative physics model and the null hypothesis,
\begin{equation}
    	\mathcal{B}_{1 0} = \frac{\int d\vec\theta' d\vec\eta'~\mathcal{L}_\texttt{\tiny{1}}(\vec\theta',\vec\eta')\cdot\Pi_\texttt{\tiny{1}}(\vec\theta',\vec\eta')}{\int d\vec\theta d\vec\eta~\mathcal{L}_\texttt{\tiny{0}}(\vec\theta, \vec\eta)\cdot\Pi_\texttt{\tiny{0}}(\vec\theta,\vec\eta)},
    \label{eq:bayes_factor}
\end{equation}
where the evidence is the average value of the likelihood function with respect to the prior distribution over all model parameters.
To compute the model evidence we use the \texttt{MultiNest} package~\cite{Feroz:2013hea}.
Given a Bayes factor it is customary to assign a qualitative description.
For this we use Jeffreys' scale~\cite{jeffreys1998theory}.

In order to evaluate the likelihood function, it is necessary to compute the expected number of events in each observable bin given the model parameters.
This expectation is obtained through Monte Carlo simulation of the detector.
The IceCube Monte Carlo is computationally expensive at high energies, so much so that it is prohibitive to produce background MC such that the statistical fluctuations of the MC are much smaller than the data fluctuations for the atmospheric muon background.
To avoid making incorrect statements due to the large MC statistical uncertainty in some bins, the analyses described in \refsec{sec:diffuse} use a modified Poisson likelihood function, $\likeSAY$, that incorporates additional uncertainty from the limited MC sample size~\cite{Arguelles:2019izp}.
This treatment produces similar results to others available in the literature~\cite{Chirkin:2013lya,Glusenkamp:2017rlp,Glusenkamp:2019uir}, but provides improved coverage properties, is numerically more stable, and is computationally more efficient~\cite{Arguelles:2019izp}.
The likelihood for this analysis is given by
\begin{align}
\begin{split}
	    \like(\vec\theta, \vec\eta) ={}& \prod_j^{n} \likeSAY(\mu_j(\vec\theta,\vec\eta), \sigma_j(\vec\theta,\vec\eta); d_j), \label{eq:likelihood}
\end{split}
\end{align}
and the priors or constraints, depending on the context, are given by
\begin{align}
\begin{split}
	    \Pi(\vec\theta, \vec\eta)={}& \left(\prod_r \Pi_r(\theta_r)\right) \cdot \left(\prod_s \Pi_s(\eta_s) \right),\label{eq:priors}
\end{split}
\end{align}
where $j$ refers to the bin number, $r$ indexes the parameters of interest, and $s$ indexes the nuisance parameters.
The variables $\theta_r$ and $\eta_r$ denote the parameters of interest and nuisance parameters, respectively.
The arguments of the likelihood $\mu_j$ and $\sigma_j$ are the expected number of events and MC statistical uncertainty of that quantity, respectively, while $d_j$ is the number of observed data events in that bin.
The parameters $\vec\theta$ and $\vec\eta$ have priors or constraints which are represented in Eq.~\eqref{eq:priors} by $\Pi_r(\theta_r)$ and $\Pi_s(\eta_s)$, respectively, and are enumerated in \reftab{tbl:priors}.
For parameters with improper uniform priors in the Bayesian treatment, we apply no external constraint in the frequentist treatment; otherwise, the prior and constraint are the same.
This results in equivalent functional forms of the product $\like(\vec\theta, \vec\eta) \Pi(\vec\theta, \vec\eta)$ for the frequentist and Bayesian analyses.

\begin{table*}
\begin{center}
\begin{tabular}{l l l l l l}
\toprule
Morphology & Observable & Bin Edge Condition & Bin Edge & Binning Minimum & Binning Maximum \\
\midrule
\multirow{2}{*}{Cascades} & Energy & $\log_{10} ( E^{i+1}_\texttt{bin edge}/E^{i}_\texttt{bin edge}) = 0.111$ & $E^0_\texttt{bin edge} = \SI{60}\TeV$ & $E^\texttt{min}=\SI{60}\TeV$ & $E^\texttt{max} = \SI{10}\PeV$ \\
& Zenith & $\cos\theta^{i}_{z,\texttt{bin edge}} - \cos\theta^{i+1}_{z,\texttt{bin edge}} = 0.2$ & $\cos\theta_{z,\texttt{bin edge}}^0 = 0$ & $\cos\theta_z^\texttt{min} = -1$ & $\cos\theta_z^\texttt{max} = 1$ \\
\midrule
\multirow{2}{*}{Tracks} & Energy & $\log_{10} ( E^{i+1}_\texttt{bin edge}/E^{i}_\texttt{bin edge}) = 0.111$ & $E^0_\texttt{bin edge} = \SI{60}\TeV$ & $E^\texttt{min}=\SI{60}\TeV$ & $E^\texttt{max} = \SI{10}\PeV$ \\
& Zenith & $\cos\theta^{i}_{z,\texttt{bin edge}} - \cos\theta^{i+1}_{z,\texttt{bin edge}} = 0.2$ & $\cos\theta_{z,\texttt{bin edge}}^0 = 0$ & $\cos\theta_z^\texttt{min} = -1$ & $\cos\theta_z^\texttt{max} = 1$ \\
\midrule
\multirow{2}{*}{Double Cascades} & Energy & $\log_{10} ( E^{i+1}_\texttt{bin edge}/E^{i}_\texttt{bin edge}) = 0.111$ & $E^0_\texttt{bin edge} = \SI{60}\TeV$ & $E^\texttt{min}=\SI{60}\TeV$ & $E^\texttt{max} = \SI{10}\PeV$ \\
& Length & $\log_{10} (l^{i+1}_\texttt{bin edge}/l^{i}_\texttt{bin edge}) = 0.1$ & $l^0 = \SI{10}\meter$ & $l^\texttt{min}=\SI{10}\meter$ & $l^\texttt{max} = \SI{1000}\meter$ \\
\bottomrule
\end{tabular}
\end{center}
\caption{\textit{\textbf{Binning of observable quantities.}}
The conditions used to construct the bin edges in each observable for each morphology are presented in this table.
An initial bin edge is given, and other bin edges are defined using a relationship between bin edges.
The lowest and highest bins are truncated if their bin edges extend beyond the defined boundaries.
For each inferred morphology, the overall binning is defined by the Cartesian product of binning in two separate observables.
This gives  210, 210, and 420 bins for cascades, tracks, and double cascades, respectively, for a total of 840 bins.}
\label{tbl:binning}
\end{table*}

The bin widths of the analysis histogram are chosen to be comparable to the detector resolution.
There are $840$ bins in observable quantities used for the analysis.
Events are first separated by their inferred morphology, and then binned in two observables.
Tracks and cascades are binned in reconstructed energy and reconstructed zenith angle, whereas double cascades are binned in reconstructed energy and the reconstructed separation between cascades.
Details of how the bin edges are defined are given in \reftab{tbl:binning}.

\section{Characterization of the astrophysical neutrino flux\label{sec:diffuse}}
\noindent
\textit{The astrophysical component observed in HESE is well described by a single power law with a spectral index of $\SPLFreqWilksIndexSummary$.
Other generic parameterizations of the astrophysical flux are also considered, but none represent a significant improvement over the single power-law hypothesis.
In these generic models, the preferred regions of parameter space are degenerate with the single power-law model.
We also introduce a generic parameterization of the flux, comprised of a set of flux segments, and report the segment normalizations and uncertainties.
In this sample, we find no evidence of an atmospheric neutrino flux from the decay of charmed hadrons and find that prompt normalizations greater than $\sim 13$ times the baseline prompt model (BERSS~\cite{Bhattacharya:2015jpa}) are strongly disfavored; the obtained limit is weaker than existing constraints obtained from other samples.
Finally, we study proposed source models, testing whether they are preferred compared to a baseline scenario.
We find that no source model scenario is strongly favored with respect to the baseline single power-law model of the astrophysical neutrino flux.
}
\newline

IceCube has reported evidence of neutrino emission associated with a blazar~\cite{IceCube:2018dnn,IceCube:2018cha}, as well as a $2.9\sigma$ excess with respect to background from a starburst galaxy~\cite{Aartsen:2019fau}.
However, these specific associations represent approximately $\SI{1}\percent$ of the astrophysical neutrino flux above $\SI{200}\TeV$~\cite{IceCube:2018cha}, leaving the origin of the vast majority of the flux still unassociated with sources~\cite{Albert:2017ohr,Aartsen:2018ywr,Aartsen:2019epb}.
Therefore, we take here a two-pronged approach to characterize the astrophysical spectrum.
Section~\ref{sec:generic_models} considers generic forms for the spectrum, which could arise from numerous physical scenarios, while \refsec{sec:specific_models} tests a small sample of specific spectra from the literature.
Separate from the astrophysical analyses, \refsec{sec:prompt} explores the atmospheric flux of neutrinos from charmed hadrons.
We provide the data, Monte Carlo, and tools necessary for these tests in the data release outlined in \refappsec{sec:release} and encourage readers to perform their own tests of model compatibility~\cite{HESEdatarelease}.

In addition to the spectral models chosen, three key assumptions are made about the astrophysical neutrino flux in this section: the flux incident on Earth is isotropic, it is the same between neutrinos and anti-neutrinos, and the same for each neutrino flavor.
An isotropic flux is expected in models where the dominant contribution is from distant sources.
We focus on the isotropic flux hypothesis as it is compatible with the available neutrino data.
Finally, IceCube is insensitive to the differences between neutrino and anti-neutrino interactions on an event-by-event basis for most energies.
So differences between the neutrino and anti-neutrino flux content do not modify the expectation of detected events in this analysis for most energies.
The one region where this does not hold is for electron anti-neutrinos near $\SI{6.3}\PeV$, where their resonant interaction with atomic electrons, called the Glashow Resonance (GR)~\cite{Glashow:1960zz,Loewy:2014zva}, can occur.
This interaction enhances the expectation of detected down-going events near the resonance energy because of the larger interaction probability in the detection volume but reduces the expectation for up-going events because of increased absorption in the Earth~\cite{Barger:2014iua,Garcia:2020jwr}.
However, because the spectrum is steeply falling, as observed in previous analyses~\cite{Aartsen:2014gkd,Aartsen:2016xlq,Aartsen:2018vez,Stettner:2019tok,Aartsen:2020aqd}, the expected number of GR events is small in comparison to the rest of the sample; approximately $\SI{3.6}\percent$ above $\SI{60}\TeV$, assuming the best-fit spectrum from \refsec{sec:spl}.

As described in \refsec{sec:backgrounds}, backgrounds from cosmic-ray showers produced in the Earth's atmosphere are a small but non-negligible contribution to the events observed in the sample; quantitatively above $\SI{60}\TeV$, $\SI{1.5}\percent$ atmospheric muons and $\SI{16}\percent$ atmospheric neutrinos.
The background model used throughout this section includes atmospheric neutrinos from pions, kaons, and charmed hadrons, as well as atmospheric muons.
Table~\ref{tbl:priors} describes the parameters of the fit for the single power-law model and the priors (or constraints) associated with them.
In addition to the astrophysical model and backgrounds with their respective nuisance parameters, detector systematics are also included.
Only the astrophysical neutrino flux parameters differ between the models described in this section; the background models and detector systematics remain the same.

\subsection{Generic models\label{sec:generic_models}}
Many well-motivated models of the astrophysical neutrino flux come in the form of power laws.
This commonality stems from the possibility that astrophysical neutrinos and cosmic rays may share a common origin, and that we observe cosmic rays at Earth with a power-law spectrum.
To examine the possibility of a power-law-like flux, we study a few generic scenarios for the astrophysical flux in the following sections.

The first scenario, called the ``single power law'' (SPL), is an unbroken power law across all energies with a freely varied normalization and spectral index.
This is the simplest model as it only has two free parameters for the astrophysical spectrum: spectral index and normalization.
It is also motivated by Fermi acceleration, which predicts a power-law energy spectrum~\cite{gaisser2016cosmic,Fermi:1949ee}.

The second scenario, called the ``double power law'' (DPL) is the sum of two unbroken power-law spectra, both with freely varying normalizations and spectral indices.
This potentially describes scenarios in which there are two populations of sources, two production mechanisms for high-energy astrophysical neutrinos that produce different power-law fluxes, or where neutrinos and anti-neutrinos have different spectra~\cite{Nunokawa:2016pop,Shoemaker:2015qul}.

A variety of source production models predict a high-energy cutoff in the neutrino spectrum, whether from limitations of the source energetics, a drop in pion production efficiency, energy losses of secondary pions and muons, or other mechanisms.
To accommodate this possibility in a functionally simple way, we define a model with a single power-law astrophysical flux that has an exponential suppression at high energies.
This third scenario, called the ``exponential cutoff,'' has an additional parameter describing the cutoff's energy scale.

The fourth scenario, called the ``log parabola'' (LP), is a simple extension to the power law that adds a changing spectral index.
This model is often used to describe gamma-ray spectra across many orders of magnitude of gamma-ray energy, which could otherwise be described as power laws in smaller energy ranges~\cite{TheFermi-LAT:2017pvy}.

\subsubsection{Single power-law flux\label{sec:spl}}
\noindent
\textit{
Our main finding is that a single power law, with a spectral index of $\SPLFreqWilksIndexSummary$, is a good description of the observed data.
This result has been robust under the improved systematic treatment of this analysis and is softer than previously reported results predominantly due to a relative excess of low-energy events in the last $\SI{4.5}\year$ of data.
We have made an extensive study of this relative excess and found no hardware, software, or calibration changes that could explain this shift.
The introduction of an additional prompt neutrino flux is only weakly correlated with the measured astrophysical spectral index, and it primarily affects the uncertainty of the astrophysical normalization measurement.
}
\newline

\begin{figure*}
    \centering
    \includegraphics[width=0.45\linewidth]{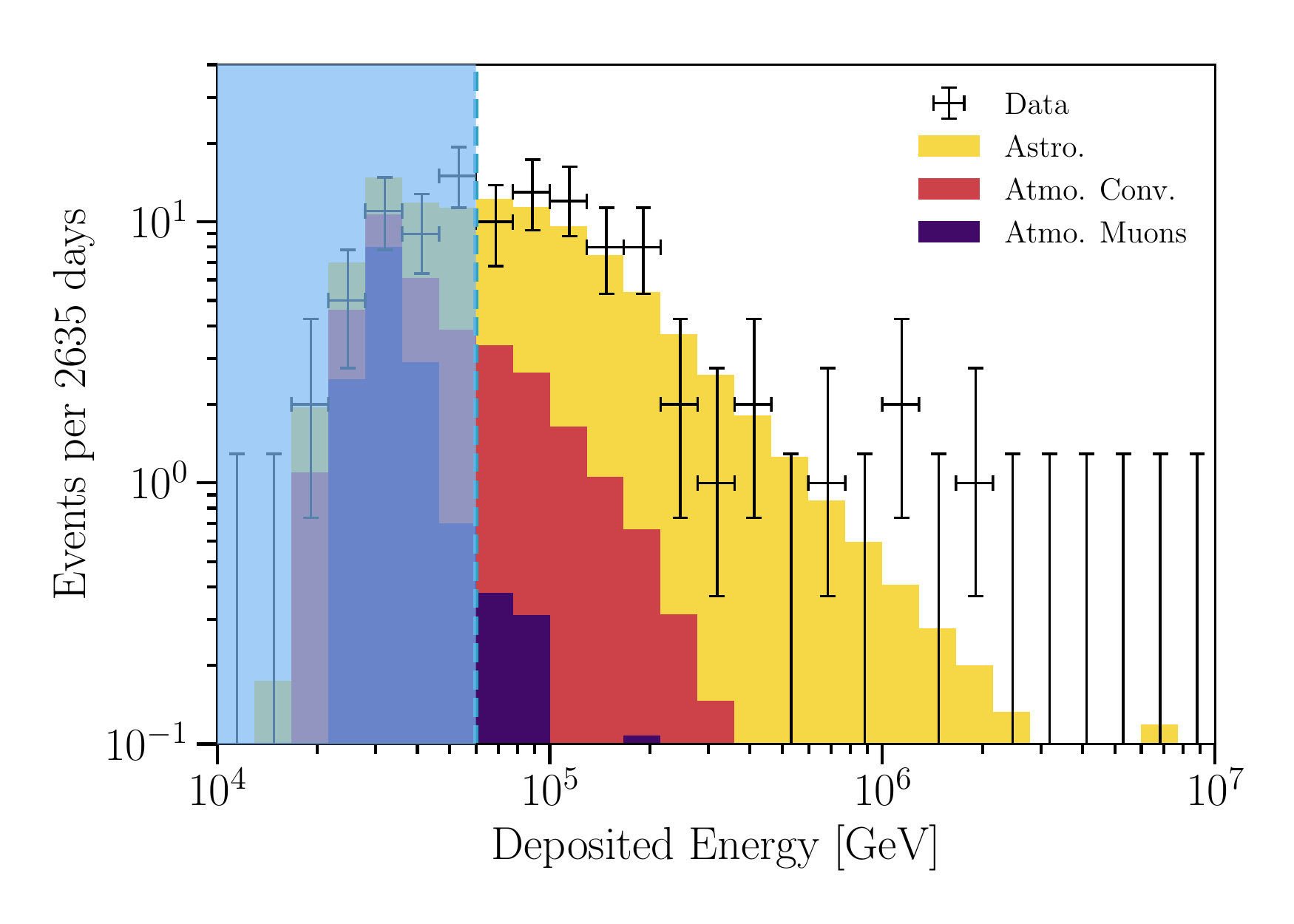}
    \includegraphics[width=0.45\linewidth]{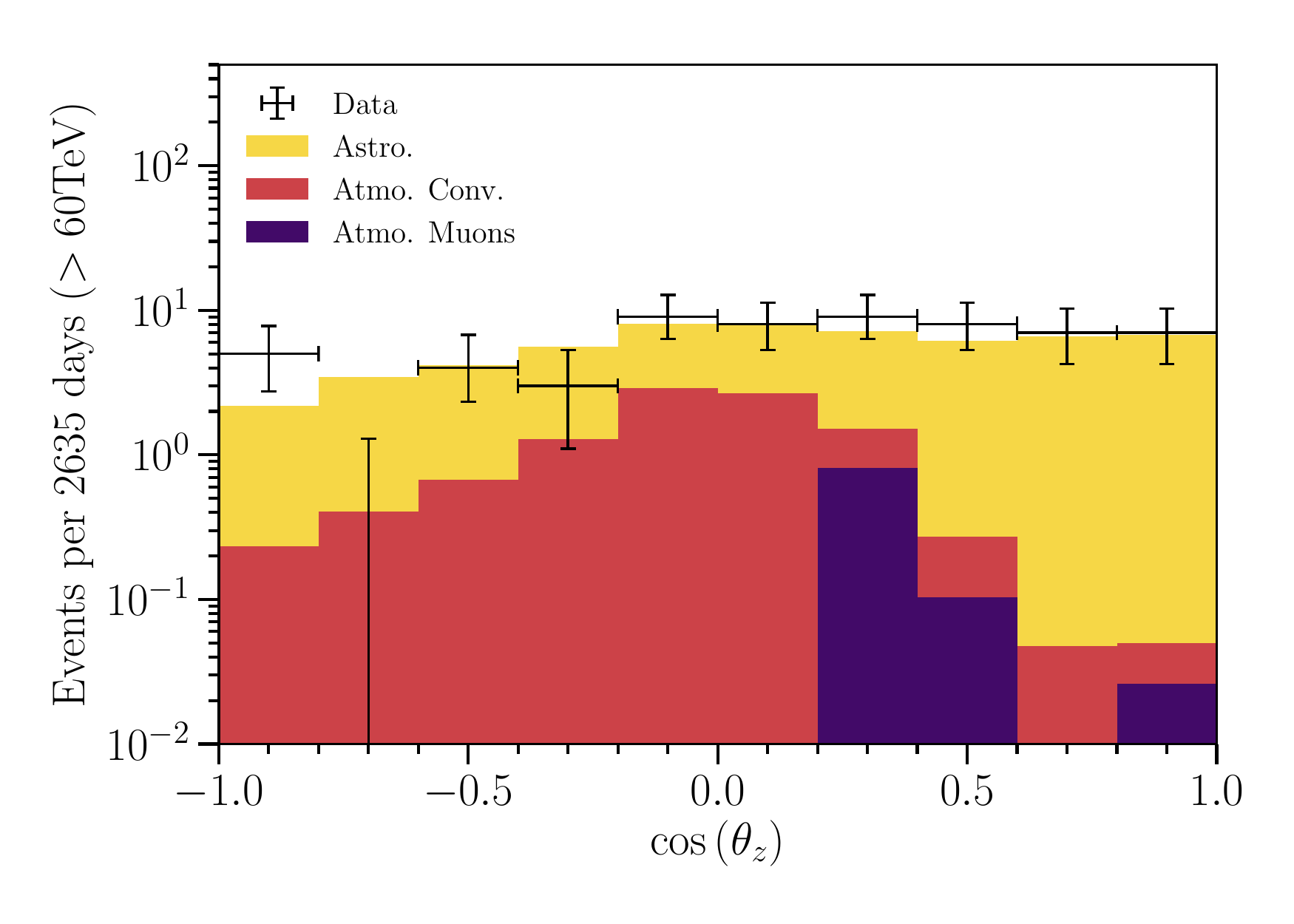}
    \caption{\textbf{\textit{Deposited energy and reconstructed $\cos\theta_z$ distributions.}}
    In these panels, the data is shown as crosses and the best-fit expectation as a stacked histogram with each color specifying a given flux component: astrophysical neutrinos (golden), conventional atmospheric neutrinos (red), and penetrating atmospheric muons (purple).
    Left: distributions of events and expected event count assuming best-fit parameters as a function of the deposited energy; events below $\SI{60}\TeV$ (light blue vertical line) are ignored in the fit.
    Right: distribution of events with energy greater than $\SI{60}\TeV$ in the cosine of their reconstructed zenith angle.
    Up-going events are on the left side of this panel and down-going events on the right.
    The expected number of events is split by components and displayed as a stacked histogram.
    The normalization of the prompt atmospheric neutrino component fits to zero, and so is not shown in the stacked histogram.
    The distribution of data events appears to be largely flat as a function of cosine zenith with a small decline towards the up-going region.
    The lower event rate in the up-going region is expected as a result of the Earth's absorption of the neutrino flux, and appears to be compatible with the Monte Carlo expectation.}\label{fig:energy-zenith}
\end{figure*}

For the single power-law-flux scenario, an isotropic flux of astrophysical neutrinos is assumed incident on the Earth with a total differential all-flavor neutrino-plus-anti-neutrino spectrum given by
\begin{gather}
\begin{split}
    \frac{d\Phi_{6\nu}}{dE} =&{}~\Phi_\texttt{astro}{\left(\frac{E_\nu}{\SI{100}\TeV}\right)}^{-\gamma_\texttt{astro}} \\
    & \cdot 10^{-18}~\textmd{GeV}^{-1}\textmd{cm}^{-2}\textmd{s}^{-1}\textmd{sr}^{-1},
\end{split}
\label{eq:spl_flux}
\end{gather}
where $\Phi_{6\nu}$ is the flux of the six neutrino species combined, $\astronorm$ is the normalization, and $\astrodeltagamma$ is the common spectral index.
These two parameters are incorporated as arguments of the likelihood, according to \refsec{sec:statistics}.
To better understand the relationship between the data and the neutrino flux contributions, we first look at projections in the two observables most different between neutrino fluxes (zenith, and energy) and compare data to the expectation from Monte Carlo assuming the nuisance parameters from the best-fit $(\hat{\vec\theta},\hat{\vec\eta})=\argmax_{\vec\theta, \vec\eta} \like(\vec\theta,\vec\eta)\cdot\Pi(\vec\theta,\vec\eta)$.
The right panel of \reffig{fig:energy-zenith} shows the data and expected number of events in bins of the cosine of the reconstructed zenith angle.
In the down-going region, the data are well described with the addition of an isotropic astrophysical neutrino flux.
Atmospheric components alone cannot describe the data well because the atmospheric neutrino components are suppressed in the down-going region by accompanying muons; see \refsec{sec:prompt} for details.
The null hypothesis (namely the atmospheric only scenario) is rejected with respect to the alternative hypothesis, including an astrophysical component at greater than $5\sigma$ with this sample.
The left panel of \reffig{fig:energy-zenith} shows the data and expected number of events in bins of reconstructed deposited energy.
The region below $\SI{60}\TeV$ is not included in the analysis because of larger background uncertainties.
However, we present the data to MC comparison in this region to demonstrate the level of agreement below the cut.
From the stacked histogram it is clear that the sample is dominated by the astrophysical component above $\SI{60}\TeV$.

\begin{figure}
    \centering
    \includegraphics[width=\linewidth]{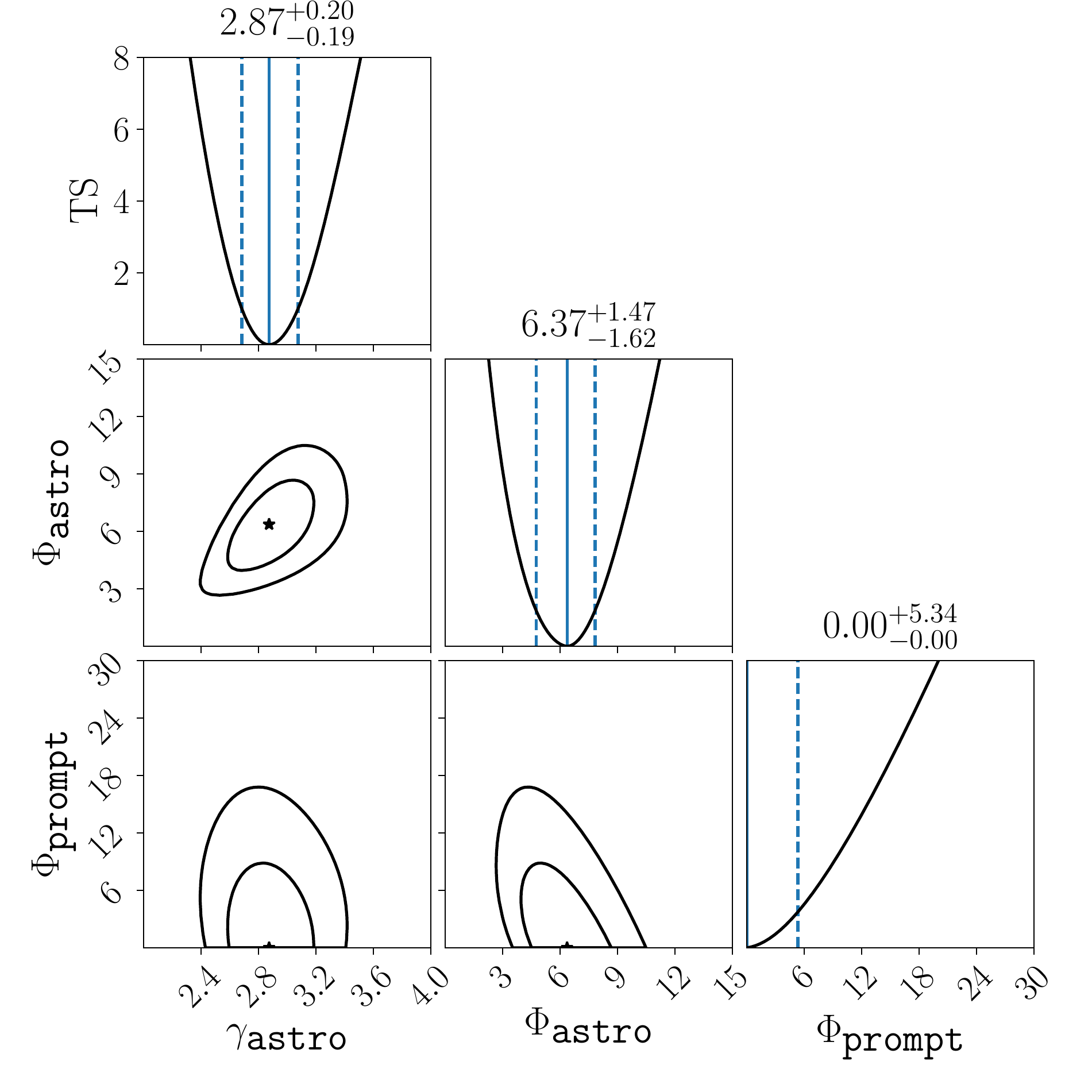}
    \caption{\textbf{\textit{Single power-law profile likelihood.}}
    Diagonal panels show the $\TS$, as a function of different model parameters, and the one sigma intervals assuming Wilks' theorem.
    Other panels show the best-fit point and two-dimensional contours.
    Solid (dashed) contours represent the $\SigmaOne$ ($\SigmaTwo$) confidence regions assuming Wilks' theorem.
    The parameter $\astrodeltagamma$ is the single power-law spectral index, $\astronorm$ is a scaling factor of the astrophysical flux at $\SI{100}\TeV$, and $\promptnorm$ is a scaling factor of the BERSS prompt neutrino flux calculation~\cite{Bhattacharya:2015jpa}; further descriptions of these parameters are provided in \refsec{tbl:priors}, Eq.~\eqref{eq:spl_flux}, and Eq.~\eqref{eq:atm_flux_equation}}\label{fig:SPL_profile}
\end{figure}

\begin{figure}
    \centering
    \includegraphics[width=\linewidth]{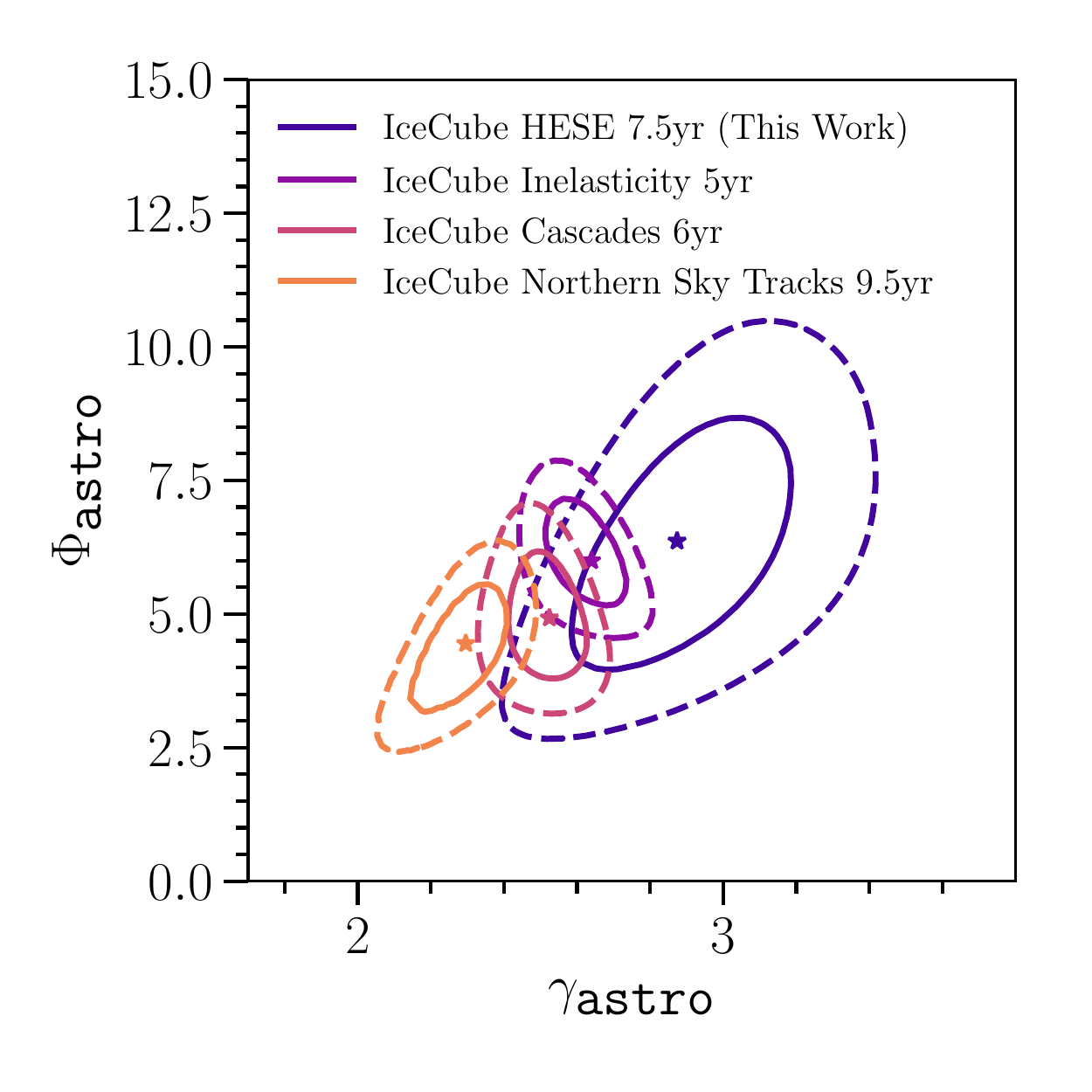}
    \caption{\textbf{\textit{Comparison of single power-law parameters from different analyses.}} Assuming an unbroken single power-law model for the astrophysical neutrino flux, results from different IceCube samples are shown.
    The horizontal axis is the spectral index of the model and the vertical axis is six-neutrino flux normalization at $\SI{100}\TeV$ given as a dimensionless multiplicative factor relative to $10^{-18}~\textmd{GeV}^{-1} \textmd{sr}^{-1} \textmd{s}^{-1} \textmd{cm}^{-2}$.
    The stars denote the different best-fit points, solid contours show the $\SigmaOne$ confidence region using the asymptotic approximation given by Wilks' theorem, and dashed contours show the $\SigmaTwo$ confidence regions.
    Blue represents results from this work, while the purple shows results from IceCube's 5yr inelasticity measurement~\cite{Aartsen:2018vez}, salmon shows results from IceCube’s 6yr cascade sample~\cite{Aartsen:2020aqd}, and orange shows IceCube’s 9.5yr Northern track sample preliminary result~\cite{Stettner:2019tok}.
    The differing preferred regions of parameter space for the astrophysical flux between the samples suggest a level of discrepancy, however a small region of parameter space is compatible with all samples at the $\SigmaTwo$ level.
    Many checks have been performed for possible explanations of the discrepancy without definitive conclusions.}\label{fig:SPL_frequentist}
\end{figure}

\begin{figure}
    \centering
    \includegraphics[width=\linewidth]{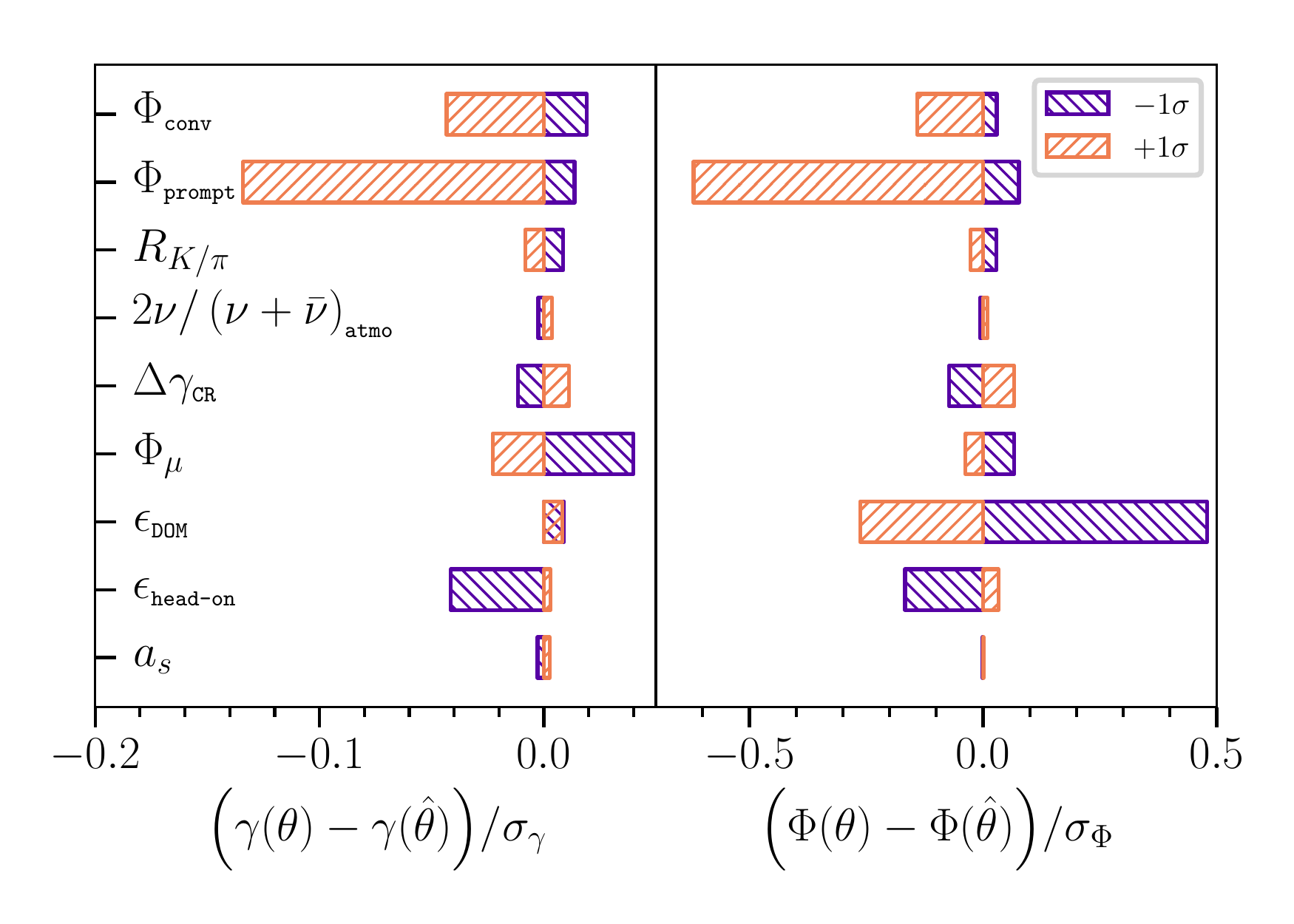}
    \caption{\textbf{\textit{Impact of systematic uncertainties on the single power-law parameters.}}
    Each panel shows the impact of the systematic on the astrophysical spectral index (left panel) and normalization (right panel).
    The impact, horizontal axis, is defined as the change in the parameter of interest relative to its uncertainty when modifying one systematic nuisance parameter.
    Orange bars indicate the effect of increasing the value of the nuisance parameter from its maximum {\it{}a posteriori} (MAP) value by $1\sigma$ as defined by the nuisance parameter's $\SigmaOne$ highest posterior density region, while purple bars indicate the corresponding reduction of the parameter.
    Systematic parameter values are given in \reftab{tbl:spl_parameters}.
    The prompt normalization ($\promptnorm$) and DOM efficiency ($\domeff$) have the largest effect on the astrophysical parameters.
    All other systematics pull the astrophysical parameters by significantly less than $0.5\sigma$.}
    \label{fig:SPL_impacts}
\end{figure}

The frequentist analysis of the single power law gives a best-fit point across all parameters, one-dimensional confidence intervals of each parameter, and two-dimensional confidence regions for the astrophysical normalization and spectral index.
The one-dimensional results are summarized in \reftab{tbl:spl_parameters}, and are obtained by assuming that the $\TS$ is $\chi^2$ distributed with one degree of freedom.
We obtain a best-fit spectral index of $\astrodeltagamma=\SPLFreqWilksIndexSummary$.
\reffig{fig:SPL_profile} shows the one-dimensional $\TS$ for $\astrodeltagamma$, $\astronorm$, and $\promptnorm$ on the diagonal panels as well as the bounds of the one-dimensional $\SigmaOne$ confidence regions plotted as vertical lines.
\reffig{fig:SPL_frequentist} and the non-diagonal panels of \reffig{fig:SPL_profile} show the $\SigmaOne$ and $\SigmaTwo$ confidence regions for the two variables on the horizontal and vertical axes assuming two degrees of freedom.
The impact of the systematics on the parameters of this model are shown in \reffig{fig:SPL_impacts}.
The most relevant systematic affecting the astrophysical normalization is the DOM efficiency and the relative contribution of neutrinos from charmed hadrons.
The astrophysical spectral index is more weakly affected by these systematics, but the normalization of the neutrino flux from charmed hadrons has the largest effect.

Our results agree with a previous iteration of this analysis~\cite{Aartsen:2014gkd} within the $2\sigma$ confidence regions of the astrophysical power-law parameters.
The previous analysis obtained a best-fit spectral index of $\astrodeltagamma^{\SI{3}\year}={2.3}^{+0.3}_{-0.3}$, compared to $\astrodeltagamma^{\SI{7.5}\year}=\SPLFreqWilksIndexSummary$ in this analysis.
This difference is primarily driven by a higher number of low-energy events observed in the latter $\SI{4.5}\year$ compared to the first $\SI{3}\year$.
A smaller contribution comes from the extension of the analysis energy range from $\SI{3}\PeV$ to $\SI{10}\PeV$, shifting the spectral index to a softer flux by $\sim 0.1$.
Further extension of the analysis energy range produces negligible changes.

To investigate the shift in spectral index between analysis iterations, an {\it{}a posteriori} analysis of the data's time dependence was performed.
Specifically, we compared a null hypothesis of a constant flux to a time-dependent spectrum with different astrophysical spectra for each of the two data partitions (first $\SI{3}\year$ and latter $\SI{4.5}\year$), where each spectrum is modeled as a single power law.
We performed a likelihood ratio based model comparison test, which disfavors the null hypothesis with a p-value of $\sim0.13$.
We conclude that there is no evidence for time dependence in this data sample.

Additionally, we tested the effect of different systematics on the fit.
We found that the inclusion or exclusion of any individual systematic or tested combination of systematics did not appreciably affect the fit result or uncertainties.

Other crosschecks were performed with the sample: comparing the spectrum of tracks and cascades, looking for differences between the up-going and down-going spectra, examining the summer and winter spectra, comparing the spectra from events in different regions of the detector, checking the charge distributions of events across many categorizations, looking for differences between charge calibrations, and checking for pulls resulting from reconstruction and simulation changes.
None of these checks showed any statistically significant differences.

\begin{figure}
    \centering
    \includegraphics[width=\linewidth]{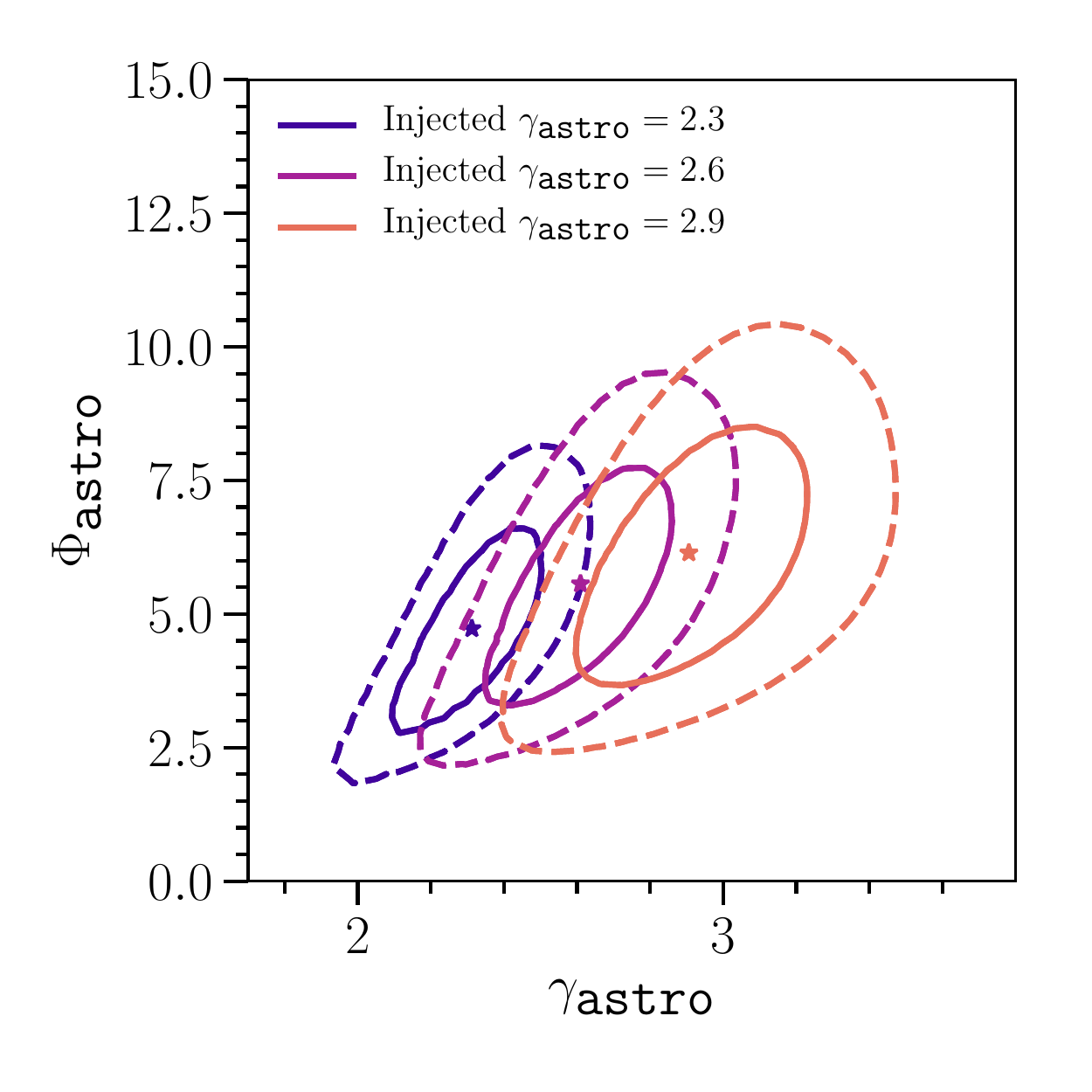}
    \caption{\textbf{\textit{Astrophysical parameter sensitivity to injected spectra.}}
    The three colors from blue to salmon show the expected astrophysical parameter $\SigmaOne$ (solid lines) and $\SigmaTwo$ (dashed lines) uncertainty for three injected astrophysical spectra $\astrodeltagamma\in\{2.3, 2.6, 2.9\}$.
    In each of these cases the expected number of events for $\SI{7.5}\year$ of livetime is injected as data using nominal values for nuisance parameters and holding the number of injected astrophysical events to be equal between the spectra.
    As expected, the spectral index uncertainty grows with softer spectra from $\sim0.14$ to $\sim0.21$ when changing $\astrodeltagamma$ from $2.3$ to $2.9$.
    }
    \label{fig:SPL_asimov_sensitivity}
\end{figure}

Although the uncertainty on $\astrodeltagamma$ is numerically similar between this analysis and the $\SI{3}\year$ analysis, this is not the result of any additional systematic uncertainty or analysis change.
This is a direct result of the change in the best-fit spectral index.
With the same amount of data, harder spectra can be measured with less uncertainty than softer spectra.
This effect is shown in \reffig{fig:SPL_asimov_sensitivity}, where we plot the uncertainty for different injected spectra ($\astrodeltagamma=\{2.3, 2.6, 2.9\}$) that have the same number of expected events in the sample.

Plotted in \reffig{fig:SPL_frequentist} are the confidence regions for other IceCube analyses.
The orange contours show the results of a single power-law fit to IceCube's up-going muon neutrino data sample~\cite{Stettner:2019tok}, the salmon contours show results from IceCube's 6yr cascade sample~\cite{Aartsen:2020aqd,hansthesis}, the purple contours show results from IceCube's 5yr inelasticity measurement~\cite{Aartsen:2018vez}, and the blue contour show results from this work.
Assuming a continuous single power law across all energies, the large values of $\astrodeltagamma$ in the preferred regions of this analysis are disfavored by the through-going muon and cascade sample results.
While these differences may be statistical, other explanations have been explored.
A thorough examination of possible detector systematics and physics systematics has not revealed a systematic cause for the differences in single power-law best-fit parameters between samples.
However, these samples cover different energies, flavors, regions of the sky, and are susceptible to different systematics and physical effects.
Differences due to these factors could help to explain the different spectral measurements and have been tested for within the samples, although presently, we have not found evidence of a primary cause.
Tests performed with the cascade sample reveal a preference for spectral softening in the tens to hundreds of $\si\TeV$ energy range~\cite{Aartsen:2020aqd}.
The flux inferred for the overlapping energy range is well consistent with the results reported here.
We briefly describe the samples for the sake of comparison.

The up-going muon neutrino sample~\cite{Stettner:2019tok}, collected over $\SI{9.5}\year$, consists of well-reconstructed muon tracks with zenith angle $\theta_z \geq \SI{85}\degree$ that also pass a boosted decision-tree based cut designed to select for through-going muon neutrino events while removing down-going muon and cascade backgrounds~\cite{Aartsen:2016xlq}.
This sample, which has negligible overlap with the sample presented in this work, contains muons of energy between $\sim\SI{100}\GeV$ and $\sim\SI{10}\PeV$, with the energy distribution peaked at $\sim\SI{1}\TeV$.
Atmospheric neutrinos dominate the sample, comprising $>\SI{99}\percent$ of events in it.
The signal of astrophysical events is only apparent at the sample's high-energy range, where the atmospheric spectrum falls below the astrophysical component.
At $\sim\SI{20}\TeV$ in reconstructed muon energy the astrophysical component is $\sim1/10\textmd{th}$ the atmospheric component.
The components are equal in flux at $\sim\SI{200}\TeV$, and the atmospheric component is $\sim1/10\textmd{th}$ of the astrophysical component at $\sim\SI{1}\PeV$.
Events in this sample with neutrino energy between $\SI{40}\TeV$ and $\SI{3.5}\PeV$ contribute $\SI{90}\percent$ of the total observed likelihood ratio between the best-fit and the atmospheric-only hypothesis.
As a function of the zenith angle, the signal-to-background ratio is lower at the horizon than for up-going events by almost an order of magnitude because of the enhanced atmospheric neutrino production at the horizon in the sensitive energy range.
This sample benefits from better control of atmospheric flux systematic parameters due to the large population of atmospheric neutrinos.
Additionally, this sample is substantially less affected by uncertainties related to muons from cosmic-ray air-showers than others because of the cuts on the reconstructed zenith angle.

The cascade neutrino sample, collected over six years, consists of cascade-like events from all directions in the sky that have neutrino energies between $\sim\SI{1}\TeV$ and $\sim\SI{10}\PeV$~\cite{Aartsen:2020aqd,hansthesis}.
Above $\SI{60}\TeV$ about $\SI{60}\percent$ of events in this sample are not contained in the HESE sample.
As the sample selects for cascade-like events, it predominantly contains electron and tau neutrinos, but also contains neutral current events from all neutrino flavors and a fraction of misidentified muon neutrinos.
The sample has a sensitive energy range from $\SI{16}\TeV$ to $\SI{2.6}\PeV$, which is defined as the smallest neutrino energy range for which a non-zero astrophysical flux is consistent with data at the $\SI{90}\percent$ confidence level.
The distribution of the signal-to-background ratio for this sample has additional features compared to the same distribution for the up-going muon neutrino sample.
These features are partly due to the rejection of atmospheric neutrino events by accompanying muons, which depends both on the neutrino energy and zenith angle.
The signal-to-background ratio ranges from 1:100 at $\si\TeV$ energies to 1:1 at $\sim\SI{20}\TeV$ to 1000:1 at $\si\PeV$ energies.
The sample is least pure near the horizon with a factor of 10 to 100 less signal per background compared to the up-going and down-going regions.
Like the up-going muon neutrino sample, this sample also benefits from better control of atmospheric flux systematic parameters due to the large population of atmospheric neutrinos, although to a lesser extent.

The sample used for the inelasticity measurement, collected over five years, consists of track and cascade events with their interaction vertex contained within the detector~\cite{Aartsen:2018vez}.
The sample is optimized to facilitate the measurement of the neutrino interaction inelasticity distribution, using both a veto and a boosted decision tree to select neutrino events while removing atmospheric muons.
The sample is sensitive in the $\SI{1}\TeV$ to $\SI{1}\PeV$ energy range with the bulk of events below $\SI{10}\TeV$.
Signal-to-background ratios of 10:1 are achieved for tracks close to $\SI{1}\PeV$ and cascades above $\SI{100}\TeV$.
Up-going track events in this sample are a factor of 10 to 1000 purer than down-going track events and a factor of 10 to 100 purer for cascades.

In contrast to these samples, the HESE selection which is the focus of this work has a similar effective area for all neutrino flavors and a signal-to-background profile with features closer to the cascade sample.
Events in this sample with neutrino energy between $\SI{69.4}\TeV$ and $\SI{1.9}\PeV$ contribute $\SI{90}\percent$ of the total observed likelihood ratio between the best-fit and the atmospheric-only hypothesis.
This signal-to-noise ratio in this sample is comparable to that of the cascade sample above $\SI{60}\TeV$ and follows a similar dependence on zenith and energy.
Above $\SI{60}\TeV$ deposited energy, the sample has 60 events with a signal-to-background ratio greater than 1:10, 59 events with a signal-to-background ratio greater than 1:1, 24 events with a signal-to-background ratio greater than 10:1, and one event with a signal-to-background ratio greater than 10000:1.
This variation in signal-to-background ratio stems from the different spectra of the fluxes and the veto's varying rejection power with respect to the zenith angle.

The through-going muon neutrino sample is $\sim\SI{50}\percent$ more sensitive to the energy spectrum under the single power-law assumption when one accounts for the parameter-space differences between the best-fit spectral indices as demonstrated in \reffig{fig:SPL_asimov_sensitivity} and the difference in sample livetime.
The HESE sample suffers from a small sample size but benefits from high astrophysical purity, while the through-going muon neutrino sample benefits from a large sample size but suffers from lower purity and worse energy resolution of tracks.
The cascade and inelasticity selections have comparable spectral sensitivity to each other and are more sensitive than the other two samples.
Both samples benefit from large sample sizes and compared to the through-going muon sample, the cascade sample and inelasticity sample benefit from the better energy resolution of cascades and starting tracks, respectively.
The cascade sample benefits from lower atmospheric neutrino contamination at high energies and improved $\nu_e$ effective area with respect to the HESE selection.

\begin{figure}
    \centering
    \includegraphics[width=\linewidth]{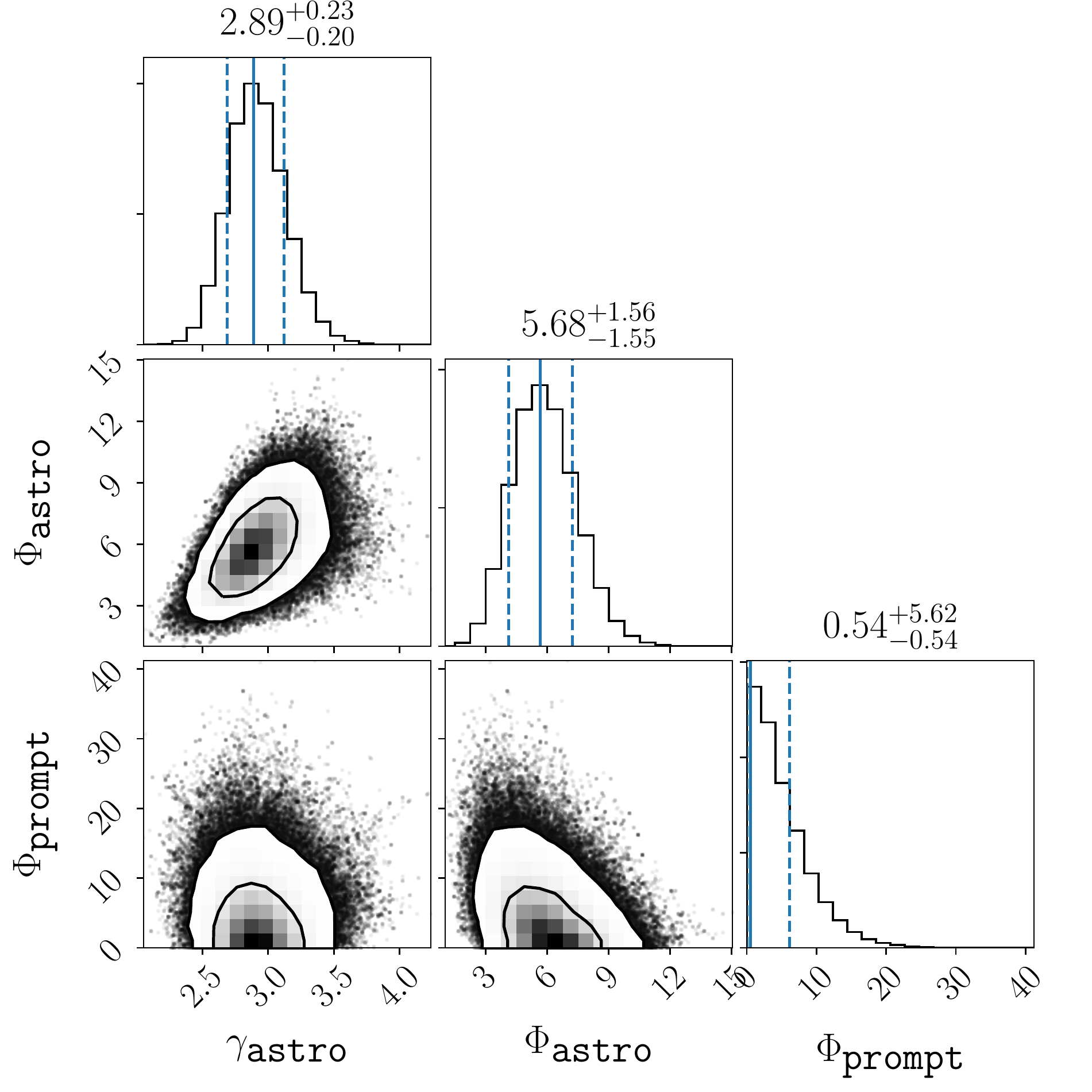}
    \caption{\textbf{\textit{Single power-law parameters posteriors.}}
    Diagonal panels show the one-dimensional posterior distribution of the parameters (joint distribution integrated over all other parameters), where the horizontal axis of the panel is the same as the horizontal axis at the bottom of the column, and the vertical axis of the panel is the probability density in an arbitrary scale.
    The solid blue lines denote the MAP estimator for each parameter, and the dashed lines denote the bounds of the one-dimensional $\SigmaOne$ HPD region; these numbers are also listed above the diagonal panels.
    Non-diagonal panels show the two-dimensional posterior distribution of the parameters, where the horizontal and vertical axes of the panel correspond to the horizontal axis at the bottom of the column and the vertical axis at the far left of the row, respectively.
    The innermost contours show the two-dimensional $\SigmaOne$ HPD region, and the outermost contours show the $\SigmaTwo$ HPD region.
    The grayscale of the histogram within the contours shows the probability density in an arbitrary scale.
    The points outside the contours show individual points from the MCMC.}\label{fig:SPL_posterior}
\end{figure}

A Bayesian analysis was also performed assuming a single power-law model for the astrophysical spectrum.
The marginal posterior distribution is used for statistical inference, as described in \refsec{sec:statistics}.
Table~\ref{tbl:spl_parameters} reports the maximum {\it{}a posteriori} (MAP) estimation of each parameter, as well as the $\SigmaOne$ highest posterior density (HPD) region.
Changing the prior of the SPL model parameters from linear-uniform to log-uniform has a much smaller effect than the reported errors of these parameters, implying the MAP is dominated by the data rather than the priors of the SPL parameters.
The Bayesian analysis finds a most likely spectral index of $\astrodeltagamma=\SPLBayesIndexSummary$, which is very similar to the frequentist estimation of these parameters.
Figure~\ref{fig:SPL_posterior} shows one-dimensional projections of the posterior distribution on the diagonal, integrated over the other variables, with the $\SigmaOne$ HPD region bounds plotted as vertical lines.
The non-diagonal panels of \reffig{fig:SPL_posterior} show contours of the $\SigmaOne$ and $\SigmaTwo$ HPD regions of the two-dimensional posterior distribution projection.
A histogram of the probability is displayed within the contours.
Outside the contours, individual samples from the MCMC are plotted.
One can see the correlation between $\astrodeltagamma$ and $\astronorm$; this correlation arises because the overall normalization of events must be roughly preserved to match the data well.
Additionally, the inverse correlation of $\astronorm$ and $\promptnorm$ is apparent in this figure, which also arises from a conserved total number of events.
Both the Bayesian and frequentist analyses show $\promptnorm$ to be compatible with zero.
The parameter estimations and corresponding errors for the conventional atmospheric normalization are in good agreement with measurements of the atmospheric neutrino flux~\cite{Aartsen:2012uu}.

\begin{table*}[thb]
    \centering
    \begin{tabular}{l rr rr}
        \toprule
        & \multicolumn{2}{c}{Frenquentist Analysis} & \multicolumn{2}{c}{Bayesian Analysis} \\
        \midrule
        Parameter & Best-fit value & $\SigmaOne$ C.L. & Most-likely value & $\SigmaOne$ H.P.D. \\
        \midrule
        \multicolumn{1}{l }{\textbf{Astrophysical neutrino flux:}} & & & &\\
        $\astronorm$ & $\SPLFreqBFNorm$ & $[\SPLFreqWilksLowerNorm,\SPLFreqWilksUpperNorm]$ & $\SPLBayesMAPNorm$ & $[\SPLBayesHPDLowerNorm,\SPLBayesHPDUpperNorm]$\\
        $\astrodeltagamma$ & $\SPLFreqBFIndex$ & $[\SPLFreqWilksLowerIndex,\SPLFreqWilksUpperIndex]$ & $\SPLBayesMAPIndex$ & $[\SPLBayesHPDLowerIndex,\SPLBayesHPDUpperIndex]$\\
        \midrule
        \multicolumn{1}{l }{\textbf{Atmospheric neutrino flux:}} & & & &\\
        $\convnorm$ & $\SPLFreqBFConvNorm$ & $[\SPLFreqWilksLowerConvNorm,\SPLFreqWilksUpperConvNorm]$ & $\SPLBayesMAPConvNorm$ & $[\SPLBayesHPDLowerConvNorm,\SPLBayesHPDUpperConvNorm]$\\
        $\promptnorm$ & $\SPLFreqBFPromptNorm$ & $[\SPLFreqWilksLowerPromptNorm,\SPLFreqWilksUpperPromptNorm]$ & $\SPLBayesMAPPromptNorm$ & $[\SPLBayesHPDLowerPromptNorm,\SPLBayesHPDUpperPromptNorm]$\\
        $\pik$ & $\SPLFreqBFKPi$ & $[\SPLFreqWilksLowerKPi,\SPLFreqWilksUpperKPi]$ & $\SPLBayesMAPKPi$ & $[\SPLBayesHPDLowerKPi,\SPLBayesHPDUpperKPi]$\\
        $\atmonunubar$ & $\SPLFreqBFAtmoNuRatio$ & $[\SPLFreqWilksLowerAtmoNuRatio,\SPLFreqWilksUpperAtmoNuRatio]$ & $\SPLBayesMAPAtmoNuRatio$ & $[\SPLBayesHPDLowerAtmoNuRatio,\SPLBayesHPDUpperAtmoNuRatio]$\\
        \midrule
        \multicolumn{1}{l }{\textbf{Cosmic ray flux:}} & & & & \\
        $\crdeltagamma$ & $\SPLFreqBFCRDeltaGamma$ & $[\SPLFreqWilksLowerCRDeltaGamma,\SPLFreqWilksUpperCRDeltaGamma]$ & $\SPLBayesMAPCRDeltaGamma$ & $[\SPLBayesHPDLowerCRDeltaGamma,\SPLBayesHPDUpperCRDeltaGamma]$\\
        $\muonnorm$ & $\SPLFreqBFMuonNorm$ & $[\SPLFreqWilksLowerMuonNorm,\SPLFreqWilksUpperMuonNorm]$ & $\SPLBayesMAPMuonNorm$ & $[\SPLBayesHPDLowerMuonNorm,\SPLBayesHPDUpperMuonNorm]$\\
        \midrule
        \multicolumn{1}{l }{\textbf{Detector:}} & & & &\\
        $\domeff$ & $\SPLFreqBFDOMEff$ & $[\SPLFreqWilksLowerDOMEff,\SPLFreqWilksUpperDOMEff]$ & $\SPLBayesMAPDOMEff$ & $[\SPLBayesHPDLowerDOMEff,\SPLBayesHPDUpperDOMEff]$\\
        $\holeice$ & $\SPLFreqBFHoleIce$ & $[\SPLFreqWilksLowerHoleIce,\SPLFreqWilksUpperHoleIce]$ & $\SPLBayesMAPHoleIce$ & $[\SPLBayesHPDLowerHoleIce,\SPLBayesHPDUpperHoleIce]$\\
        $a_s$ & $\SPLFreqBFAniScale$ & $[\SPLFreqWilksLowerAniScale,\SPLFreqWilksUpperAniScale]$ & $\SPLBayesMAPAniScale$ & $[\SPLBayesHPDLowerAniScale,\SPLBayesHPDUpperAniScale]$\\
        \bottomrule
    \end{tabular}
    \caption{\textbf{\textit{Single power-law model parameters.}} The frequentist analysis column shows the best-fit parameters and their corresponding $\SigmaOne$ C.L. interval according to Wilks' theorem for the single power-law model.
    The Bayesian analysis column shows the most-likely values of the parameters as well as the $\SigmaOne$ highest probability density (HPD) interval.
    Parameter name descriptions and priors(constraints) are given in \reftab{tbl:priors}.}
    \label{tbl:spl_parameters}
\end{table*}

\subsubsection{Double power-law flux\label{sec:dpl}}
\noindent
\textit{
This model can be parameterized as hard and soft power-law components.
We find that the preferred value of the hard component spectral index is close to that of the single power-law result, \textit{i.e.} $\gamma_\texttt{hard} \sim 2.8$ and $\gamma_\texttt{astro} \sim 2.9$, while the soft component spectral index most-likely value is $\gamma_\texttt{soft} \sim 3.1$.
The latter is poorly constrained, and values as large as $\gamma_\texttt{soft}=3.4$ are contained within the $\SigmaOne$ highest probability density region.
The two components' normalizations are highly correlated, with either equal to zero allowed within the two-dimensional $\SigmaOne$ highest probability density region.
This observation is aligned with our conclusion that there is no indication of an additional power-law component in this sample.
}
\newline

\begin{figure}
    \centering
    \includegraphics[width=\linewidth]{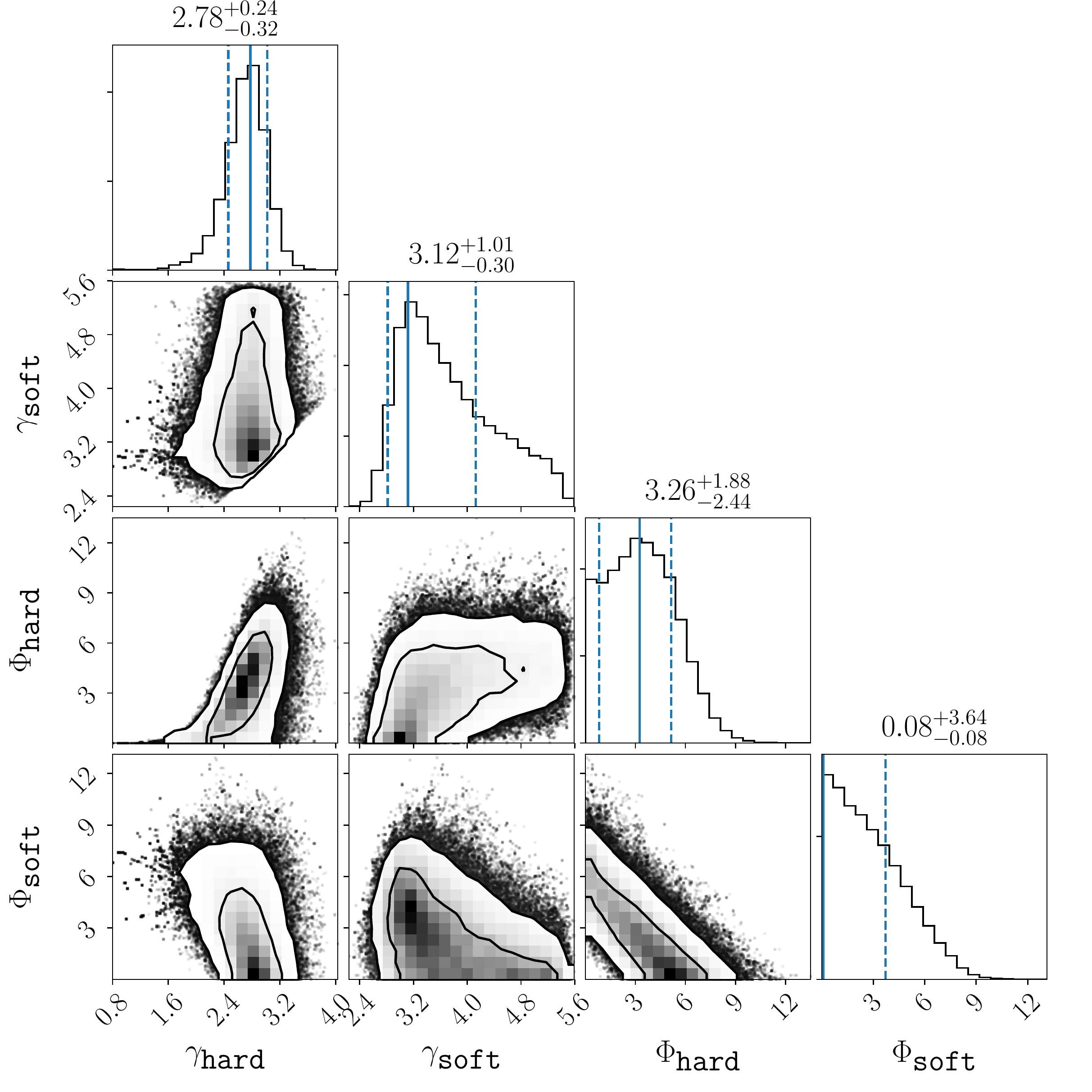}
    \caption{\textbf{\textit{Double power-law model hard/soft parameters posterior distribution.}}
    Results derived from the posterior distribution of the model are shown in the same style as \reffig{fig:SPL_posterior}.
    The figure shows the one- and two-dimensional posterior distributions for the parameters of the hard and soft components of the astrophysical neutrino flux.
    The diagonal panels show the one-dimensional posterior with the parameter MAP estimation and $\SigmaOne$ HPD region indicated, while the non-diagonal panels show the two-dimensional posterior with $\SigmaOne$ and $\SigmaTwo$ regions indicated.}\label{fig:dpl_posterior}
\end{figure}

The double power law is an extension to the single power law, which introduces a second power-law component with duplicated free parameters.
In this model the astrophysical differential flux is described as
\begin{equation}
\begin{split}
    &\frac{d\Phi_{6\nu}}{dE} = \\
&\left(\Phi_\texttt{hard}{\left(\frac{E_\nu}{\SI{100}\TeV}\right)}^{-\gamma_\texttt{hard}}\right. + \left. \Phi_\texttt{soft}{\left(\frac{E_\nu}{\SI{100}\TeV}\right)}^{-\gamma_\texttt{soft}}\right) \\
    & \cdot 10^{-18}~\textmd{GeV}^{-1}\textmd{cm}^{-2}\textmd{s}^{-1}\textmd{sr}^{-1},
\end{split}
\label{eq:dpl_flux}
\end{equation}
where $\gamma_\texttt{hard} \leq \gamma_\texttt{soft}$.
In \reffig{fig:dpl_posterior}, the normalizations of the two components are seen to be anti-correlated as the data require an astrophysical component, but both normalizations are compatible with zero within the two-dimensional $\SigmaOne$ HPD region of the bi-normalization posterior distribution.
This behavior indicates that two power-law components are not significantly preferred over a single power-law component.
A second point in the parameter space, where the two spectral indices are equal, reduces to the single power-law scenario as well.
When the flux is parameterized in terms of the difference between the spectral parameters, it is clear that this region is well within the $\SigmaOne$ HPD region.
Since the $\SigmaOne$ HPD regions are compatible with a single power law, a meaningful estimate for the ``critical energy,'' where the two power laws are equal, cannot be obtained.

\subsubsection{Single power law with spectral cutoff\label{sec:cutoff}}
\noindent
\textit{
A spectral cutoff is connected to the highest energies that the sources producing neutrinos can achieve.
We have performed a model comparison analysis between the single power-law model without a cutoff and one with a cutoff.
We find that models that include a cutoff with values smaller than $\BayesFactorStrongCutoff$ are strongly disfavored with respect to the no-cutoff hypothesis due to a large number of lower energy events.
Models with cutoff energy above $\BayesFactorEqualCutoff$ have evidence close to or greater than the no-cutoff hypothesis but are not substantially favored as the softer spectra fit by the data do not expect many events above $\sim\si\PeV$ energies.
}
\newline

The flux of the astrophysical component is given as
\begin{equation}
\begin{split}
    \frac{d\Phi_{6\nu}}{dE}={}&\Phi_\texttt{astro}{\left(\frac{E_\nu}{\SI{100}\TeV}\right)}^{-\gamma_\texttt{astro}} \\ & \cdot e^{-\frac{E_{\nu}}{E_\texttt{cutoff}}} \\ & \cdot 10^{-18}\textmd{GeV}^{-1}\textmd{cm}^{-2}\textmd{s}^{-1}\textmd{sr}^{-1}.
\end{split}\label{eq:cutoff_flux}
\end{equation}

\begin{figure}
    \centering
    \includegraphics[width=\linewidth]{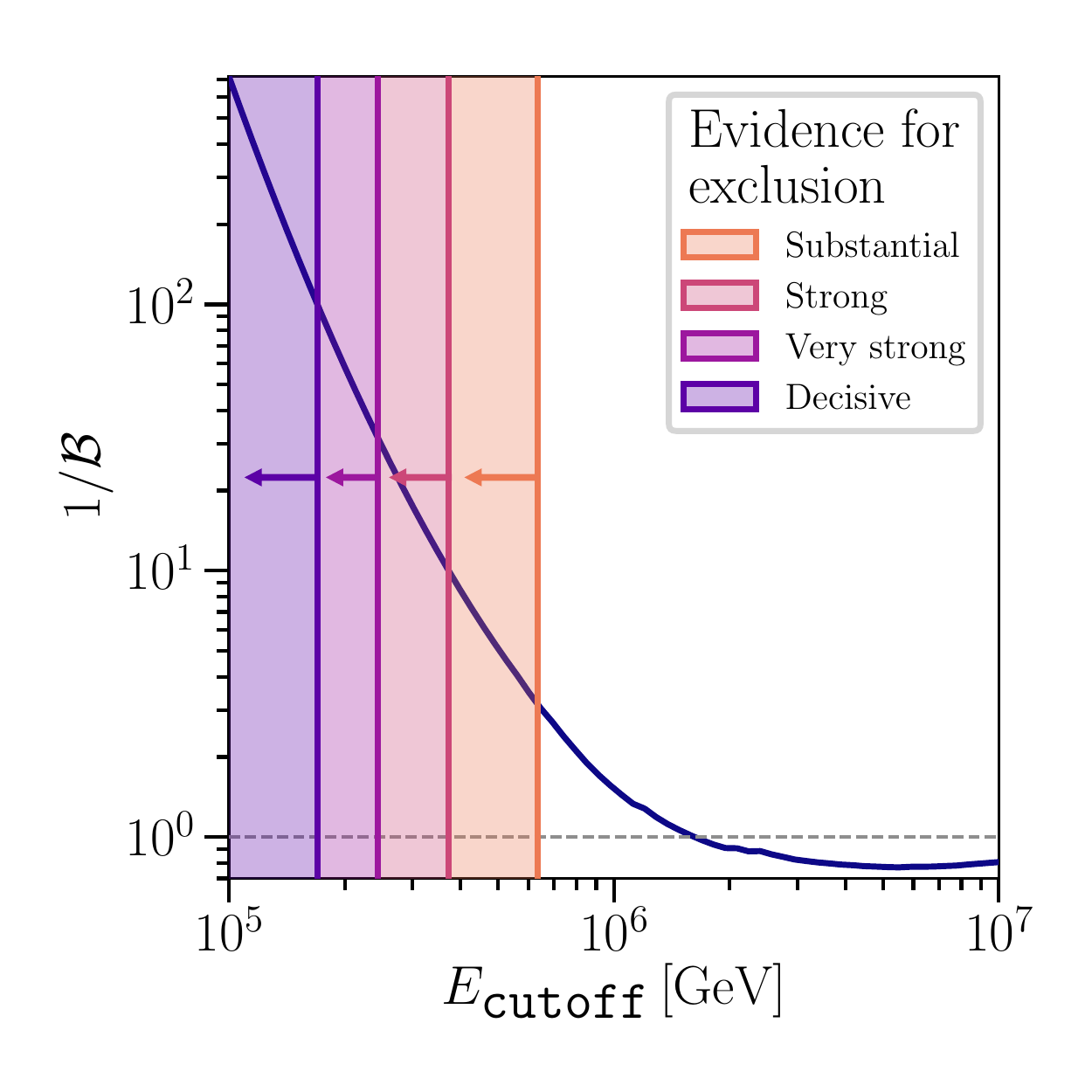}
    \caption{\textbf{\textit{Spectral cutoff Bayes factors and regions of exclusion.}}
    The inverse of the Bayes factor, $\mathcal{B}$, is plotted in the blue curve as a function of the cutoff energy assumed in the alternative hypothesis.
    Regions of the cutoff energy shown by the shaded regions are disfavored with respect to the null hypothesis with varying degrees of certainty according to Jeffreys' scale.
    The Bayes factor is computed for each value of the cutoff energy with the single power-law as a null hypothesis.
    The gray dashed line indicates where the evidence of the cutoff model and the single power-law model are equal.}\label{fig:cutoff_bayes}
\end{figure}

To study the preference between cutoff scenarios, we compute the Bayes factor for many values of the cutoff energy $E_\texttt{cutoff}$, where the null hypothesis is the single power-law model, and the alternative hypothesis is the cutoff model.
The Bayes factor in this case is defined as
\begin{equation}
\begin{split}
	\mathcal{B}(E_\texttt{\small{cutoff}}) ={}& \frac{\int d\vec\eta~\mathcal{L}_\texttt{\tiny{cutoff}}(E_\texttt{\tiny{cutoff}},\vec\eta)\cdot\Pi(\vec\eta)}{\int d\vec\eta~\mathcal{L}_\texttt{\tiny{SPL}}(\vec\eta)\cdot\Pi(\vec\eta)},
\end{split}\label{eq:cutoff_bayes_factor}
\end{equation}
where $\mathcal{L}_\texttt{\tiny{cutoff}}$ is the likelihood of the cutoff model, $\mathcal{L}_\texttt{\tiny{SPL}}$ is the likelihood of the single power-law model, and $\Pi$ is the set of priors given in \reftab{tbl:priors}.
\reffig{fig:cutoff_bayes} shows the inverse of the Bayes factor $\mathcal{B}$ as a function of $E_\texttt{cutoff}$.
For most values of $E_\texttt{cutoff}$, the Bayes factor is less than one; this implies that the data in this sample favors a model with no cutoff in most cases.
We can exclude values of the cutoff with some level of certainty for regions where $\mathcal{B} < 1$
\reffig{fig:cutoff_bayes} shows excluded regions of the cutoff chosen according to Jeffreys' scale.

\begin{figure}
    \centering
    \includegraphics[width=\linewidth]{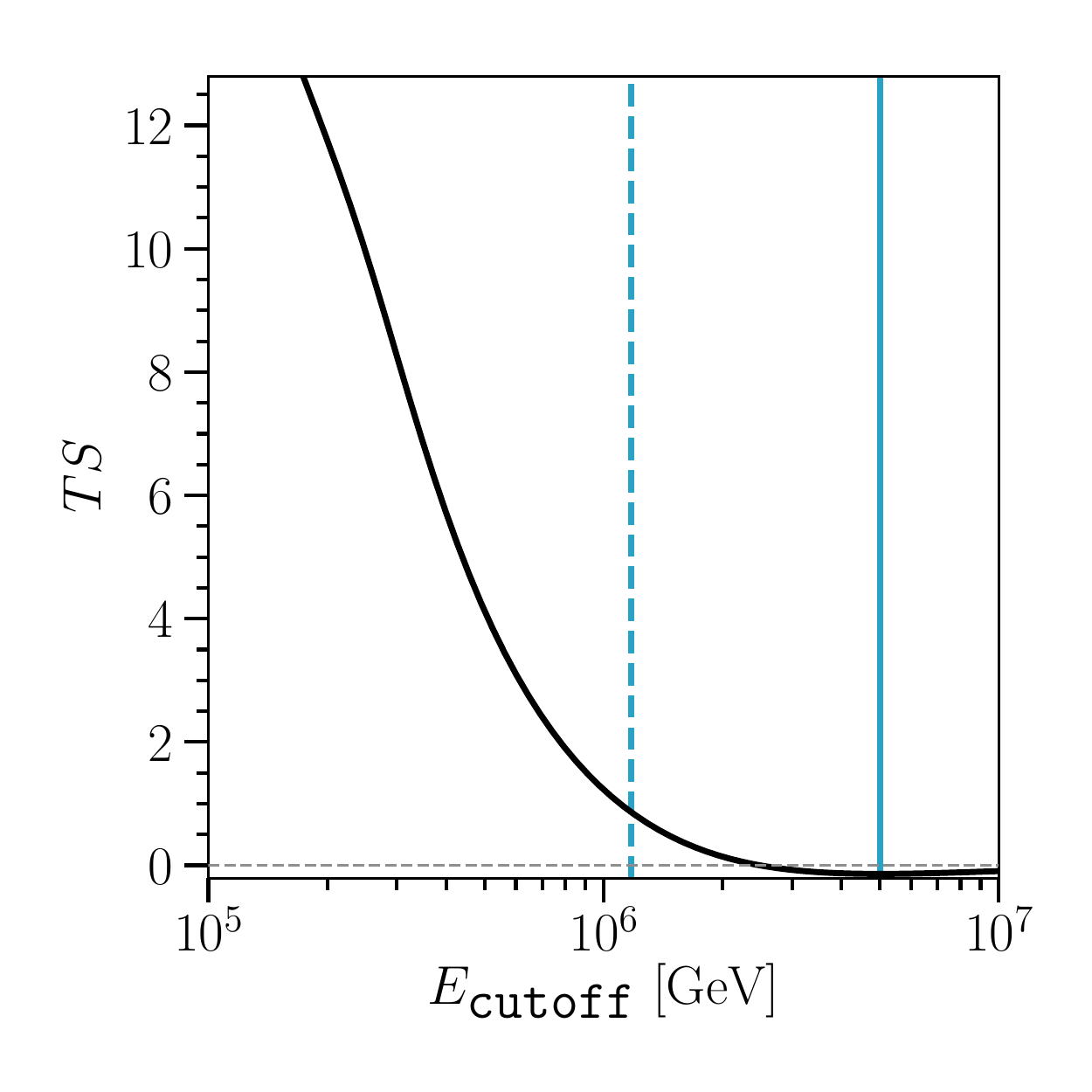}
    \caption{\textbf{\textit{Cutoff model comparison using frequentist test-statistic.}}
    The test-statistic comparing the cutoff hypothesis and the single power-law hypothesis is plotted as the black curve.
    The gray dashed line indicates where the cutoff model's test-statistic and the single power-law model are equal; there is a slight preference for the cutoff scenario in the several $\si\PeV$ region, although this is not statistically significant.
    We also report the results of parameter estimation for the cutoff model in this plot.
    The best-fit value for the cutoff energy, $E_\texttt{cutoff}$, is shown as the solid blue line and the dashed blue line indicates the boundary of the $\SigmaOne$ confidence interval.}\label{fig:cutoff_freq}
\end{figure}

In addition to the Bayes factor treatment described above, we also perform a test using a frequentist test-statistic, defined as
\begin{equation}
\begin{split}
	\TS(E_\texttt{\small{cutoff}}) ={}& -2\log\left({\frac{\max_{\vec\eta}\mathcal{L}_\texttt{\tiny{cutoff}}(E_\texttt{\tiny{cutoff}},\vec\eta)\cdot\Pi(\vec\eta)}{\max_{\vec\eta}\mathcal{L}_\texttt{\tiny{SPL}}(\vec\eta)\cdot\Pi(\vec\eta)}}\right),
\end{split}\label{eq:cutoff_ts}
\end{equation}
in order to compare to other IceCube measurements of a spectral cutoff.
As before, the null hypothesis is the single power-law model, and the alternative hypothesis is the spectral cutoff model with the cutoff energy as a free parameter.
This model-comparison test obtains a p-value of $\CutoffPValue$ and best-fit cutoff energy of $\CutoffBFCutoff$.
To further visualize the cutoff energy parameter space favored or disfavored, we plot the test statistic as a function of the cutoff energy in \reffig{fig:cutoff_freq}.
Cutoff energies below $\CutoffWilksLowerCutoff$ are disfavored at more than the $\SigmaOne$ confidence level while cutoff energies above this, including the no cutoff scenario, are compatible within the $\SigmaOne$ confidence level.

\subsubsection{Log-parabola flux\label{sec:log_parabola}}
\noindent
\textit{
This spectral model has two relevant parameters: the spectral index at the $\SI{100}\TeV$ pivot point and the spectral index rate of change.
In this model, the spectral index at the pivot point and the spectral index rate of change are simultaneously compatible with the single power-law spectral index and zero, respectively.
This result implies that in the measured energy range, we observe no indication of log-linear spectral change.
}
\newline

In log-energy log-flux space, the single power law can be represented as a line.
A simple, functional extension is to add curvature to this line.
This gives the log-parabola model which has the form

\begin{equation}
\begin{split}
    \frac{d\Phi_{6\nu}}{dE}={}&\Phi_\texttt{astro}{\left(\frac{E_\nu}{\SI{100}\TeV}\right)}^{-\left(\alpha+\beta \mathop{\log_{10}}(\frac{E_\nu}{\SI{100}\TeV})\right)} \\ & \cdot 10^{-18}\textmd{GeV}^{-1}\textmd{cm}^{-2}\textmd{s}^{-1}\textmd{sr}^{-1},
\end{split}\label{eq:logparabola_flux}
\end{equation}
where $\alpha$ is the spectral index at $\SI{100}\TeV$, and $\beta$ governs how the effective spectral index changes with energy.
In the Bayesian analysis of this model, we have chosen improper uniform priors for both $\alpha$ and $\beta$.
At $\SI{100}\TeV$ the most-likely spectral index ($\alpha=2.78$) is still soft, and compatible with the most-likely SPL spectral index ($\astrodeltagamma=\SPLBayesMAPIndex$) within the $\SigmaOne$ HPD region of $\alpha$.
There is one region of the parameter space where the log-parabola model becomes the same as a single power law when $\beta=0$.
This region of parameter space is within the $\SigmaOne$ HPD region of $\beta$, informing us that the data is most compatible with a model that is close to a single power law rather than a model with larger curvature.

\begin{figure}
    \centering
    \includegraphics[width=\linewidth]{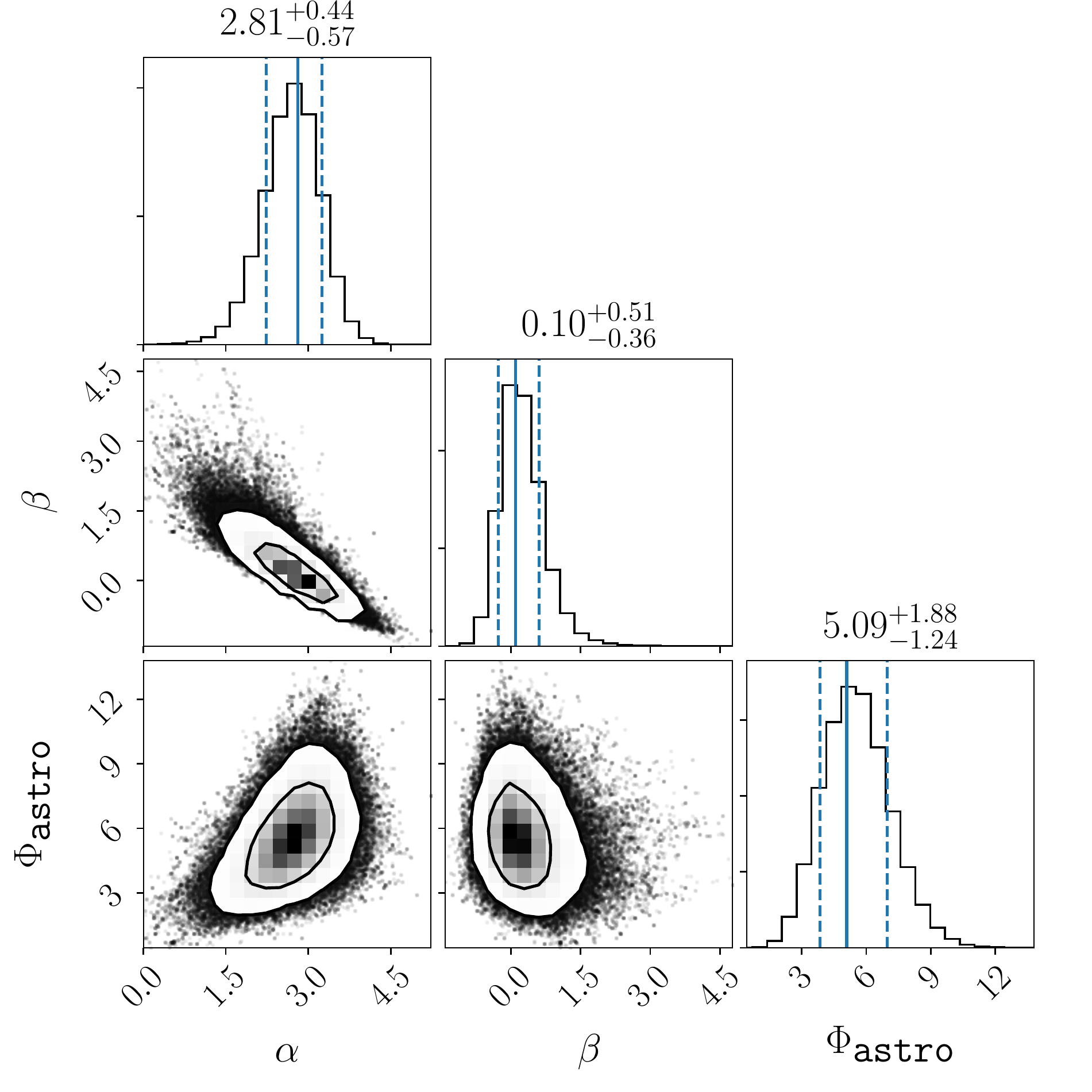}
    \caption{\textbf{\textit{Log-parabola astrophysical model parameters posterior distribution.}}
    Results derived from the posterior distribution of the model are shown in the same style as \reffig{fig:SPL_posterior}.
    The figure shows the one- and two-dimensional posterior distributions for the astrophysical flux normalization, $\astronorm$; the spectral index at $\SI{100}\TeV$, $\alpha$; and the change in spectral index, $\beta$.
    The diagonal panels show the one-dimensional posterior with the parameter MAP estimation and $\SigmaOne$ HPD region indicated, while the non-diagonal panels show the two-dimensional posterior with $\SigmaOne$ and $\SigmaTwo$ regions indicated.
    As we can see from the posterior distribution, $\beta$ is compatible with zero which implies that an unbroken power law is a good fit to the data under these model constraints.}\label{fig:log_parabola_powerlaw_corner}
\end{figure}

\subsubsection{Segmented power-law flux\label{sec:unfolding}}
\noindent
\textit{The neutrino spectrum can be generically parameterized as a set of narrow $E^{-2}$ power-law segments.
Here, the normalization of each of these segments and their uncertainties are reported.
It is notable that the two lowest energy segments' energy content is higher than the five highest energy segments by a factor of $\sim 2-3$.
The origin of this increase also drives the soft spectrum observed in the single power law model.
Without a break in the spectrum above $\sim\SI{50}\TeV$, the flux normalization measured around $\SI{100}\TeV$ is incompatible with a gamma-ray transparent source model given current gamma-ray measurements~\cite{Murase:2015xka,Ando:2015bva,Bechtol:2015uqb,Meszaros:2017fcs,Capanema:2020rjj}.
}
\newline

The models explored in previous sections restrict the spectrum to be described by an unbroken power-law-like model across the entire energy range.
In this section, a more general parameterization of the astrophysical flux is introduced.
The neutrino energy spectrum is split into segments equally spaced in $\log E_\nu$, assumed to behave as $E^{-2}$ within each segment, and then the normalizations of each segment are allowed to vary independently.
While not entirely general, this model can describe a wide variety of fluxes with the current detector energy resolution.
The astrophysical neutrino flux within each segment is given by
\begin{align}
\begin{split}
    \frac{d\Phi_{6\nu}}{dE} ={}& \Phi_{i} {\left(\frac{E_\nu}{E_{c,i}}\right)}^{-2}\\
    & \cdot 10^{-18}\textmd{GeV}^{-1}\textmd{cm}^{-2}\textmd{s}^{-1}\textmd{sr}^{-1},
\label{eq:segmented_flux}
\end{split}
\end{align}
where $\Phi_i$ is the normalization constant for each bin, $E_{c,i}$ is the log-center of each bin.

\begin{figure*}[ht]
    \centering
    \includegraphics[width=0.45\linewidth]{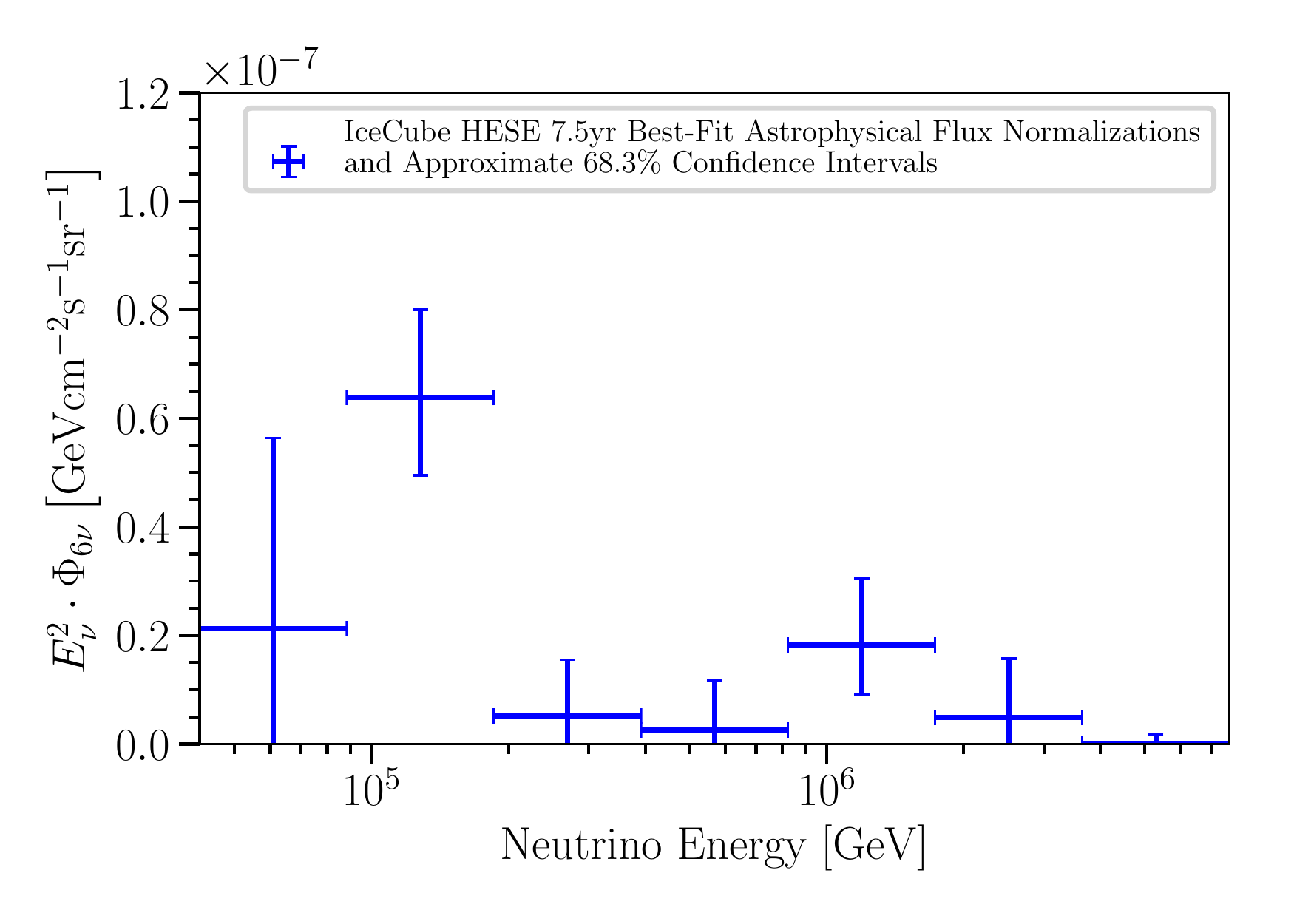}
    \includegraphics[width=0.45\linewidth]{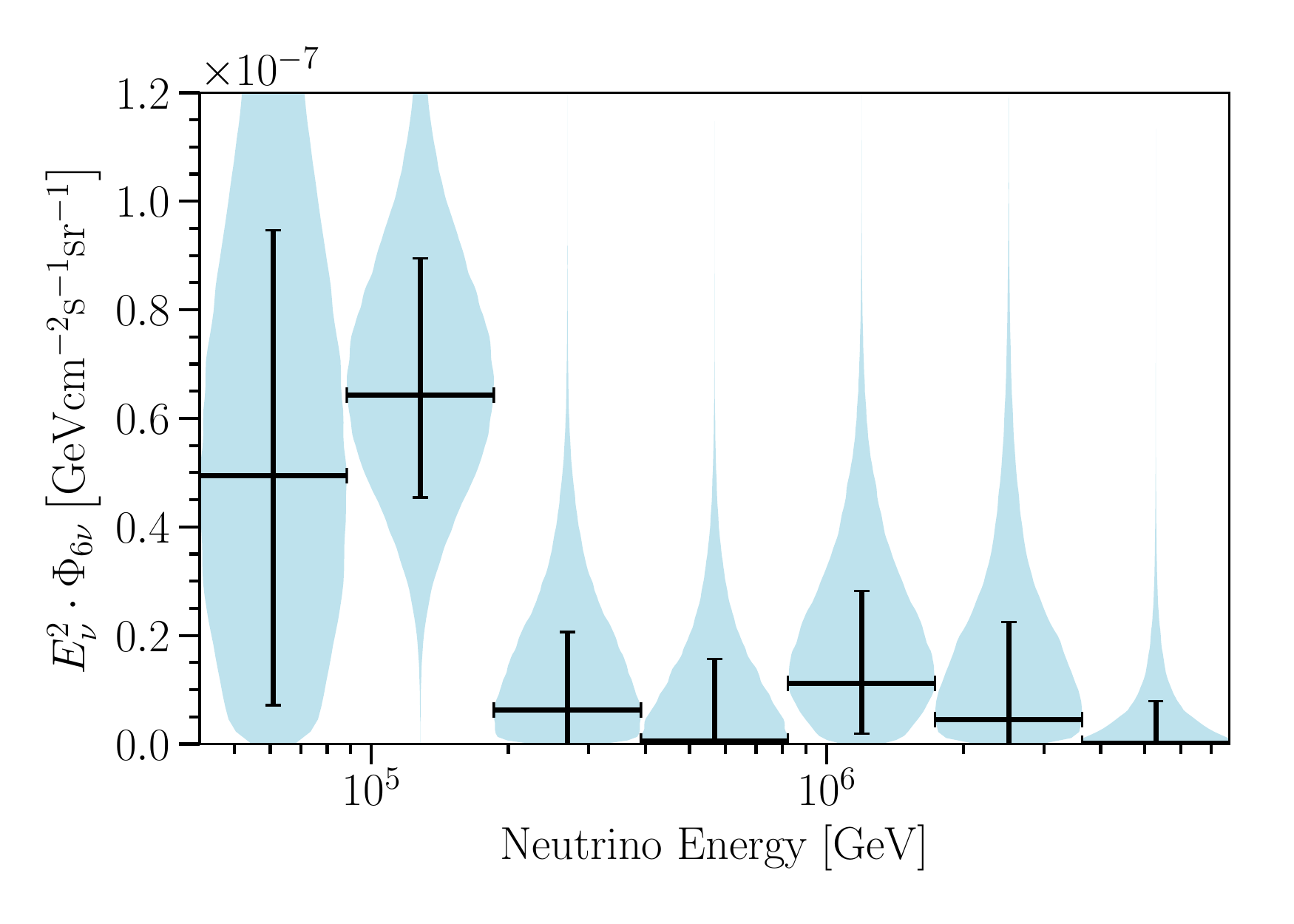}
    \caption{\textbf{\textit{Segmented power-law fit.}}
    The differential flux as obtained by fitting the normalizations of independent $E^{-2}$ segments defined in true neutrino energy.
    Left: in this plot each error bar shows the region in which $\Delta\like\leq0.5$ while holding all other parameters fixed, providing an approximate $\SigmaOne$ confidence level interval for the astrophysical normalization in that particular segment.
    Right: in this plot each error bar shows the $\SigmaOne$ highest probability density credible interval for the astrophysical normalization in that particular segment, assuming a uniform prior on the normalizations.
    The width of the turquoise bands is proportional to the posterior density of the normalization in that segment, and the horizontal scales are arbitrary.}\label{fig:segmented}
\end{figure*}

This model is analyzed in the same way as previous analyses~\cite{Aartsen:2014gkd,Aartsen:2015zva,Aartsen:2017mau}.
Namely, the best-fit point for the normalizations and their one-dimensional errors are obtained, which are plotted for seven energy segments (enumerated in \reftab{tbl:segment_normalizations}) in the left panel of \reffig{fig:segmented}.
Other energy segments are profiled over and not shown because they are poorly constrained by the data and do not provide meaningful information.
To compare with previous work, we estimate errors by fixing all parameters except one normalization and find the range of this normalization for which $\Delta\max_{\vec\theta}\like\cdot\Pi\leq 0.5$; this procedure produces smaller errors than the profile likelihood technique.

As a complement to the aforementioned frequentist approach, this model is also analyzed in a Bayesian way.
Assuming improper positive uniform priors for the normalizations of the power-law segments, we sample the model's posterior distribution.
The right panel of \reffig{fig:segmented} shows the one-dimensional MAP estimation of each normalization independently for the same seven energy segments as before; the remaining energy segments have been marginalized over as the data does not significantly constrain them.
Errors of each normalization are constructed by integrating the joined distribution over all other parameters and then computing the $\SigmaOne$ HPD region of that segment.
The one-dimensional posterior density is also plotted as a turquoise band to demonstrate the shape of the distribution, although the relative scale between bands is arbitrary.
Finally, in \reftab{tbl:segment_normalizations}, the segments' normalizations are reported for both the frequentist and Bayesian analysis.
We suggest readers use the supplied data release to draw accurate conclusions about the spectrum rather than the errors reported here as correlations exist between the parameters.

The most notable feature of the segmented power-law fit, reported in this section, is the large neutrino flux at the lower energy range: between $\SI{60}\TeV$ and $\SI{200}\TeV$.
This lower-energy contribution to the observed flux is what drives the soft spectral index reported in \refsec{sec:spl} for the single power-law scenario.
Under the assumption that the astrophysical flux sources are transparent to gamma rays, we would expect a correspondingly large gamma-ray flux to be observed.
The flux measured in the second-lowest bin around $\sim\SI{100}\TeV$ is not incompatible with current gamma-ray measurements on its own, but may not be compatible depending on the underlying neutrino spectrum.
For this data to remain compatible with a gamma-ray transparent source model, a spectral break is needed above $\sim\SI{50}\TeV$, such that the spectrum is harder below the break.
However, the cascade sample has measured a normalization of $E_\nu^2 \Phi_{6\nu}=\sim\SI{1.3e-7}{\GeV\per\square\cm\per\s\per\sr}$ around $\sim\SI{30}\TeV$, implying that the required spectral break does not occur above $\sim\SI{50}\TeV$.
This would suggest the existence of gamma-ray opaque sources that dominate the neutrino flux at the lowest energies in this analysis~\cite{Murase:2015xka,Ando:2015bva,Bechtol:2015uqb,Meszaros:2017fcs,Capanema:2020rjj}.

Pionic gamma rays accompany high-energy neutrinos at the site of production, in fact their emission rates are intimately related by~\cite{10.3389/fspas.2019.00032}
\begin{equation}
    \frac{1}{3} \sum_{\alpha} E_{\nu}^{2} Q_{\nu_{\alpha}}\left(E_{\nu}\right) \simeq \frac{K_{\pi}}{4}\left[E_{\gamma}^{2} Q_{\gamma}\left(E_{\gamma}\right)\right]_{E_{\gamma}=2 E_{\nu}},
\end{equation}
where $Q$ is the source rate function of neutrinos or gamma-rays, $K_{\pi}$ accounts for the ratio of charged-to-neutral pion production via proton-gas interactions ($pp$) or photo-hadronic interactions ($p\gamma$).
However, these pionic gamma rays interact with the extragalactic background light (EBL) and cascade to lower energies~\cite{Berezinsky:1975zz} to contribute to the IGRB.
The high intensity of the neutrino flux at energies below $\SI{100}\TeV$ compared to the IGRB flux measured by Fermi satellite may indicate that sources responsible for their production cannot be transparent to very-high-energy gamma rays.
In this scenario, the gamma rays produced interact with lower energy photons in the source, initiating electro-magnetic cascades that could be visible at lower frequencies, and therefore, would not overshoot the measured IGRB flux.
We should also note that photo-hadronic interactions should be the dominant channel of neutrino production in sources responsible for the high-intensity flux, as hadro-nuclear interactions would result in an overwhelming gamma ray flux at lower energies that cannot be tolerated by the measured IGRB~\cite{Murase:2013rfa, Bechtol:2015uqb,Capanema:2020rjj}.

\subsection{Atmospheric flux from charmed hadrons\label{sec:prompt}}
\noindent
\textit{
In this analysis, the astrophysical component and the ``prompt'' neutrino flux can be distinguished by their energy and angular distributions.
We find no evidence for a prompt component of the atmospheric neutrino flux; normalizations greater than 13 times the baseline model are strongly disfavored with respect to the no-charmed-hadron neutrino flux hypothesis.
Additionally, we explore an astrophysical flux free hypothesis and find that this is rejected at greater than $5\sigma$ with respect to a single power-law astrophysical plus atmospheric flux hypothesis.
}
\newline

Although most neutrinos in cosmic-ray air showers are produced from the decay of muons, pions, and kaons, at high energies, the decay of charmed hadrons can also produce neutrinos.
Due to the short decay length of charmed hadrons compared to their interaction length, they mostly decay without losing energy, yielding a harder spectrum of neutrinos than the conventional component~\cite{gaisser2016cosmic}, that is more uniform across the sky.
Two ingredients are needed to compute the flux from charmed hadrons: the production cross section of charmed hadrons and the cosmic-ray flux.
For this work, the relevant part of the production cross section has not been measured at collider experiments as it is only accessible in the far-forward region very close to the beam-line where there is little to no instrumentation.
The charmed hadron production cross section can be computed by means of perturbative QCD~\cite{Garzelli:2015psa,Gauld:2015kvh,Gauld:2015yia,Bhattacharya:2016jce,Garzelli:2016xmx} and non-perturbative techniques such as dipole-model interactions~\cite{Goncalves:2006ch,Enberg:2008te,Arguelles:2015wba}.
In the region of interest, from $\SI{10}\TeV$ to $\SI{10}\PeV$, the expected flux has an uncertainty of at least a factor of two in normalization due to re-normalization and factorization scale uncertainties, as well as uncertainties in the cosmic-ray composition, and charm mass uncertainties according to~\cite{Abelev:2012vra,Garzelli:2016xmx,Zenaiev:2019ktw}.
These uncertainty estimations do not include the possibility of additional non-perturbative contributions, {\it e.g.} intrinsic charm~\cite{Laha:2016dri}.
The prompt flux shape variation in this region of interest arises primarily from changes in the cosmic-ray models.
In this analysis, we use the BERSS calculation~\cite{Bhattacharya:2015jpa} with passing fractions from~\cite{Arguelles:2018awr} to predict the baseline prompt contribution to the data sample.

\begin{figure*}
    \centering
    \includegraphics[width=0.45\linewidth]{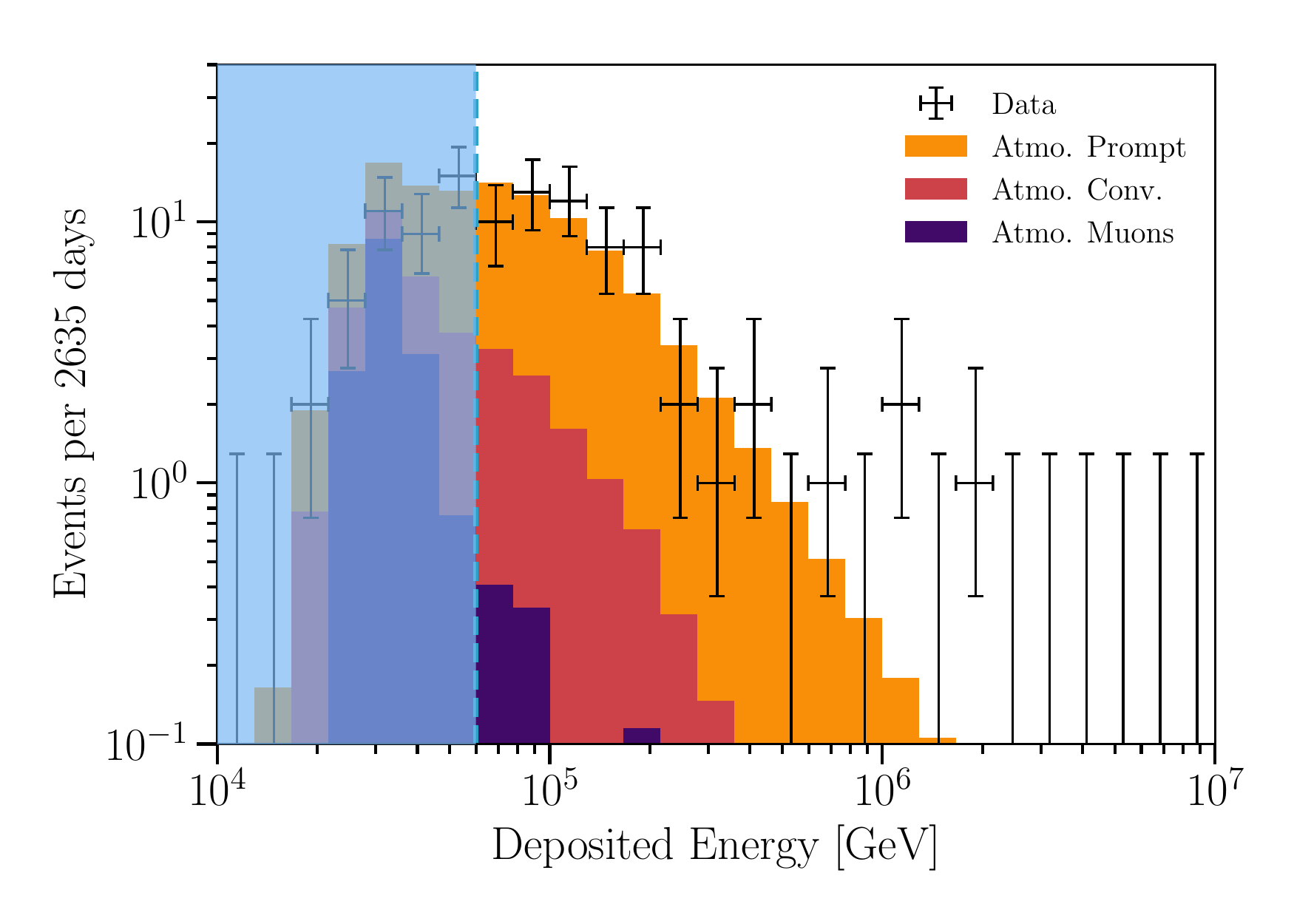}
    \includegraphics[width=0.45\linewidth]{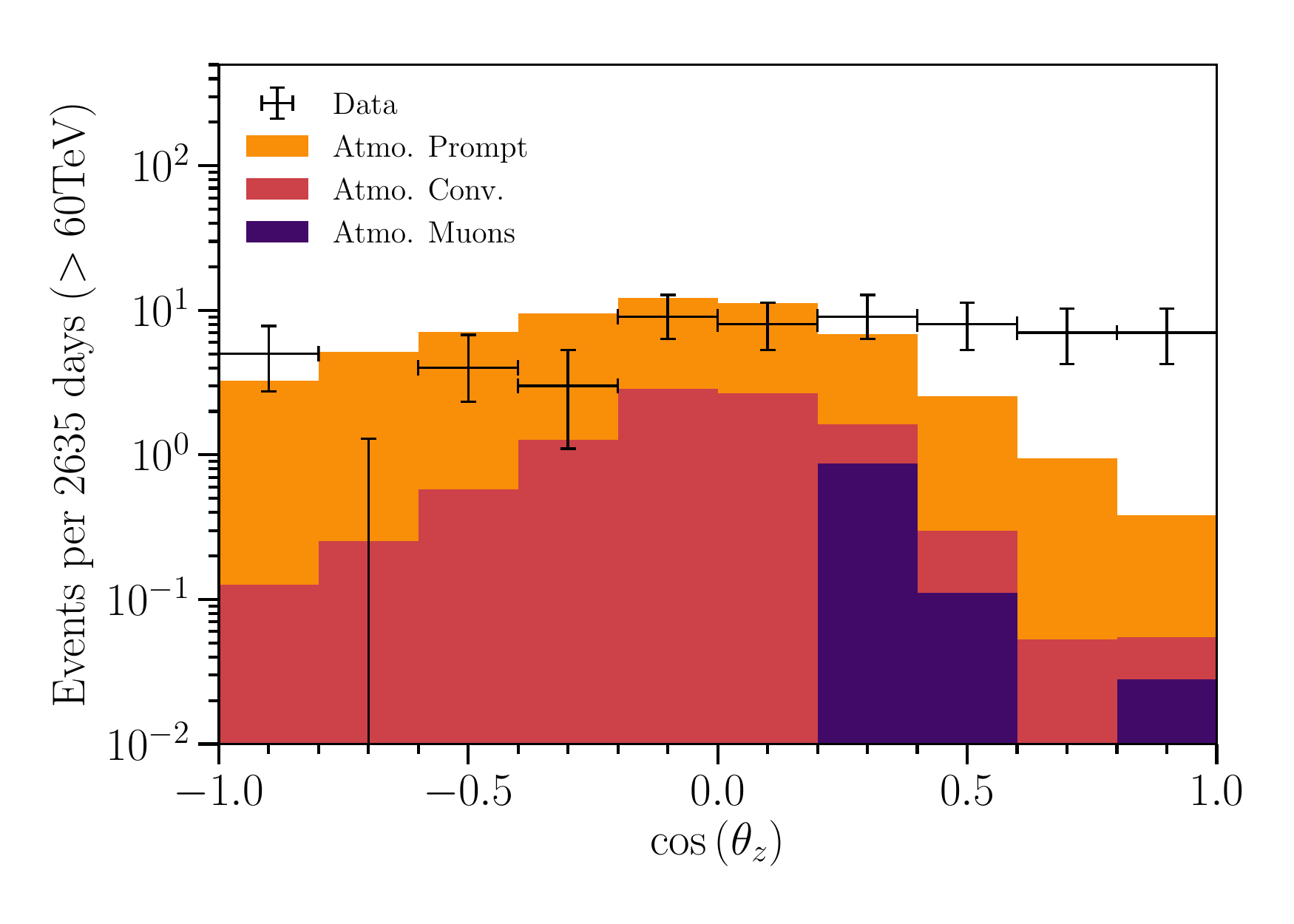}
    \caption{\textbf{\textit{Atmospheric-background-only fit to the data.}} In these figures we present the best fit in the absence of an astrophysical component.
    The left panel shows the deposited energy distribution and the right panel the angular distribution.
    As can be seen in the right panel, the angular distribution is in tension with the expectation in several bins.
    This amounts to a greater than $5\sigma$ difference with respect to the best-fit astrophysical model.
    The colors are the same as in \reffig{fig:energy-zenith}.}\label{fig:prompt_no_astro}
\end{figure*}

An analysis of the HESE sample can be performed considering only the atmospheric muon and neutrino components.
This results in a best-fit prompt normalization of $\NoAstroPromptNorm$ times the baseline model and is shown in \reffig{fig:prompt_no_astro}.
As can be seen from \reffig{fig:prompt_no_astro}, the predicted angular distribution in this background-only fit fails to explain the southern-sky event rate.
The atmospheric only hypothesis is disfavored by greater than $5\sigma$ in comparison to the single power-law astrophysical plus atmospheric flux hypothesis.

Compared to constraints on the prompt normalization from other IceCube samples, the best-fit prompt normalization obtained in the background-only fit is in tension with these results~\cite{Aartsen:2013eka,Aartsen:2016xlq,Aartsen:2014muf}.
Some constraints have been obtained when considering a single-power-law astrophysical component, and are thus dependent on this model assumption.
However, the constraints from~\cite{Aartsen:2013eka} predate the observation of high-energy extraterrestrial neutrinos and are conservative because this scenario is equivalent to zero contribution from the astrophysical flux.
The latter results in a constraint of the prompt normalization of $\ICFiftyNinePromptUpperLimit$ times the ERS calculation~\cite{Enberg:2008te} at $\SI{90}\percent$ C.L.; a model which is approximately $2.5$ times larger than the benchmark model used in this analysis.

When we allow for the existence of a single power-law astrophysical component, the best-fit prompt component normalization is zero.
In this same scenario, using the frequentist statistical construction assuming Wilks' theorem with one degree of freedom, a $\SigmaOne$ C.L. prompt normalization upper bound of $\SPLFreqWilksUpperPromptNorm$ and a $\SI{90}\percent$ upper limit of $\SPLFreqWilksNinetyUpperLimit$ is obtained.
This result is in agreement with the results summarized in \reftab{tbl:prompt}.

\begin{figure}
    \centering
    \includegraphics[width=\linewidth]{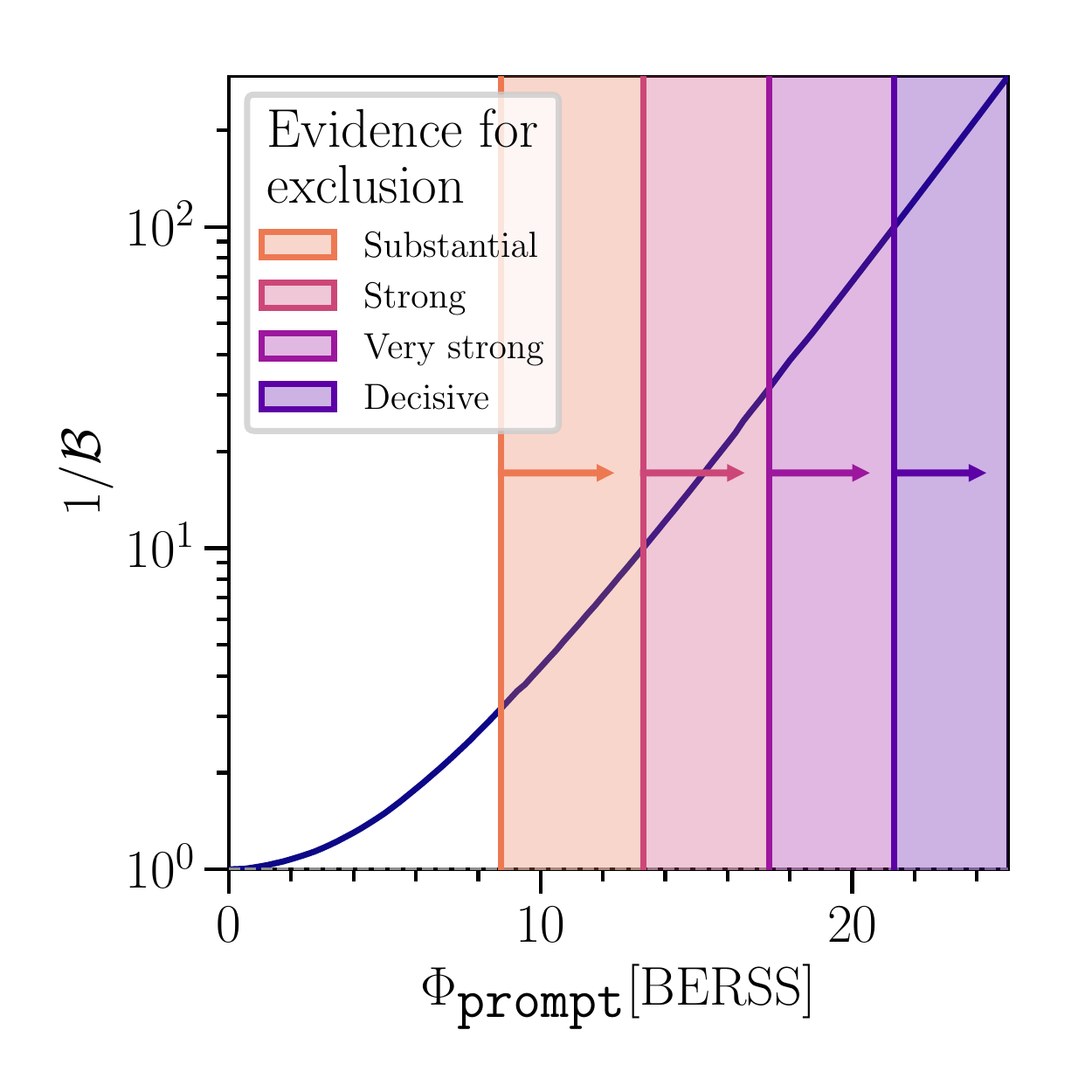}
    \caption{\textbf{\textit{Prompt neutrino normalization constraints.}}
    The horizontal axis shows the size of the prompt normalization with respect to the baseline BERSS model discussed in \refsec{sec:backgrounds} with passing fractions given in~\cite{Arguelles:2018awr}.
    The vertical axis gives the reciprocal of the Bayes factor; decreasing Bayes factors imply a more disfavored prompt normalization.
    The shaded regions denote exclusions of the prompt normalization according to Jeffreys' scale from substantial to decisive.}\label{fig:prompt_bayes}
\end{figure}

Additionally, in the Bayesian framework, the most-likely value of the prompt normalization is $\SPLBayesPromptNormSummary$ when assuming an improper uniform prior for the prompt normalization.
In this case, the prompt normalization posterior distribution strongly depends on the prior choice.
For this reason, we report our Bayesian prompt normalization results in terms of the Bayes factor between the no-prompt hypothesis and a given prompt normalization; see \refsec{sec:statistics} for details.
In \reffig{fig:prompt_bayes} we show the Bayes factor obtained assuming a uniform prior on the astrophysical neutrino normalization and spectral index.
According to Jeffreys' scale, we find that prompt normalizations greater than $\BayesFactorStrongPromptNorm$ are disfavored at the strong level, compared to the no-prompt scenario.

\begin{table*}
\begin{center}
\begin{tabular}{l c c}
\toprule
 & Frequentist upper limit ($\SI{90}\percent$ C.L.)& Bayesian model rejection (strong) \\
\midrule
    Northern sky muons IC59~\cite{Aartsen:2013eka} & $\ICFiftyNinePromptUpperLimit \times \phi_{ERS}$ & --  \\
    Northern sky muons IC86~\cite{Aartsen:2016xlq} & $1.06 \times \phi_{ERS}$ & -- \\
    All-sky medium-energy starting cascades~\cite{Aartsen:2014muf} & $1.52 \times \phi_{ERS}$ & -- \\
    HESE $\SI{7.5}\year$ (this work) & $\SPLFreqWilksNinetyUpperLimit \times \phi_{BERSS}$ & $ \BayesFactorStrongPromptNorm \times \phi_{BERSS}$ \\
\bottomrule
\end{tabular}
\end{center}
\caption{\textit{\textbf{Summary of constraints on the flux of charmed mesons.}} The benchmark $\phi_{ERS}$~\cite{Enberg:2008te} and $\phi_{BERSS}$~\cite{Bhattacharya:2015jpa}, are such that $\phi_{ERS}(\SI{100}\TeV) \approx 2.5 \cdot \phi_{BERSS}(\SI{100}\TeV)$.}
\label{tbl:prompt}
\end{table*}

\subsection{Source-specific models\label{sec:specific_models}}
\noindent
\textit{Several models are selected from the literature that predict the neutrino flux from a variety of sources: AGN, low-luminosity AGN, choked jets in core-collapse SN, star burst galaxies, low-luminosity BLLacs, and GRBs.
These models are tested against a baseline single power-law astrophysical flux model by considering two types of alternative hypotheses.
In the first, a source model comprises the entire astrophysical neutrino flux; in the second, a source model and a single power law both contribute to the flux.
These tests fix the model parameters to nominal values, and so do not have the power to exclude them for the full allowed range of parameters.
None of these tested alternative scenarios are strongly favored compared to the single power-law hypothesis.
Of the scenarios tested, some are disfavored with the model parameters' nominal values, see \reftab{tbl:diffuse_models} for more information.
}
\newline

\begin{figure}
    \centering
    \includegraphics[width=\linewidth]{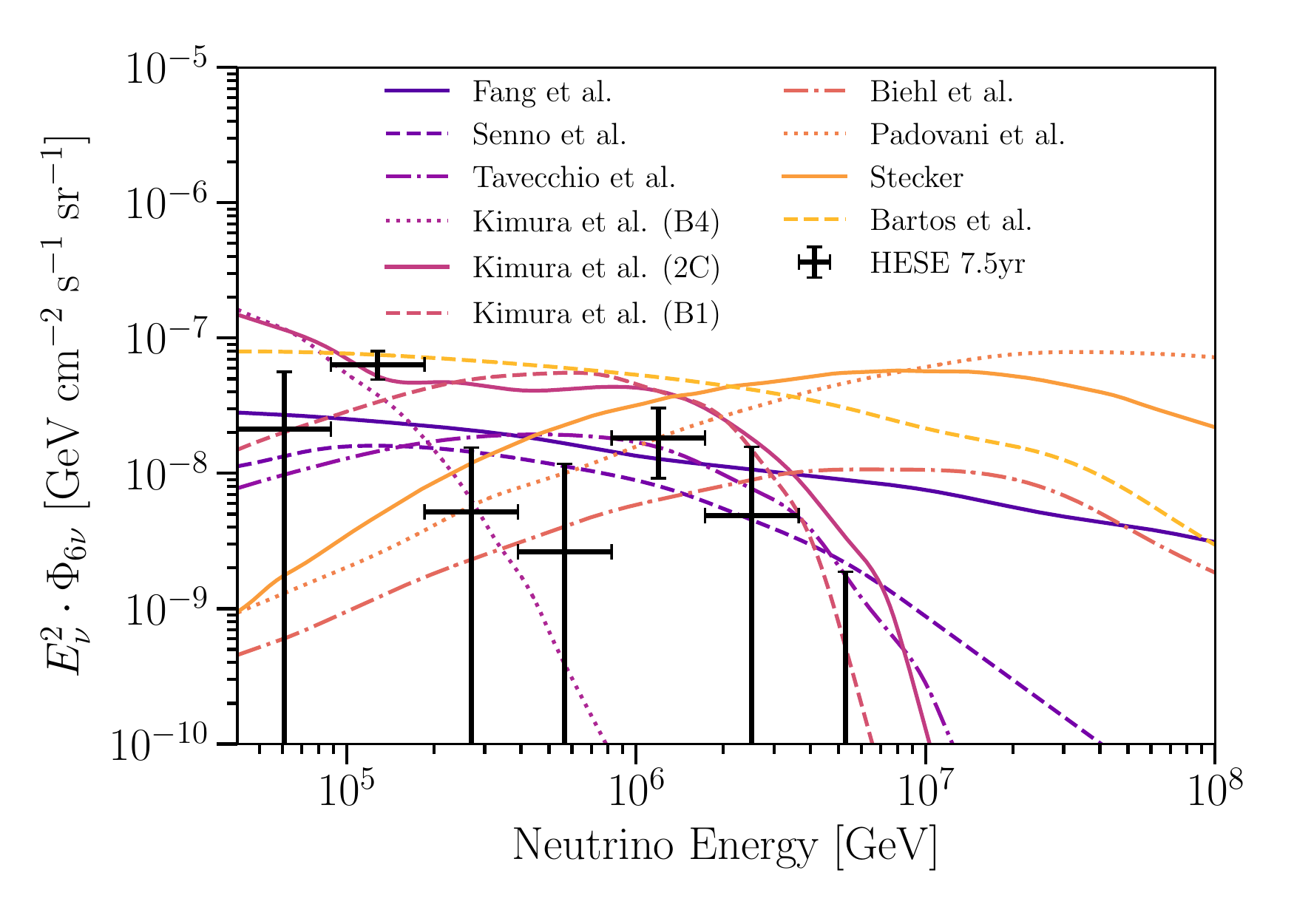}
    \caption{\textbf{\textit{Source-specific models tested in this work together with the segmented-fit outcome.}}
    Models discussed in \refsec{sec:specific_models} and listed in \reftab{tbl:diffuse_models} are shown as lines.
    The HESE segmented power-law model, described in \refsec{sec:unfolding}, best-fit normalizations are shown as black crosses.
    Models are ordered in the color scale from largest (darkest color) to smallest (lightest color) evidence as reported in \reftab{tbl:diffuse_models}.}
    \label{fig:source_models_fluxes}
\end{figure}

Section~\ref{sec:generic_models} characterizes the observed astrophysical neutrino events employing generic models.
These studies show that a single power law is a good fit to the data.
Nevertheless, in this section, we study the compatibility of the observed events with specific source predictions of the astrophysical neutrino flux proposed in the literature.
The specific source fluxes used in this analysis, together with the result of the segmented power-law fit can be seen in \reffig{fig:source_models_fluxes}.
These models were chosen because they have a significant flux contribution in the energy range that this analysis is sensitive to.
Thus, {\it e.g.} we do not test cosmogenic~\cite{Halzen:1992cz} neutrino flux models, which predict neutrinos from cosmic rays interacting with the cosmic microwave background since they are expected to contribute at higher energies where dedicated IceCube searches exist~\cite{Aartsen:2018vtx}; see~\cite{Safa:2019ege} for a recent discussion on the expected rate of cosmogenic flux in this analysis energy range.

Astrophysical neutrino flux predictions have, in principle, many parameters that may modify the expected flux.
For simplicity, this analysis does not study the models' internal parameters but is limited to some nominal values of their parameters, fixing the source model to a single flux.
This analysis means that the tests do not cover these models in their entirety, and so the results presented here should not be interpreted as categorically excluding any of the models tested.
The analysis takes the form of a model selection test with two non-nested alternatives, which we assume to be equally likely in our interpretation of the results.

In the primary analysis, the two alternatives are: a single power law and a specific source model.
The Bayes factor of these two scenarios is used as a criterion for model selection; see \refsec{sec:statistics} for details.
Thus, the single power law serves as a benchmark model to compare against.
In the case of the single power-law model the evidence is calculated by marginalizing over the two model parameters, normalization and spectral index, assuming uniform priors in a compact region defined as $(\astronorm,\astrodeltagamma) \in [0, 25]\times [2,4]$, as well as the analysis nuisance parameters with priors given in \reftab{tbl:priors}.
Boundaries of this uniform box prior are chosen to encompass recent measurements of the astrophysical flux and the bulk of the posterior distribution mass.
To first order, expansion of this box prior changes the evidence by a factor proportional to the parameter space's size.
The alternative scenario has no free parameters, and thus the posterior integral is only over the nuisance parameters.
Figure~\ref{fig:source_models_fluxes} shows the specific source fluxes with a color scale that orders scenarios by their evidence.
In this study, the single power law is penalized due to additional model complexity with respect to the specific source scenario.

\begin{table*}
    \centering
    \begin{tabular}{l l r r r r}
        \toprule
        \multirow{2}{*}{\makecell[l]{}Source class} & \multirow{2}{*}{\makecell[l]{Model}} & \multirowcell{2}{Model only \\ Bayes factor} & \multirowcell{2}{Model + SPL \\ Bayes factor} & \multirowcell{2}{Most-likely \\ SPL $\astrodeltagamma$} & \multirowcell{2}{Most-likely \\ SPL $\astronorm$} \\
        & & & & & \\ \midrule
        \SteckerTableSummary \\ \midrule
        \FangTableSummary \\ \midrule
        \KimuraBOneTableSummary \\ \midrule
        \KimuraBFourTableSummary \\ \midrule
        \KimuraTwoCompTableSummary \\ \midrule
        \MariaBLLacsTableSummary \\ \midrule
        \MurasechockedJetsTableSummary \\ \midrule
        \SBGminBmodelTableSummary \\ \midrule
        \TavecchilowPowerTableSummary \\ \midrule
        \TDEWinterBiehlTableSummary \\
        \bottomrule
    \end{tabular}
    \caption{\textbf{\textit{Astrophysical neutrino flux model comparison test results.}} Each row shows the source-specific scenario tested, the Bayes factor of the model on its own, the Bayes factor of the model in conjunction with a power-law component, the most likely spectral index of the accompanying power-law component with corresponding $\SigmaOne$ HPD region, and the most likely normalization of the accompanying power-law component with corresponding $\SigmaOne$ HPD region.
    Assuming the two alternatives are equally likely, we can interpret small Bayes factors (those less than one) as indicating preference for the single power-law model and large Bayes factors (those greater than one) as indicating preference for the alternative scenario.}\label{tbl:diffuse_models}
\end{table*}

Since some of these models are not intended to explain the whole astrophysical neutrino spectrum, a secondary analysis is also performed.
In this analysis, two models are considered: a single power law on its own and a single power law together with a source model.
In this case, when comparing the null and alternative hypotheses, the constant parameter-space factor that results from any choice of the single power-law flux parameter boundaries will cancel.
So we can use improper uniform priors without introducing an arbitrary scaling factor.
The same prior dependence in the other Bayesian analyses still remains, as the two models' likelihoods can peak in different regions of parameter space.

The results of both analyses are shown in \reftab{tbl:diffuse_models}.
For each model, the ``Model only Bayes factor'' is reported as the primary analysis result.
For the secondary analysis, the ``Model plus single power-law Bayes factor'' is reported together with the most-likely spectral index and normalization and their errors.

\begin{figure*}
    \centering
    \subfloat{\includegraphics[width=0.45\linewidth]{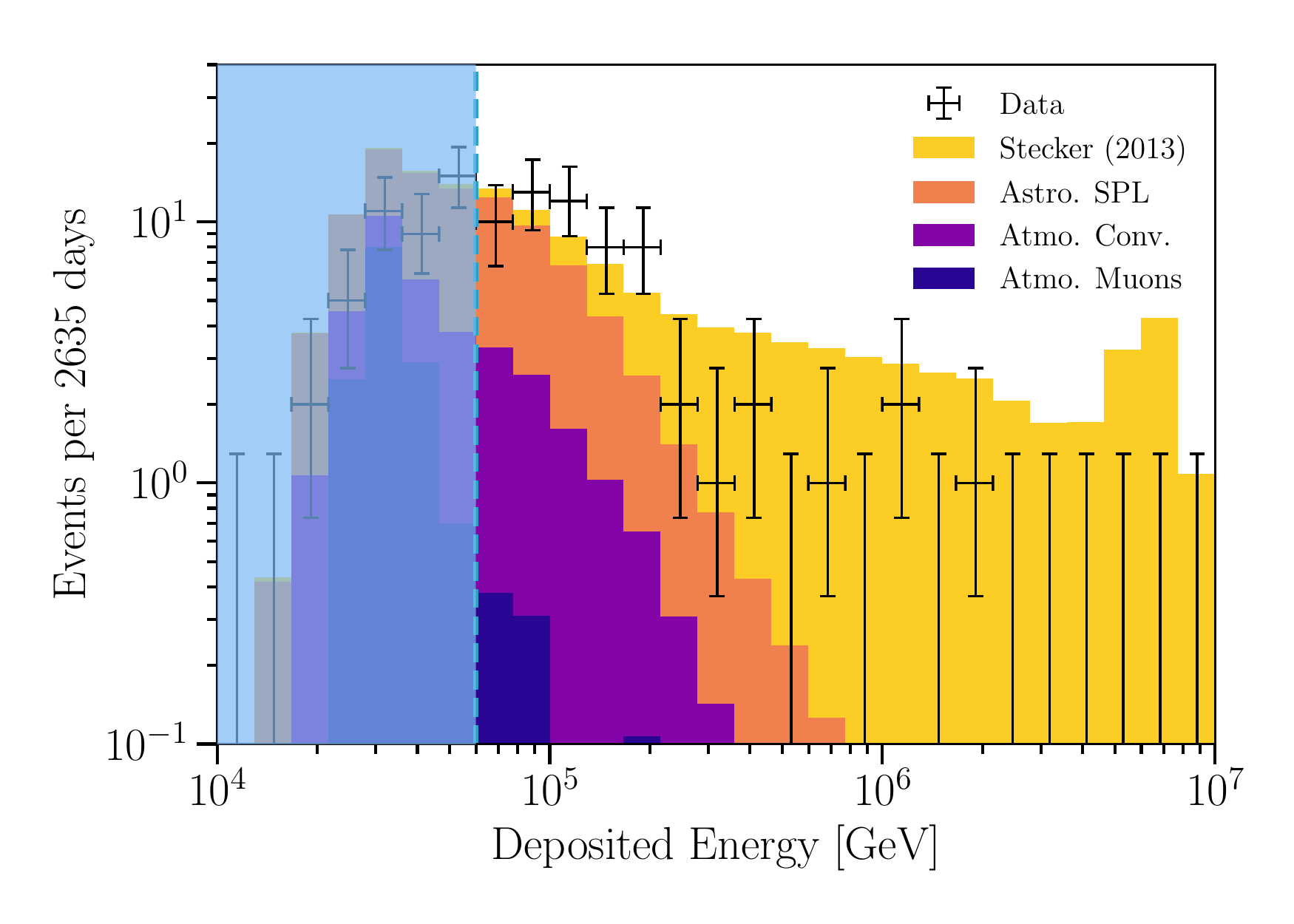}}
    \subfloat{\includegraphics[width=0.45\linewidth]{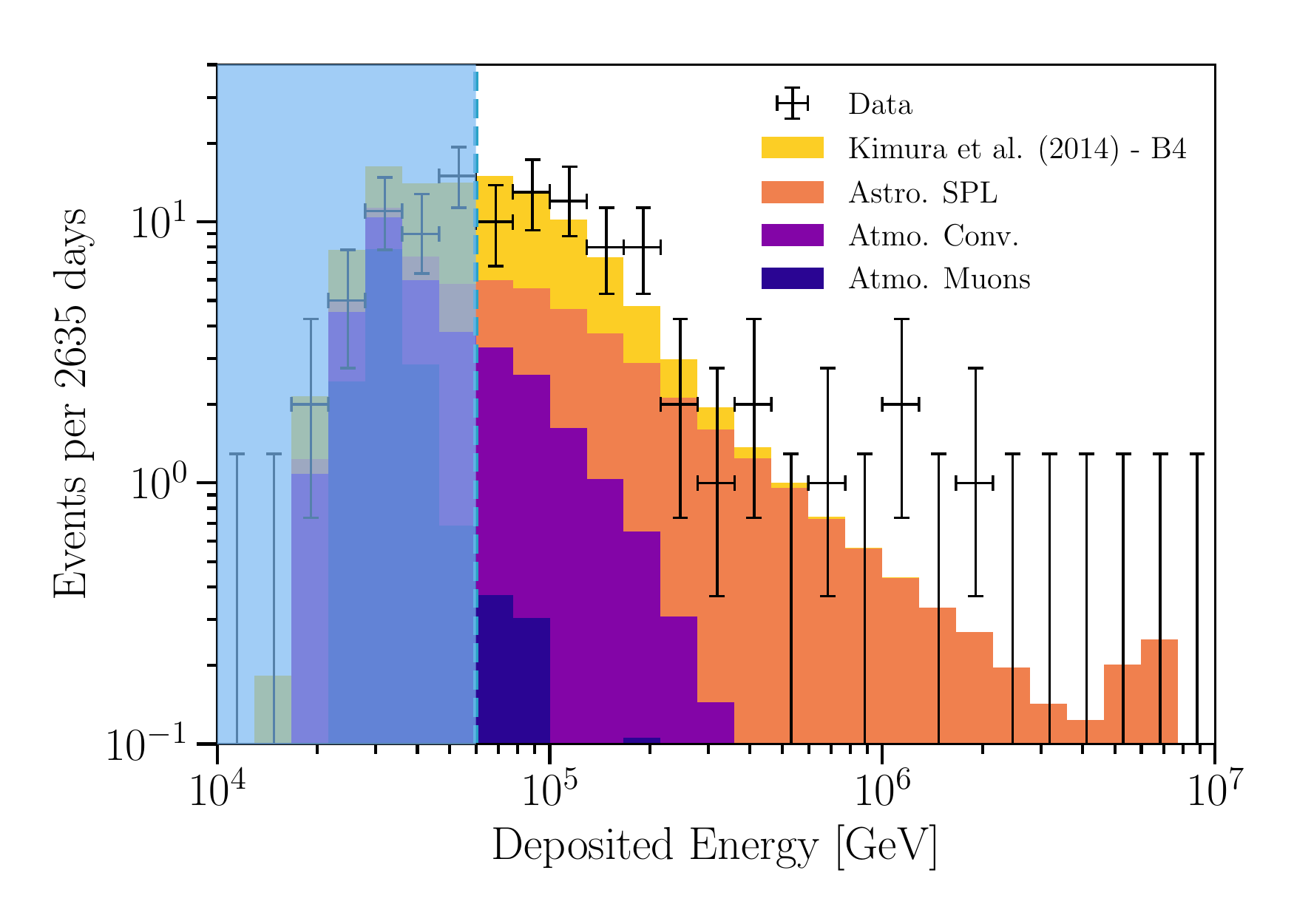}}
    \caption{\textbf{\textit{Source-specific model energy distributions.}}
    Different panels show the predicted energy distribution compare to the data for a subset of the models considered in \reftab{tbl:diffuse_models}.
    In the cases shown above there are two astrophysical components: a specific source model and an additional single power-law component.
    All the components are shown as a stacked histogram at the best-fit value of the components normalizations.
    Left: The \Stecker~model is shown as an example of a case where the Bayes factor assuming equally likely alternatives indicates significant preference for the single power-law model over the two component scenario.
    Right: The \KimuraBFour~model is shown as an example of a case where an additional single power law is needed to explain the distribution of events.}\label{fig:source_models}
\end{figure*}

Given the obtained Bayes factors models can be organized into two categories:
\begin{itemize}
    \item Models with Bayes factors much less than one for both analyses.
    In this case, the single power law is a better description of the data and the addition of an additional single power law to the source model does not alter this conclusion.
    Figure~\ref{fig:source_models} (left) provides an example of this category.
    \item Models with Bayes factor much less than one when compared to the single power law, but that improve when introducing an additional single power-law component.
    Models in this category can only describe part of the flux and require the existence of a second component to be compatible with the data.
    Figure~\ref{fig:source_models} (right) provides an example of this category.
\end{itemize}
To conclude, in this section, we have studied models of astrophysical neutrinos proposed in the literature with nominal parameters and compare them to our baseline parameterization, the single power-law spectrum.
No tested scenario with nominal parameters is substantially preferred over the baseline model, and some scenarios -- those with Bayes factors much smaller than unity -- are disfavored; see \reftab{tbl:diffuse_models}.
More extensive tests of models in the literature using the information provided in~\cite{HESEdatarelease} are encouraged.

\section{Conclusions\label{sec:conclusion}}
We have updated the previous analysis by improving the description of the atmospheric background (\refsec{sec:backgrounds}) and incorporating an additional $\SI{4.5}\year$ of data taking.
The largest change to the atmospheric background description is the update of the atmospheric passing fractions to use the calculation from~\cite{Arguelles:2018awr} and the change in the baseline neutrino flux from charmed hadrons to the flux calculated in~\cite{Bhattacharya:2015jpa}.
The detector uncertainties and description have also been improved.
The detector single-photo-electron distributions have been re-calibrated, resulting in a $\sim\SI{4}\percent$ change in the inferred charge on average.
Uncertainties on the absolute detector efficiency and angular acceptance are now included.
Finally, reconstruction uncertainties are now accounted for in a more detailed manner.

The atmospheric-only scenario requires a prompt neutrino normalization $\sim 20$ times larger than the baseline model (\refsec{sec:prompt}).
Such a scenario is ruled out in this work and by prompt normalization upper limits from complimentary IceCube analyses.
The strongest of these limits constrains the normalization to be less than $\sim2.65$ times the BERSS flux at $\SI{100}\TeV$~\cite{Aartsen:2016xlq} at the $\SI{90}\percent$ C.L, and the weakest constrains the normalization to be less than $\sim9.5$ times the BERSS flux at $\SI{100}\TeV$~\cite{Aartsen:2013eka} at the $\SI{90}\percent$ C.L. which is conservative because of the zero observed astrophysical flux.
In an analysis using only this sample, the atmospheric-only solution is disfavored with a decisive criterion according to Jeffreys' scale compared to a model that incorporates a single power-law astrophysical component.
This constraint stems from the inability of an atmospheric-only model to reproduce the observed zenith distribution.
These results reinforce the conclusions of previous analyses~\cite{Aartsen:2013jdh,Aartsen:2014gkd,Aartsen:2015zva} regarding the astrophysical neutrino flux.
Although the prompt component has a distinct angular signature in the HESE sample, the component's normalization is far too small for this analysis to be sensitive without orders of magnitude more data.

The observed data is compared to generic astrophysical neutrino model assumptions using frequentist and Bayesian statistical prescriptions (\refsec{sec:statistics}).
We report the preferred parameter space for a simple single power-law model $\SPLFreqWilksIndexSummary$ (\refsec{sec:spl}) and find this is in agreement with previous results after accounting for the differences in analysis design and the possibility of statistical fluctuations.
The shift in the spectral index comes partially from the extension of the analysis energy range, and as a result of many additional cascade events in the low-energy region of the analysis observed in the latter $\SI{4.5}\year$.
The significance and uncertainties of this astrophysical measurement remain robust in the face of additional systematics.
Results from other IceCube samples differ in their best-fit parameters, but overlap in their $\SigmaTwo$ C.L. regions.

Comparisons of the data to generic astrophysical neutrino models have also been presented, such as a double power law (\refsec{sec:dpl}), a spectral cutoff (\refsec{sec:cutoff}), and log-parabola (\refsec{sec:log_parabola}).
No preference is found for any of these models compared to the single power law; in particular, the existence of a cutoff is constrained to be at energies greater than $\BayesFactorStrongCutoff$ with a strong criterion according to Jeffreys' scale and similarly excluded below $\CutoffWilksLowerCutoffNinety$ at the $\SI{90}\percent$ C.L.

As a new part of this analysis, the preference for some specific source scenarios proposed in the literature~\cite{Stecker:2013fxa,Fang:2017zjf,Kimura:2014jba,Padovani:2015mba,Senno:2015tsn,Bartos:2015xpa,Tavecchio:2014eia,Biehl:2017hnb} is quantified assuming nominal parameters of the models with respect to the single power law using the Bayes factor.
Some models provide a compatible description of the data, while others fail to explain the observations even with the addition of a power-law component.
No scenario tested provides a substantially improved description of the data compared to the single power-law, which describes the data well.
Models that do not describe the data well with nominal parameters may do so with other model parameters, and other models not considered may provide better descriptions of the data.
On this point, readers are encouraged to use the data release~\cite{HESEdatarelease} to perform more detailed tests.

Other measurements have been performed with this sample in a manner consistent with the analysis techniques presented here.
These are the measurement of the neutrino flavor composition~\cite{HESETAU}, searches for additional neutrino interactions~\cite{HESEFLV} and dark matter in the galactic core~\cite{HESEDM}, and a measurement of the neutrino cross-section~\cite{HESEXS}.

We conclude that, given the available data in this sample's sensitive energy range, the astrophysical neutrino flux is well described by a single power law, and there is no evidence for additional spectral structure in this sample.
Despite this, many models remain compatible with the data, and larger samples will be required to differentiate between the different proposed spectra.
The spectral index uncertainty has been reduced by adding an additional $\SI{4.5}\year$ of data; however, this addition has also shifted the measurement to a softer spectrum, away from other measurements.
These other measurements also have reduced uncertainties with the addition of more data.
Although a purely statistical explanation cannot be ruled out, differences between these spectral measurements remain unexplained and will require further study to resolve.

\begin{acknowledgments}
The IceCube collaboration acknowledges the significant contributions to this manuscript from Carlos Arg\"uelles, Austin Schneider, Juliana Stachurska, and Tianlu Yuan.

We acknowledge the support from the following agencies:
USA {\textendash} U.S. National Science Foundation-Office of Polar Programs,
U.S. National Science Foundation-Physics Division,
Wisconsin Alumni Research Foundation,
Center for High Throughput Computing (CHTC) at the University of Wisconsin{\textendash}Madison,
Open Science Grid (OSG),
Extreme Science and Engineering Discovery Environment (XSEDE),
Frontera computing project at the Texas Advanced Computing Center,
U.S. Department of Energy-National Energy Research Scientific Computing Center,
Particle astrophysics research computing center at the University of Maryland,
Institute for Cyber-Enabled Research at Michigan State University,
and Astroparticle physics computational facility at Marquette University;
Belgium {\textendash} Funds for Scientific Research (FRS-FNRS and FWO),
FWO Odysseus and Big Science programmes,
and Belgian Federal Science Policy Office (Belspo);
Germany {\textendash} Bundesministerium f{\"u}r Bildung und Forschung (BMBF),
Deutsche Forschungsgemeinschaft (DFG),
Helmholtz Alliance for Astroparticle Physics (HAP),
Initiative and Networking Fund of the Helmholtz Association,
Deutsches Elektronen Synchrotron (DESY),
and High Performance Computing cluster of the RWTH Aachen;
Sweden {\textendash} Swedish Research Council,
Swedish Polar Research Secretariat,
Swedish National Infrastructure for Computing (SNIC),
and Knut and Alice Wallenberg Foundation;
Australia {\textendash} Australian Research Council;
Canada {\textendash} Natural Sciences and Engineering Research Council of Canada,
Calcul Qu{\'e}bec, Compute Ontario, Canada Foundation for Innovation, WestGrid, and Compute Canada;
Denmark {\textendash} Villum Fonden, Danish National Research Foundation (DNRF), Carlsberg Foundation;
New Zealand {\textendash} Marsden Fund;
Japan {\textendash} Japan Society for Promotion of Science (JSPS)
and Institute for Global Prominent Research (IGPR) of Chiba University;
Korea {\textendash} National Research Foundation of Korea (NRF);
Switzerland {\textendash} Swiss National Science Foundation (SNSF);
United Kingdom {\textendash} Department of Physics, University of Oxford.
United Kingdom {\textendash} Science and Technology Facilities Council (STFC), part of UK Research and Innovation.
\end{acknowledgments}

\bibliography{hesebiblio}

\clearpage
\newpage
\onecolumngrid
\appendix

\begin{center}
\textbf{\large Appendix}
\end{center}

\section{Expected number of events table\label{sec:events_table}}
In this appendix, we report the expected number of events for each component of the high energy starting event flux for events with greater than $\SI{60}\TeV$ deposited energy in \reftab{tbl:spl_bf_events_high}.
These are computed assuming the single power-law astrophysical flux hypothesis best-fit point given in \reftab{tbl:spl_parameters}.
The components reported are: number of expected atmospheric muons ($N_\mu$), atmospheric neutrinos from the conventional component ($N_{\texttt{conv}}$), atmospheric neutrinos from charmed hadrons ($N_{\texttt{prompt}}$), and from astrophysical neutrinos ($N_{\texttt{astro}}$).
We report these components not only for the analysis energy range, but also split by morphology (cascade, tracks, and double cascades), and split by reconstructed direction (up and down).

\begin{table}[h!]
    \begin{minipage}{\linewidth}
    \begin{tabular}{l c c c c c c}
        \toprule
        $E > \SI{60}\TeV$ & $N_\mu$ & $N_{\texttt{conv}}$ & $N_{\texttt{prompt}}$ & $N_{\texttt{astro}}$ & Total & Data \\
        \midrule
        Total Events	& 0.9 & 9.1 & 0.0 & 48.4 & 58.4 & 60.0 \\
        \midrule
        Up	& 0.0 & 5.5 & 0.0 & 18.1 & 23.5 & 21.0 \\ 
        Down	& 0.9 & 3.6 & 0.0 & 30.3 & 34.9 & 39.0 \\ 
        \midrule
        Cascade	& 0.0 & 4.4 & 0.0 & 38.0 & 42.4 & 41.0 \\ 
        Track	& 0.9 & 4.5 & 0.0 & 8.2 & 13.7 & 17.0 \\ 
        Double Cascade	& 0.0 & 0.2 & 0.0 & 2.1 & 2.3 & 2.0 \\ 
        \bottomrule
    \end{tabular}
    \end{minipage}
    \begin{minipage}{\linewidth}
    \caption{\textbf{\textit{Single power-law best-fit event expectations between $\SI{60}\TeV$ and $\SI{10}\PeV$.}}
    The left-most column indicates the event category, which corresponds to a particular choice of morphology, or a direction.
    The right-most column shows the number of data events observed for a given category.
    Each intermediate column corresponds to the expected number of events in the sample for a given source category: atmospheric muons ($N_\mu$), conventional atmospheric neutrinos ($N_{\texttt{conv}}$), prompt atmospheric neutrinos ($N_{\texttt{prompt}}$), astrophysical neutrinos ($N_{\texttt{astro}}$), and total number of events from all source categories.}
    \label{tbl:spl_bf_events_high}
    \end{minipage}
\end{table}
\FloatBarrier

\section{Sideband distributions\label{sec:sidebands}}
\begin{figure}[hbt!]
    \centering
    \includegraphics[width=0.5\linewidth]{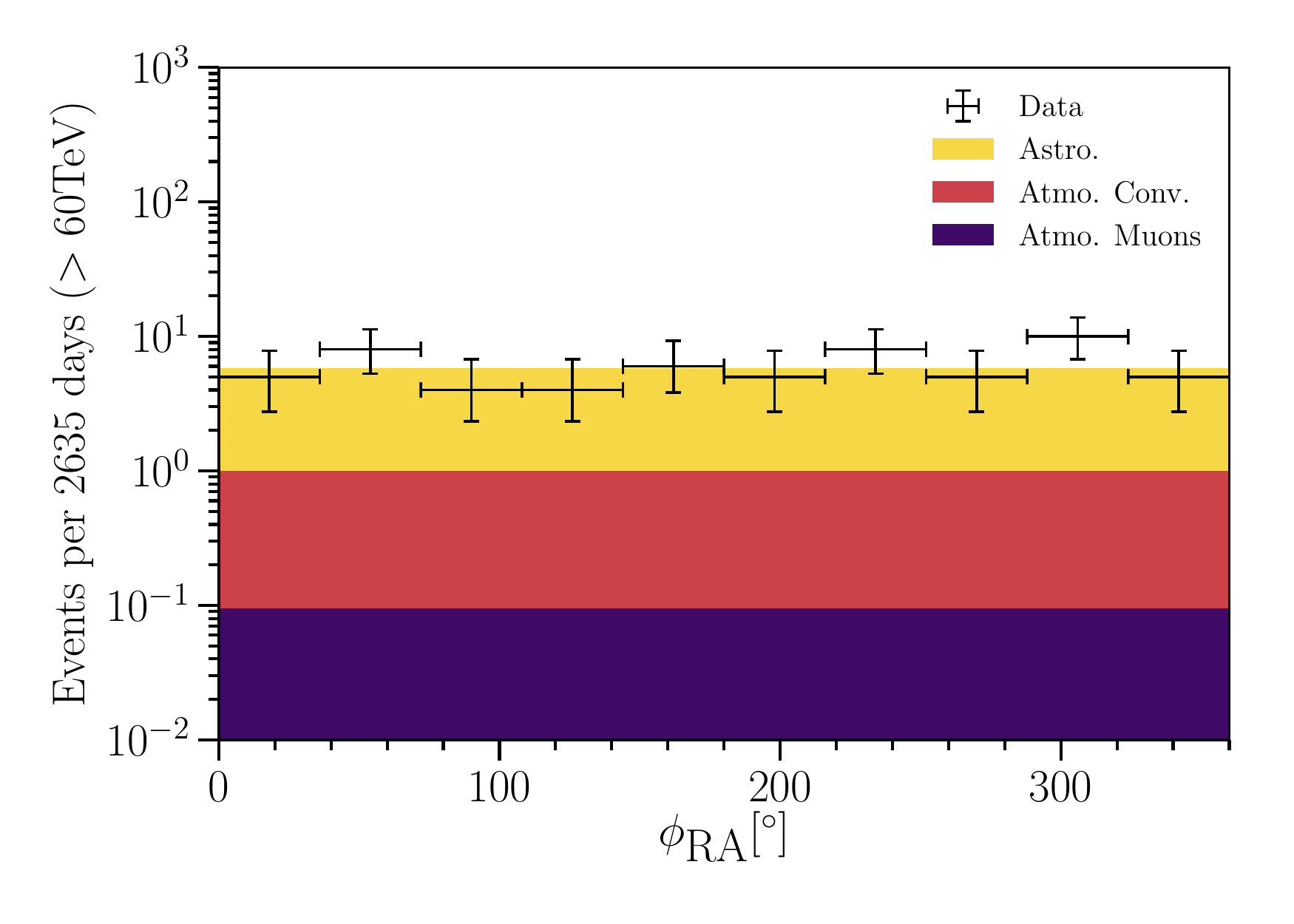}
    \caption{\textbf{\textit{Right ascension distribution.}}
    Events above $\SI{60}\TeV$ are shown together with the stacked expectation of different components using simulation weighted with the best-fit parameters for the single power-law astrophysical model.
    This figure has the same format as \reffig{fig:energy-zenith}}\label{fig:events_ra}
\end{figure}

The energy, zenith, and morphologies of the events are used to measure the atmospheric and astrophysical neutrino components.
For a dominant astrophysical isotropic component the right ascension distribution should be uniform.
During the unblinding process of this sample, this distribution was used as a control distribution.
This distribution is shown in \reffig{fig:events_ra}.

\FloatBarrier
\section{Effects of systematics in analysis distributions\label{sec:extendedsys}}
In this section we describe in more detail the implementation of the detector and ice systematic uncertainties discussed in \refsec{sec:detector_systematics}.
The most relevant detector and ice systematics have been implemented in the analysis via continuous parameterizations.
These are the absolute DOM efficiency ($\domeff$), the head-on DOM efficiency ($\holeice$), and ice anisotropy scale ($\anisotropy$).
The effect, of each systematic parameter, is different for each of the morphology bins in the analysis.
In this section, we show the effect of the systematics across all morphologies weighted at the best-fit point composition.

\begin{figure*}[!ht]
    \centering
    \subfloat{\includegraphics[width=0.3\linewidth]{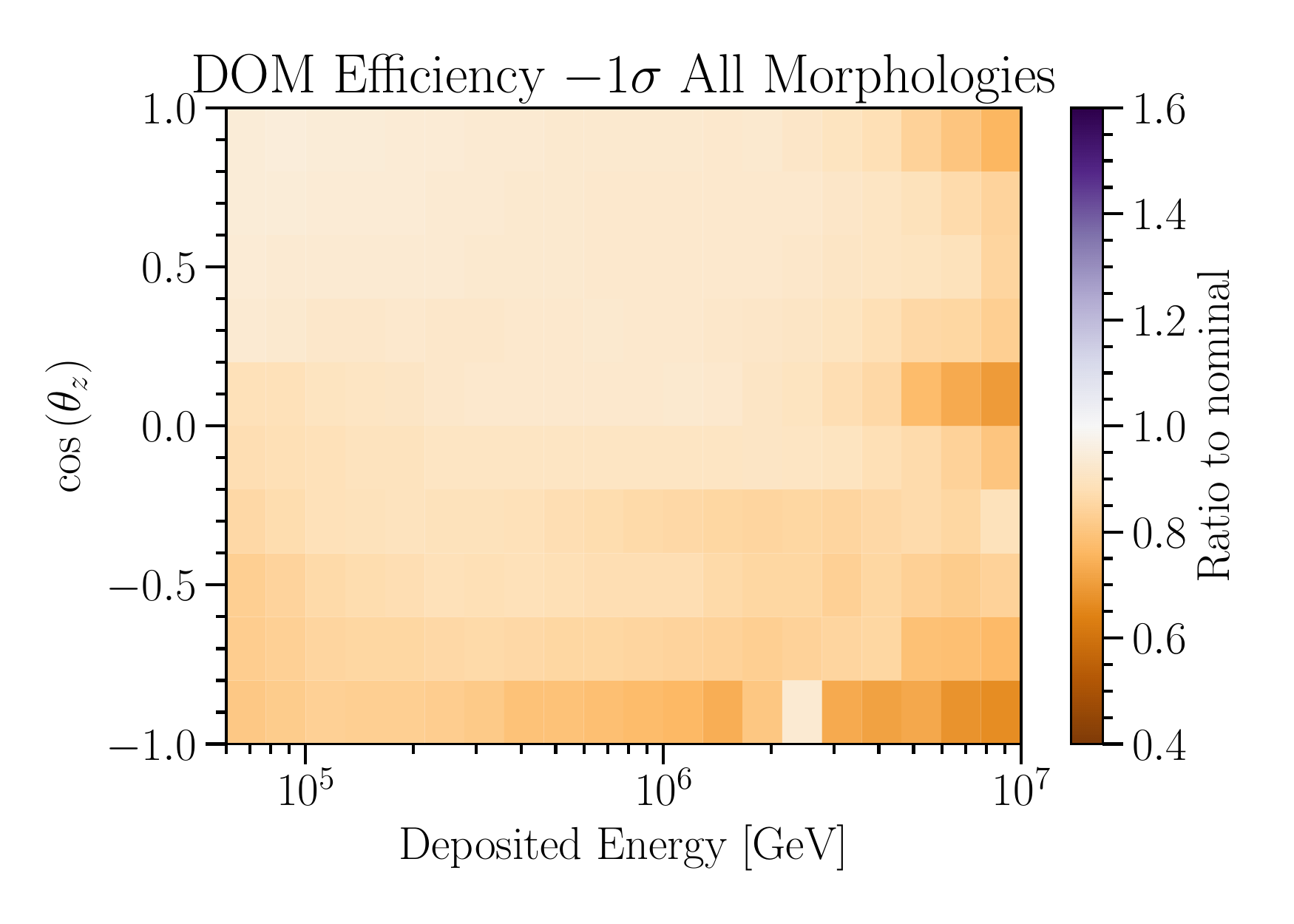}}
    \subfloat{\includegraphics[width=0.3\linewidth]{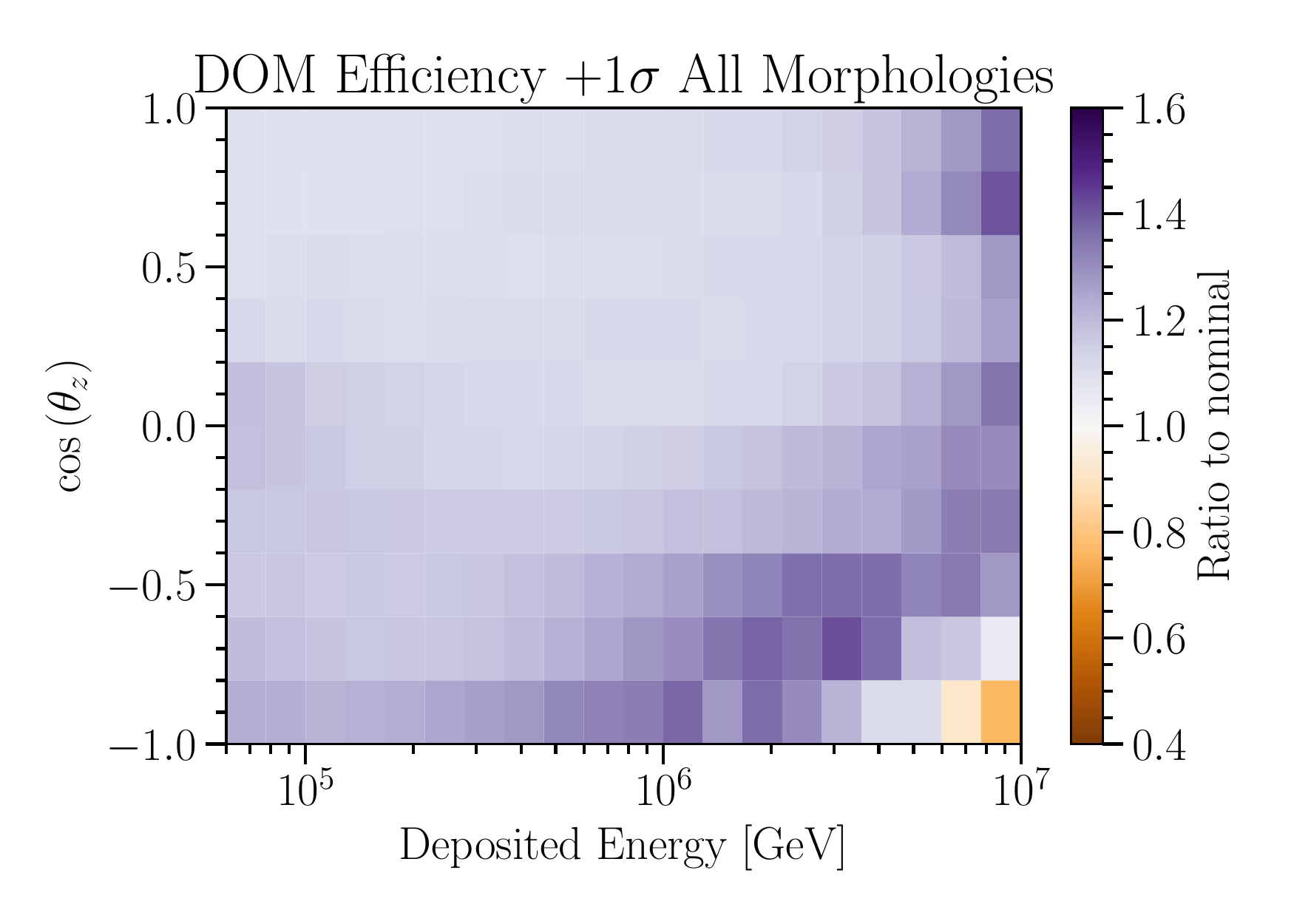}}
    \subfloat{\includegraphics[width=0.3\linewidth]{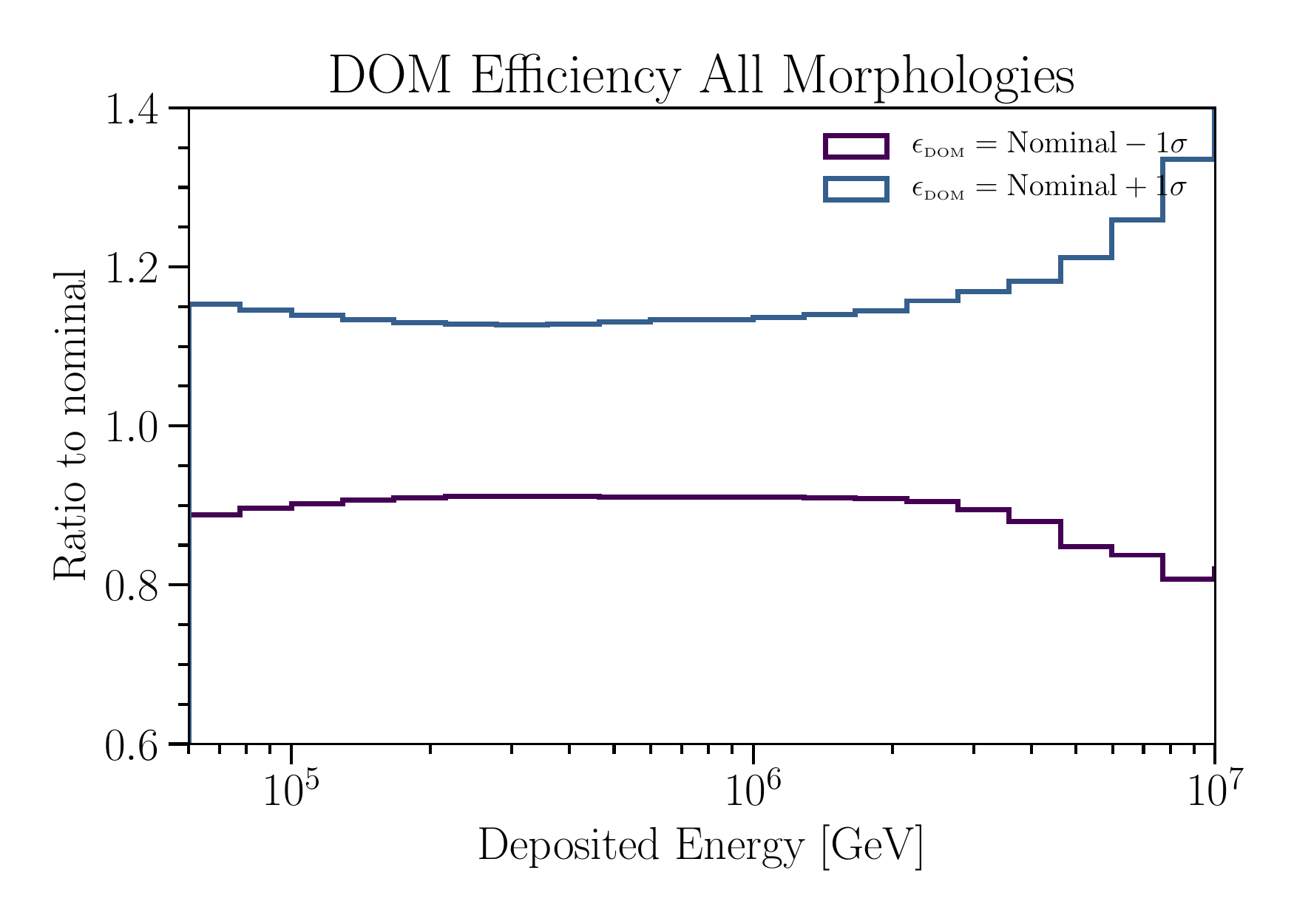}}
    \caption{\textbf{\textit{Effect of changing the DOM efficiency on the sample.}}
    Left and center panels show the ratio of the systematic expectation with respect to the nominal as a function of the deposited energy and the cosine of the zenith angle for decreasing and increasing the efficiency by one sigma respectively.
    The color scale is proportional to the change and less saturated colors correspond to lesser change.
    The right-most panel shows the ratio of energy distributions, when decreasing and increasing the efficiency, as dark and light lines as a function of the deposited energy.
    In these three panels all morphologies are considered.}\label{fig:domeff}
\end{figure*}

\reffig{fig:domeff} shows the change in event rate, with respect to the analysis nominal simulation, in the deposited energy and zenith distributions when shifting the $\domeff$ by one sigma (the prior width).
The main effect of this parameter is to increase or decrease the rate in the sample, having minor deposited energy and zenith dependence.
Thus, this systematic has the largest impact in the absolute normalization of the reported fluxes.

\begin{figure*}[!ht]
    \centering
    \subfloat{\includegraphics[width=0.3\linewidth]{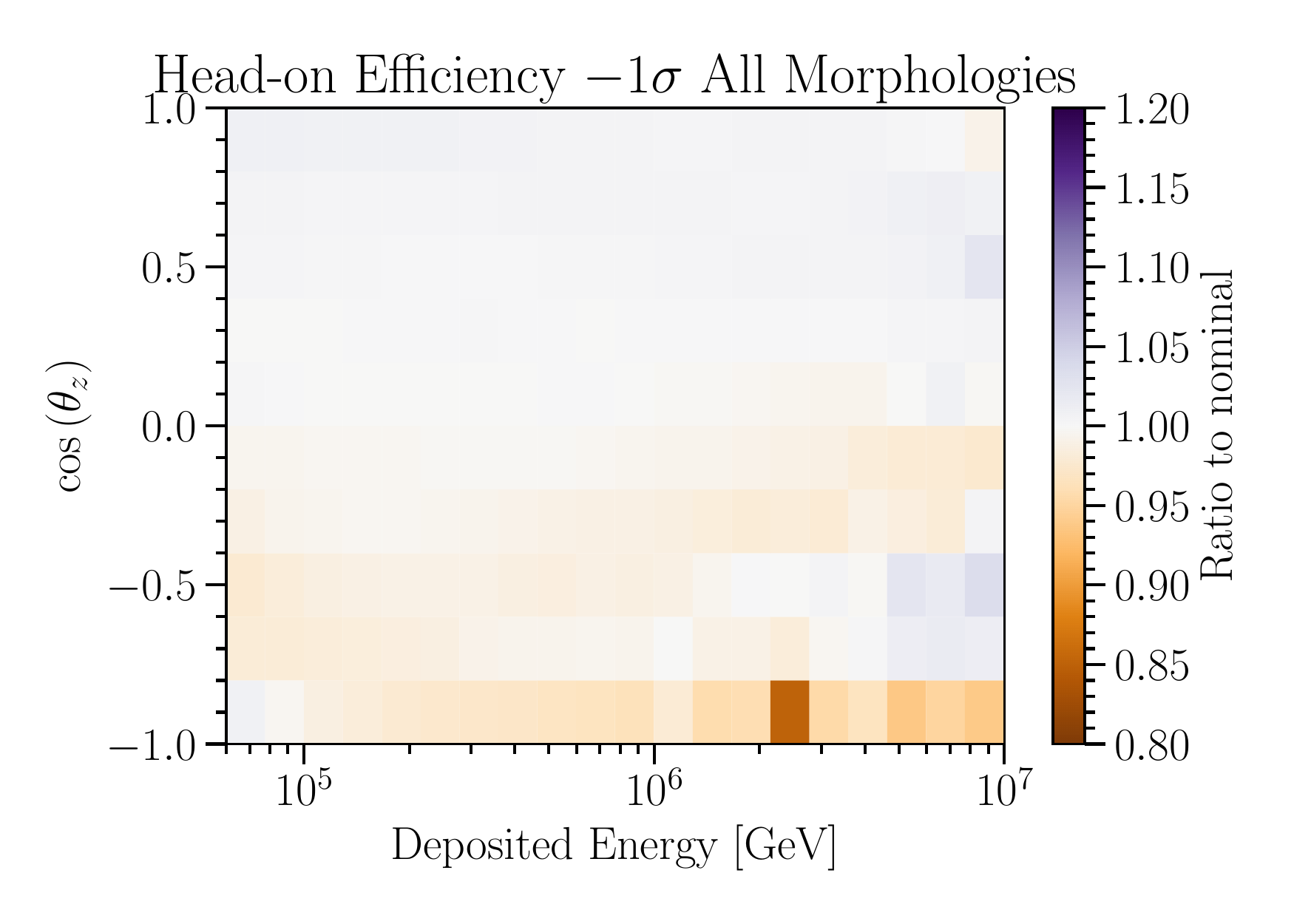}}
    \subfloat{\includegraphics[width=0.3\linewidth]{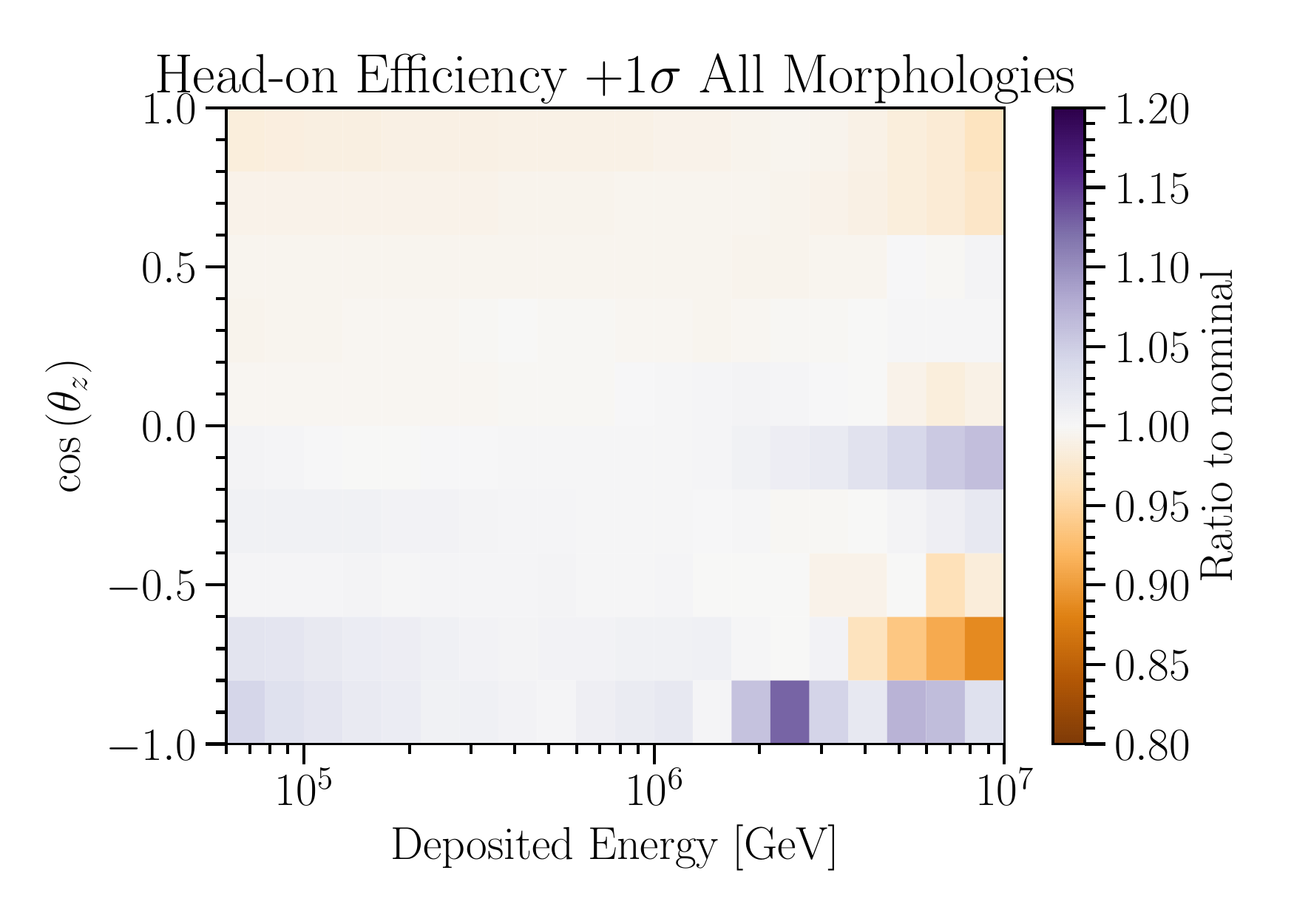}}
    \subfloat{\includegraphics[width=0.3\linewidth]{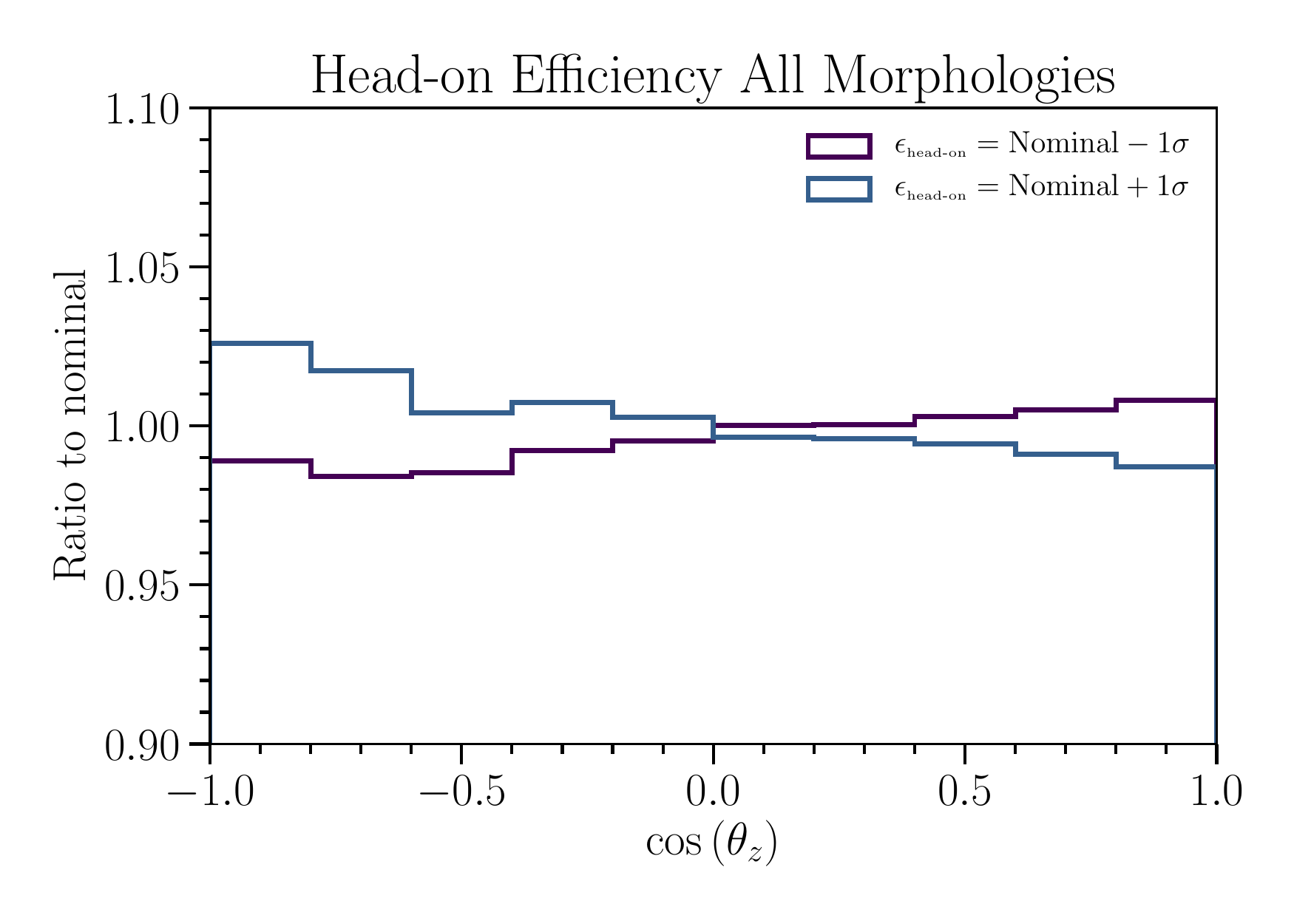}}
    \caption{\textbf{\textit{Effect of changing the hole ice on the sample.}}
    The layout and colors have the same meaning as in \reffig{fig:domeff}, and show the effect of decreasing and increasing the head-on efficiency.
    Since this effect is predominantly an effect in the angular distribution, the right most panel shows the ratio to nominal as a function of the cosine of the zenith angle.
    In these three panels all morphologies are considered.}\label{fig:holeice}
\end{figure*}

\reffig{fig:holeice} shows the relative rate change when modifying the DOM head-on efficiency, $\holeice$, by one sigma (the prior width).
In this case, the effect of the systematic is mostly energy dependent, but -- as expected for a change in the DOM angular response -- it introduces a change in the angular distribution.

\begin{figure*}[!ht]
    \centering
    \subfloat{\includegraphics[width=0.3\linewidth]{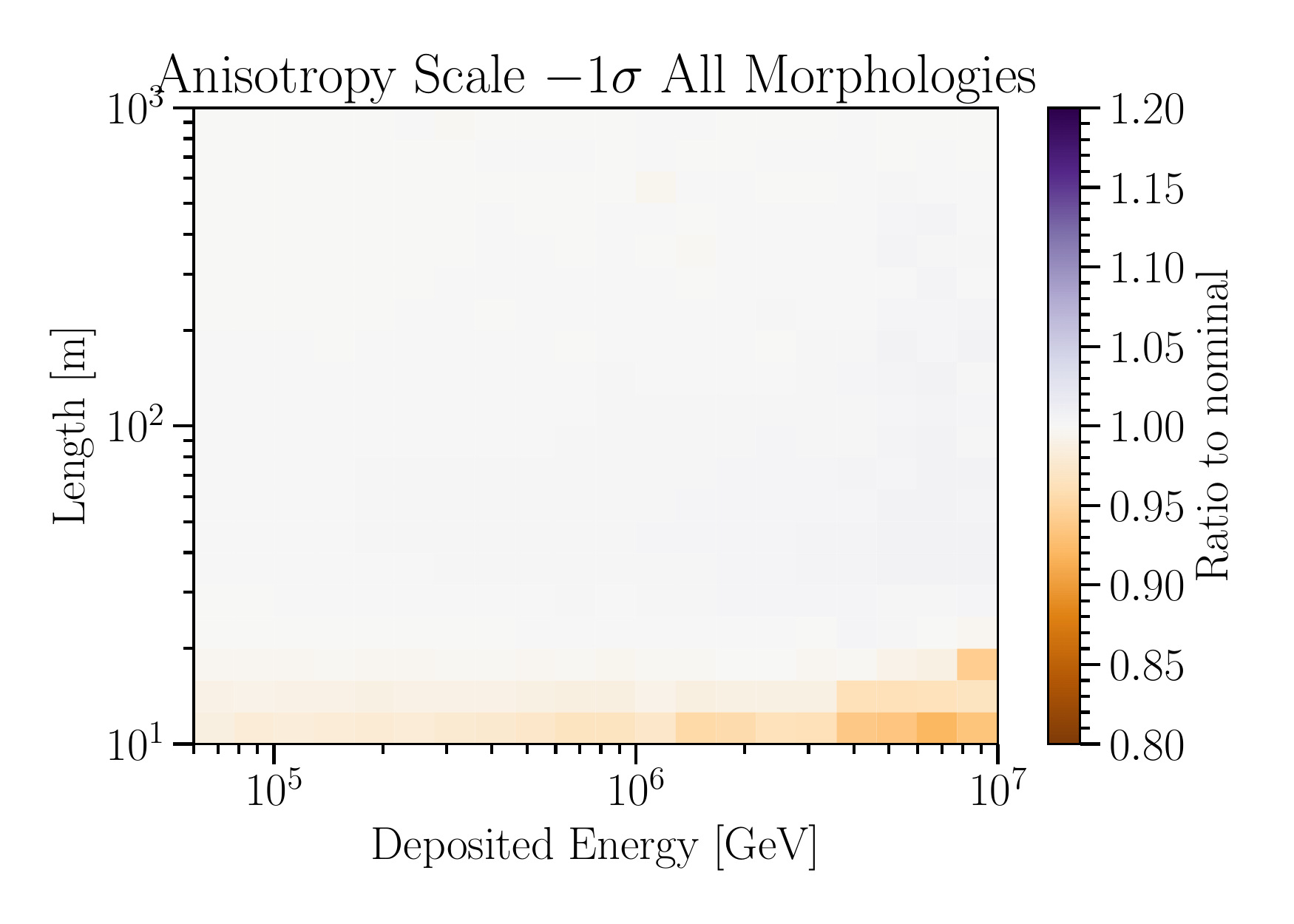}}
    \subfloat{\includegraphics[width=0.3\linewidth]{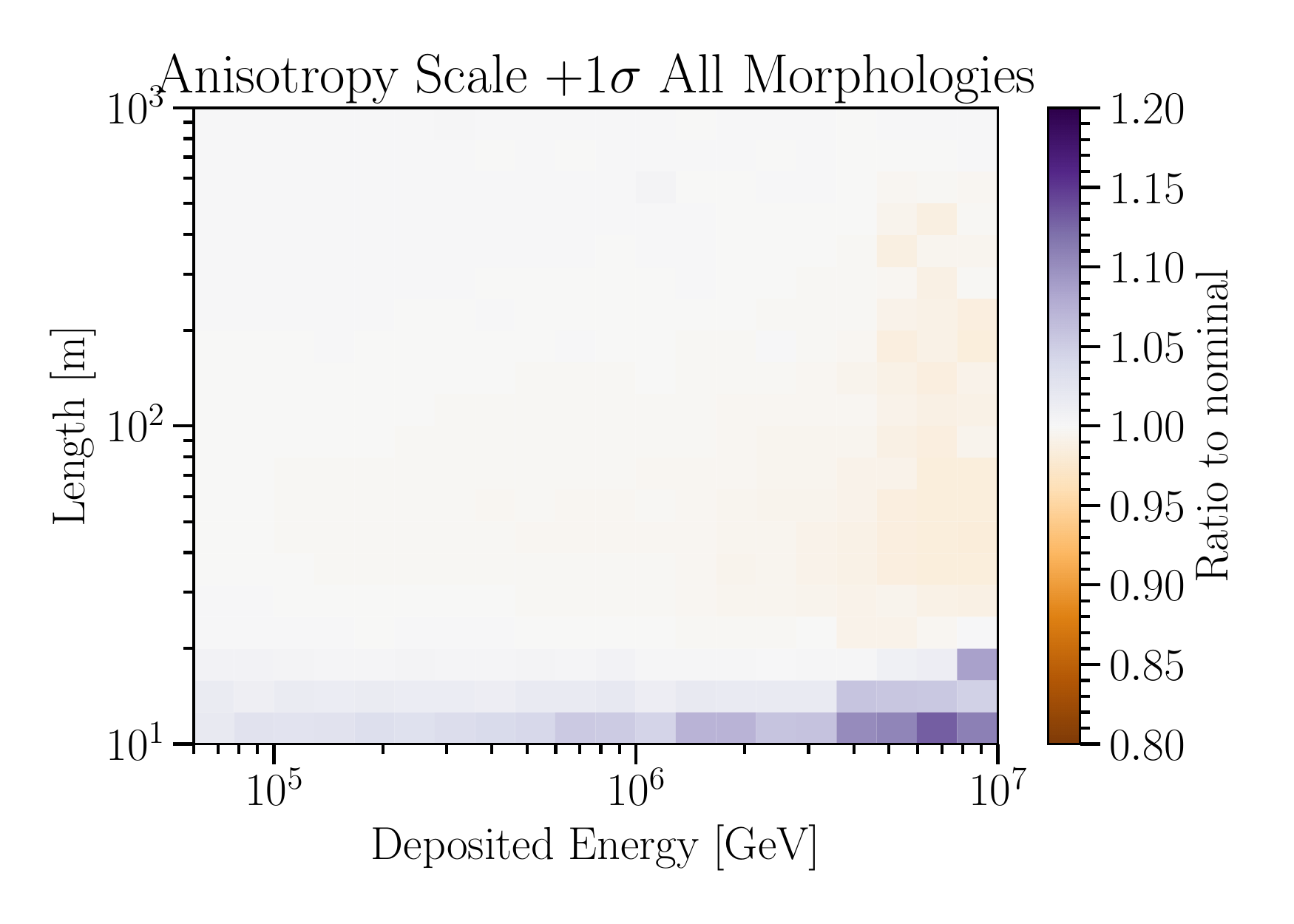}}
    \subfloat{\includegraphics[width=0.3\linewidth]{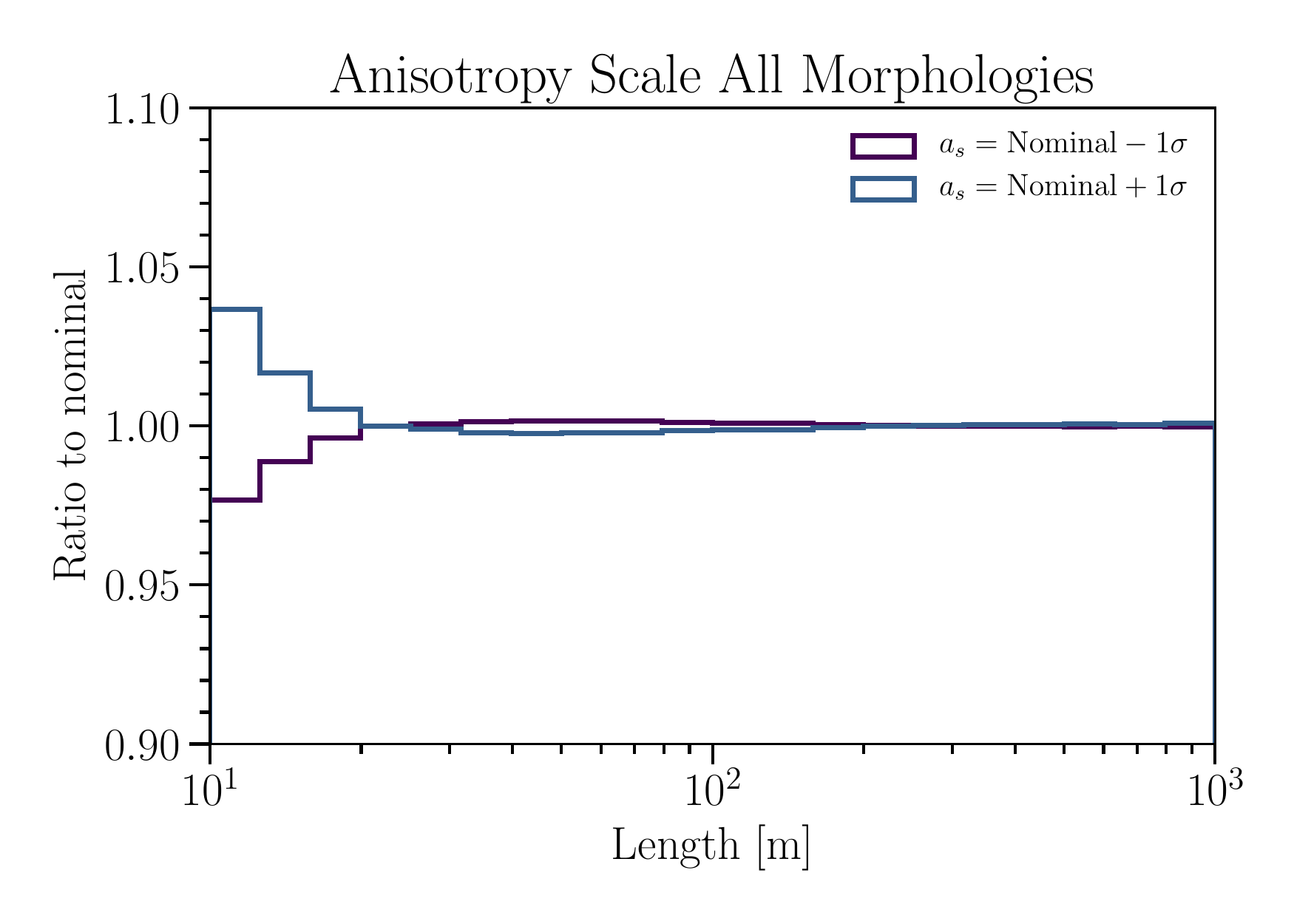}}
    \caption{\textbf{\textit{Effect of changing ice anisotropy on length distribution in the sample.}}
    The layout and colors have the same meaning as in \reffig{fig:domeff}, and show the change relative the baseline to ice model anisotropy.
    The right most panel shows the ratio of length distributions, when decreasing and increasing the anisotropy, as dark and light lines as a function of the reconstructed length.
    In these three panels all morphologies are considered.}\label{fig:anisotropy}
\end{figure*}

\reffig{fig:anisotropy} shows the effect of changing the bulk ice anisotropy parameter by one sigma (the prior width) in the length and deposited energy distributions.
This systematic has a very small effect on the deposited energy distribution, but a significant one in the double-cascade vertex separation distribution.
Close to the major anisotropy axis, events reconstructed as single cascades become more elongated with increasing anisotropy scale.
The elongation of these cascades can cause some events to migrate to the double cascade classification.
Double cascades that lie along the major axis have the same elongation behavior with increasing anisotropy scale, causing some events to migrate to bins of larger reconstructed length.
This behavior is nearly reversed along the minor anisotropy axis.

\FloatBarrier

\section{Single photo-electron charge distribution calibration\label{sec:charge_calibration}}
Each of IceCube's digital optical modules uses a discriminator circuit to trigger on the PMT voltage level and a selection of digitizers with different time windows, digitization rates, and dynamic range to record the PMT output after a trigger.
To perform accurate physics measurements we must be able to convert the digitized waveforms into deposited charge time stamps.
This is achieved with {\it in situ} measurements of a variety of calibration constants for each DOM.
The voltage and timing calibration is described in detail by~\cite{Aartsen:2016nxy}.
The PMTs in the DOMs operate at a gain of $10^7$.
In the PMT operation that results in this gain, the process of producing knock-off electrons is highly stochastic, meaning that a single photo-electron can potentially give rise to a broad distribution of charges at the final stage.
To achieve the target gain of $10^7$, the PMT gain as a function of high voltage is calibrated using the response to background ``dark-noise'' photons.
For single photo-electrons (SPEs) the charge before amplification $e$ is known.
By integrating the voltage samples (as a function of time, and dividing by the load resistance) of many single photo-electron waveforms, a histogram of response charges is formed for each high voltage setting.
The histogram of response charges above a certain charge is fit with the sum of an exponential and Gaussian distribution.
The peak of the Gaussian component (SPE peak) is used to set the PMT high voltage such that the gain at this point in the distribution is $10^7$.
The discriminator threshold is then set to be $0.25$ times the voltage that corresponds to the Gaussian peak.
This calibration procedure is repeated at the start of each IceCube season.
The stability of the DOM operating conditions makes this calibration frequency sufficient.

After the digitizer readout and gain of the PMT have been calibrated, the interpretation of the digitized waveform readout as deposited charge then relies on the calibration of the single photo-electron charge distribution (SPE curve).
Previous IceCube analyses used the same sum of exponential and Gaussian distributions that were fit during the calibration procedure to interpret the readout waveforms.
However, in 2015 further investigation revealed that a more accurate procedure than that used for initial calibration resulted in SPE peaks that were on average $\SI{4.3}\percent$ higher than expected.
This realization prompted a detector wide re-calibration effort for all seasons of IceCube data, internally referred to as ``Pass2.''
The offset in the initial calibration procedure is in part due to contamination from multi-photo-electron pulses and failures of the fitting procedure to match the data well.
The Pass2 re-calibration procedure uses raw waveform information collected via an unbiased random filter designed to capture detector noise and background muon events.
In a manner similar to the calibration procedure, the integrated charges from each digitizer are first deconvolved from electronics effects and then binned so that a function can be fit to the distribution.
Depending on the available information and fit quality an exponential plus Gaussian distribution or the sum of an exponential and two Gaussian distributions is fit to each histogram above the 5th percentile of the data.
This procedure is performed on a per season basis for each digitizer on every DOM.
Once the new SPE curves have been obtained, the data is reinterpreted accordingly and processed with the filters used for the 2017 data taking season in order to provide a uniform set of filters for all the re-processed years of data.

The changes introduced by the new filters are small for most IceCube analyses.
However, the re-calibration of the SPE curves has resulted in a change to inferred deposited charge, and therefore energy, of approximately $\SI{-4}\percent$ on average.
This systematic bias was partially accounted for in other IceCube analyses that consider the uncertainty in the absolute photon efficiency of the DOMs.
This analysis and others forthcoming use the new Pass2 calibration to obtain more accurate estimations of the deposited charge and energy of events in the detector.
Notably in this analysis, the reconstructed deposited energy has changed in comparison to previous analyses of this sample due to the charge re-calibration.
As the selection uses cuts on the charge of events which have not been altered, some events have been removed from the sample after re-calibration.
Further calibration efforts of the DOM response are ongoing and have since yielded more accurate measurements that will be used in future analyses~\cite{IceCube:2020nwx}.
\FloatBarrier

\section{Detailed likelihood description\label{sec:likelihood}}
The analysis likelihood is given by
\begin{align}
    \begin{split}
	    \like(\vec\theta, \vec\eta) ={}& \left(\prod_j^{n} \likeSAY(\mu_j(\vec\theta,\vec\eta), \sigma_j(\vec\theta,\vec\eta); d_j) \right) \cdot \left(\prod_r \Pi_r(\theta_r)\right) \cdot \left(\prod_s \Pi_s(\theta_s) \right),\label{eq:likelihood_appendix}
    \end{split}
\end{align}
where $1\leq j\leq n$ refers to the bin number, $r$ indexes the parameters of interest, $s$ indexes the nuisance parameters, $\theta_r$ are the different parameters of interest, $\eta_s$ are the different nuisance parameters, $\Pi_r$ are the priors on the parameters of interest, $\Pi_s$ are the priors on the nuisance parameters, $\mu_j$ refers to the expectation in bin $j$, $\sigma_j$ refers to the standard deviation of the expectation in bin $j$, $d_j$ refers to the number of data events in that bin, $\vec\theta$ refers to all the parameters of interest, $\vec\eta$ refers to all the nuisance parameters, and $\likeSAY$ is the effective likelihood described in~\cite{Arguelles:2019izp}.
The arguments of the likelihood are then
\begin{align}
    \begin{split}
        \mu_j(\vec\theta,\vec\eta) = \sum_i w^j_i(\vec\theta,\vec\eta)~\textmd{and}~\sigma_j(\vec\theta,\vec\eta) = \sqrt{\sum_i {\left(w^j_i(\vec\theta,\vec\eta)\right)}^2},
    \end{split}
\end{align}
where $w_i^j$ are the weights in bin $j$.
The event weights have contributions from each flux component, which we enumerate as $\texttt{conv}$, $\texttt{prompt}$, $\texttt{muon}$, and $\texttt{astro}$ for the conventional atmospheric neutrino, prompt atmospheric neutrino, atmospheric muon, and astrophysical neutrino fluxes, respectively.
We also split the weights into their flux dependence, $\alpha_i^\texttt{component}$, and systematic corrections, $\beta_i^\texttt{component}$.
As neutrino events and atmospheric muons are simulated separately, $w_i^\texttt{muon}$ is zero if the other weight components are non-zero and vice-versa.
The conventional and prompt systematic corrections are applied in the same manner regardless of neutrino type, whereas the astrophysical systematic corrections are applied on a per-flavor basis.
In symbolic notation, this can be written as:
\begin{align}
	\begin{split}
        w_i ={}& w^\texttt{conv}_i + w^\texttt{prompt}_i + w^\texttt{muon}_i + w^\texttt{astro}_i,
	\end{split}\\
	\begin{split}
        w_i^\texttt{conv} ={}& \alpha^\texttt{conv}_i \beta^\texttt{conv}_i,
	\end{split}\\
	\begin{split}
        w_i^\texttt{prompt} ={}& \alpha^\texttt{prompt}_i \beta^\texttt{prompt}_i,
	\end{split}\\
    \begin{split}
        w_i^\texttt{muon} ={}& \alpha^\texttt{muon}_i,
    \end{split}\\
	\begin{split}
        w_i^\texttt{astro} ={}& \alpha^\texttt{astro}_i \beta^{\texttt{astro}, \|p_i\|}_i,
	\end{split}
\end{align}
where $p_i$ denotes the particle type, and $\|p_i\|$ denotes the particle type irrespective of whether it is an anti-particle or not.

To correct for differences between the simulated event generation probability distributions and those of another hypothesis given by $\vec\theta$ and $\vec\eta$, we must concern ourselves with the true simulated properties of each simulation-event's primary particle.
We denote $p_i$ as the particle type of the event's primary particle, $E_i$ as the primary particle's initial energy, $\theta_i^{z}$ as the primary's zenith angle, and $d_i$ as the primary's depth at the first intersection with a cylinder centered around and containing the detector.
The main correction is between the generated distribution of neutrinos and a baseline flux of neutrinos for a particular livetime.
This correction necessitates a weighting factor $(\Phi \cdot T) / (N \cdot P)$, where $\Phi$ is the differential flux, $T$ the livetime of the sample, $N$ is the number of generated events, and $P$ is the probability density of event generation.
Additional corrections can account for deviations from the baseline model, and effects not simulated.

The conventional component considers the neutrino flux from pions ($\pi$) and kaons (K) separately.
For a neutrino of type $p_i$, of energy $E_i^\nu$, and with zenith angle $\theta_i^{\nu,z}$, the differential flux of such neutrinos for pions and kaons is $\Phi_\texttt{HONDA}^{\pi, p_i}(E_i^\nu, \theta_i^{\nu, z})$ and $\Phi_\texttt{HONDA}^{K, p_i}(E_i^\nu, \theta_i^{\nu, z})$, respectively~\cite{Honda:2006qj}.
The overall normalization of the conventional atmospheric neutrino flux is modified via the parameter $\convnorm$ and the relative normalizations of the components from pions and kaons are controlled by the parameter $\pik$.
To modify the slope of the conventional atmospheric neutrino spectrum the parameter $\crdeltagamma$ acts as a spectral index correction about a $\SI{2020}\GeV$ pivot point.
The proportion of atmospheric neutrinos and anti-neutrinos is allowed to vary via the parameter ${2\nu/\left(\nu+\bar{\nu}\right)}_\texttt{atmo}$ which can range from zero (all anti-neutrinos) to two (all neutrinos), where the nominal value of one gives the relative neutrino to anti-neutrino content specified in the chosen atmospheric models.
Finally, as we only simulate single neutrinos and not all the products of the cosmic-ray air showers from which they originate, a correction factor is needed to account for the probability that a neutrino event may be rejected by the presence of an accompanying muon.
To account for this the weight is multiplied by the probability that the neutrino is not rejected by an accompanying muon $P_\texttt{passing}^{\texttt{conv},p_i}(E_i^\nu, \theta_i^{\nu, z}, D_i^\nu)$.
In symbolic notation, this is given by:
\begin{align}
    \begin{split}
        \alpha_i^{\texttt{conv}} ={}& \Phi_\texttt{conv} \left(\frac{\Phi_\texttt{HONDA}^{\pi, p_i}(E_i^\nu, \theta_i^{\nu, z})T_\texttt{sample}}{N_\texttt{gen} P_\texttt{gen}^{p_i}(E_i^\nu, \theta_i^{\nu, z})} + R_{K/\pi} \frac{\Phi_\texttt{HONDA}^{K, p_i}(E_i^\nu, \theta_i^{\nu, z})T_\texttt{sample}}{N_\texttt{gen} P_\texttt{gen}^{p_i}(E_i^\nu, \theta_i^{\nu, z})}\right) \cdot {\left(\frac{E_i^\nu}{\SI{2020}\GeV}\right)}^{-\Delta\gamma_\texttt{CR}} \\
        &\cdot P_\texttt{passing}^{\texttt{conv},p_i}(E_i^\nu, \theta_i^{\nu, z}, D_i^\nu)
        \cdot \left\{ \begin{array}{ll} {2\nu/\left(\nu+\bar{\nu}\right)}_\texttt{atmo} & \quad p_i\texttt{ is }\nu \\ {2-2\nu/\left(\nu+\bar{\nu}\right)}_\texttt{atmo} & \quad p_i\texttt{ is }\bar\nu \end{array} \right. .
    \end{split}
\end{align}

The prompt neutrino component considers the flux of atmospheric neutrinos from charmed hadrons $\Phi_\texttt{BERSS}^{p_i}(E_i^\nu, \theta_i^{\nu, z})$~\cite{Bhattacharya:2015jpa}.
Similar corrections as those defined for the conventional component are used for the prompt component.
In this case the normalization of the prompt neutrino flux is controlled by $\promptnorm$.
The parameters $\crdeltagamma$ and ${2\nu/\left(\nu+\bar{\nu}\right)}_\texttt{atmo}$ serve the same purpose in the prompt neutrino component, except that the pivot point is chosen to be $\SI{7887}\GeV$.
A term must also be included to account for the probability of the neutrino being rejected due to accompanying muons: $P_\texttt{passing}^{\texttt{prompt},p_i}(E_i^\nu, \theta_i^{\nu, z}, D_i^\nu)$.
In symbolic notation, this is given by:
\begin{align}
    \begin{split}
        \alpha_i^{\texttt{prompt}} ={}& \Phi_\texttt{prompt}\left(\frac{\Phi_\texttt{BERSS}^{p_i}(E_i^\nu, \theta_i^{\nu, z})T_\texttt{sample}}{N_\texttt{gen} P_\texttt{gen}^{p_i}(E_i^\nu, \theta_i^{\nu, z})}\right) \cdot {\left(\frac{E_i^\nu}{\SI{7887}\GeV}\right)}^{-\Delta\gamma_\texttt{CR}} \\
        &\cdot P_\texttt{passing}^{\texttt{prompt},p_i}(E_i^\nu, \theta_i^{\nu, z}, D_i^\nu)
        \cdot\left\{ \begin{array}{ll} {2\nu/\left(\nu+\bar{\nu}\right)}_\texttt{atmo} & \quad p_i\texttt{ is }\nu \\
        {2-2\nu/\left(\nu+\bar{\nu}\right)}_\texttt{atmo} & \quad p_i\texttt{ is }\bar\nu \end{array}\right. .
    \end{split}
\end{align}

The flux of atmospheric muons from cosmic-ray air showers is modelled by a parameterization of muons from air showers simulated with the \CORSIKA{} package assuming the Hillas-Gaisser H4a~\cite{Gaisser:2013bla} cosmic-ray flux model and SIBYLL 2.1~\cite{Ahn:2009wx} hadronic model.
This parameterized flux is denoted by $\Phi_\texttt{GaisserH4a}(E_i^\nu, \theta_i^{\nu,z},d_i^\nu)$.
As for other fluxes, a normalization factor $\muonnorm$ is included.
The only other correction is to shift the baseline flux to the center of the data derived prior.
This is accomplished with the $2.1\cdot N_\texttt{tagged}^\mu / N_\MUONGUN$ factor.
Where 2.1 is the ratio of detection volumes for the full and reduced volume event selections (not accounting for differences in efficiency), $N_\texttt{tagged}^\mu$ is the number of tagged muons, and $N_\MUONGUN$ is the number of expected events in the baseline atmospheric muon model before re-scaling.
Namely,
\begin{align}
    \begin{split}
        \alpha_i^\texttt{muon} ={}& \Phi_\mu \frac{\Phi_\texttt{GaisserH4a}(E_i^\mu, \theta_i^{\mu,z},d_i^\mu) T_\texttt{sample}}{N_\texttt{gen} P_\texttt{gen}^{p_i}(E_i^\mu, \theta_i^{\mu,z},d_i^\mu)} \cdot \frac{2.1 \cdot N^\mu_\texttt{tagged}}{N_\MUONGUN}.
    \end{split}
\end{align}

The astrophysical component is modeled with a single power law as a baseline, although this flux can be replaced with other models as has been done in \refsec{sec:diffuse}.
The factors accounting for the generation are the same as for the other neutrino fluxes, and a normalization factor $\astronorm$ is included.
The baseline normalization is chosen to be $10^{-18} [\textmd{GeV}^{-1}\textmd{cm}^{-2}\textmd{s}^{-1}\textmd{sr}^{-1}]$ at $\SI{100}\TeV$.
The factor of $2 \pi$ stems from the uniform azimuthal distribution of the astrophysical flux, and $\astrodeltagamma$ governs the index of the spectrum.
In symbolic notation, this is given by:
\begin{align}
    \begin{split}
        \alpha_i^{\texttt{astro}} ={}& \Phi_\texttt{astro} \left(\frac{10^{-18} [\textmd{GeV}^{-1}\textmd{cm}^{-2}\textmd{s}^{-1}\textmd{sr}^{-1}]2 \pi T_\texttt{sample}}{N_\texttt{gen} P_\texttt{gen}^{p_i}(E_i^\nu, \theta_i^{\nu, z})}\right) \cdot {\left(\frac{E_i^\nu}{\SI{100}\TeV}\right)}^{-\gamma_\texttt{astro}}.
    \end{split}
\end{align}

Similarly to true simulated properties, the reconstructed event properties are also needed for the purpose of binning and systematic corrections.
We denote $m_i^R$ as the inferred morphology of event $i$, $E_i^R$ as the reconstructed deposited energy of the event, $\theta_i^{R,z}$ as the reconstructed event zenith angle, and $l_i^R$ as the reconstructed distance between energy depositions in the double cascade reconstruction.
The effects of detector systematic parameters $\epsilon_\texttt{DOM}$, $\epsilon_\texttt{head-on}$, and $a_s$ are assumed to be independent, and are each accounted for by corrections to the expectation stored in b-splines.
For each systematic the corrections for tracks and cascades are applied using the same combination of observables, while double cascades differ.
These are organized as:
\begin{align}
    \begin{split}
        \beta_i^x ={}& \left\{\begin{array}{ll} \beta_i^{\texttt{dc}, x} & \quad m_i^R = \texttt{double cascade} \\ \beta_i^{\texttt{t/c}, x} & \quad m_i^R = \texttt{track}~\textmd{or}~m_i^R = \texttt{cascade} \end{array}\right.
    \end{split}
\end{align}

Parameterized expectations for each systematic $s$ are denoted by $f_s$, and corrections are ratios of $f_s$ evaluated at a specific systematic parameter value to $f_s$ evaluated at the nominal systematic parameter value denoted by $s_0$.
These correction factors are explicitly given by:
\begin{align}
    \begin{split}
        \beta_i^{\texttt{dc}, x} ={}& \left( \frac{f_{\epsilon_\texttt{DOM}}^{x, \texttt{dc}}(E_i^R, \theta_i^{R, z}, \epsilon_\texttt{DOM})}{f_{\epsilon_\texttt{DOM}}^{x, \texttt{dc}}(E_i^R, \theta_i^{R, z}, \epsilon_{\texttt{DOM}, 0})} \right) \cdot \left(\frac{f_{\epsilon_\texttt{head-on}}^{x, dc}(E_i^R, l_i^R, \epsilon_\texttt{head-on})}{f_{\epsilon_\texttt{head-on}}^{x, dc}(E_i^R, l_i^R, \epsilon_{\texttt{head-on},0})} \right) \cdot \left(\frac{f_{a_s}^{x, dc}(l_i^R, a_s)}{f_{a_s}^{x, dc}(l_i^R, a_{s,0})} \right),
    \end{split}\\
    \begin{split}
        \beta_i^{\texttt{t/c}, x} ={}& \left( \frac{f_{\epsilon_\texttt{DOM}}^{x, m_i^R}(E_i^R, \theta_i^{R, z}, \epsilon_\texttt{DOM})}{f_{\epsilon_\texttt{DOM}}^{x, m_i^R}(E_i^R, \theta_i^{R, z}, \epsilon_{\texttt{DOM}, 0})} \right) \cdot \left(\frac{f_{\epsilon_\texttt{head-on}}^{x, m_i^R}(E_i^R, \theta_i^{R, z}, \epsilon_\texttt{head-on})}{f_{\epsilon_\texttt{head-on}}^{x, m_i^R}(E_i^R, \theta_i^{R, z}, \epsilon_{\texttt{head-on},0})}\right) \cdot \left(\frac{f_{a_s}^{x, dc}(E_i^R, a_s)}{f_{a_s}^{x, dc}(E_i^R, a_{s,0})} \right).
    \end{split}
\end{align}
\FloatBarrier

\section{Data release for additional characterization of the astrophysical neutrino flux\label{sec:release}}
The different astrophysical and atmospheric models explored in this paper represent only a small portion of the theoretical model space.
To facilitate better model tests and the combination of results from different experiments and data sets, we provide the data and simulated event information used in the analyses described in \refsec{sec:diffuse}.
The data release information is provided in ~\cite{HESEdatarelease}.

The data used for the analyses described in \refsec{sec:diffuse} is provided in a \texttt{json} formatted file, which contains the 102 data events that pass the selection described in \refsec{sec:selection}.
For each data event we provide the following variables:
\begin{itemize}
    \item \texttt{recoDepositedEnergy} - The reconstructed deposited energy of the event, given in $\si\GeV$.
    \item \texttt{recoMorphology} - The inferred morphology of the event, where \texttt{0, 1, 2} correspond to cascades, tracks, and double cascades, respectively.
    \item \texttt{recoZenith} - The reconstructed zenith direction of the event, given in radians. 
    \item \texttt{recoLength} - The reconstructed length of the event, given in meters.
\end{itemize}

A second \texttt{json} formatted file contains the MC events used to compute the expected data event rates.
For each event we provide the following variables:
\begin{itemize}
    \item \texttt{primaryType} - The simulated initial particle flavor, given in the Monte Carlo numbering scheme outlined by the Particle Data Group~\cite{PhysRevD.98.030001}.
    \item \texttt{primaryEnergy} - The simulated true energy of the initial particle (neutrino or muon), given in $\si\GeV$.
    \item \texttt{primaryZenith} - The simulated true zenith direction of the initial particle (neutrino or muon), given in radians.
    \item \texttt{trueLength} - The simulated true length between cascades in a double cascade the event, given in meters.
    \item \texttt{interactionType} - The simulated neutrino interaction of the initial particle. Values of \texttt{1, 2, 3} correspond to CC, NC, and GR interactions, respectively. For simulated atmospheric muons a value of \texttt{0} is given.
    \item \texttt{weightOverFluxOverLivetime} - The MC weight of the neutrino event, divided by the simulated flux and detector livetime, given in units of $\si\GeV~\si\sr~\si\cm^{2}$. This is set to zero for atmospheric muon events.
    \item \texttt{muonWeightOverLivetime} - The MC weight for each muon event, divided by the detector livetime. This is set to zero for neutrino events.
    \item \texttt{pionFlux} - The nominal conventional atmospheric neutrino flux from pion decay, as described in \refsec{sec:backgrounds}, given in units of $\si\GeV^{-1}\si\s^{-1}\si\sr^{-1}\si\cm^{-2}$.
    \item \texttt{kaonFlux} - The nominal conventional atmospheric neutrino flux from kaon decay, as described in \refsec{sec:backgrounds}, given in units of $\si\GeV^{-1}\si\s^{-1}\si\sr^{-1}\si\cm^{-2}$.
    \item \texttt{promptFlux} - The nominal prompt atmospheric flux neutrino flux, as described in \refsec{sec:backgrounds}, given in units of $\si\GeV^{-1}\si\s^{-1}\si\sr^{-1}\si\cm^{-2}$.
    \item \texttt{conventionalSelfVetoCorrection} - The veto passing fraction for conventional neutrinos, computed as described in \refsec{sec:backgrounds}. 
    \item \texttt{promptSelfVetoCorrection} - The veto passing fraction for prompt neutrinos, computed as described in \refsec{sec:backgrounds}.
    \item \texttt{recoDepositedEnergy} - The reconstructed deposited energy of the event, given in $\si\GeV$.
    \item \texttt{recoMorphology} - The inferred morphology of the event. \texttt{0, 1, 2} correspond to cascades, tracks, and double cascades, respectively.
    \item \texttt{recoLength} - The reconstructed length of the event, given in meters.
    \item \texttt{recoZenith} - The reconstructed zenith angle of the event, given in radians.
\end{itemize}

\begin{figure}[hbt!]
    \centering
    \includegraphics[width=0.5\linewidth]{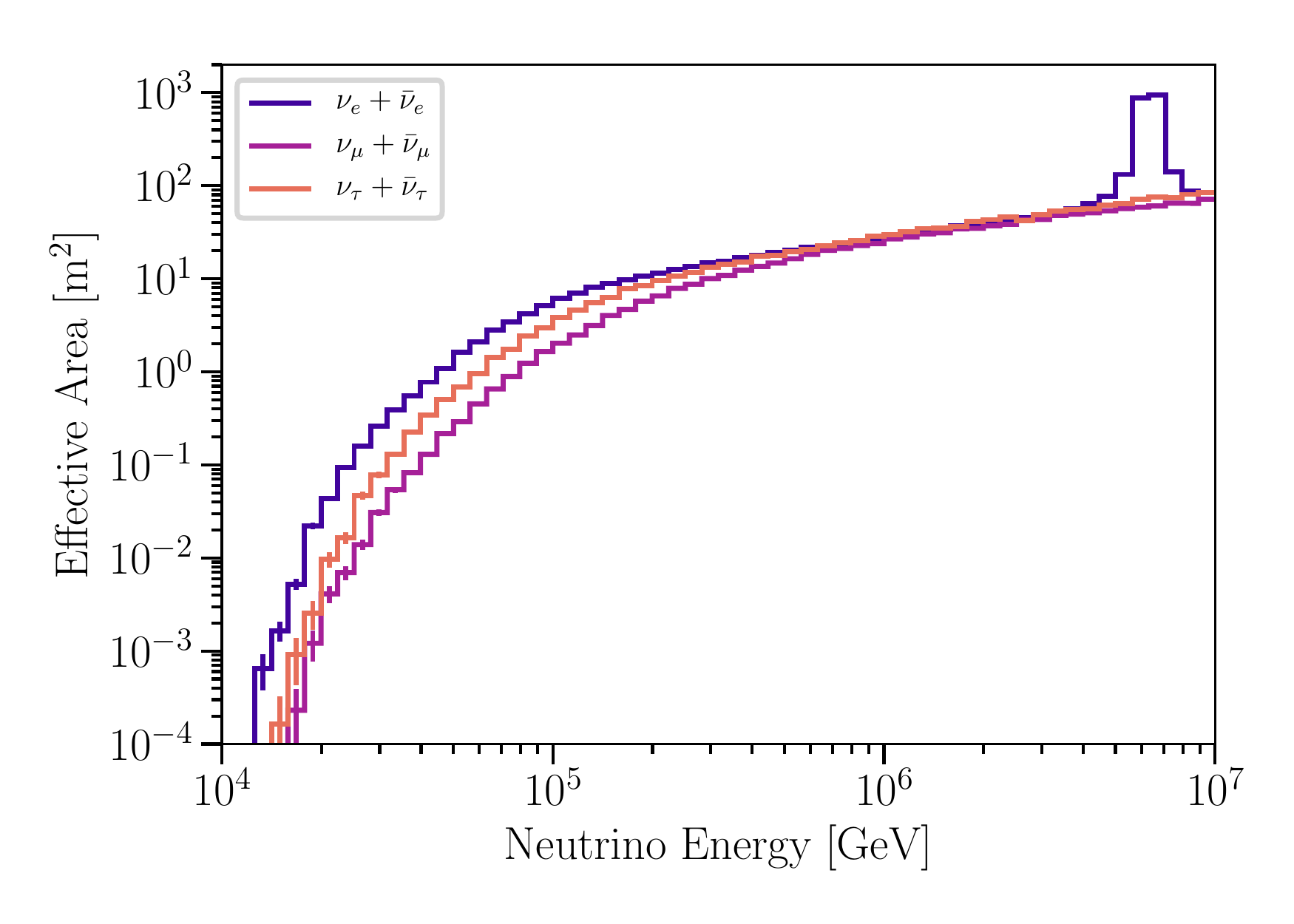}
    \caption{\textbf{\textit{All-sky energy dependent astrophysical neutrino effective area.}}
    The direction averaged effective area for the three neutrino flavors is shown as a function of the neutrino energy incident on Earth.
    This effective area takes into account the effects of absorption in the Earth.
    To obtain the expected number of astrophysical neutrinos, this effective area can be multiplied by the all-sky astrophysical neutrino flux.
    On the other hand, to obtain the expected number of atmospheric events, the atmospheric passing fraction must be included in the calculation of the effective area.}\label{fig:effective_area}
\end{figure}

We provide a \texttt{python3} example code of using the MC information, where the energy dependent neutrino effective area of the selection is computed and the corresponding distributions are plotted.
The output of this code is shown in \reffig{fig:effective_area}.

The interpolating b-splines used to perform the systematic corrections, described in Sections~\ref{sec:detector_systematics} and~\ref{sec:likelihood}, are provided as \texttt{fits} files which can be read with the \texttt{PHOTOSPLINE} software package~\cite{photospline}.

The final component of this data release is a sample code, also written in \texttt{python3}, which reproduces the fit of the data to a single power-law astrophysical flux in the same manner as described in \refsec{sec:spl}. The primary goal of these scripts are to provide a working example utilizing the information provided in the data files, and we encourage readers to use these files as a jumping point into their own analyses. A more detailed description of the files in this example are provided in the accompanying \texttt{README} file in \cite{HESEdatarelease}.
\FloatBarrier

\section{Table of segmented power-law parameters\label{sec:segmented_table}}
The segmented power-law splits the astrophysical neutrino energy spectrum into independent $E^-2$ segments the normalizations of which are allowed to vary.
Additional discussion of this model is found in~\refsec{sec:unfolding}.
The best fit point and approximate $\SigmaOne$ confidence intervals of this astrophysical model, as well as the MAP estimators and $\SigmaOne$ credible regions, are provided in the table below for the purpose of producing additional visualizations.
Readers seeking to use this event sample in other analyses should refer to the data release for the most accurate information as the errors reported here do not describe correlations between the parameters.

\begin{table*}[thb]
    \centering
    \begin{tabular}{l r r}
        \toprule
        & \multicolumn{1}{c}{Frequentist Analysis} & \multicolumn{1}{c}{Bayesian Analysis} \\
        \midrule
        Energy Range & Best-fit value and $\SigmaOne$ C.L. & Most-likely value and $\SigmaOne$ H.P.D. \\
        \midrule
        $[4.20\cdot 10^{4}, 8.83\cdot 10^{4}]$ & ${5.7 }_{-5.7 }^{+9.5 }\times10^{-18}$ & ${1.3}_{-1.1}^{+1.2}\times10^{-17}$ \\
        $[8.83\cdot 10^{4}, 1.86\cdot 10^{5}]$ & ${3.89}_{-0.87}^{+0.99}\times10^{-18}$ & ${3.9}_{-1.1}^{+1.5}\times10^{-18}$ \\
        $[1.86\cdot 10^{5}, 3.91\cdot 10^{5}]$ & ${7.1 }_{-7.1 }^{+1.4 }\times10^{-20}$ & ${8.6}_{-8.6}^{+20 }\times10^{-20}$ \\
        $[3.91\cdot 10^{5}, 8.23\cdot 10^{5}]$ & ${8.1 }_{-8.1 }^{+28  }\times10^{-21}$ & ${1.9}_{-1.9}^{+47 }\times10^{-21}$ \\
        $[8.23\cdot 10^{5}, 1.73\cdot 10^{6}]$ & ${1.3 }_{-0.64}^{+0.86}\times10^{-20}$ & ${7.9}_{-6.5}^{+12 }\times10^{-21}$ \\
        $[1.73\cdot 10^{6}, 3.64\cdot 10^{6}]$ & ${7.7 }_{-7.7 }^{+17  }\times10^{-22}$ & ${7.2}_{-7.2}^{+28 }\times10^{-22}$ \\
        $[3.64\cdot 10^{6}, 7.67\cdot 10^{6}]$ & ${0.0 }_{-0.0 }^{+6.7 }\times10^{-23}$ & ${5.4}_{-5.4}^{+280}\times10^{-24}$ \\
    \bottomrule
    \end{tabular}
    \caption{\textbf{\textit{Segmented power-law model normalizations.}}
    The left-most column shows the energy range in $\si\GeV$ of each segment, while the other columns show the six-neutrino flux at the center of each bin in units of $[{\rm GeV}^{-1} {\rm s}^{-1} {\rm sr}^{-1} {\rm cm}^{-2}]$ for the frequentist and Bayesian analyses.
    The frequentist analysis column shows the best-fit parameters and their approximate $\SI{68.3}\percent$ confidence interval.
    The Bayesian analysis column shows the most-likely values of the parameters, as well as the $\SI{68.3}\percent$ highest probability density interval (HPD).}
    \label{tbl:segment_normalizations}
\end{table*}
\section{High-energy astrophysical neutrino source searches\label{sec:sources}}
\noindent
\textit{In this section, we report the results of searches for neutrino sources performed with this sample and describe the reconstruction choices made for these analyses that differ from those used in \refsec{sec:diffuse}.
We do not find a significant indication of a point-like neutrino source or correlation with the galactic plane in these searches.
The most-significant source location is found to have a null-hypothesis post-trials p-value of 0.092.
}
\newline

\subsection{Directional reconstruction for neutrino source searches\label{sec:sourcesreco}}

In the search for neutrino sources, a conservative choice is made to model the background with data-derived distributions in an effort to avoid potential bias from mismodelling of the backgrounds.
Related to this choice, data below the $\SI{60}\TeV$ cut is included in this analysis.
As Monte-Carlo (MC) simulation is not used in this approach, we can use a more accurate but computationally more expensive directional reconstruction for these analyses, which would otherwise be prohibitive to use on a MC sample.
Events in data are reconstructed by fully re-simulating cascades or tracks with in-ice photon propagation~\cite{Chirkin:2013dfi}.
A re-simulation of the light propagation is performed for each considered combination of direction and interaction vertex.
This re-simulated information is re-used when considering different times and energies for the event.
We first perform a localized random search in the reconstruction parameter space to find the minimum of the test-statistic described in~\cite{Chirkin:2013lya}, comparing re-simulated waveforms to data.
This provides a localized starting point in the high-dimensional parameter space for efficiently estimating the directional posterior distribution.
An ``approximate Bayesian computation'' (ABC)~\cite{Marjoram15324} is performed around the estimated minimum using the same test-statistic in order to sample the posterior distribution of the reconstruction parameters, assuming a uniform prior in direction and vertex position.
This method allows us to sample the test-statistic posterior distribution despite the non-deterministic nature of the re-simulation step.
After the ABC step, we marginalize over other parameters to obtain the event direction's posterior distribution.
We then parameterize the directional posterior distribution with an eight-parameter Fisher-Bingham distribution~\cite{Yuan:2020}, which is used in the analyses described later in this section.

The errors reported for these events with this reconstruction method are near twice the angular size on average than those quoted previously for the same events.
This stems from the different treatment of uncertainties.
Previously, re-simulations of the events used variations of the scattering and absorption for individual ice layers.
This previously used method accounts for statistical variations and uncertainties in the ice layers but neglects other known detector uncertainties.
On the other hand, the reconstruction now used for the source searches explicitly includes calibration uncertainty on the simulated distribution of observed charge in its test-statistic.
This new method of introducing the detector uncertainties covers the known uncertainties in the photon arrival distribution from calibration and is more conservative than that previously used.
This change in the treatment of uncertainty results in larger angular errors, which are consistent with other more complete treatments of the calibration uncertainty~\cite{Aartsen:2019jcj}.
Therefore, the previously reported errors were likely underestimated.
This ice model used in the new reconstruction also incorporates more calibration information than the model used previously, a change that improves the accuracy of the reconstructed direction for cascades.
The inferred directional probability distribution of some events has changed substantially with respect to what was inferred with the original reconstruction.
The extent of these changes is expected based on studies of simulated events, given that the previous errors undercover and the modelling of the ice properties used for reconstruction has been improved significantly.
We note that the directional information inferred from cascade events, particularly in the detector's azimuthal dimension, is highly dependent on the modelling of the ice, which continues to improve.
The ice model used for the source searches reconstruction in this work adds information about the azimuthal anisotropy of the scattering and absorption of photons based on~\cite{Williams:2014era}.
Ice models in development will eventually add information about the birefringent crystalline structure of the ice~\cite{Chirkin:2019vyq,Chirkin:2013lpu}, changing again what we infer about the cascade directional probability distributions.

Except for four events classified as double cascades, data are reconstructed based on the classification described above.
Since algorithms to estimate the double cascade angular uncertainties were not available, the four double cascades are reconstructed, assuming a cascade hypothesis for the source searches.

\subsection{Neutrino source searches}

In a simple, spatial-only, analysis, we searched for clustering consistent with either point-like or spatially-extended galactic emission.
In each case, we used an unbinned likelihood function, described in~\cite{Braun:2008bg}, given by
\begin{align}
	    \like(n_s;\vec{\psi}_s) &= \prod_i^N
	    \left[
	    \frac{n_s}{N}\cdot \pdfSxi
	    + \left( 1 - \frac{n_s}{N} \right)\cdot B
	    \right],
	    \label{eq:pslikelihood}
\end{align}
where $N$ is the total number of observed events, $n_s$ is the expected number of signal events, $\vec{\psi}_s$ is the source position in the sky, $\vec{x}_i$ represents the properties of event $i$, $B=1/(4\pi)$ is the background spatial distribution which we take to be uniform, and $\pdfSxi=P_i(\vec{\psi}_s)$ is the spatial distribution expected from the signal population which is taken to be equal to the posterior density of the event direction $P_i(\vec{x}_s)$ at the source position.
We neglect the energy information in this search.
For point-like sources, $\pdfSxi$ is determined entirely by the posterior distributions described in \refsec{sec:reconstruction}; for spatially extended emission, those must be convolved with the assumed spatial distribution of the source.
In order to quantify preference for the alternative (presence of a neutrino source) versus the null (isotropic scenario) hypothesis, we use the following test statistic
\begin{equation}
    \TS=-2\ln[\like(n_s=0)/\like(\hat{n}_s)],
\end{equation}
where $\hat{n}_s$ is the value that maximizes the likelihood.
The null-hypothesis p-value is determined by comparing the observed test statistic value, $\TS_\texttt{obs}$, with the $\TS$ distribution expected given background only correcting for the number of source hypotheses considered.
This background $\TS$ distribution is determined by repeating the experiment using modified versions of the dataset, where we have randomized the right ascension of all events.

\begin{figure}
    \centering
    \includegraphics[width=\linewidth]{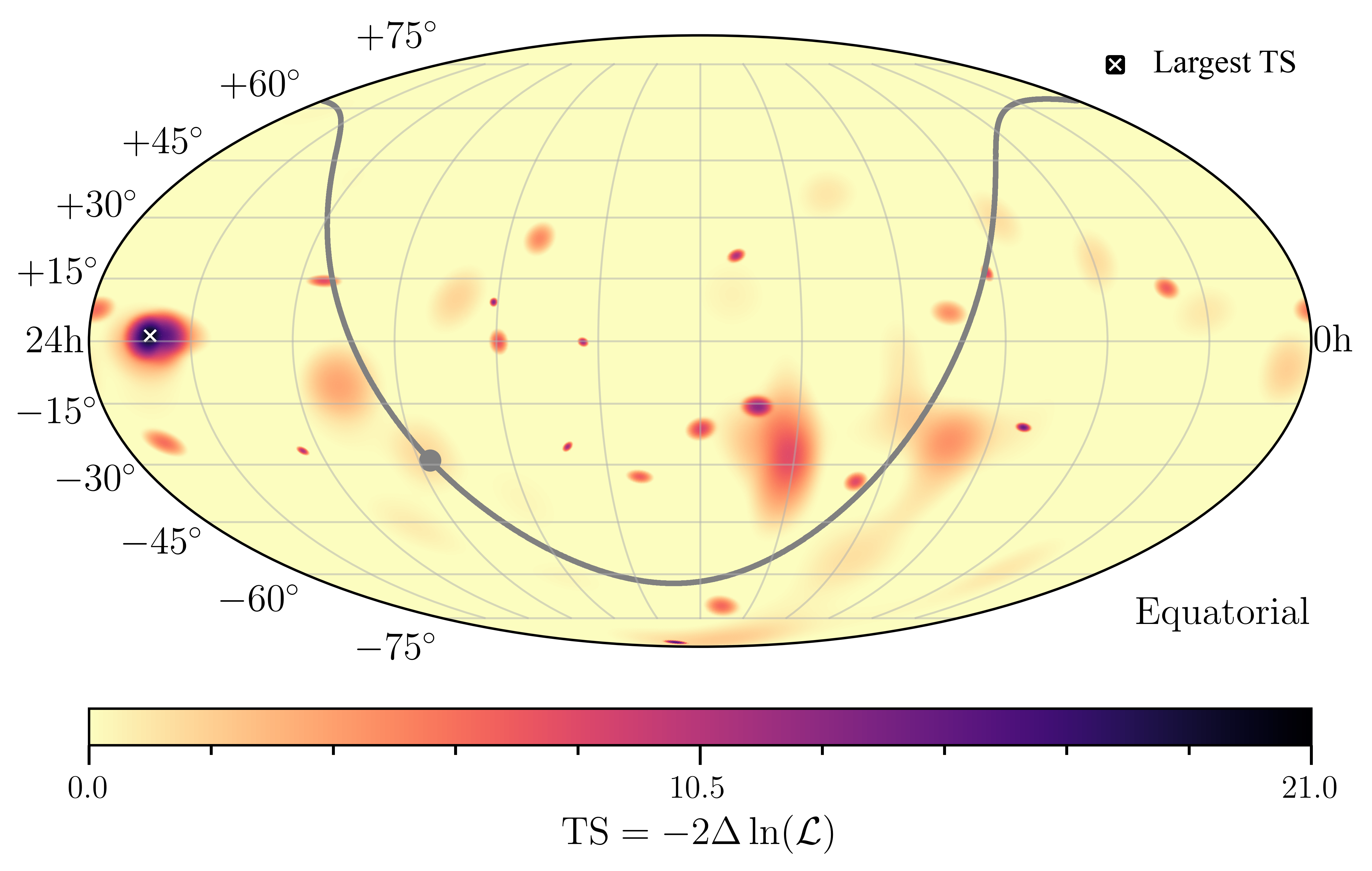}
    \caption{\textbf{\textit{Point source $\TS$ map.}}
    The $\TS$ at each point on the sky is indicated by the blue color scale.
    The Galactic center and Galactic plane are indicated by the gray dot and gray curve, respectively.
    Unlike in previous analyses~\cite{Aartsen:2013jdh,Aartsen:2014gkd,Aartsen:2015zva,Aartsen:2017mau}, the most significant position is well separated from the Galactic plane, at $(\alpha,\delta)=(342.1^\circ, 1.3^\circ)$, indicated by the white $\times$.
    } \label{fig:ps_skymap}
\end{figure}

The results of testing for point-like emission in the sky from many directions are shown in \reffig{fig:ps_skymap}.
The maximum test statistic is found at equatorial coordinates $(\alpha,\delta)=(\SI{342.1}\degree, \SI{1.3}\degree)$, with a null-hypothesis post-trials p-value of 0.092.
The hottest spot no longer correlates with the galactic plane as was the case in previous analyses.
In an unweighted hottest-spot test restricted to a predefined source list consisting of 74 source candidates -- which has been studied in previous iterations of this analysis~\cite{Aartsen:2014gkd}, with through-going tracks~\cite{Aartsen:2016oji}, and with contained cascades~\cite{Aartsen:2019epb} -- no significant emission was found; the null-hypothesis post-trials p-value is 0.76.
This list of 74 sources is enumerated in the supplementary material of \cite{Aartsen:2014gkd}.

We performed two searches for spatially-extended Galactic emission.
First, as in past work~\cite{Aartsen:2014gkd,Aartsen:2015zva,Aartsen:2017mau}, we tested for uniform emission within a Galactic plane region expressed in terms of Galactic latitude as $|b|<b_\text{max}$, scanning over possible values $b_\text{max}\in[2.5^\circ,30^\circ]$.
Here, the greatest pre-trial significance ($\text{p}=0.02)$ was found for $b_\text{max}=5^\circ$.
After accounting for multiple, partially correlated tests over the range of the $b_\text{max}$ scan, we find a post-trials p-value of $0.062$.

\Fermi-LAT has inferred a diffuse flux of gamma rays originating in decays of neutral pions produced by interactions between cosmic rays and Galactic gas and dust~\cite{Ackermann:2012pya}.
Although this inference is based on measurements made at energies far below the threshold of this analysis ($E_\gamma \lesssim\SI{100}\GeV$), it is possible that this gamma ray flux extends to higher energies and is accompanied by high energy neutrinos produced in decays of charged pions.
Therefore, in a second, more model-dependent search similar to that performed in~\cite{Aartsen:2019epb,Aartsen:2017ujz}, we test for emission following the spatial profile of the \Fermi-LAT best-fit result on Galactic hadronic emission below $\SI{100}\GeV$, using the method described in~\cite{Aartsen:2017ujz}.
Here we find a p-value of $0.089$.

To conclude, this analysis's new reconstruction techniques are more conservative and generally increase the reported angular error.
A detailed accounting of these changes is found in \refsec{sec:comparison}.
With these searches, we find no evidence of clustering in the sample or correlation with tested sources.
Other, more sensitive, searches for neutrino sources have been performed with IceCube data and provide better constraints on the scenarios they test~\cite{Aartsen:2019fau,Aartsen:2019epb}.
In the future, improved results will be based on other, dedicated samples.

\FloatBarrier

\section{Event comparison\label{sec:comparison}}
\begin{longtable}[e]{@{\extracolsep{\fill}}l c c l l l}
\toprule
\makecell{Event\\ID} & HPD Area (7.5yr) [sr] & HPD Area (prev.) [sr] & HPD Overlap Area [sr] \\
\midrule
\endfirsthead
\makecell{Event\\ID} & HPD Area (7.5yr) [sr] & HPD Area (prev.) [sr] & HPD Overlap Area [sr] \\
\midrule
\endhead
\endfoot
\bottomrule \\*
\caption{\textbf{\textit{Comparison of event reconstructions.}}
As the reconstruction used for the neutrino source searches has changed in this analysis as opposed to previous analyses, this table provides a comparison of the reconstructed event properties.
The angular size of the $\SI{50}\percent$ and $\SI{90}\percent$ directional highest posterior density regions, and the overlap of these regions are given for each reconstruction.
Only events with previously released properties are shown in the table; the reconstructed properties of all events, including full directional posterior distributions can be found in the data release.
The highest posterior density regions of the directions can be directly computed for the reconstruction used in this analysis, and so are used directly in this comparison.
For the previously used reconstructions, the reconstruction likelihood as a function of direction is converted to a directional probability density with a smoothing representative of the average angular resolution of each morphology.
This probability density is then re-calibrated to match the angular resolution obtained from the re-simulations of the events, and the highest posterior density regions computed in this way are used for the comparison.
Events with no properties listed were determined to be coincident muons.
Events without properties listed from this analysis were removed from the sample for reasons discussed in \refsec{sec:selection}.}
\label{tbl:comparison}
\endlastfoot
\makecell[tc]{1} & \makecell[tl]{$\SI{50}\percent$: $0.564$\\$\SI{90}\percent$: $1.71$} & \makecell[tl]{$\SI{50}\percent$: $0.253$\\$\SI{90}\percent$: $0.826$} & \makecell[tl]{$\SI{50}\percent$: $0.116$\\$\SI{90}\percent$: $0.584$} \\
\makecell[tc]{2} & \makecell[tl]{$\SI{50}\percent$: $0.940$\\$\SI{90}\percent$: $3.02$} & \makecell[tl]{$\SI{50}\percent$: $0.605$\\$\SI{90}\percent$: $2.75$} & \makecell[tl]{$\SI{50}\percent$: $0.226$\\$\SI{90}\percent$: $1.17$} \\
\makecell[tc]{3} & \makecell[tl]{$\SI{50}\percent$: $1.26\times10^{-3}$\\$\SI{90}\percent$: $4.19\times10^{-3}$} & \makecell[tl]{$\SI{50}\percent$: $1.81\times10^{-3}$\\$\SI{90}\percent$: $6.06\times10^{-3}$} & \makecell[tl]{$\SI{50}\percent$: $0.00$\\$\SI{90}\percent$: $7.31\times10^{-4}$} \\
\makecell[tc]{4} & \makecell[tl]{$\SI{50}\percent$: $0.0421$\\$\SI{90}\percent$: $0.141$} & \makecell[tl]{$\SI{50}\percent$: $0.0481$\\$\SI{90}\percent$: $0.160$} & \makecell[tl]{$\SI{50}\percent$: $0.00$\\$\SI{90}\percent$: $0.0413$} \\
\makecell[tc]{5} & - & \makecell[tl]{$\SI{50}\percent$: $1.33\times10^{-3}$\\$\SI{90}\percent$: $4.43\times10^{-3}$} & - \\
\makecell[tc]{6} & - & \makecell[tl]{$\SI{50}\percent$: $0.0913$\\$\SI{90}\percent$: $0.302$} & - \\
\makecell[tc]{7} & \makecell[tl]{$\SI{50}\percent$: $1.10$\\$\SI{90}\percent$: $3.18$} & \makecell[tl]{$\SI{50}\percent$: $0.545$\\$\SI{90}\percent$: $1.76$} & \makecell[tl]{$\SI{50}\percent$: $0.00$\\$\SI{90}\percent$: $0.195$} \\
\makecell[tc]{8} & \makecell[tl]{$\SI{50}\percent$: $1.87\times10^{-3}$\\$\SI{90}\percent$: $6.22\times10^{-3}$} & \makecell[tl]{$\SI{50}\percent$: $1.65\times10^{-3}$\\$\SI{90}\percent$: $5.51\times10^{-3}$} & \makecell[tl]{$\SI{50}\percent$: $3.64\times10^{-4}$\\$\SI{90}\percent$: $3.09\times10^{-3}$} \\
\end{longtable}
\FloatBarrier

\section{Source searches reconstructions\label{sec:reco}}
To also support analyses of the source search variety, we provide the directional posterior distributions used in the analyses described in \refsec{sec:sources}.
These posterior distributions are parameterized with the Fisher-Bingham eight-parameter distribution (FB8).
The directional probability density of the FB8 distribution can be written as 
\begin{equation}
    f_8(\vec{x})=c_8(\kappa, \beta, \eta, \vec{\nu})^{-1}e^{\kappa \vec{\nu}\cdot\Gamma^T\vec{x}+\beta\left[\left(\vec{\gamma_2}\cdot\vec{x}\right)^2+\eta\left(\vec{\gamma_3}\cdot\vec{x}\right)^2\right]}.
\end{equation}
\label{eq:fb8}
In this parameterization $\Gamma$ is a rotation matrix which also specifies $\vec{\gamma_2}$ and $\vec{\gamma_3}$, which are the second and third columns of $\Gamma$ respectively.
The unit vector $\vec{\nu}$ specifies a direction.
The parameters $\kappa$, $\beta$, and $\eta$, are scalars that satisfy the conditions $\kappa\geq0$, $\beta\geq0$, and $|\eta|\leq1$.
The term $c_8(\kappa,\beta,\eta)$ normalizes the distribution in direction and in this work is computed using the method described in~\cite{Yuan:2020}.
In total these can be specified by eight parameters; the matrix by three parameters, the vector by two parameters, and the scalars by three parameters.
For each event in the sample we provide the parameters of the corresponding FB8 distribution, as well as a $\texttt{HEALPix}$ compatible $\texttt{fits}$ file that contains an estimate of the integrated probability within each of the $786432$ considered pixels.
\begin{longtable}[e]{@{\extracolsep{\fill}}l l l l l l}
\toprule
\makecell[cl]{Event\\ID} & HPD Area (7.5yr) [sr] & $\Gamma:(\theta, \phi, \psi)$ [rad] & $(\kappa, \beta, \eta)$ & $\vec\nu:(\alpha, \rho)$ [rad] & MAP : (RA, DEC) \\
\midrule
\endfirsthead
\makecell[cl]{Event\\ID} & HPD Area (7.5yr) [sr] & $\Gamma:(\theta, \phi, \psi)$ [rad] & $(\kappa, \beta, \eta)$ & $\vec\nu:(\alpha, \rho)$ [rad] & MAP : (RA, DEC) \\
\midrule
\endhead
\endfoot
\bottomrule \\*
\caption{\textbf{\textit{Source searches event reconstructions.}}
This table summarizes the reconstruction parameterization used for the source searches in \refappsec{sec:sources}.
The left-most column gives the solid-angle of the $\SI{50}\percent$ and $\SI{90}\percent$ highest-posterior-density regions to provide a sense of scale for the angular errors.
The central three columns provide the parameters of the FB8 distribution that parameterize the reconstruction errors of the event; these values are given to an arbitrary precision that is not necessarily indicative of the precision to which these values can be determined.
The right-most column lists the maximum {\it a posteriori} estimator of the event direction in equatorial coordinates, which is equivalent to the mode of the FB8 distribution.
Events without properties listed are not included in this sample for reasons discussed in \refsec{sec:selection} but are listed here to remain consistent with the numbering scheme of previous analyses.}
\label{tbl:source_searches_reco}
\endlastfoot
\belowbaseline[0pt]{\makecell[tl]{$1$}} & \belowbaseline[0pt]{\makecell[tl]{$\begin{aligned}\SI{50}\percent&{}{:}~0.564\\[-5pt]\SI{90}\percent&{}{:}~1.71\end{aligned}$}} & \belowbaseline[0pt]{\makecell[tl]{$\begin{aligned}\theta&{}{:}~0.45826837\\[-5pt]\phi&{}{:}~0.91828843\\[-5pt]\psi&{}{:}~1.17643995\end{aligned}$}} & \belowbaseline[0pt]{\makecell[tl]{$\begin{aligned}\kappa&{}{:}~19.57268893\\[-5pt]\beta&{}{:}~15.57446717\\[-5pt]\eta&{}{:}~-0.79817007\end{aligned}$}} & \belowbaseline[0pt]{\makecell[tl]{$\begin{aligned}\alpha&{}{:}~0.27093361\\[-5pt]\rho&{}{:}~-1.48396204\end{aligned}$}} & \belowbaseline[0pt]{\makecell[tl]{$\begin{aligned}\textrm{RA}&{}{:}~\SI{58.54}\degree\\[-5pt]\textrm{DEC}&{}{:}~\SI{9.59}\degree\end{aligned}$}} \\
\belowbaseline[0pt]{\makecell[tl]{$2$}} & \belowbaseline[0pt]{\makecell[tl]{$\begin{aligned}\SI{50}\percent&{}{:}~0.940\\[-5pt]\SI{90}\percent&{}{:}~3.02\end{aligned}$}} & \belowbaseline[0pt]{\makecell[tl]{$\begin{aligned}\theta&{}{:}~1.55358028\\[-5pt]\phi&{}{:}~-0.93835829\\[-5pt]\psi&{}{:}~1.54259543\end{aligned}$}} & \belowbaseline[0pt]{\makecell[tl]{$\begin{aligned}\kappa&{}{:}~4.96233247\\[-5pt]\beta&{}{:}~3.70300725\\[-5pt]\eta&{}{:}~1.00\end{aligned}$}} & \belowbaseline[0pt]{\makecell[tl]{$\begin{aligned}\alpha&{}{:}~0.60313619\\[-5pt]\rho&{}{:}~-1.79160961\end{aligned}$}} & \belowbaseline[0pt]{\makecell[tl]{$\begin{aligned}\textrm{RA}&{}{:}~\SI{246.80}\degree\\[-5pt]\textrm{DEC}&{}{:}~\SI{-8.23}\degree\end{aligned}$}} \\
\end{longtable}
\FloatBarrier


\end{document}